%% file: RamsesRT.tex
\documentclass[usenatbib,usegraphicx,useAMS]{mn2e}
\pdfoutput=1
\usepackage{graphicx}
\usepackage{subfig}
\usepackage{verbatim}
\usepackage[usenames,dvipsnames]{color}
\usepackage{booktabs}
\usepackage{dsfont}
\usepackage{listings} 
\usepackage{amsmath}
\usepackage{soul}
\usepackage{amssymb}
\allowdisplaybreaks[1]

\newcommand{\App}[1]{Appendix~\ref{#1}}

\newcommand{\eq}[1]{(\ref{#1})}
\newcommand{\Eq}[1]{Eq.~\ref{#1}}

\newcommand{\Fig}[1]{Fig.~\ref{#1}}
\newcommand{\Sec}[1]{\S\ref{#1}}
\newcommand{\Tab}[1]{Table~\ref{#1}}
\newcommand{\Pse}[1]{Listing~\ref{#1}}

\def\art  {{\tt ART}}
\def\aton  {{\tt ATON}}
\def \Capreole{{\tt{Capreole}}}
\def \C2R{{\tt{C$^2$-Ray}}}
\def \CC2R{{\tt{Capreole+C$^2$-Ray}}}

\def\coral  {{\tt Coral}}
\def\crash  {{\tt Crash}}
\def\flash  {{\tt Flash}}
\def\flashhc  {{\tt Flash-HC}}
\def\ffte  {{\tt FFTE}}
\def\ift  {{\tt IFT}}
\def\licorice  {{\tt Licorice}}

\def\rhod    {{\tt RH1D}}
\def\ramses    {{\tt RAMSES}}
\def\ramsesrt  {{\tt RAMSES-RT}}
\def\rsph  {{\tt RSPH}}
\def\simplex  {{\tt SimpleX}}
\def \TC2R{{\tt{TVD+C$^2$-Ray}}}
\def\zeusmp  {{\tt Zeus-MP}}

\def\hei    {{\rm{He\textsc{i}}}}
\def\heii   {\rm{He\textsc{ii}}}
\def\heiii   {\rm{He\textsc{iii}}}
\def\hi    {{\rm{H\textsc{i}}}}
\def\hii   {\rm{H\textsc{ii}}}
\def\heisub {\rm{He \scriptscriptstyle I}}
\def\heiisub {\rm{He \scriptscriptstyle II}}
\def\heiiisub {\rm{He \scriptscriptstyle III}}
\def\hisub {\rm{H \scriptscriptstyle I}}
\def\hiisub {\rm{H \scriptscriptstyle II}}
\def\nh {n_{\rm{H}}}
\def\nel {n_{\rm{e}}}
\def\nhe {n_{\rm{He}}}
\def\nhei {n_{\heisub}}
\def\nheii {n_{\heiisub}}
\def\nheiii {n_{\heiiisub}}
\def\nhi {n_{\hisub}}
\def\nhii {n_{\hiisub}}
\def\xhi {x_{\rm{H \scriptscriptstyle I}}}
\def\xhii {x_{\rm{H \scriptscriptstyle II}}}
\def\xhei {x_{\rm{He \scriptscriptstyle I}}}
\def\xheii {x_{\rm{He \scriptscriptstyle II}}}
\def\xheiii {x_{\rm{He \scriptscriptstyle III}}}
\def\cci {{\rm{cm}}^{-3}}   
\def\ccitwod {{\rm{cm}}^{-2}}   
\def\cs {\rm{cm}^{2}}       
\def\cmmone {\rm{cm}^{-1}}  
\def\sm {\rm{s}^{-1}}       
\def\ccs {\rm{cm}^{3} \, \rm{s}^{-1}}                           
\def\coolU {\rm{erg}  \, \rm{cm}^{3} \, \rm{s}^{-1}}         
\def\ergs {\rm{erg} \, \rm{s}^{-1}}                         
\def\flux {{\rm{s}}^{-1} \, {\rm{cm}}^{-2}}                 
\def\phflux {{\rm{photons}} \; {\rm{s}}^{-1} \, {\rm{cm}}^{-2}}
\def\emrate {\rm{photons} \; \rm{s}^{-1}}                   
\def\age {{\tau}}           
\def\bF {{\bf{F}}}          
\def\bn {{\bf{n}}}          
\def\bx {{\bf{x}}}          
\def\CH {{\Lambda}}         
\def\CHp {{\Lambda^\prime}} 
\def\Cool {{\mathcal{L}}}   
\def\cred {{c_r}}           
\def\cse {{\sigma}^{\rm{E}}} 
\def\csn {{\sigma}^{\rm{N}}} 
\def\dt {\Delta t}          
\def\dtrt {\Delta t_{\rm{RT}}}    
\def\dttc {\Delta t_{\rm{TC}}}    
\def\egy {\bar{\epsilon}}   
\def\edens {E}  
\def\etherm {{\varepsilon}}    
\def\Et {{\bf{\mathds{D}}}} 
\def\eTconv {\frac{(\gamma-1)\mh}{\rho \kb}}
\def\fc {{f_c}}             
\def\Fphot {F}              
\def\Fpr {{\dot{F}}}        
\def\Heat {{\mathcal{H}}}   
\def\ident {\mathbf{I}}     
\def\lmax {\ell_{\rm{max}}} 
\def\lmin {\ell_{\rm{min}}} 
\def\Lbox {L_{\rm{box}}}    
\def\Lum {L}                
\def\kb {k_{\rm{B}}}        
\def\mh {m_{\rm{H}}}        
\def\Nphot {N}              
\def\Npr {{\dot{N}}}        
\def\nui {{\nu_{i \scriptscriptstyle 0}}} 
\def\nuf {{\nu_{i \scriptscriptstyle 1}}} 
\def\pa{{\partial}}         
\def\pt{{\partial t}}       
\def\Pt {{\bf{\mathds{P}}}} 
\def\rate {{\mathcal{S}}}   
\def\recA {\alpha^{\rm{A}}} 
\def\recB {\alpha^{\rm{B}}} 
\def\recAH {\alpha^{\rm{A}}_{\rm{H \scriptscriptstyle II}}}
\def\recBH {\alpha^{\rm{B}}_{\rm{H \scriptscriptstyle II}}}
\def\rs {r_{\rm{S}}}        
\def\ri {r_{\rm{I}}}        
\def\sage {{\tau_{\star}}}  
\def\state {{\mathcal{U}}}  
\def\staten {{\widetilde{\mathcal{U}}}}  
\def\stateF {{\mathcal{F}}} 
\def\Tmu {T_{\mu}}          
\def\tcross {t_{\rm{cross}}}
\def\trec {t_{\rm{rec}}}    
\def\vel {{\bf{u}}}         
\def\vi {v_{\rm{I}}}        
\def\wsim {w_{\rm{sim}}}    
\def\Ila {Il06} 
\def\Ilb {Il09} 
\def\AT {AT08}  

\def\lya {{\rm{Ly}$\alpha$}}
\def\vsk{\vskip 0.4cm} 

\mathchardef\mhyphen="2D

\long\def\symbolfootnote[#1]#2{\begingroup%
\def\thefootnote{\fnsymbol{footnote}}\footnote[#1]{#2}\endgroup}

\def\joki{}

\begin{document}

\title[RAMSES-RT]{RAMSES-RT: Radiation hydrodynamics in the
  cosmological context} \author[Rosdahl et al.]
{J.~Rosdahl,$^{1,2}$\thanks{E-mail: joki@strw.leidenuniv.nl}
  J.~Blaizot,$^{1}$ D. Aubert,$^{3}$ T. Stranex,$^{4}$
  and R. Teyssier$^{4,5}$ \\
  $^1$Universit\'e de Lyon, Lyon, F-69003, France ; \\
  \; Universit\'e Lyon 1, Observatoire de Lyon, 9 avenue Charles
  Andr\'e, Saint-Genis Laval,
  F-69230, France ; \\
  \; CNRS, UMR 5574, Centre de Recherche Astrophysique de Lyon \\
  $^2$Leiden Observatory, Leiden University, P.O. Box 9513, 2300 RA,
  Leiden, The Netherlands \\
  $^3$Observatoire Astronomique de Strasbourg,
  Universite de Strasbourg, CNRS UMR 7550, \\
  \; \; 11 rue de l'Universite,
  F-67000 Strasbourg, France \\
  $^4$Institute of Theoretical Physics, University of Z\"urich,
  Winterthurerstrasse 190, 8057 Z\"urich — Switzerland \\
  $^5$IRFU/SAp, CEA Saclay, F-91191 Gif-sur-Yvette Cedex, France }

\maketitle
\begin{abstract}
  We present a new implementation of radiation hydrodynamics (RHD) in
  the adaptive mesh refinement (AMR) code \ramses{}. The multi-group
  radiative transfer (RT) is performed on the AMR grid with a
  first-order Godunov method using the M1 closure for the Eddington
  tensor, and is coupled to the hydrodynamics via non-equilibrium
  thermochemistry of hydrogen and helium. This moment-based approach
  has the large advantage that the computational cost is independent
  of the number of radiative sources - it can even deal with
  continuous regions of emission such as bound-free emission from
  gas. As it is built directly into \ramses{}, the RT takes natural
  advantage of the refinement and parallelization strategies already
  in place. Since we use an explicit advection solver for the
  radiative transport, the time step is restricted by the speed of
  light - a severe limitation that can be alleviated using the
  so--called ``reduced speed of light" approximation. We propose a
  rigorous framework to assess the validity of this approximation in
  various conditions encountered in cosmology and galaxy formation.
  We finally perform with our newly developed code a complete suite of
  RHD tests, comparing our results to other RHD codes.  The tests
  demonstrate that our code performs very well and is ideally suited
  for exploring the effect of radiation on current scenarios of
  structure and galaxy formation.
\end{abstract}
\begin{keywords}
  methods: numerical,
  radiative transfer
\end{keywords}

\section{Introduction} \label{Intro.sec} With the surging interest in
reionization and the first sources of light in the Universe, and also
thanks to a steadily increasing computational power, cosmological
simulation codes have begun to include ionizing radiative transfer
(RT) in the last decade or so. This is generally seen as a
second-order component in most astrophysical processes, but important
nonetheless, and is obviously very important in the context of
simulating reionization. Due to the challenges involved, most
implementations have started out with the post-processing of ionizing
radiation on simulations including only dark matter, but a few have
begun doing {\it coupled} radiation hydrodynamics (RHD), which model
the interplay of radiation and gas.

It is highly desirable to follow self-consistently, with RHD
simulations, the time-evolution and morphology of large-scale
intergalactic medium (IGM) reionization and at the same time the
smaller scale formation of the presumed sources of reionization; how
galaxy formation is regulated by the ionizing radiation being
released, how much of the radiation escapes from the galaxies to
ionize the IGM, how first generation stars are formed in a metal-free
environment and how radiative and supernovae feedback from those stars
affect the inter-galactic medium. The galaxies and the IGM are indeed
inter-connected via the ionizing radiation: the photons released from
the galaxies affect the state of the surrounding gas via ionization
and heating and may even prevent it from falling in or condensing into
external gravitational potentials, especially small ones
\citep[e.g.][]{Wise:2008bq,2011MNRAS.417L..93O}, which can then in
turn significantly alter the ionization history.

The importance of RT and RHD is of course not limited to the epoch of
reionization. Stars keep emitting ionizing radiation after this epoch
and their radiative feedback likely has an effect on the
post-reionization regulation of star-formation
\citep[e.g.][]{Pawlik:2009jv,Hopkins:2011fk}, the mass distribution of
stellar populations \citep{2012ApJ...754...71K} and even gas outflows
\citep{2012MNRAS.427..968H}.

Radiation hydrodynamics are complex and costly in simulations.  The
inclusion of coupled radiative transfer in hydrodynamical codes in
general is challenging mainly because of the high dimensionality of
radiative transfer (space-, angular-, and frequency dimensions) and
the inherent difference between the typical timescales of radiative
transfer and non-relativistic hydrodynamics.  Simulating the
interaction between small and large scales (so relevant to the epoch
of cosmic reionization) makes things even worse: one wants to
simulate, in a statistically significant region of the Universe
(i.e. of the order of 100 comoving Mpc across) the condensation of
matter in galaxy groups on Mpc scales, down to individual galaxies on
kpc scales, followed by the formations of stellar nurseries in those
galaxies on pc scales, and ultimately the formation of stars on sub-pc
scales and then the effect of radiation from those stars back to the
large scale IGM.  This cycle involves size differences of something
like 9 to 10 orders of magnitude -- which is too much for the most
advanced codes and computers today, actually even so without the
inclusion of radiative transfer.

Due to these challenges, simulations typically focus on only a subset
of these scales; either they consider reionization on large scales and
apply sub-resolution recipes to determine stellar luminosities and UV
escape fractions, or they ignore the cosmological context and focus on
star formation and escape fractions in isolated galaxies or even
isolated stellar nurseries.

A number of large scale 3D radiative transfer simulations of
reionization have been carried out in recent years
\citep[e.g.][]{Gnedin:1997fq, MiraldaEscude:2000iz, Gnedin:2000hp,
  Ciardi:2003dz, Sokasian:2004ex, Iliev:2006cl, Zahn:2007gg,
  Croft:2008if, Baek:2010el, Aubert:2010if, Petkova:2011jh}, though
they must all to some degree use subgrid recipes for star formation
rates, stellar luminosities and UV escape fractions, none of which are
well constrained. The ionization history in these simulations thus
largely depends on these input parameters and resolution -- some in
fact use the observational constraints of the ionization history to
derive constraints on these free parameters
\citep[e.g.][]{Sokasian:2004ex, Croft:2008if, Baek:2010el,
  Aubert:2010if, Petkova:2011jh}.  Furthermore, most of these works
have used a post-processing RT strategy instead of RHD, which neglects
the effect the ionizing radiation has on the formation of luminous
sources.

The primary driver behind this work is the desire to understand the
birth of galaxies and stars during the dark ages, and how they link
with their large scale environment. We have thus implemented a RHD
version of the widely used cosmological code \ramses{}
\citep{Teyssier:2002fj}, that we call \ramsesrt{}, with the goal of
running cosmological RHD simulations, optimized for \textit{galactic}
scale radiation hydrodynamics. \ramses{} is an adaptive mesh
refinement (AMR) code, which greatly cuts costs by adaptively allowing
the resolution to follow the formation of structures. The RHD
implementation takes full advantage of the AMR strategy, allowing for
high resolution simulations that can self consistently model the
interplay of the reionizing Universe and the formation of the first
galaxies.

Some of the goals we will be able to tackle with this implementation
are:
\begin{itemize}
\item Study radiative feedback effects in primordial galaxies. These
  galaxies are by definition young and small, and the first stars are
  thought to be gigantic and very bright due to the lack of
  metals. The ionizing radiation from these first stars is likely to
  have a dramatic effect on the galaxy evolution.  This is closely
  associated with the formation of molecules, needed to form the first
  stars, which is sensitive to the radiation field. Radiative feedback
  effects also appear to be relevant in lower-redshift galaxies, and
  likely have a considerable impact on the initial mass function of
  stellar populations \citep[][]{2012ApJ...754...71K}.
\item Investigate the escape of ionizing photons from early galaxies,
  how it affects the ionization history and external structure
  formation, e.g. the formation of satellite galaxies.
\item Study the emission and absorption properties of galaxies and
  extended structures. Observable properties of gas are highly
  dependent on its ionization state, which in turn depends on the
  local radiation field \citep[e.g.][]{Oppenheimer:2013vd}.  To
  predict it correctly, and to make correct interpretations of
  existing observations, one thus needs to model the ionization state
  consistently, for which RHD simulations are needed.
\item Improve sub-resolution recipes: of course we have not
  implemented a miracle code, and we are still nowhere near simulating
  simultaneously the 9 to 10 orders of magnitude in scale needed
  for fully self-consistent simulations of reionization. Sub-resolution
  strategies are still needed, and part of the objective is to improve
  those via small-scale simulations of stellar feedback (SNe,
  radiation, stellar winds).
\end{itemize}

It is useful here to make clear the distinction between continuum and line
radiative transfer: our goal is to study the interplay of ionizing
radiation, e.g. from stellar populations and AGN, and the
interstellar/intergalactic gas. We consider \textit{continuum}
radiation, because the spectra of stars (and AGN) are smooth enough
that emission and absorption processes are not sensitive to subtle
rest-frame frequency shifts, be they due to local gas velocities or
cosmological expansion.

On the other side is \textit{line} transfer, i.e. the propagation of
radiation over a narrow frequency range, usually corresponding to a
central frequency that resonates with the gas particles. An important
example is the propagation of \lya{} photons. Here, one is interested
in the complex frequency and direction shifts that take place via
scattering on the gas particles, and gas velocities and subtle
frequency shifts are vital components. Line transfer is mostly done to
interpret observational spectra, e.g. from \lya{} emitting/absorbing
galaxies \citep[e.g.][]{Verhamme:2006jt}, and is usually run in
post-processing under the assumption that the line radiation has a
negligible effect on the gas dynamics \citep[through this assumption
is not neccessarily true; see][]{Dijkstra:2009ku}.

There is a bit of a grey line between those two regimes of continuum
and line radiation - some codes are even able to do both
\citep[e.g.][]{Baek:2009gm,Pierleoni:2009kr,Yajima:2012fo}. Our
implementation deals strictly with continuum radiation though, as do
most RHD implementations, for the sake of speed and memory
limitations. We do approximate multi-frequency, but only quite
coarsely, such that simulated photons represent an average of photons
over a relatively wide frequency range, and any subtle frequency
shifts and velocity effects are ignored.

\subsection{Radiative transfer schemes and existing implementations}
Cosmological hydrodynamics codes have traditionally been divided into
two categories: Smoothed Particle Hydrodynamics (SPH) and AMR.  The
drawbacks and advantages of each method have been thoroughly explored
\citep[e.g.][]{Agertz:2007gd, Wadsley:2008if, Tasker:2008cu} and we
now believe that both code types agree more or less on the final
result {\it if they are used carefully with recently developed fixes
  and improvements\joki{, and if applied in their regimes of
  validity}}. On the radiation side, it is quite remarkable that we
have the same dichotomy between ray-tracing codes and moment-based
codes. Comparative evaluations of both methods have been performed in
several papers \citep[][]{Iliev:2006jz, Altay:2008cg, Aubert:2008jj,
  Pawlik:2008kk, Iliev:2009kn, Maselli:2009cw, Petkova:2009fx,
  Cantalupo:2011ke, Pawlik:2011ei, Petkova:2011dz, Wise:2011iw}, and
here again, each method has its own specific advantage over the other
one. Comparing both methods in the coupled case (RHD) within the more
challenging context of galaxy formation, such as in the recent Aquila
comparison project, \citep[][]{Scannapieco:2012iy}, remains to be
done.

\subsubsection{Ray-based schemes}
Here the approximation is made that the radiation field is dominated
by a limited number of sources. This allows one to approximate the
local intensity of radiation, $I_{\nu}$, as a function of the optical
depth $\tau$ along \textit{rays} from each source.

The simplest solution is to cast rays, or \textbf{long
  characteristics} from each source to each cell (or volume element)
and sum up the optical depth at each endpoint. With the optical depths
in hand, $I_{\nu}$ is known everywhere and the rates of
photoionization, heating and cooling can be calculated. While this
strategy has the advantage of being simple and easy to parallelize
(each source calculation is independent from the other), there is a
lot of redundancy, since any cell which is close to a radiative source
is traversed by many rays cast to further-lying cells, and is thus
queried many times for its contribution to the optical depth. The
parallelization is also not really so advantageous in the case of
multiprocessor codes, since rays that travel over large lengths likely
need to access cell states over many CPU nodes, calling for a lot of
inter-node communication. Furthermore, the method is expensive: the
computational cost scales linearly with the number of radiative
sources, and each RT timestep has order $\mathcal{O}(N_{\rm{sources}}
\ N_{\rm{cells}})$ operations, where $N_{\rm{sources}}$ is the number
of radiative sources and $N_{\rm{cells}}$ is the number of volume
elements. Implementation examples include \cite{{Abel:1999ch}},
\cite{{Cen:2002gr}} and \cite{Susa:2006tz}.

\textbf{Short characteristics} schemes overcome the redundancy problem
by not casting separate rays for each destination cell. Instead, the
calculation of optical depths in cells is propagated outwards from the
source, and is in each cell based on the entering optical depths from
the inner-lying cells. Calculation of the optical depth in a cell thus
requires some sort of interpolation from the inner ones. There is no
redundancy, as only a single ray segment is cast through each cell in
one time-step. However, there is still a large number of operations
and the problem has been made inherently serial, since the optical
depths must be calculated in a sequence which follows the radiation
ripple away from the source. Some examples are
\cite{{Nakamoto:2001cf}}, \cite{{Mellema:2006cl}},
\cite{{Whalen:2006ke}} and \cite{{Alvarez:2006kr}}.

\textbf{Adaptive ray tracing} \citep[e.g.][]{Abel:2002kt,
  Razoumov:2005fn, Wise:2011iw} is a variant on long characteristics,
where rays of photons are integrated outwards from the source,
updating the ray at every step of the way via absorption. To minimize
redundancy, only a handful of rays are cast from the source, but they
are split into sub-rays to ensure that all cells are covered by them,
and they can be merged again if need be.

\textbf{Cones} are a variant on short characteristics, used in
conjunction with SPH \citep{{Pawlik:2008kk},{Pawlik:2011ei}} and the
moving-mesh AREPO code \citep{{Petkova:2011dz}}. The angular dimension
of the RT equation is discretized into tesselating cones that can
collect radiation from multiple sources and thus ease the
computational load and even allow for the inclusion of continuous
sources, e.g. gas collisional recombination.

A \textbf{hybrid method} proposed by \cite{{Rijkhorst:2006ke}}
combines the long and short characteristics on patch-based grids (like
AMR), to get rid of most of the redundancy while keeping the parallel
nature. Long characteristics are used inside patches, while short
characteristics are used for the inter-patches calculations.

\textbf{Monte-Carlo} schemes do without splitting or merging of rays,
but instead reduce the computational cost by sampling the radiation
field, typically both in the angular and frequency dimensions, into
photon packets that are emitted and traced away from the source. The
cost can thus be adjusted with the number of packets emitted, but
generally this number must be high in order to minimize the noise
inherent to such a statistical method. Examples include
\cite{{Ciardi:2001is}}, \cite{{Maselli:2003ex}},
\cite{{Altay:2008cg}}, \cite{{Baek:2009gm}}, and
\cite{{Cantalupo:2011ke}}. An advantage of the Monte-Carlo approach of
tracking individual photon packets is that it naturally allows for
keeping track of the scattering of photons. For line radiation
transfer, where doppler/redshift effects in resonant photon scattering
are important, Monte-Carlo schemes are the only feasible way to go --
though in these cases, post-processing RT is usually sufficient
\citep[e.g.][]{{Cantalupo:2005hq}, {Verhamme:2006jt},
  {Laursen:2007kl}, {Pierleoni:2009kr}}.

Ray-based schemes in general assume infinite light speed, i.e. rays
are cast from source to destination instantaneously. Many authors note
that this only affects the initial speed of ionization fronts
(I-fronts) around points sources (being faster than the light speed),
but it may also result in an over-estimated I-front speed in
underdense regions (see \Sec{Ila4.sec}), and may thus give incorrect
results in reionization experiments where voids are re-ionized too
quickly. Some ray schemes \citep[e.g.][]{{Pawlik:2008kk},
  {Petkova:2011dz},{Wise:2011iw}} allow for finite light speed, but
this adds to the complexity, memory requirement and computational
load. \joki{With the exception of the cone-based methods (and to some
  degree the \cite{Wise:2011iw} implementation), which can combine
radiation from many sources into single rays, ray-based schemes share
the disadvantage that the computational load increases linearly with
the number of radiative sources. Moment methods can naturally
  tackle this problem, though other limitations appear instead.}

\subsubsection{Moment-based radiative transfer}
An alternative to ray-tracing schemes is to reduce the angular
dimensions by taking angular moments of the RT equation
(Eq. \ref{th_RT.eq}). Intuitively this can be thought of as switching
from a beam description to that of a field or a fluid, where the
individual beams are replaced with a ``bulk" direction that represents
an average of all the photons crossing a given volume element in
space. This infers useful simplifications: two angular dimensions are
eliminated from the problem, and the equations take a form of
conservation laws\joki{, like the Euler equations of
  hydrodynamics. They} are thus rather easily coupled to these
equations, and can be solved with numerical methods designed for
hydrodynamics. \joki{Since radiation is not tracked individually from
  each source, but rather just added to the radiation field, the
  computation load is naturally independent of the number of sources.}

The main advantage is also the main drawback: the directionality is
largely lost in the moment approximation and the radiation becomes
somewhat diffusive, which is generally a good description in the
optically thick limit, where the radiation scatters a lot, but not in
the optically thin regime where the radiation is
free-streaming. Radiation has a tendency to creep around corners with
moment methods. Shadows are usually only coarsely approximated, if at
all, though we will see e.g. in section \Sec{iliev3.sec} that sharp
shadows can be maintained with idealized setups and a specific solver.

The large value of the speed of light is also an issue. Moment methods
based on an explicit time marching scheme have to follow a Courant
stability condition that basically limits the radiation from crossing
more than one volume element in one time-step.  This requires to
perform many time-steps to simulate a light crossing time in the
free-streaming limit, or, as we will see later, to reduce artificially
the light speed.  Implicit solvers can somewhat alleviate this
limitation, at the price of inverting large sparse matrices which are
usually ill-conditioned and require expensive, poorly parallelized,
relaxation methods.

The frequency dimension is also reduced, via integration over
frequency bins: in the grey (single group) approximation the integral
is performed over the whole relevant frequency range, typically from
the hydrogen ionization frequency and upwards. In the multigroup
approximation, the frequency range is split into a handful of bins, or
photon groups, (rarely more than a few tens due to memory and
computational limitations) and the equations of radiative transfer can
be solved separately for each group. \joki{Ray-tracing schemes also
  often discretize into some number of frequency bins, and they are
  usually more flexible in this regard than moment-based schemes:
  while the spectrum of each source can be discretized individually in
  ray-tracing, the discretization is fixed in space in moment-based
  schemes, i.e. the frequency intervals and resulting averaged photon
  properties must be the same everywhere, due to the field
  approximation.}

In the simplest form of moment-based RT implementations, so-called
flux limited diffusion (FLD), only the zero-th order moment of the
radiative transfer equation is used\joki{, resulting in an elliptic
  set of conservation laws.} A closure is provided in the form of a
local diffusion relation, which lets the radiation flow in the
direction of decreasing gas internal energy (i.e. in the direction
opposite of the energy gradient). This is realistic only if the medium
is optically thick, and shadows cannot be modelled. The FLD method has
been used by e.g. \cite{Krumholz:2007kp}, \cite{{Reynolds:2009kj}} and
\cite{{Commercon:2011eq}}, mainly for the purpose of studying the
momentum feedback of infrared radiation onto dusty and optically thick
gas, rather than photoionization of hydrogen and helium.

\cite{{Gnedin:2001cw}} and \cite{{Petkova:2009fx}} used the optically
thin variable Eddington tensor formalism (OTVET), in which the
direction of the radiative field is composed on-the-fly in every point
in space from all the radiative sources in the simulation, assuming
that the medium between source and destination is transparent (hence
optically thin). This calculation is pretty fast, given the number of
relevant radiative sources is not overburdening, and one can neglect
these in-between gas cells.  \cite{{Finlator:2009jb}} take this
further and include in the calculation the optical thickness between
source and destination with a long characteristics method, which makes
for an accurate but slow implementation. \joki{A clear disadvantage
  here is that in using the radiation sources to close the moment
  equations and compute the flux direction, the scaling of the
  computational load with the number of sources is re-introduced,
  hence negating one of the main advantages of moment-based RT.}

\cite{{Gonzalez:2007gd}}, \AT{} and \cite{{Vaytet:2010tv}} -- and now
us -- use a different closure formalism, the so-called M1 closure,
which can establish and retain bulk directionality of photon flows,
and can to some degree model shadows behind opaque obstacles. The M1
closure is very advantageous in the sense that it is purely local,
i.e. it requires no information which lies outside the cell, which is
not the case for the OTVET approximation.

As shown by \cite{{Dubroca:1999gk}}, the M1 closure has the further
advantage that it makes the system of RT equations take locally the
form of a hyperbolic system of conservations laws, where the
characteristic wave speeds can be calculated explicitly and are
usually close, but always smaller than the speed of light $c$.
Hyperbolic systems of conservation laws are mathematically well
understood and thoroughly investigated, and a plethora of numerical
methods exist to deal with them \citep[e.g.][]{Toro99}. In fact, the
Euler equations are also a hyperbolic system of conservation laws,
which implies we have the RT equations in a form which is well suited
to lie alongside existing hydrodynamical solvers, e.g. in \ramses{}.

\subsection{From \aton{} to \ramsesrt{}}
\label{aton.sec}

The \aton{} code \citep[\AT{},][]{{Aubert:2010if}} uses graphical
processing units, or GPUs, to post-process the transfer of
monochromatic photons and their interaction with hydrogen gas. GPUs
are very fast, and therefore offer the possibility to use the correct
(very large) value for the speed of light and perform hundreds to
thousands of radiation sub-cycles at a reasonable cost, but only if
the data is optimally structured in memory, such that volume elements
that are close in space are also close in memory.  It is ideal for
post-processing RT on simulation outputs that are projected onto a
Cartesian grid, but hard to couple directly with an AMR grid in order
to play an active part in any complex galaxy formation
simulation. Even so, we have in the newest version of the \aton{} code
included the possibility to perform fully coupled RHD simulations
using a Cartesian grid only (this usually corresponds to our coarser
grid level in the AMR hierarchy), where RT is performed using the ATON
module on GPUs.

In our \ramsesrt{} implementation we use the same RT method as \aton{}
does -- the moment method with the M1 Eddington tensor closure.  The
biggest difference is that \ramsesrt{} is built directly into the
\ramses{} cosmological hydrodynamics code, allowing us to perform RHD
simulations directly on the AMR grid, without any transfer of data
between different grid structures. Furthermore, we have expanded the
implementation to include multigroup photons to approximate
multifrequency, and we have added the interactions between photons and
helium. We explicitly store and advect the ionization states of
hydrogen and helium, and we have built into \ramsesrt{} a new
non-equilibrium thermochemistry model that evolves these states along
with the temperature and the radiation field through chemical
processes, photon absorption and emission. Finally, for realistic
radiative feedback from stellar populations, we have enabled
\ramsesrt{} to read external spectral energy distribution (SED) models
and derive from them luminosities and UV ``colors" of simulated
stellar sources.

We have already listed a number of RT implementations, two of which
even function already in the \ramses{} code
\citep[\AT{},][]{{Commercon:2011eq}}, and one might ask whether
another one is really needed.  To first answer for the \aton{}
implementation, it is optimized for a different regime than
\ramsesrt{}. As discussed, \aton{} prefers to work with structured
grids, but it cannot deal well with adaptive refinement. This, plus
the speed of \aton{}, makes it very good for studying large scale
cosmological reionization, but not good for AMR simulations of
individual halos/galaxies, e.g. cosmological zoom simulations, where
the subject of interest is the effect of radiative feedback on the
formation of structures and galaxy evolution, and escape fractions of
ionizing radiation.  The \cite{Commercon:2011eq} implementation is
on the opposite side of the spectrum. Being based on the FLD method,
it is optimized for RHD simulations of optically thick protostellar
gas. It is a monogroup code that doesn't track the ionization state of
the gas. Furthermore, it uses a rather costly implicit solver, which
makes it hard to adapt to multiple adaptive time stepping usually used
in galaxy formation problems.

A few codes have been used for published 3D cosmological RHD
simulations with ionizing radiation. \joki{As far as we can see these
  are \cite{Gnedin:2000hp}, \cite{Kohler:2007kp} (both in {$\tt
    SLH-P^3M$}), \cite{Shin:2008dm}, \cite{{Petkova:2009fx}} (in {\tt
    GADGET}), \cite{{Wise:2011iw}} (in {\tt ENZO}),
  \cite{Finlator:2011ji} (in {\tt GADGET}), \cite{Hasegawa:2013fv}
  ({\tt START}), and \cite{Pawlik:2013jz} (in {\tt GADGET})}. A few
others that have been used for published astrophysical (ionizing) RHD
simulations but without a co-evolving cosmology are
\cite{{Mellema:2006cl}}, \cite{{Susa:2006tz}}, \cite{{Whalen:2006ke}},
and \cite{{Baek:2009gm}}. The rest apparently only do post-processing
RT, aren't parallel or are otherwise not efficient enough. Many of
these codes are also optimized for cosmological reionization rather
than galaxy-scale feedback.

Thus there aren't so many cosmological RHD implementations out there,
and there should be room for more.  The main advantage of our
implementation is that our method allows for an unlimited number of
radiative sources and can even easily handle continuous sources, and
is thus ideal for modelling e.g. the effects of radiative feedback in
highly resolved simulations of galaxy formation, UV escape fractions,
and the effects of self-shielding on the emission properties of gas
and structure formation, e.g. in the context of galaxy formation in
weak gravitational potentials.

The structure of the paper is as follows: in \Sec{Method.sec} we
present the moment based RT method we use. In \Sec{Impl.sec} we
explain how we inject and transport radiation on a grid of gas cells,
and how we calculate the thermochemistry in each cell, that
incorporates the absorption and emission of radiation. In \Sec{Dt.sec}
we present two tricks we use to speed up the RHD code, namely to
reduce the speed of light, and to ``smooth" out the effect of operator
splitting. In \Sec{RHD.sec} we describe how the radiative transfer
calculation is placed in the numerical scheme of \ramses{}, and
demonstrate that the radiation is accurately transported across an AMR
grid. In \Sec{tests.sec}, we present our test suite, demonstrating
that our code performs very well in coupled radiation hydrodynamics
problems and finally, \Sec{Discussion.sec} summarizes this work and
points towards features that may be added in the future. Details of
the thermochemistry and additional code tests are described in the
appendix.

\section{Moment-based radiative transfer with the M1 closure} 
\label{Method.sec}
Let $I_{\nu}(\bx,\bn,t)$ denote the radiation specific intensity at
location $\bx$ and time $t$, such that
\begin{equation}
  I_{\nu} \ d\nu \ d\Omega \ dA \ dt
\end{equation}
is the energy of photons with frequency over the range $d\nu$ around
$\nu$ propagating through the area $dA$ in a solid angle $d\Omega$
around the direction $\bn$.

The equation of radiative transfer \citep[e.g.][]{{Mihalas:1984vm}}
describes the local change in $I_{\nu}$ as a function of propagation,
absorption and emission, \begin{equation}\label{th_RT.eq}
  \frac{1}{c} \frac{\partial I_{\nu}}{\partial t}
  + \bn \cdot \nabla I_{\nu}
  = -\kappa_{\nu}I_{\nu} + \eta_{\nu},
\end{equation}
where $c$ is the speed of light, $\kappa_{\nu}(\bx,\bn,t)$ is an
absorption coefficient and $\eta_{\nu}(\bx,\bn,t)$ a source
function. 

By taking the zeroth and first angular moments of \eq{th_RT.eq}, we
can derive the moment-based RT equations that describe the
time-evolution of photon number density $\Nphot_\nu$ and flux
$\bF_\nu$ (see e.g. \AT):

\begin{align}
  \frac{\partial \Nphot_{\nu}}{\partial t}+\nabla \cdot \bF_{\nu} 
  &= -\sum_j^{\hi,\hei,\heii}
  {n_j \sigma_{\nu j}cN_{\nu}}
   + \Npr^{\star}_{\nu} + \Npr^{rec}_{\nu}
  \label{RT_N_phys.eq}\\
  \frac{\partial \bF_{\nu}}{\partial t}
  +{c^2}\nabla \cdot \Pt_{\nu}
  &= -\sum_j^{\hi,\hei,\heii}
  {n_j \sigma_{\nu j}c\bF_{\nu}}, \label{RT_N_phys2.eq} 
\end{align}
where $\Pt_{\nu}$ is the radiative pressure tensor that remains to be
determined to close the set of equations. Here we have split the
absorption coefficient into constituent terms, $n_j \sigma_{\nu j}$,
where $n_j$ is number density of the photo-absorbing species $j$
(=\hi{}, \hei{}, \heii{}), and $\sigma_{\nu j}$ is the ionization
cross section between $\nu$-frequency photons and species $j$.
Furthermore we have split the source function into (e.g. stellar,
quasar) injection sources, $\Npr^{\star}_{\nu}$, and recombination
radiation from gas, $\Npr^{rec}_{\nu}$. Here we only consider the
photo-absorption of hydrogen and helium, which is obviously most
relevant in the regime of UV photons. However, other absorbers can
straightforwardly be added to the system.

Eqs. \eq{RT_N_phys.eq}-\eq{RT_N_phys2.eq} are continuous in $\nu$, and
they must be discretized to be usable in a numerical code. \AT{}
collected all relevant frequencies into one bin, so the equations
could be solved for one \emph{group} of photons whose attributes
represent averages over the frequency range. For a rough approximation
of multifrequency, we split the relevant frequency range into a number
of photon groups, defined by
\begin{equation}\label{nuintegral_multi.eq}
  \Nphot_{i} = \int_{\nui}^{\nuf}{\Nphot_{\nu} \; d\nu}, \ \ \ \ \ \ \ \ \
  \bF_{i} = \int_{\nui}^{\nuf}{\bF_{\nu} \; d\nu},
\end{equation}
where $(\nui,\nuf)$ is the frequency interval for group $i$. In the
limit of one photon group, the frequency range is
$(\nui,\nuf)=(\nu_{\hi},\infty)$; with $M>1$ groups, the frequency
intervals should typically be mutually exclusive and set up to cover
the whole H-ionizing range:
\begin{displaymath}
  [\nu_{00},\nu_{01}:\nu_{10},\nu_{11}:...:\nu_{M0},\nu_{M1}]
  =[\nu_{\hi},\infty[.
\end{displaymath}

Integrating the RT equations (\ref{RT_N_phys.eq}) and
(\ref{RT_N_phys2.eq}) over each frequency bin corresponding to the
group definitions yields $M$ sets of four equations:
\begin{align}
  \frac{\partial \Nphot_{i}}{\partial t}+\nabla \cdot \bF_{i} &= 
  -\sum_j^{\hi,\hei,\heii}{n_jc\csn_{ij}N_i}
  \ \ + \Npr^{\star}_{i} + \Npr^{rec}_{i}, \label{RTfin1.eq}
  \\
  \frac{\partial \bF_{i}}{\partial t}
  +{c^2}\nabla \cdot \Pt_{i} &=
  -\sum_j^{\hi,\hei,\heii}{n_jc\csn_{ij}\bF_i}, \label{RTfin2.eq}
\end{align}
where $\csn_{ij}$ represent average cross sections between each group
$i$ and species $j$, defined by\footnote{here we assume the spectral
  shape of $\bF_{\nu}$ to be identical, within each group, to that of
  $\Nphot_{\nu}$.}
\begin{equation}\label{csn_multi0.eq}
  \csn_{ij}=
  \frac{\int_{\nui}^{\nuf}{\sigma_{\nu j} N_{\nu} \ d\nu}}
  {\int_{\nui}^{\nuf}{N_{\nu} \ d\nu}}.
\end{equation}
We simplify things however by defining the group cross sections as
global quantities, assuming a frequency distribution of energy
$J(\nu)$ for the radiative sources (e.g. a black-body or some
sophisticated model). The cross sections are thus in practice
evaluated by
\begin{equation}\label{csn_multi.eq}
  \csn_{ij}=
  \frac{\int_{\nui}^{\nuf}{\sigma_{\nu j} J(\nu) /h\nu \ d\nu}}
  {\int_{\nui}^{\nuf}{J(\nu)  /h\nu \ d\nu}},
\end{equation}
where $h\nu$ is photon energy (with $h$ the Planck constant).
Likewise, average photon energies within each group are evaluated by
\begin{equation}\label{egy_multi.eq}
  \egy_{i}=
  \frac{\int_{\nui}^{\nuf}{J(\nu) \ d\nu}}
  {\int_{\nui}^{\nuf}{J(\nu) / h\nu \ d\nu}},
\end{equation}
and furthermore, for the calculation of photoionization
heating\footnote{see \Eq{PHeat2.eq}}, energy weighted cross sections
are stored for each group - absorbing species couple:
\begin{equation}\label{cse_multi.eq}
  \cse_{ij}=
  \frac{\int_{\nui}^{\nuf}{\sigma_{\nu j} J(\nu) \ d\nu}}
  {\int_{\nui}^{\nuf}{ J(\nu) \ d\nu}}.
\end{equation}
In \ramsesrt{}, $\csn_{ij}$, $\cse_{ij}$ and $\egy_{i}$ can be either
set by hand or evaluated on-the-fly from spectral energy distribution
tables as luminosity weighted averages from in-simulation stellar
populations, using the expressions from \cite{{Verner:1996dm}} for
$\sigma_{\nu, \hi}$, $\sigma_{\nu, \hei}$ and $\sigma_{\nu, \heii}$.

For each photon group, the corresponding set of equations
\eq{RTfin1.eq}-\eq{RTfin2.eq} must be closed with an expression for
the pressure tensor $\Pt$. This tensor is usually described as the
product of the photon number density and the so-called Eddington
tensor $\Et$ (see Eq.~\ref{edd2.eq}), for which some meaningful and
physical expression is desired. Some formalisms have been suggested
for $\Et_{\nu}$.  \cite{{Gnedin:2001cw}}, \cite{{Finlator:2009jb}},
and \cite{{Petkova:2009fx}} have used the so called optically thin
Eddington tensor formalism (OTVET), in which $\Pt$ is composed
on-the-fly from all the radiation sources, the main drawback being the
computational cost associated with collecting the positions of every
radiative source relative to every volume element. Instead, like \AT{}
(and \citealt{{Gonzalez:2007gd}} before them), we use the M1 closure
relation \citep{{Levermore:1984cs}}, which has the great advantages
that it is purely local, i.e. evaluating it in a piece of space only
requires local quantities, and that it can retain a directionality
along the flow of the radiative field. In our frequency-discretized
form, the pressure tensor is given in each volume element for each
photon group by
\begin{equation}\label{edd2.eq}
  \Pt_i = \Et_i\Nphot_i
\end{equation}
where the Eddington tensor is
\begin{equation}\label{eddTens.eq}
  \Et_i = 
    \frac{1-\chi_i}{2}\ident + \frac{3\chi_i-1}{2} 
  \bn_i \otimes \bn_i
\end{equation}
and
\begin{equation}\label{Udir.eq}
  \bn_i = \frac{\bF_i}{|\bF_i|}, \ \ \ \ \
  \chi_i = \frac{3+4 f_i^2}{5+2\sqrt{4-3 f_i^2}}, \ \ \ \ \
  f_i = \frac{ |\bF_i|}{c\Nphot_i},
\end{equation}
are the unit vector pointing in the flux direction, the Eddington
factor and the reduced flux, respectively.  The reduced flux describes
the directionality of the group $i$ radiation in each point, and must
always have $0 \le f_i \le 1$. A low value means the radiation is
predominantly isotropic, and a high value means it is predominantly
flowing in one direction. \joki{Photons injected into a point (via an
  increase in photon density only) initially have zero reduced flux
  and thus are isotropic. Away from the source, the moment equations
  and M1 closure develop a preferred outwards direction, i.e. the
  reduced flux tends towards one. Beams can be injected by imposing
  unity reduced flux on the injected photons. In this case, the M1
  closure correctly maintains unity reduced flux (and $\chi=1$) along
  the beam (see demonstrations in \Fig{hllglfcomp.fig} and Sections
  \ref{amr_transport.sec}, \ref{iliev3.sec}, and \ref{iliev7.sec}).}
For the arguments leading to these expressions and a
general discussion we point the reader towards
\cite{{Levermore:1984cs}} and \cite{{Gonzalez:2007gd}} and \AT{}.

\section{The radiative transfer implementation} \label{Impl.sec} We
will now describe how pure radiative transfer is solved on a grid --
without yet taking into consideration the hydrodynamical coupling. The
details here are not very specific to \ramsesrt{} and are much like
those of \AT{}.

In addition to the usual hydrodynamical variables stored in every grid
cell in \ramses{} (gas density $\rho$, momentum density $ \rho\bf{u}$,
energy density $E$, metallicity $Z$), \ramsesrt{} has the following
variables: First, we have the $4 \times M$ variables describing photon
densities $\Nphot_i$ and fluxes $\bF_{i}$ for the $M$ photon
groups. Second, in order to consistently treat the interactions of
photons and gas, we track the non-equilibrium evolution of hydrogen
and helium ionization in every cell, stored in the form of passive
scalars which are advected with the gas, namely
\begin{align}
  \xhii   = \frac{\nhii}{\nh},   \ \ \ \ \ 
  \xheii  = \frac{\nheii}{\nhe}, \ \ \ \ \ 
  \xheiii = \frac{\nheiii}{\nhe}. \label{x_fractions.eq}
\end{align}

For each photon group, we solve the set of equations
\eq{RTfin1.eq}-\eq{RTfin2.eq} with an operator splitting strategy,
which involves decomposing the equations into three steps that are
executed in sequence over the same time-step $\Delta t$, which has
some pre-determined length. The steps are:

\begin{enumerate}
\item \textit{Photon injection step}, where radiation from stellar and
  other radiative sources (other than gas recombinations) is injected
  into the grid. This corresponds to the $\Npr^{\star}_{i}$ term in
  \eq{RTfin1.eq}.
\item \textit{Photon transport step}, where photons are propagated in
  space. This corresponds to solving \eq{RTfin1.eq}-\eq{RTfin2.eq}
  with the RHS being equal to zero.
\item \textit{Thermochemistry step}, where the rest of the RHS of
  \eq{RTfin1.eq}-\eq{RTfin2.eq} is solved. This is where the photons
  and the gas couple, so here we evolve not only the photon densities
  and fluxes, but also the ionization state and temperature of the
  gas.
\end{enumerate}

\subsection{The injection step}\label{Injection.sec}
The equations to solve in this step are very simple,
\begin{equation}\label{inj.eq}
  \frac{\partial N_i}{\partial t}=\Npr^{\star}_i,
\end{equation}
where $\Npr^{\star}_i$ is a rate of photon injection into photon group
$i$, in the given cell. Normally, the injected photons come from
stellar sources, but they could also include other point sources such
as AGN, and also pre-defined point sources or even continuous
``volume" sources\footnote{In \cite{{Rosdahl:2012bt}}, we emitted UV
  background radiation from cosmological void regions, under the
  assumption that they are transparent to the radiation.}.

Given the time $t$ and time-step length $\dt$, the discrete update in
each cell done for each photon group is the following sum over all
stellar particles situated in the cell:
\begin{align} \label{injnum.eq}
  N^{n+1}_i &=N^{n}_i \\
  &+\frac{f_{\rm{esc}}}{V} \sum^{\rm{cell \ stars}}_{\star} 
    m_{\star} \left[ 
      \Pi_i(\sage^{n+1},Z_{\star}) -
      \Pi_i(\sage^{n},Z_{\star}) \right], \nonumber
\end{align}
where $n$ denotes the time index ($n=t$ and $n+1=t+\Delta t$),
$f_{\rm{esc}}$ is an escape fraction, $V$ is the cell volume,
$m_{\star}$, $\sage$ and $Z_{\star}$ are mass, age and metallicity of
the stellar particles, respectively, and $\Pi_i$ is some model for the
accumulated number of group $i$ photons emitted per solar mass over
the lifetime (so far) of a stellar particle. The escape fraction,
$f_{\rm{esc}}$ is just a parameter that can be used to express the
suppression (or even boosting) of radiation from processes that are
unresolved inside the gas cell.

\ramsesrt{} can read stellar energy distribution (SED) model tables to
do on-the-fly evaluation of the stellar particle luminosities,
$\Pi_i$. Photon cross sections and energies can also be determined
on-the-fly from the same tables\joki{, to represent
  luminosity-weighted averages of the stellar populations in a
  simulation}. Details are given in \App{SEDs.sec}.

\subsection{The transport step}\label{Transport.sec}
The equations describing free-flowing photons are
\begin{align}
  \frac{\partial N}{\partial t}+\nabla \cdot \bF=0, 
  \label{transport.eq} \\
  \frac{\partial \bF}{\partial t}+c^2\nabla \cdot \Pt=0, 
  \label{transport2.eq}
\end{align}
i.e. \eq{RTfin1.eq}-\eq{RTfin2.eq} with the RHS $=0$. Note that we
have removed the photon group subscript, since this set of equations
is solved independently for each group over the time-step.

\begin{figure*}
  \centering
  \subfloat{\includegraphics[width=0.22\textwidth]
    {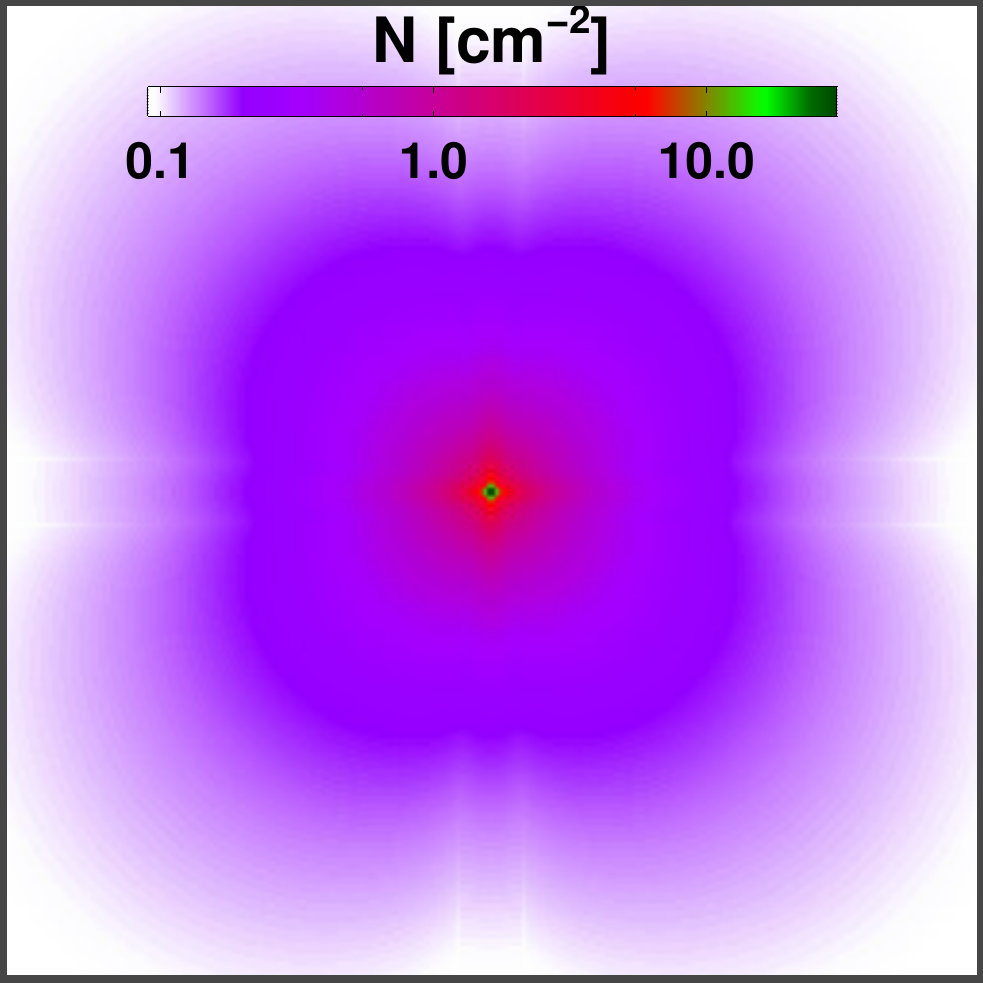}}
  \subfloat{\includegraphics[width=0.22\textwidth]
    {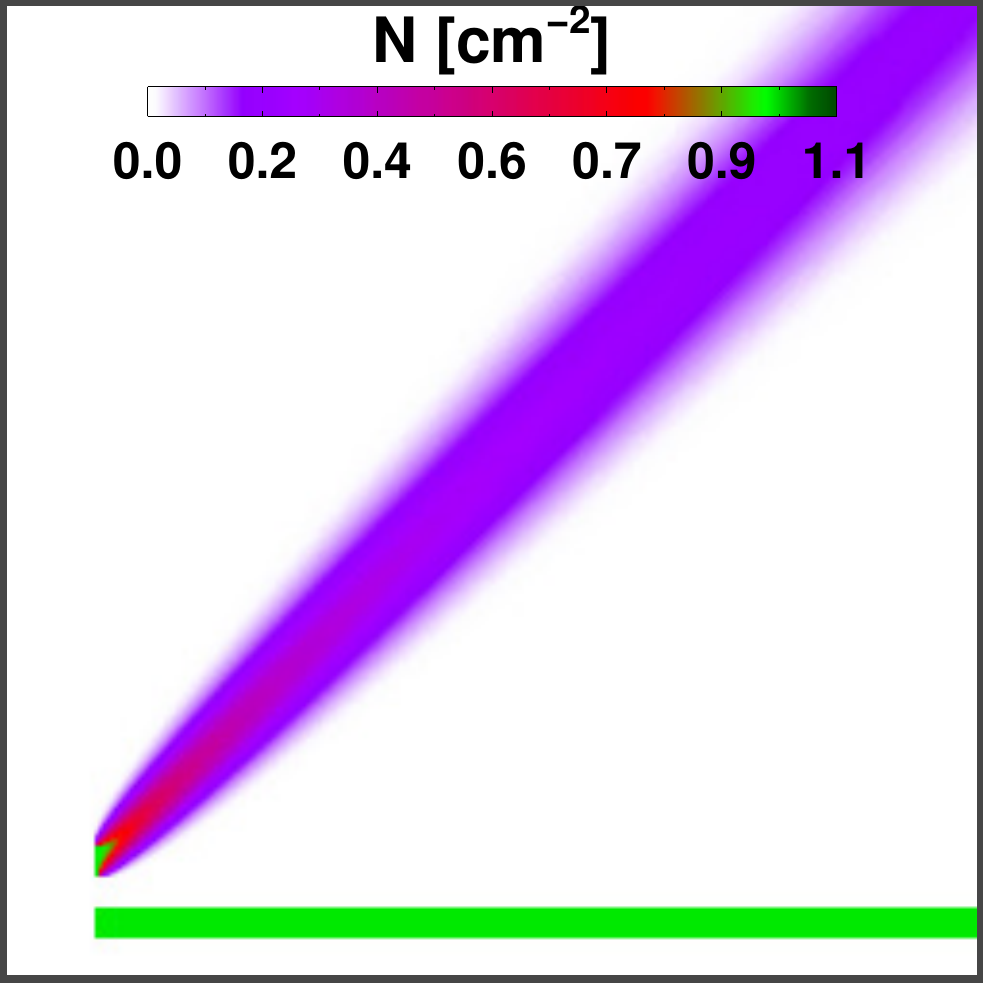}}
  \subfloat{\includegraphics[width=0.22\textwidth]
    {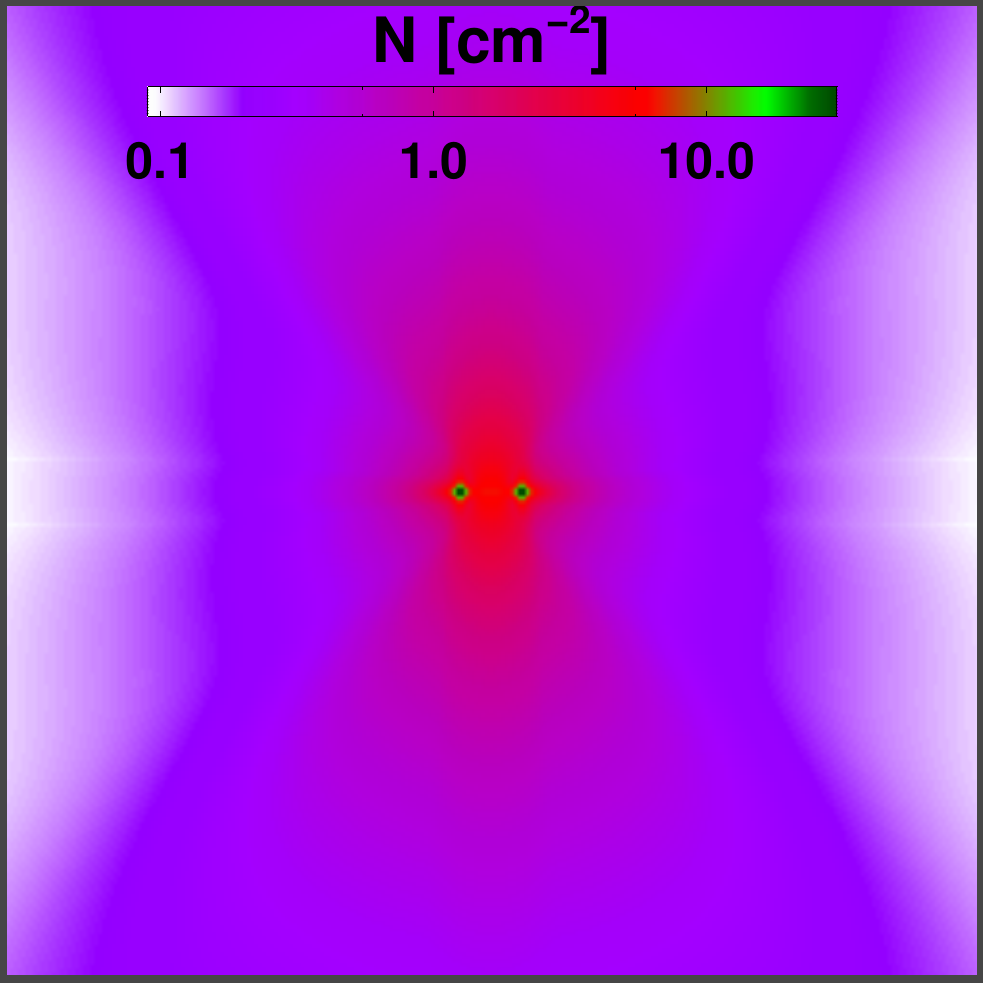}}
  \subfloat{\includegraphics[width=0.22\textwidth]
    {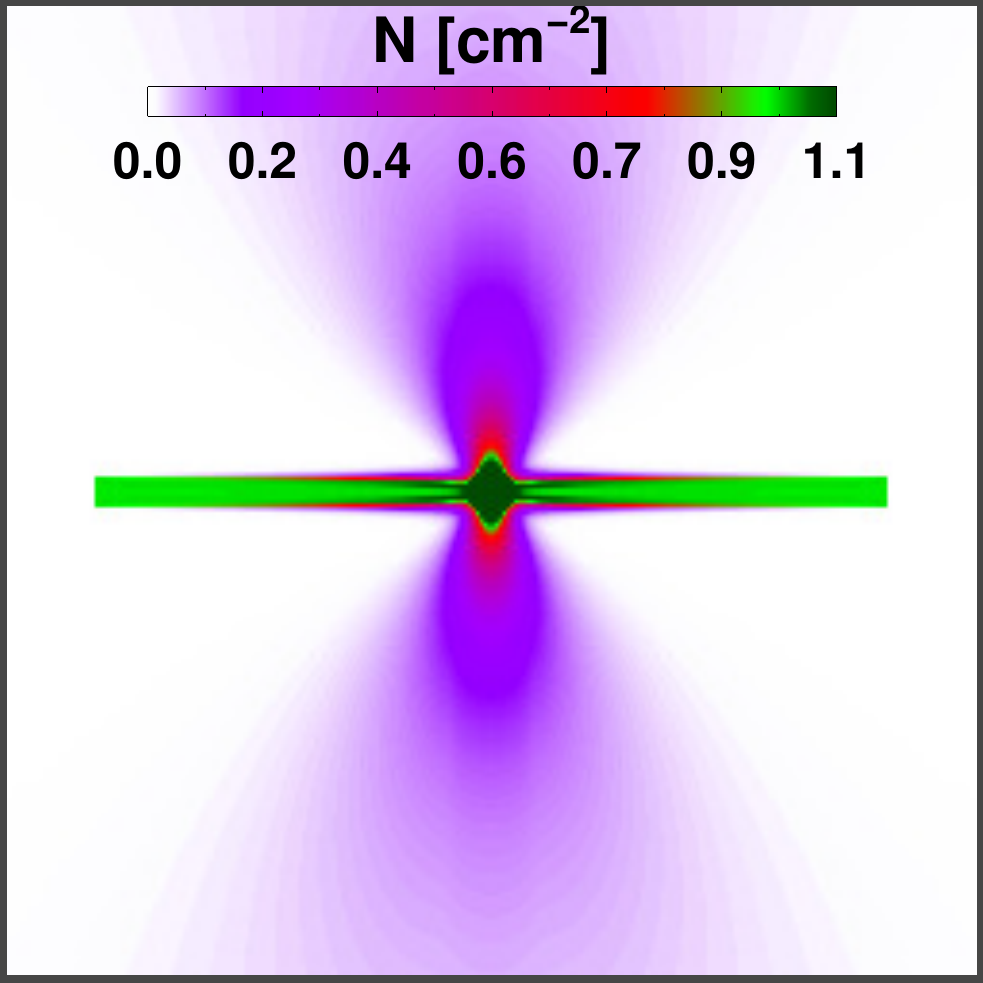}}
  \vspace{-4.mm}
  \subfloat{\includegraphics[width=0.22\textwidth]
    {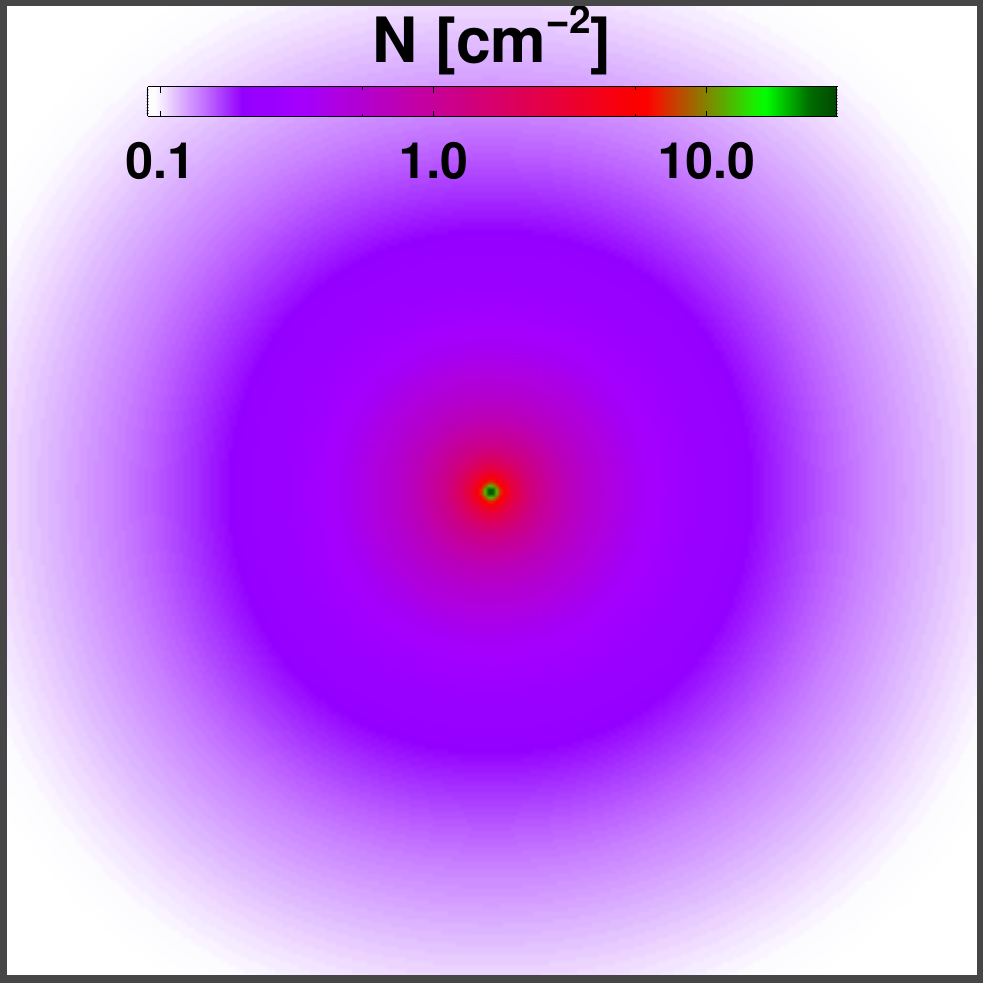}}
  \subfloat{\includegraphics[width=0.22\textwidth]
    {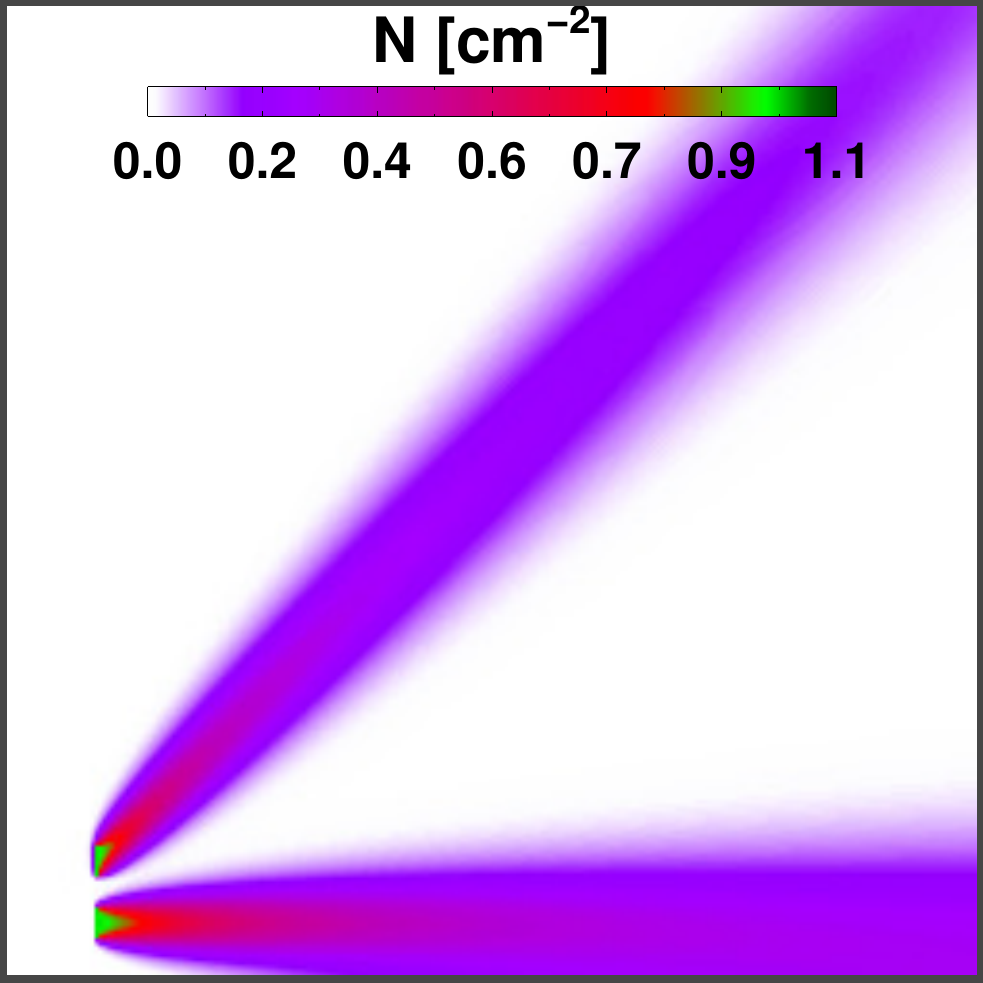}}
  \subfloat{\includegraphics[width=0.22\textwidth]
    {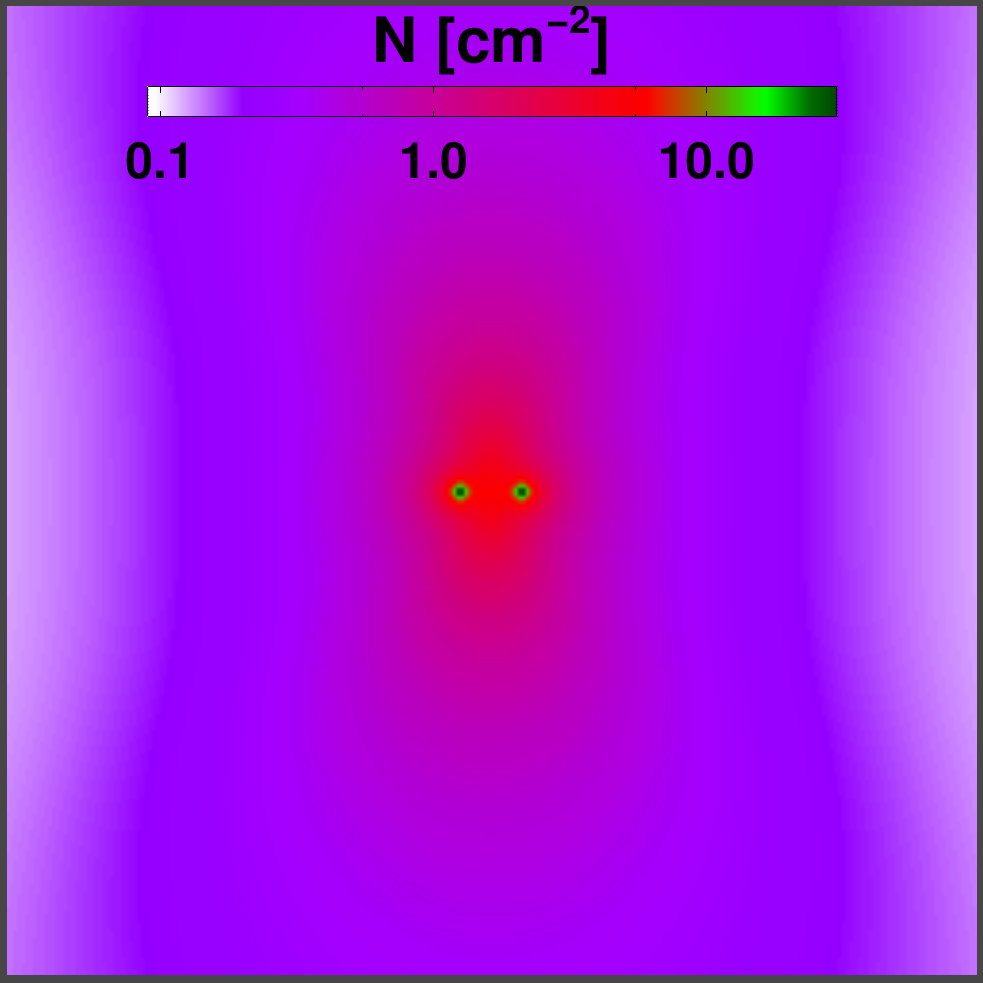}}
  \subfloat{\includegraphics[width=0.22\textwidth]
    {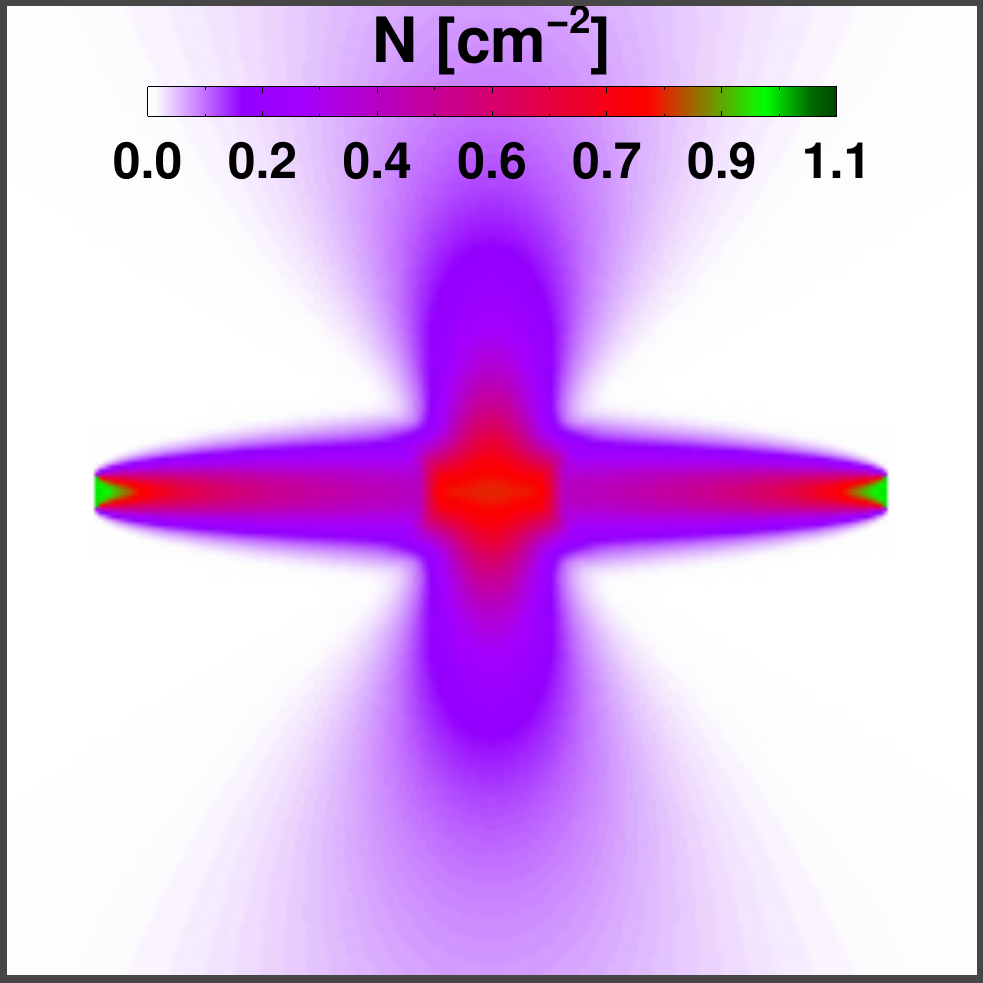}}
  \caption
  {\label{hllglfcomp.fig}Comparison of radiative transport in 2D
    \ramsesrt{} runs (no photon-gas interaction) with the HLL
    (\textbf{top}) and GLF (\textbf{bottom}) flux functions, using
    isotropic point sources and beams. The box width in all runs is
    $1$ cm and the resolution is $256^2$ cells. For each isotropic
    point source, $10^{10}$ photons $\sm$ are injected continuously,
    and for each beam a constant photon density of $N=1 \ \ccitwod$, with
    a unity reduced flux, is imposed on a region of one cell width and
    eight cell heights at the beam origin. \joki{The snapshots are
      taken at $t=5.2 \times 10^{-11}$ s (a bit less than two light
      crossing times, long enough that a static configuration has been
      reached).} \textbf{Far left} frames show single isotropic point
    sources. \textbf{Middle left} frames show attempts at creating
    horizontal and diagonal beams (with $\bF=(cN,0)$ and
    $\bF=(cN,cN)/\sqrt{2}$, respectively).  \textbf{Middle right}
    frames show two isotropic point sources and how the photons behave
    between them. \textbf{Far right} frames show two beams of opposing
    directions and how a spurious weak perpendicular radiation source
    forms where they meet.}
\end{figure*}

We can write the above equations in vector form 
\begin{equation}\label{dUdt.eq}
  \frac{\partial \state}{\partial t}+\nabla \stateF(\state)=0,
\end{equation}
where $\state=[N,\bF]$ and $\stateF(\state)=[\bF,c^2\Pt]$. To solve
\eq{dUdt.eq} over time-step $\Delta t$, we use an explicit
conservative formulation, expressed here in 1D for simplicity,
\begin{equation}\label{rt_transport.eq}
 \frac{\state_{l}^{n+1}-\state_{l}^{n}}{\Delta t}
 +\frac{\stateF_{l+1/2}^n-\stateF_{l-1/2}^n}{\Delta x}=0,
\end{equation}
where $n$ again denotes time index and $l$ denotes cell index along
the x-axis. $\stateF_{l+1/2}$ and
$\stateF_{l-1/2}=\stateF_{(l-1)+1/2}$ are intercell fluxes evaluated
at the cell interfaces. Simple algebra gives us the updated cell
state,
\begin{equation}\label{dUdt_sol.eq}
  \state_{l}^{n+1}= \state_{l}^{n} +
  \frac{\Delta t}{\Delta x}
  \left(\stateF_{l-1/2}^n-\stateF_{l+1/2}^n \right),
\end{equation}
and all we have to do is determine expressions for the intercell
fluxes.

Many intercell flux functions are available for differential equations
of the form \eq{dUdt.eq} which give stable results in the form of
\eq{dUdt_sol.eq} \citep[see e.g.][]{Toro99}, as long as the Courant
time-step condition is respected (see \Sec{dt.sec}). Following \AT{}
and \cite{{Gonzalez:2007gd}} we implement two flux functions which can
be used in \ramsesrt{}.

One is the Harten-Lax-van Leer (HLL) flux function
\citep{{Harten:1983um}},
\begin{equation}\label{hll.eq}
  \left( \stateF_{\rm{HLL}} \right)_{l+1/2}^{n}
  = \frac{\lambda^+ \stateF_l^n - \lambda^- \stateF_{l+1}^n
  + \lambda^+ \lambda^- \left( \state_{l+1}^n - \state_{l}^n \right)}
  {\lambda^+ - \lambda^-},
\end{equation}
where
\begin{align*}
\lambda^+&=\max(0,\lambda^{\rm{max}}_{l},\lambda^{\rm{max}}_{l+1}), \\
\lambda^-&=\min(0,\lambda^{\rm{min}}_l,\lambda^{\rm{min}}_{l+1})
\end{align*}
are maximum and minimum eigenvalues of the Jacobian $\partial \stateF
/ \partial \state$. These eigenvalues mathematically correspond to
wave speeds, which in the case of 3D radiative transfer depend only on
the magnitude of the reduced flux $f$ \eq{Udir.eq} and the angle of
incidence of the flux vector to the cell interface. This dependence
has been calculated and tabulated by \cite{{Gonzalez:2007gd}}, and we
use their table to extract the eigenvalues.

The other flux function we have implemented is the simpler Global Lax
Friedrich (GLF) function,
\begin{equation}\label{glf.eq}
  \left( \stateF_{\rm{GLF}} \right)_{l+1/2}^{n}
  = \frac{\stateF_l^n + \stateF_{l+1}^n}{2}
  - \frac{c}{2}\left( \state_{l+1}^n - \state_{l}^n \right),
\end{equation}
which corresponds to setting the HLL eigenvalues to the speed of
light, i.e. $\lambda^-=-c$ and $\lambda^+=c$, and has the effect of
making the radiative transport more diffusive. Beams and shadows are
therefore better modelled with the HLL flux function than with the GLF
one, whereas the inherent directionality in the HLL function results
in radiation around isotropic sources (e.g. stars) which is noticeably
asymmetric, due to the preference of the axis directions.

\Fig{hllglfcomp.fig} illustrates the difference between the two flux
functions in some idealized 2D \ramsesrt{} tests, where we shoot off
beams and turn on isotropic sources. \joki{Here the photon-gas
  interaction is turned off by setting all photoionization cross
  sections to zero ($\csn_{j}=\cse_{j}=0$ for any species $j$).} It
can be seen that the HLL flux function fails to give isotropic
radiation (far left) and that the GLF function gives more diffusive
beams (second from left). Note also how the diffusivity of beams with
the HLL flux function is direction-dependent. A horizontal or vertical
beam is perfectly retained while a diagonal one ``leaks" to the sides
almost as much as with the GLF function, which has the advantage of
being fairly consistent on whether the beam is along-axis or
diagonal. The right frames of the figure give an idea of how the
radiative transport behaves in the case of multiple sources, i.e. with
opposing beams and neighboring isotropic sources. The two opposing
beams example is a typical configuration where the M1 closure relation
obviously fails, creating a spurious source of radiation,
perpendicular to the beam direction: \joki{Since opposing fluxes
  cannot cross each other in a single point in the moment
  approximation, the radiation is ``squeezed'' into those
  perpendicular directions.} \joki{It is unclear to us how much of a
  problem this presents in astrophysical contexts. Beams, which
  clearly represent the worst case scenario, are not very relevant,
  but multiple nearby sources are.}  We generally prefer to use the
GLF flux function, since we mostly deal with isotropic sources in our
cosmological/galactic simulations, but the choice of function really
depends on the problem. There is no noticeable difference in the
computational load, so if shadows are important, one should go for
HLL. \AT{} have compared the two flux functions in some of the
benchmark RT tests of \cite{{Iliev:2006jz}} and found that they give
very similar results. We do likewise for the test we describe in
\Sec{iliev6.sec}, and come to the same conclusion.
 
\subsection{The thermochemical step}
Here we solve for the interaction between photons and gas. This is
done by solving \eq{RTfin1.eq} and \eq{RTfin2.eq} with zero divergence
and stellar injection terms.

Photon absorption and emission have the effect of heating and cooling
the gas, so in order to self-consistently implement these
interactions, we evolve along with them the thermal energy density
$\etherm$ of the gas and the abundances of the species that interact
with the photons, here \hi{}, \hei{} and \heii{} via photoionizations
and \hii{}, \heii{} (again) and \heiii{} via recombinations. We follow
these abundances in the form of the three ionization fractions
$\xhii$, $\xheii$ and $\xheiii$, that we presented in
Eqs. \eq{x_fractions.eq}. The set of non-equilibrium thermochemistry
equations solved in \ramsesrt{} consists of:
\begin{align}
  \frac{\partial \Nphot_{i}}{\partial t} &=
  -\sum_j^{\hi,\hei,\heii}{n_jc\csn_{ij}N_i} \label{thchem_N.eq}\\*
  &+ \sum_j^{\hii, \heii, \heiii}  
    b^{rec}_{ji} \, [  \recA_{j} - \recB_{j} ]\, n_j \, n_e,
    \nonumber \\
  \frac{\partial \bF_{i}}{\partial t} &=
  -\sum_j^{\hi,\hei,\heii}{n_jc\csn_{ij}\bF_i}, 
  \label{thchem_F.eq} \\*
  \frac{\partial \etherm}{\partial t} &=
  \Heat + \Cool 
  \label{thchem_T.eq} \\
  \nh \frac{\pa \xhii}{\pt} &=
  \nhi \left( \beta_{\hi} \ \nel 
  + \sum_{i=1}^M \csn_{i \hi} c \Nphot_i \right) 
  \label{thchem_xhii.eq} \\*
  &- \nhii \ \recAH \ \nel, \nonumber \\
  \nhe \frac{\pa \xheii}{\pt} &=
  \nhei \left( \beta_{\hei} \ \nel
    + \sum_{i=1}^M \csn_{i \hei} c \Nphot_i \right) 
  \label{thchem_xheii.eq} \\*
  &+ \nheiii \ \recA_{\heiii} \ \nel \nonumber \\*
  &- \nheii \left( \beta_{\heii} \nel 
  +  \recA_{\heii} \nel
  +  \sum_{i=1}^M \csn_{i \heii} c \Nphot_i \right)
   \nonumber \\
  \nhe \frac{\pa \xheiii}{\pt} 
  &= \nheii \left( \beta_{\heii} \ \nel 
  + \sum_{i=1}^M \csn_{i \heii} c \Nphot_i
\right) \label{thchem_xheiii.eq} \\*
  &- \ \nheiii \ \recA_{\heiii} \ \nel  \nonumber
\end{align}

In the photon density and flux equations, \eq{thchem_N.eq} and
\eq{thchem_F.eq}, we have replaced the photon emission rate
$\Npr^{rec}_{i}$ with the full expression for recombinative emissions
from gas. Here, $\recA_{j}(T)$ and $\recB_{j}(T)$ represent case A and
B recombination rates for electrons combining with species $j$
($=\hii, \heii, \heiii$).  The $b^{rec}_{ji}$ factor is a boolean
(1/0) that states which photon group $j$-species recombinations emit
into, and $\nel$ is electron number density (a direct function of the
H and He ionization states, neglecting the contribution from metals).

The temperature-evolution, \eq{thchem_T.eq} is greatly simplified here
(see \App{neq_cooling.sec} for details). Basically it consists of two
terms: the photoheating rate
$\Heat(\Nphot_i,\xhii,\xheii,\xheiii,\nh)$ and the radiative cooling
rate $\Cool(T,\Nphot_i,\xhii,\xheii,\xheiii,\nh)$.

The $\xhii{}$ evolution \eq{thchem_xhii.eq} consists of, respectively
on the RHS, \hi{} collisional ionizations, \hi{} photo-ionizations,
and \hii{} recombinations. Here, $\beta(T)$ is a rate of collisional
ionizations. The $\xheii$ evolution \eq{thchem_xheii.eq} consists of,
from left to right, \hei{} collisional ionizations, \heiii{}
recombinations, \hei{} photo-ionizations, and \heii{} collisional
ionizations, recombinations, and photoionizations. Likewise, the
$\xheiii$ evolution \eq{thchem_xheiii.eq} consists of \heii{}
collisional ionizations and photoionizations, and \heiii{}
recombinations. The expressions we use for rates of recombinations and
collisional ionizations are given in \App{rates.sec}.

The computational approach we use to solving Equations
\eq{thchem_N.eq}-\eq{thchem_xheiii.eq} takes inspiration from
\cite{{Anninos:1997in}}. The basic premise is to solve the equations
over a sub-step in a specific order (the order we have given),
explicitly for those variables that remain to be solved (including the
current one), but implicitly for those that have already be solved
over the sub-step. Eqs. \eq{thchem_N.eq} and \eq{thchem_F.eq} are thus
solved purely explicitly, using the backwards-in-time (BW) values for
all variables on the RHS. Eq. \eq{thchem_T.eq} is partly implicit in
the sense that it uses forward-in-time values for $\Nphot$ and $\bF$,
but BW values for the other variables. And so on, ending with
Eq. \eq{thchem_xheiii.eq}, which is then implicit in every variable
except the one solved for ($\xheiii$). We give details of the
discretization of these equations in \App{neq_cooling.sec}.

\subsubsection{The 10\% thermochemistry rule}\label{10prule.sec}
For accuracy, each thermochemistry step is restricted by a local
cooling time which prohibits any of the thermochemical quantities to
change by a substantial fraction in one time-step.  We therefore
sub-cycle the thermochemistry step to fill in the global RT time-step
(see next section), using what can be called \textit{the $10\%$ rule}:
In each cell, the thermochemistry step is initially executed with the
full RT time-step length, and then the fractional update is
considered. If any of the evolved quantities ($\Nphot_i, \Fphot_i$,
$\etherm$, ionization fractions) have changed by more than $10\%$, we
backtrack and do the same calculation with half the time-step
length. Conversely, if the greatest fractional change in a sub-step is
$<5\%$, the timestep length is doubled for the next sub-step (without
the backtracking).

\joki{Together, the quasi-implicit approach used in solving the
  thermochemistry, and the $10\%$ rule, infer that photons are in
  principle conserved only at the $10\%$ level\joki{\footnote{\joki{As
        discussed in \Sec{amr_transport.sec}, the photon transport
        accurately conserves photons, so thermochemistry errors are
        the sole source of non-conservation}}}. This is because the
  thermochemistry solver is explicit in the photon density updates
  (i.e. uses before-timestep values of ionization fractions), but the
  following ionization fraction updates are implicit in the photon
  densities (i.e. they use after-timestep values for the photon
  densities). Thus, in the situation of a cell in the process of being
  photoionized, the ionization fractions are underestimated at the
  photon density updates and the photon densities are underestimated
  at the ionization fraction updates. Conversely, if the cell gas is
  recombining, the recombination-photon emission is slightly
  overestimated, since before-timestep values for the ionization
  fractions are used for the emissivity. However, judging from the
  performance in RT tests (\Sec{tests.sec}) and thermochemistry tests
  (\Sec{CoolTests.sec}), this does not appear to be cause for
  concern.}

\subsubsection{The on-the-spot approximation}\label{otsa.sec}
The photon-emitting recombinative term, the second RHS sum in
\eq{thchem_N.eq}, is optionally included. Excluding it is usually
referred to as the on-the-spot approximation (OTSA), meaning that any
recombination-emitted photons are absorbed ``on the spot'' by a
near-lying atom (in the same cell), and hence these photon emissions
cancel out by local photon absorptions. If the OTSA is assumed, the
gas is thus not photoemitting, and the case A recombination rates are
replaced with case B recombination rates in
\eq{thchem_N.eq}-\eq{thchem_xheiii.eq}, i.e. photon-emitting
recombinations straight down to the ground level are not counted. The
OTSA is in general a valid approximation in the optically thick regime
but not so when the photon mean free path becomes longer than the cell
width.

It is a great advantage of our RT implementation that it is not
restricted to a limited number of point sources. The computational
load does not scale at all with the number of sources, and photon
emission from gas (non-OTSA) comes at no added cost, whereas it may
become prohibitively expensive in ray-tracing implementations.

\section{Timestepping issues}
\label{Dt.sec}
RT is computationally expensive, and we use two basic tricks to speed
up the calculation. One is to reduce the speed of light, the other is
to modify slightly the traditional operator splitting approach, by
increasing the coupling between photon injection and advection on one
hand and thermochemistry and photo-heating on the other hand.

\subsection{The RT timestep and the reduced speed of light}
\label{dt.sec}
In each iteration before the three RT steps of photon injection,
advection and thermochemistry are executed, the length of the
time-step, $\dtrt$, must be determined.

We use an explicit solver for the radiative transport
\eq{rt_transport.eq}, so the advection timestep, and thus the global
RT timestep, is constrained by the Courant condition (here in 3D),
\begin{equation}\label{dt_courant.eq}
  \dtrt <  \frac{\Delta x}{3c},
\end{equation}
where $\Delta x$ is the cell width. This time-step constraint is
severe: it results in an integration step which is typically 300 times
shorter than in non-relativistic hydrodynamical simulations, where the
speed of light is replaced by a maximum gas velocity ($\sim 1000\
km/s$) in Eq. \ref{dt_courant.eq}. In a coupled (RHD) simulation, this
would imply a comparable increase in CPU time, either because of a
global timestep reduction (as we chose to implement, see
Sec. \ref{RHD.sec}), or because of many radiative sub-steps (as is
implemented e.g. in \aton{}\footnote{But \aton{} runs on GPUs, which
  are about a hundred times faster than CPUs, whereas \ramsesrt{} runs
  on CPUs and thus can't afford such huge amount of RT
  subcycling. NB: \aton{} also increases the timestep by working
    on the coarse grid, and hence multiplying $\Delta x$ by a factor
    $\sim 2^{6-8} = 64 - 256$ in Eq. \ref{dt_courant.eq}.}). In the
case of radiative transfer with the moment equations, there are two
well-known solutions to this problem.

The first solution is to use an \textit{implicit} method instead of an
explicit one to solve the transport equation, which means using
forward-in-time intercell fluxes in \eq{rt_transport.eq}, i.e
replacing $\stateF^n\equiv\stateF^{t}$ with
$\stateF^{n+1}\equiv\stateF^{t+\dt}$. This seemingly simple change
ensures that the computation is \textit{always stable}, no matter how
big the time-step, and we can get rid of the Courant
condition. However: \textit{(i)} It doesn't mean that the computation
is accurate, and in fact we still need some time-stepping condition to
retain the accuracy, e.g. to restrain any quantity to be changed by
more than say $10\%$ in a single time-step. Furthermore, such a
condition usually must be checked by trial-and-error, i.e. one guesses
a time-step and performs a global transport step (over the grid) and
then checks whether the accuracy constraint was broken anywhere. Such
trial-and-error time-stepping can be very expensive since it is a
global process. \textit{(ii)} Replacing $\stateF^{t}$ with
$\stateF^{t+\dt}$ is actually not simple at
all. Eqs. \ref{rt_transport.eq} become a system of coupled algebraic
equations that must be solved via matrix manipulation in an iterative
process, which is complicated, computationally expensive, and of
limited scope (i.e. can't be easily applied to any problem). Due to
these two reasons we have opted out of the implicit approach. It is
absolutely a valid approach however, and used by many
\citep[e.g.][]{{Petkova:2009fx},{Commercon:2011eq}}.

The second solution, which we have chosen instead, is to keep our
solver explicit, and relax the Courant condition by changing the speed
of light to a \textit{reduced light speed} $\cred \ll c$, the payoff
being that the time-step \eq{dt_courant.eq} becomes longer. This is
generally referred to as the reduced speed of light approximation
(RSLA), and was introduced by \cite{{Gnedin:2001cw}}.  The idea of the
RSLA is that in many applications of interest, the propagation of
light is in fact limited by the much slower speed of ionizing
fronts. In such situations, reducing the speed of light, while keeping
it higher than the fastest I-front, will yield the correct solution at
a much reduced CPU cost. In the following section, we provide a
framework to help judge how accurate the RSLA may be in various
astrophysical contexts.

\subsection{A framework for setting the reduced light speed
  value} \label{reduced_c.sec} In the extremely complex framework of
galaxy formation simulations, the accuracy of the results obtained
using the RSLA can really only be assessed by convergence tests. It is
nonetheless useful to consider a simple idealized setup in order to
derive a physical intuition of where, when, and by how much one may
reduce the speed of light. In this section, we thus discuss the
expansion of an ionized region around a central source embedded in a
uniform neutral medium.

We consider a source turning on and emitting ionizing photons at a
rate $\Npr$ into a homogeneous hydrogen-only medium of number density
$\nh$. An expanding sphere of ionized gas forms around the source and
halts at the Str\"omgren radius $\rs$ within which the rate of recombinations
equals the source luminosity:
\begin{align} \label{Stromgren_radius.eq}
\rs=\left( \frac{3\Npr}{4 \pi \recB \nh^2} \right)^{1/3},
\end{align}
where $\recB \sim 2.6 \times 10^{-13} \ \ccs$ is the case-B
recombination rate at $T\sim 10^4\ K$, and where we have assumed that
the plasma within $\rs$ is fully ionized.

The relativistic expansion of the I-front to its final radius
  $\rs$ is derived in \cite{Shapiro:2006gd}, and may be expressed as:
\begin{equation}\label{Stromgr_w_rel.eq}
  w=qy - \ln (1-y^3),
\end{equation}
where $w = t / \trec$ is time in units of the recombination time
$\trec=(\nh \recB)^{-1}$, $y = \ri/\rs$ is the position of the I-front
in units of $\rs$, and the factor $q \equiv \tcross/\trec \equiv
\rs/(c \trec)$ describes the light crossing time $\tcross$ across the
Str\"omgren radius in units of the recombination time, and basically
encompasses all the free parameters in the setup (source luminosity,
gas density, and temperature). Writing $q \propto \Npr^{1/3}
\nh^{1/3}$, we see that in many astrophysical contexts, $q$ stays in
the range $\sim 10^{-3}-10^{-2}$ (see \Tab{c_red.tbl}), simply because
we are generally either interested in the effect of bright sources
(e.g. a whole galaxy) on relatively low-density gas (e.g. the IGM) or
of fainter sources (e.g. an O-star) on high-density gas (at
e.g. molecular-cloud densities).

\begin{figure}
  \centering
  \includegraphics[width=0.48\textwidth]
  {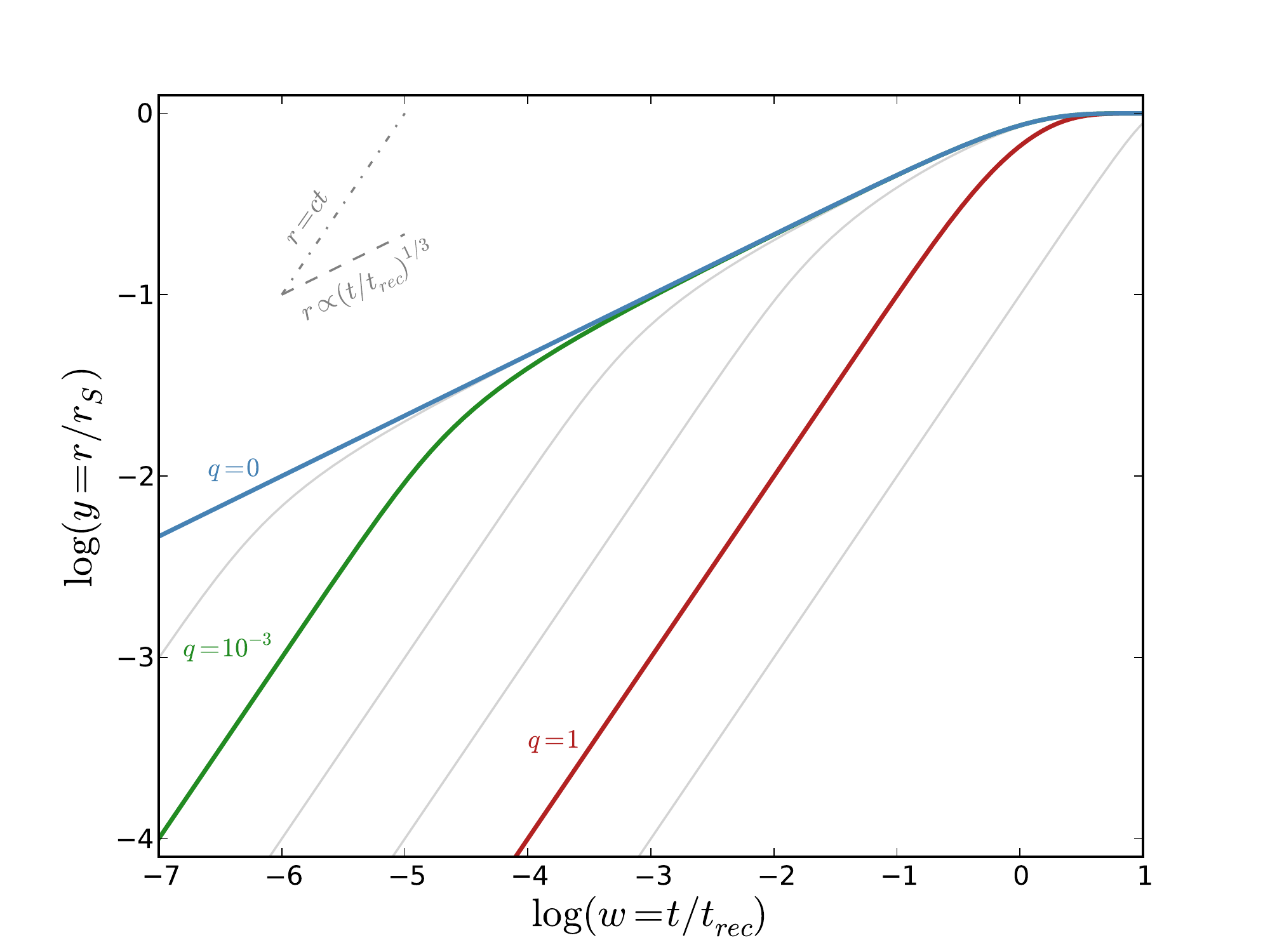}
  \caption{\label{Shapiro.fig}I-front expansion in a Stromgren sphere
    for a set of values of the dimensionless crossing time $q$. The
    blue curve shows the infinite-light-speed limit ($q=0$). The green
    curve shows a typical case with $q=10^{-3}$, and the red curve
    shows the $q=1$ case, as discussed in the text. The thin grey
    curves show other values of $q$, spanning the range $10^{-4}-10$
    in steps of one dex. The grey lines in the top-left corner of the
    plot show slopes corresponding to an expansion at the speed of
    light (dot-dashed line) or as $(t/\trec)^{1/3}$ (dashed line). For
    any $q > 0$, the I-front radius is accurately described by the
    $q=0$-limit after a crossing time.}
\end{figure}

Let us now discuss briefly the evolution of an I-front given by
  Eq. \ref{Stromgr_w_rel.eq} for illustrative values of $q$:

\begin{itemize}
\item  {\it $q = 0$ (blue curve of Fig. \ref{Shapiro.fig})}: this is the
  limiting non-relativistic case, which assumes an infinite speed of
  light ($\tcross=0$). In this case, the I-front expands
  roughly as $y \propto w^{1/3}$ (its speed decreases as
  $w^{-2/3}$) almost all the way to $\rs$, which it
  reaches after about a recombination time. 
\item {\it $q=1$ (red curve of Fig. \ref{Shapiro.fig})}: here (and for
  all $q>1$), the I-front basically expands at the speed of light all
  the way to $\rs$, which it thus reaches after a crossing time
  \joki{(which is equal to a recombination time in this case)}.
\item {\it $q=10^{-3}$ (green curve of Fig. \ref{Shapiro.fig})}: in
  this typical case, the I-front starts expanding at the speed of
  light, until $w\sim (q/3)^{3/2}$.  It then slows down and quickly
  reaches the limiting $q=0$ behavior after a crossing time (at $w\sim
  q$). The I-front then reaches $\rs$ after a recombination time (at
  $w \sim 1$).
\end{itemize}
An important feature appearing in the two latter cases is that {\it
  for any physical setup $q>0$, the I-front is always well described
  by the $q=0$ limit after a crossing time (i.e. $w \gtrsim q$).} We
can use this feature to understand the impact of reducing the speed of
light in our code. Say we have a physical setup described by a value
$q_0$. Reducing the speed of light by a factor $\fc < 1$ ($c_r = \fc
c$) implies an increase by a factor $1/\fc$ of the effective crossing
time, and the effective $q$ in our experiment becomes $q_0 / \fc$. The
solution we obtain with $c_r$ will be accurate only after an effective
crossing time, i.e. after $w = q_0 / \fc$. Before that time, the
reduced-light-speed solution will lag behind the real one.

How much one may reduce the speed of light in a given numerical
experiment then depends on the boundary conditions of the problem and
their associated timescales. Call $\tau_{\rm sim}$ the shortest
relevant timescale of a simulation. For example, if one is interested
in the effect of radiative feedback from massive stars onto the ISM,
$\tau_{\rm sim}$ can be set to the lifetime of these stars. If one is
running a very short experiment (see Sec. \ref{Ila4.sec}), the
duration of the simulation may determine $\tau_{\rm sim}$. Given this
timescale contraint $\tau_{\rm sim}$, one may reduce the speed of
light by a factor such that the I-fronts will be correctly described
after a timelapse well shorter than $\tau_{\rm sim}$, i.e. $\tcross /
\fc \ll \tau_{\rm sim}$. In other words, one may typically use $\fc =
\min(1; \sim 10\times \tcross / \tau_{\rm sim})$. We now turn to a
couple of concrete examples.

\begin{table*}
  \centering
  \caption
  {Str\"omgren sphere properties for typical cosmological regimes,
    with the inferred minimum allowed light speed fractions.}
  \label{c_red.tbl}
  \begin{tabular}{c|ccccccccc}
    \toprule
    Regime & $\nh$  [$\cci$] & $\Npr$  [$\rm{s}^{-1}$] &
    $\rs$  [kpc] & $\tcross$ [Myr] & $\trec$ [Myr]   & 
    $q$ & $\tau_{\rm sim}$ [Myr] & $\wsim$  & $f_{\rm{c,min}}$ \\
    \midrule
    MW ISM & $10^{-1}$ & $2 \ 10^{50}$ & $0.9$ & 
    $3 \ 10^{-3}$ & $1.2$ & $2 \ 10^{-3}$ & $1$ & $1$ 
    & $3 \ 10^{-2}$ \\
    MW cloud & $10^{2}$ & $2 \ 10^{48}$ & $2 \ 10^{-3}$ & 
    $6 \ 10^{-6}$ & $1 \ 10^{-3}$ & $5 \ 10^{-3}$ & $0.1$ & $80$ 
    & $6 \ 10^{-4}$ \\
    Iliev tests 1,2,5 & $10^{-3}$ & $5 \ 10^{48}$ & $5.4$ & 
    $2 \ 10^{-2}$ & $122.3$ & $1.4 \ 10^{-4}$ & $10$ & $8 \ 10^{-2}$ 
    & $2 \ 10^{-2}$ \\
    Iliev test 4 & $10^{-4}$ & $7 \ 10^{52}$ & $600$ & 
    $2$ & $1200$ & $2 \ 10^{-3}$ & $0.05$ & $4 \ 10^{-5}$ 
    & $1$ \\
    \bottomrule
  \end{tabular}
\end{table*}

\subsection{Example speed of light calculations}
In \Tab{c_red.tbl} we take some concrete (and of course very
approximate) examples to see generally what values of $\fc$ are
feasible. We consider cosmological applications from inter-galactic to
inter-stellar scales and setups from some of the RT code tests
described in \Sec{tests.sec}.

\subsubsection*{Reionization of the inter-galactic medium}
Here we are concerned with the expansion of ionization fronts away
from galaxies and into the IGM, as for example in the fourth test of
\cite{Iliev:2006jz} (hereafter \Ila{}). In this test, the IGM gas
density is typically $\nh = 10^{-4}$ cm$^{-3}$, and the sources have
$\dot{N} = 7\ 10^{52} \ s^{-1}$. In such a configuration, the
Str\"omgren radius is $\rs \sim 600$ kpc, corresponding to a crossing
time $\tcross \sim 2$ Myr. Because of the low density of the gas, the
recombination time is very long ($\gtrsim 1$ Gyr), and we are thus
close to the $q = 10^{-3}$ case discussed above (the green curve in
Fig. \ref{Shapiro.fig}).

Test 4 of \Ila{} is analyzed at
output times $\tau_{\rm sim,1} = 0.05$ Myr and $\tau_{\rm sim,2} =
0.4$ Myr (see Fig. \ref{Il4maps.fig}). In both cases, $\tau_{\rm sim}
< \tcross$, and we cannot reduce the speed of light to get an accurate
result at these times, because the expanding front has not yet reached
the $q=0$ limit. Interestingly, we cannot increase the speed of light
either, as is done in \Ila{} with C$^2$-Ray which assumes an infinite
light-speed. From Fig. \ref{Shapiro.fig}, it is clear that this
approximation (the blue curve) will over-predict the radius of the
front. We can use the analysis above to note that had the results been
compared at a later output time $\tau_{\rm sim} > 2$ Myr, the
infinite-light-speed approximation would have provided accurate
results. It is only ten times later, however, that reducing the speed
of light by a factor ten would have provided accurate results.

We conclude that propagating an I-front in the IGM at the proper speed
requires to use a value of the speed of light close to the correct
value. This is especially true in Test 4 of the \Ila{}
paper (last row of Table~\ref{c_red.tbl}). This confirms that for
cosmic reionization related studies, using the correct value for the
speed of light is very important.

\subsubsection*{Inter-stellar medium}
There is admittedly a lot of variety here, but as a rough estimate, we
can take typical densities to be $\nh\sim10^{-1} \ \cci$ in the
large-scale ISM and $\nh\sim10^{2} \ \cci$ in star-forming clouds. In
the stellar nurseries we consider single OB stars, releasing
$\Npr_{OB}\sim 2 \times 10^{48}$ photons per second, and in the
large-scale ISM we consider groups of ($\sim 100$) OB stars. The
constraining timescale is on the order of the stellar cycle of OB
stars ($\tau_{\rm sim}\sim~10$ Myr), and less for the stellar
nurseries.  In these two cases, which are representative of the dense
ISM inside galactic disks, we see in Table~\ref{c_red.tbl} that the
allowed reduction factor for the speed of light is much larger ($f_c
\simeq 10^{-4}$ to $10^{-3}$).  This is due to two effects acting
together: the gas density is higher, but the sources are fainter,
since we are now resolving individual stellar clusters, and not an
entire galaxy. Tests 1, 2 of \Ila{} and test 5 of its' RHD sequel
\citep{Iliev:2009kn} are also representative of such a favorable
regime to use the reduced speed of light approximation (second to last
row in Table~\ref{c_red.tbl}).  This rigorous analysis of the problem
at hand confirms that propagating I-front in galaxy formation
simulation can be done reliably using our current approach, while
cosmic reionization problems are better handled with GPU acceleration
and the correct speed of light.

\subsection{Smoothed RT}\label{smooth.sec}
\begin{figure}
  \centering
  \subfloat{\includegraphics[width=0.4\textwidth]
    {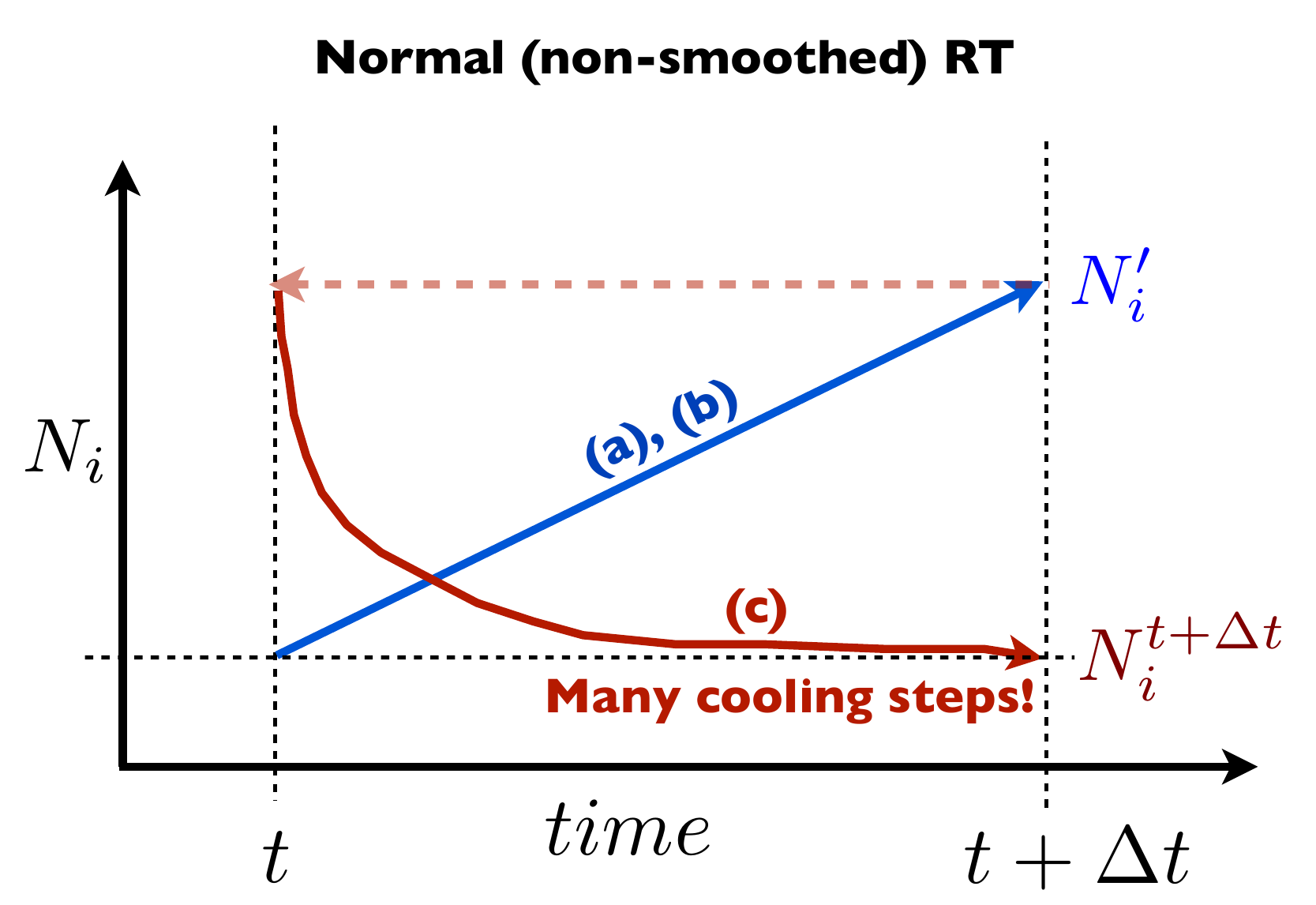}} \\
  \subfloat{\includegraphics[width=0.4\textwidth]
    {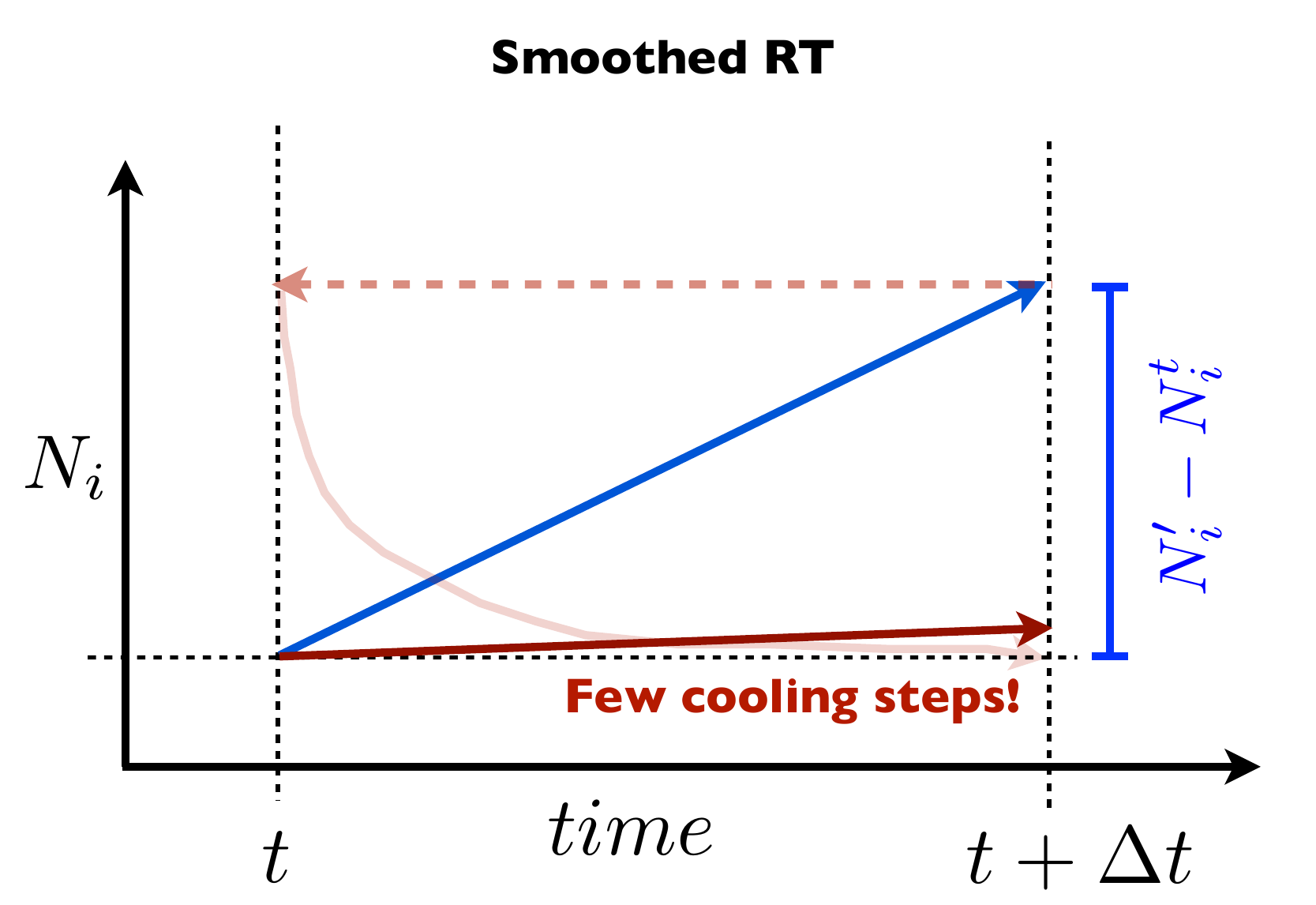}}
  \caption[A sketch explaining smoothed RT]
  {\label{smooth_a.fig}Sketch plots showing a photon density evolution
    over a global RT time-step with normal RT (\textbf{top}) and
    smoothed RT (\textbf{bottom}). In normal RT the photon density is
    updated to $\Nphot'$ during photon transport (a) and injection
    (b). This is then used as an initial state for thermochemistry
    (c). It is often the case that the photons are depleted over the
    global time-step $\dt$, in a process which takes many
    thermochemistry subcycles. In smoothed RT, the photon density
    state is not updated by the transport and injection steps, but
    rather the \textit{difference} is used to infer a photon injection
    rate for the cell, which is gradually added during each
    thermochemistry substep. This can dramatically reduce the needed
    number of chemistry substeps.}
\end{figure}

A problem we had to face, while performing \ramsesrt{} galaxy
formation runs, as well as the various test cases presented here, is
that there is often a small number of cells, usually along I-fronts,
or close to strong radiation sources, that execute a huge number of
thermochemistry subcycles in a single RT time-step. This is in part
fault of the operator-splitting approach used, where the RT equations
have been partly decoupled. Specifically, the photon density updates
happen in three steps in this approach (see \Fig{smooth_a.fig},
top). The photon injection step always increases the number of
photons, usually by a relatively large amount, and the transport step
does the same when it feeds photons into cells along these
I-fronts. The thermochemistry step in the I-front cells has the exact
opposite effect: the photon density decreases again via
absorptions. If the photon-depletion time is shorter than the Courant
time, we have a curious situation where the cell goes through an
inefficient cycle during the thermochemistry subcycles: it starts
neutral with a large abundance of photons (that have come in via the
transport and/or photon injection steps). It first requires a number
of subcycles to evolve to a (partly) ionized state, during which the
photon density is gradually decreased. It can then reach a turnaround
when the photons are depleted. If the RT time-step is not yet
finished, the cell then goes into a reverse process, where it becomes
neutral again. This whole cycle may take a large number of
thermochemical steps, yet the cell gas ends up being in much the same
state as it started.

In reality, the ionization state and photon density would not cycle
like this but would rather settle into a semi-equilibrium where the
rate of ionizations equals that of recombinations.

For the purpose of saving up on computing time and reducing the number
of thermochemistry subcycles, we have implemented an optional strategy
we call \textit{smoothed RT} that roughly corrects this
non-equilibrium effect of operator splitting (see \Fig{smooth_a.fig},
bottom). In it, the result of ($N'_i, \bF'_i$) from the transport and
injection steps in each cell is used to infer a \textit{rate} for the
thermochemistry step, rather than being set as an initial condition.
We use the pre-transport, pre-injection values of $\Nphot^t_i$ and
$\bF^t_i$ as initial conditions for the thermochemistry, but instead
update the thermochemistry equations (\ref{thchem_N.eq}) and
(\ref{thchem_F.eq}) to
\begin{align}
  \frac{\partial \Nphot_{i}}{\partial t} &= 
  -\sum_j^{\hi,\hei,\heii}{n_jc\csn_{ij}N_i}
  \ \ + \ \Npr^{rec}_{i} \ + \ \dot{\Nphot}_i, 
  \label{thchem_N_smooth.eq}
  \\
  \frac{\partial \bF_{i}}{\partial t} &= 
  -\sum_j^{\hi,\hei,\heii}{n_jc\csn_{ij}\bF_i} \ 
  + \ \dot{\bF}_i, 
  \label{thchem_F_smooth.eq}
\end{align}
where the new terms at the far right represent the \textit{rates} at
which the photon densities and fluxes changed in the transport and
injection steps, i.e.
\begin{align}
  \dot{\Nphot}_i = \frac{\Nphot_i'-\Nphot^t_i}{\dt}, \label{smoothrate1.eq} \\
  \dot{\bF}_i = \frac{\bF_i'-\bF^t_i}{\dt},          \label{smoothrate2.eq}
\end{align}
\joki{where $\Nphot^t_i$ and $\bF^t_i$ ($\Nphot_i'$ and $\bF_i'$)
  denote a cell state before (after) cell injection (Eq. \ref{inj.eq})
  and transport (Eqs. \ref{transport.eq} and \ref{transport2.eq}) have
  been solved over $\Delta t$. The injection and transport steps are
  unchanged from the normal operator splitting method, except for the
  fact that the cell states are not immediately updated to reflect the
  end results of those steps. The results of the injection step go
  only as initial conditions into the transport step, and the end
  results of the transport step are only used to calculate the photon
  density and flux rates of change via Eqs. (\ref{smoothrate1.eq}) and
  (\ref{smoothrate2.eq}).  Only after the thermochemistry step does a
  cell get a valid state that is the result of all three steps.}

The idea is that when the photons are introduced like this into the
thermochemistry step, they will be introduced \textit{gradually} in
line with the subcycling, and the photon density vs. ionization
fraction cycle will disappear as a result and be replaced with a
semi-equilibrium, which should reduce the number of subcycles and the
computational load. The total photon injection (or depletion) will
still equal $\Nphot_i'-\Nphot^t_i$, so in the limit that there are no
photoionizations or photon-emitting recombinations, the end result is
exactly the same photon density (and flux) as would be left at the end
of the transport and injection steps without smoothed RT. 

\joki{The advantage of the smoothing approach is perhaps best
  explained with an example: consider a cell with a strong source of
  radiation and gas dense and neutral enough that the timescales of
  cooling, ionization and/or recombination are much shorter than the
  global timestep length, $\Delta t$. This could either be a source
  containing a stellar particle or a cell along an ionization
  front. Without smoothing, the photoionization rate in the cell can
  change dramatically as a result of photon injection/transport. The
  thermochemistry step thus starts with a high rate of photoionzations
  which gradually goes down in the thermochemistry sub-cycling as the
  gas becomes more ionized and the photons are absorbed. With
  smoothing, this dramatic change in the photoionization rate never
  happens, thus requiring fewer thermochemistry sub-cycles to react. A
  situation also exists where the smoothing approach \textit{slows
    down} the thermochemistry: if a cell contains a strong source of
  radiation, but diffuse gas (i.e. \textit{long} timescales compared
  to $\Delta t$ for cooling, ionization and/or recombination), the
  non-smoothed approach would result in little or no thermochemistry
  sub-cycling, whereas the smoothed approach would take many
  sub-cycles just to update the radiation field and effectively reach
  the final result of the injection and transport steps.}

The gain in computational speed is thus quite dependent on the problem
at hand, and also on the reduced light speed, which determines the
size of the RT time-step, $\Delta t$. We've made a comparison on the
computational speed between using the smoothed and non-smoothed RT in
a cosmological zoom simulation from the NUT simulations suite
\citep[e.g.][]{{Powell:2011ex}} that includes the transfer of UV
photons from stellar sources. Here, smoothed RT reduces the average
number of thermochemistry subcycles by a factor of $6$ and the
computing time by a factor $3.5$. So a lot may indeed be gained by
using smoothed RT.

One could argue that the ionization states in I-fronts are better
modelled with smoothed RT, since the cycle of photon density and
ionization fraction is a purely numerical effect of operator
splitting. We have intentionally drawn a slightly higher end value of
$\Nphot_i$ in the smoothed RT than non-smoothed in \Fig{smooth_a.fig}:
whereas non-smoothed RT can completely deplete the photons in a cell,
smoothed RT usually leaves a small reservoir after the
thermochemistry, that more accurately represents the
``semi-equilibrium value".

Of course an alternative to smoothed RT, and a more correct solution,
is to attack the root of the problem and reduce the global time-step
length, i.e. also limit the transport and injection steps to the
$10\%$ rule. Reducing the global time-step length is highly
impractical though; the main reason for using operator splitting in
the first place is that it enables us to separate the timescales for
the different steps.

The same method of smoothing out discreteness that comes with operator
splitting (in the case of pure hydrodynamics) has previously been
described by \cite{{Sun:1996hq}}, where it is referred to as
``pseudo-non-time-splitting".

\section{Radiation hydrodynamics in Ramses}
\label{RHD.sec}
\ramses{} \citep{{Teyssier:2002fj}} is a cosmological adaptive mesh
refinement (AMR) code that can simulate the evolution and interaction
of dark matter, stellar populations and baryonic gas via gravity,
hydrodynamics and radiative cooling. It can run on parallel computers
using the message passing interface (MPI) standard, and is optimized
to run very large numerical experiments. It is used for cosmological
simulations in the framework of the expanding Universe, and also
smaller scale simulations of more isolated phenomena, such as the
formation and evolution of galaxies, clusters, and stars. Dark matter
and stars are modelled as collisionless particles that move around the
simulation box and interact via gravity. We will focus here on the
hydrodynamics of \ramses{} though, which is where the RT couples to
everything else.

\ramses{} employs a second-order Godunov solver on the Euler equations
of gravito hydrodynamics in their conservative form,
\begin{align}
  \frac{\partial \rho}{\partial t}+
     \nabla \cdot \left( \rho \vel \right) &= 0 \label{euler1.eq} \\
  \frac{\partial}{\partial t} \left( \rho \vel \right) 
     + \nabla \cdot \left( \rho \vel \otimes \vel \right ) 
     + \nabla P &= - \rho\nabla \phi  \label{euler2.eq} \\
  \frac{\partial \edens}{\partial t}  
     + \nabla \cdot \left( \left( \edens+P \right ) \vel \right) 
     &= -\rho \vel \cdot \nabla \phi + \Lambda(\rho,\etherm), 
     \label{consE.eq}
\end{align}
where $t$ is time, $\rho$ the gas density, $\vel$ the bulk velocity,
$\phi$ the gravitational potential, $\edens$ the gas total energy
density, $P$ the pressure, and $\Lambda$ represents radiative cooling
and heating via thermochemistry terms (resp. negative and positive),
which are functions of the gas density, temperature and ionization
state. In \ramses{}, collisional ionization equilibrium (CIE) is
traditionally assumed, which allows the ionization states to be
calculated as surjective functions of the temperature and density and
thus they don't need to be explicitly tracked in the code. $\edens$ is
divided into kinetic and thermal energy density ($\etherm$)
components:
\begin{align}
  \edens=\frac{1}{2}\rho u^2+\etherm.
\end{align}
The system of Euler equations is closed with an equation of state
which relates the pressure and thermal energy,
\begin{align}\label{hydro_closure.eq}
  P=(\gamma-1)\etherm,
\end{align}
where $\gamma$ is the ratio of specific heats. The Euler equations are
adapted to super comoving coordinates, to account for cosmological
expansion, by a simple transformation of variables (see
\Sec{cosmo.sec}).

The Euler equations are solved across an AMR grid structure. Operator
splitting is employed for the thermochemistry source terms,
i.e. $\Lambda$ is separated from the rest of the Euler equations in
the numerical implementation -- which makes it trivial to modify the
thermochemistry solver, i.e. change it from equilibrium to
non-equilibrium.

\begin{figure}
  \centering
  \includegraphics[width=0.1\textwidth]
  {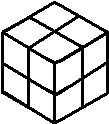}
\caption[An oct]{\label{oct.fig}An oct -- the basic grid
  element in \ramses{}.}
\end{figure}

The basic grid element in \ramses{} is an oct (\Fig{oct.fig}), which
is a grid composed of eight cubical cells. A {\it conservative} state
vector $\state=(\rho,\rho \vel,\edens,\rho Z)$ is associated with each
cell storing its hydrodynamical properties of gas density $\rho$,
momentum density $\rho \vel$, total energy density $\edens$ and metal
mass density $\rho Z$.  (One can also use the {\it primitive} state
vector, defined as ${\cal W} = (\rho,\vel,P, Z)$.) Each cell in the
oct can be recursively refined to contain sub-octs, up to a maximum
level $\ell$ of refinement. The whole \ramses{} simulation box is one
oct at $\ell=1$, which is homogeneously and recursively refined to a
minimum refinement level $\lmin$, such that the coarse (minimum) box
resolution is $2^{\lmin}$ cells on each side. Octs at or above level
$\lmin$ are then adaptively refined during the simulation run, to
follow the formation and evolution of structures, up to a maximum
refinement level $\lmax$, giving the box a maximum \textit{effective}
resolution of $2^{\lmax}$ cell widths per box width.  The cell
refinement is \emph{gradual}: the resolution must never change by more
than one level across cell boundaries.

\subsection{\ramses{} multi-stepping approach}
\begin{figure}
  \centering
  \includegraphics[width=0.47\textwidth]
    {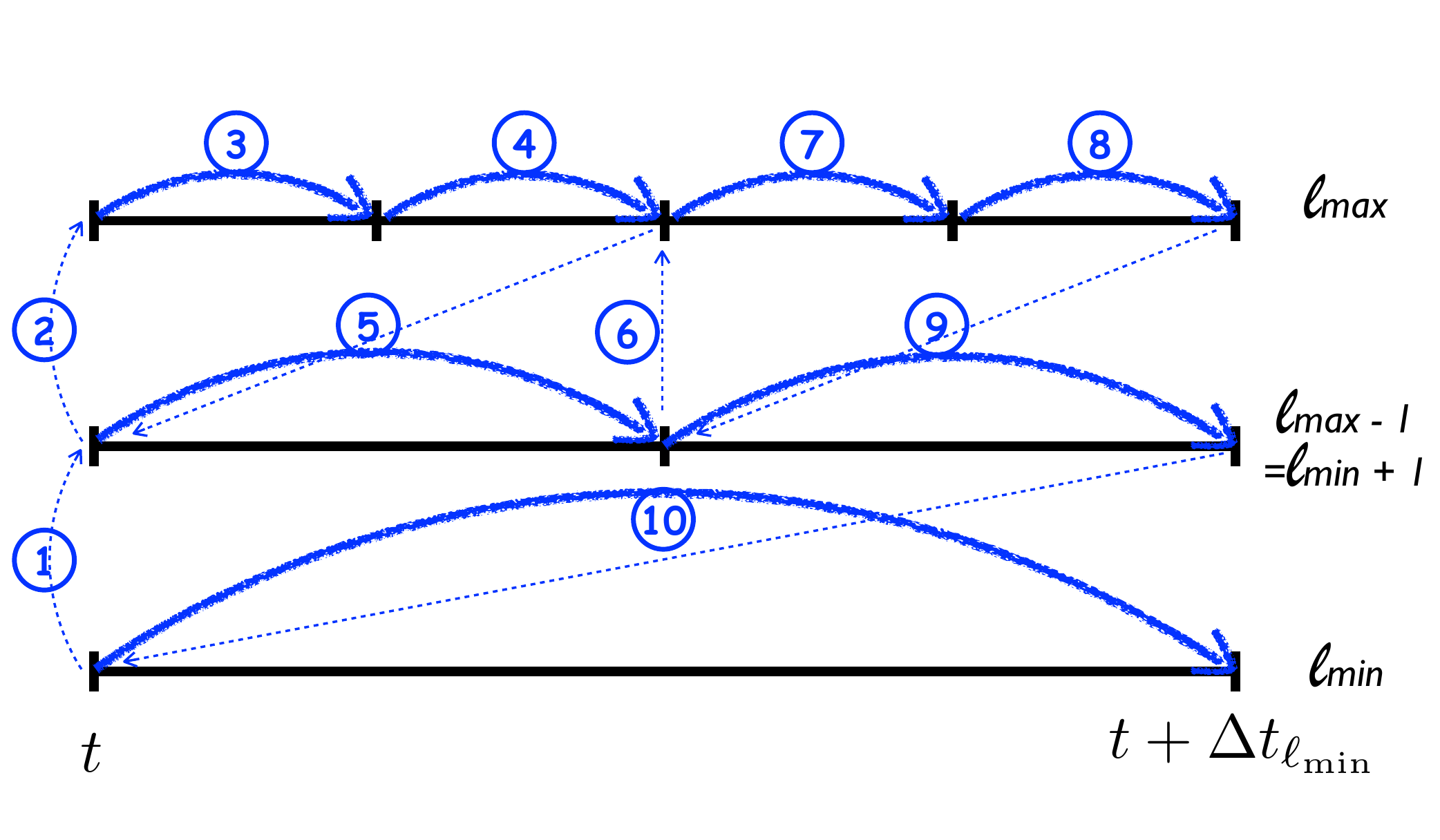}
  \caption[Recursive time-stepping in \ramses{}]
  {\label{amr_step.fig}Recursive hydro time-stepping over one coarse
    time-step in the AMR levels of \ramses{}, here shown for a
    three-level AMR structure. Each solid arrowed line represents a
    time-step which is executed for all cells belonging to the
    corresponding AMR level. The numbers indicate the order of the
    time-stepping, including the calls to finer levels (1, 2, and 6).}
\end{figure}

With \emph{AMR multi-stepping}, the resolution is not only adaptive in
terms of volume, but also in \emph{time}, with different timestep
sizes on different refinement levels. A coarse time-step, over the
whole AMR grid, is initiated at the coarse level, $\lmin$, as we show
schematically in \Fig{amr_step.fig}. First, the coarse time-step
length $\dt_{\lmin}$ is estimated via (the minimum of) Courant
conditions in all $\lmin$ cells. Before the coarse step is executed,
the next finer level, $\lmin+1$, is made to execute the same
time-step, in two substeps since the finer level Courant condition
should approximately halve the time-step length. This process is
recursive: the next finer level makes its own time-step estimate
(Courant condition, but also $\dt_{\ell}\le \dt_{\ell-1}$) and has its
next finer level to execute two substeps. This recursive call up the
level hierarchy continues to the highest available level $\lmax$,
which contains only \textit{leaf} cells and no sub-octs. Here the
first two substeps are finally executed, with step lengths
$\dt_{\lmax}\le\dt_{\lmin}/2^{\lmax-\lmin}$. When the two $\lmax$
substeps are done, the $\lmax-1$ time-step is re-evaluated to be no
longer than the sum of the two substeps just executed at $\lmax$, and
then one $\lmax-1$ step is executed. Then back to level $\lmax$ to
execute two steps, and so on. The substepping continues in this
fashion across the level hierarchy, ending with one time-step for the
coarsest level cells (with a modified time-step length $\dt_{\lmin}$).

At the heart of \ramses{} lies a recursive routine called
$\tt{amr\_step(\ell)}$ which describes a single time-step at level
$\ell$, and is initially called from the coarsest level ($\lmin$).  To
facilitate our descriptions on how the RT implementation is placed
into \ramses{}, we illustrate the routine in pseudocode format in
\Pse{ramses.ps}, where we have excluded details and bits not directly
relevant to RHD (e.g. MPI syncing and load-balancing, adaptive
refinement and de-refinement, particle propagation, gravity solver,
star formation, and stellar feedback).

First, the recursion is made twice, solving the hydrodynamics over two
sub-steps at all finer levels. Then the Euler equations are solved
over the current coarse time-step, for all cells belonging to the
current level. It is important to note here that the hydrodynamical
quantities are fully updated at the current level in the
$\tt{hydro\_solver}$, but there are also intermediate hydro updates in
all neighboring cells at the next coarser level. The coarser level
update is only \textit{partial} though, because it only reflects the
intercell fluxes across inter-level boundaries, and fluxes across
other boundaries (same level or next coarser level) will only be
accounted for when the coarser level time-step is fully
advanced. Until then, these coarser level neighbor cells have gas
states that are not well defined, since they only reflect some of
their intercell fluxes. It effectively means that at any point between
the start and finish of the primary (coarse) call to $\tt{amr\_step}$,
there are some cells in the simulation box (lying next to finer level
cells) that have ill-defined intermediate hydrodynamical states. This
point is further illustrated in \App{App_interlevel}. It is important
to keep in mind when considering the coupling of RT with the
hydrodynamics of \ramses{}.

Having put down the basics of AMR hydrodynamics, we are now in a
position to add radiative transfer.

\begin{lstlisting}[float,caption={The AMR step in
    \ramses{}.},frame=trBL, label={ramses.ps}]
recursive subroutine amr_step($\ell$):

if $\ell$ < $\lmax$ and any cells exist in $\ell+1$
  call amr_step($\ell+1$)
  call amr_step($\ell+1$)
call hydro_solver($\ell$): all $\ell$ cells and some $\ell-1$
call eq_thermochemistry($\ell$): all $\ell$ leaf cells

end
\end{lstlisting}

\subsection{\ramsesrt{}}\label{ramsesrt.sec}
In \ramsesrt{}, each cell stores some additional state variables. Here
${\state}=(\rho,\rho \vel,E,\rho Z,\rho \xhii,\rho \xheii,\rho
\xheiii,\Nphot_i,\bF_i)$, where $\xhii{}$, $\xheii{}$ and $\xheiii{}$
are the hydrogen and helium ionization fractions, which are advected
with the gas as passive scalars (in the hydro solver), and $\Nphot_i$,
$\bF_i$ represent the $4M$ variables of photon density and flux for
each of the $M$ photon groups. Note that this represents a hefty
increase in the memory requirement compared to the hydrodynamics only
of \ramses{}: the memory requirement for storing $\state$ (which is
the bulk of the total memory in most simulations) is increased by a
factor of $1.5(1+4/9M)$, where the $1.5$ represents the ionization
fractions and the parenthesis term represents the photon fluxes and
densities.  Thus, with three photon groups, the memory requirement is
increased by roughly a factor $3.5$ compared to a traditional
\ramses{} simulation.

\vsk

Given the time-scale difference between hydrodynamics and radiative
transfer, the obvious approach to performing RHD is to sub-cycle the
three radiative transfer steps (injection, advection, thermochemistry)
within the hydrodynamical step. There is, however, a major drawback to
this approach, which is that it is incompatible with AMR
multi-stepping: the RT sub-cycling must be done before/after each
hydrodynamical AMR step \emph{at the finest refinement level only},
and since light can in principle cross the whole box within the fine
level hydrodynamical timestep, the RT sub-cycling must be done over
the whole grid, over all levels. However, the partial hydrodynamical
flux between cells at level boundaries always leaves some cells
between the fine level steps with an intermediate (i.e. partially
updated) gas state. This makes the thermochemistry ill-defined in
those cells, since it needs to update the gas temperature in every
cell, and for this to work the temperature must have a well defined
and unique value everywhere. There are three ways around this:

First is to perform the RT subcycling only after a \emph{coarse}
hydrodynamical step, but here potentially thousands of fine-scale
hydro steps would be executed without taking into account the
thermochemistry. 

\begin{figure}
  \centering
  \includegraphics[width=0.47\textwidth]
  {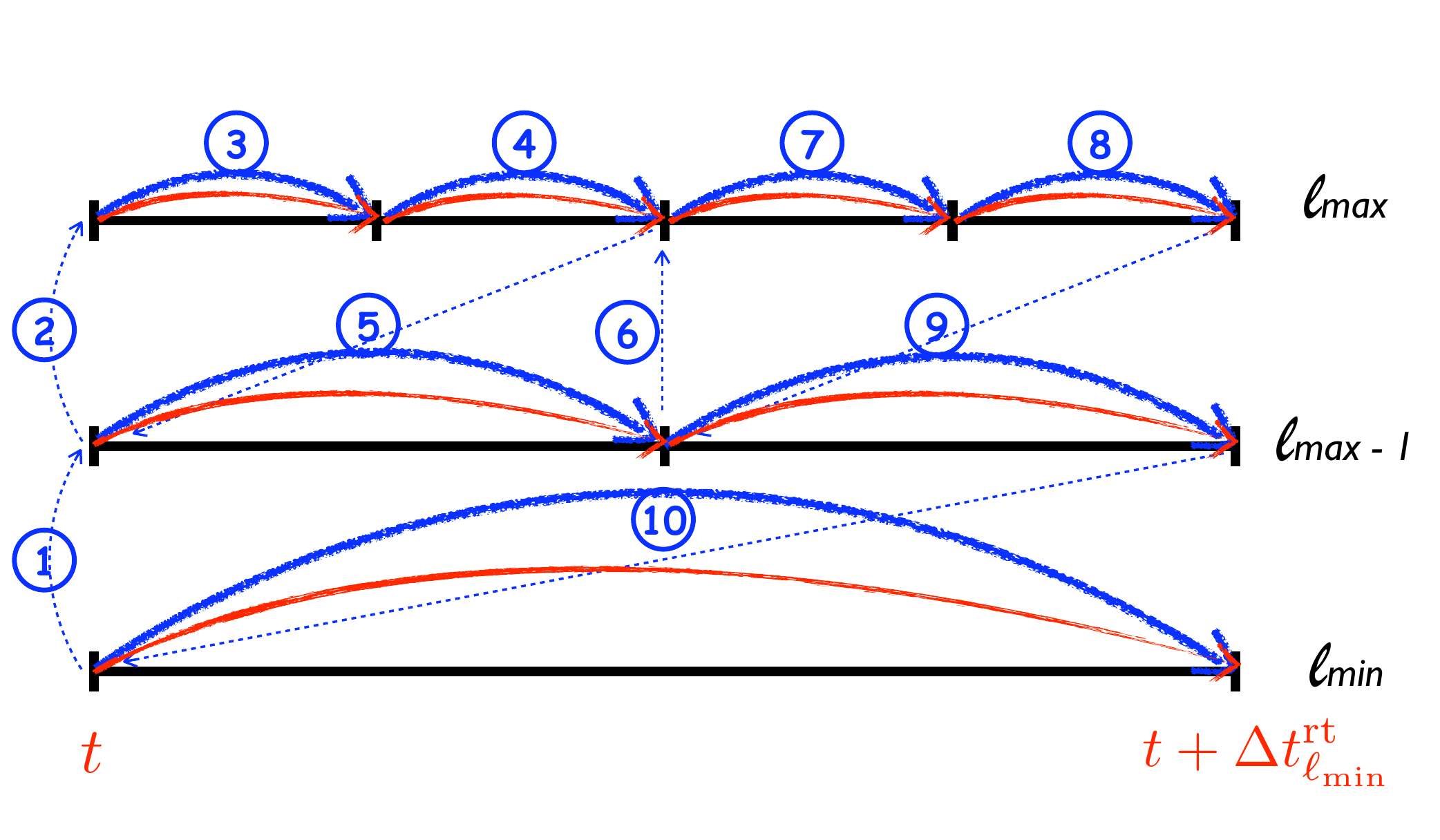} \caption{\label{rt_timestep.fig}Diagram of the
  $\tt{amr\_step}$ in \ramsesrt{}. This is much like the normal
  $\tt{amr\_step}$ in \ramses{}, except that the time-step length has
  the extra constraint of the light speed Courant condition, and each
  level $\ell$ step also performs photon injection, RT transport and
  thermochemistry over the same time-step and level.}
\end{figure}

\begin{lstlisting}[,float
  ,caption={The AMR step in \ramsesrt{}.}
  ,frame=trBL, label={RT_timestepping.ps}]
recursive subroutine amr_step($\ell$)

if $\ell$ < $\lmax$ and any cells in $\ell+1$
  call amr_step($\ell+1$)
  call amr_step($\ell+1$)
call photon_injection_step($\ell$)
call hydro_solver($\ell$): all $\ell$ cells and some $\ell-1$
call rt_transport($\ell$): all $\ell$ cells and some $\ell-1$
call neq_thermochemistry($\ell$): all $\ell$ leaf cells

end
\end{lstlisting}

Second, is to prohibit AMR multi-stepping, which makes the whole grid
well defined after each step and thus allows for RT sub-cycling over
the whole box. Multi-stepping is however one of the main advantages of
AMR, and essentially allows us to refine in time as well as space, so
this isn't really an option. 

We thus default to the third strategy, which we use in
\ramsesrt{}. Here we drop the subcycling of RT within the hydro step
and perform the two \emph{on the same timestep length}, which is the
minimum of the RT and hydro timestep. Thus, with each hydro step, at
any level, the RT steps are performed \emph{over the same level
  only}. The basic scheme is illustrated in \Fig{rt_timestep.fig}, and
the pseudocode for the updated $\tt{amr\_step}$ is shown in
\Pse{RT_timestepping.ps}. Obviously, the main drawback here is the
timescale difference, which can be something like a factor of
$100-1000$, meaning the number of hydrodynamical steps is increased
the same factor and the run-time accordingly (plus numerical diffusion
likely becomes a problem with such small hydrodynamical
time-steps). However, if we also apply a reduced speed of light, we
can shrink this factor arbitrarily, down to the limit where the
hydro-timestep is the limiting factor and the only increase in
computational load is the added advection of photons (which is
considerably cheaper for one photon group than the hydrodynamical
solver) and the non-equilibrium thermochemistry (which typically has a
computational cost comparable to the equilibrium solver of \ramses{},
provided we use RT smoothing). The question, which we have tried to
answer in \Sec{reduced_c.sec}, is then how far we are allowed to go in
reducing the light speed.

Parallelization is naturally acquired in \ramsesrt{} by simply taking
advantage of the MPI strategies already in place in \ramses{}.

\subsection{Radiation transport on an AMR grid}\label{amr_transport.sec}
In \ramsesrt{}, the radiation variables are fully incorporated into the
AMR structure of \ramses{}. The ionization fractions and photon
densities and fluxes are refined and de-refined along with the usual
hydro quantities, with a choice of interpolation schemes for newly
refined cells (straight injection or linear interpolation). The
radiative transfer, i.e. injection, transport and thermochemistry, is
multi-stepped across the level hierarchy, thus giving AMR refinement
both in space and time. Inter-level radiation transport is tackled in
the same way as the hydrodynamical advection, i.e. transport on level
$\ell$ includes partial updates of neighbouring cells on level
$\ell-1$. Update of the finer level cell RT variables over level
boundaries involves the RT variables in a coarser cell, which are
evaluated, again with the same choice of interpolation
schemes. \ramsesrt{} includes optional refinement criteria on photon
densities, ion abundances and gradients in those, in addition to the
usual refinement criteria that can be used in \ramses{} (on mass and
gradients in the hydrodynamical quantities).

\begin{figure*}
  \centering
  \subfloat{\includegraphics[width=0.3\textwidth]
    {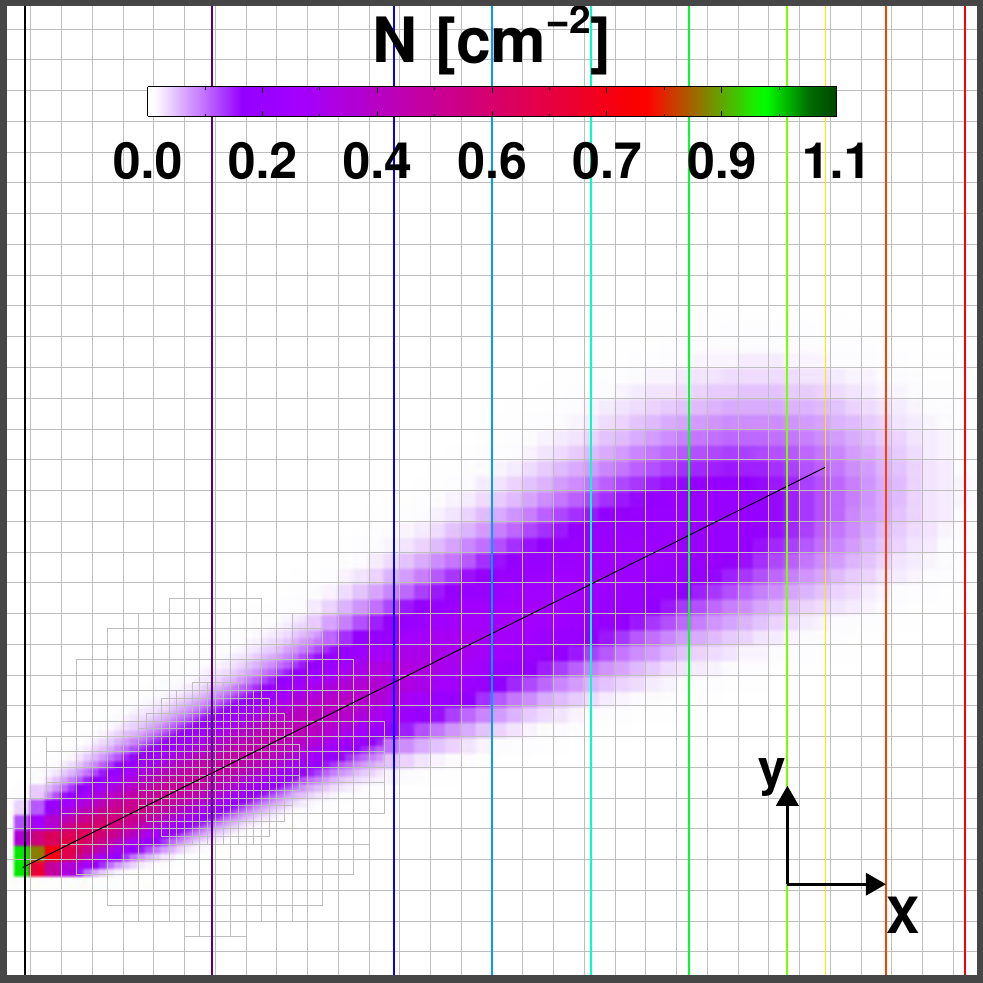}}
  \subfloat{\includegraphics[width=0.65\textwidth]
    {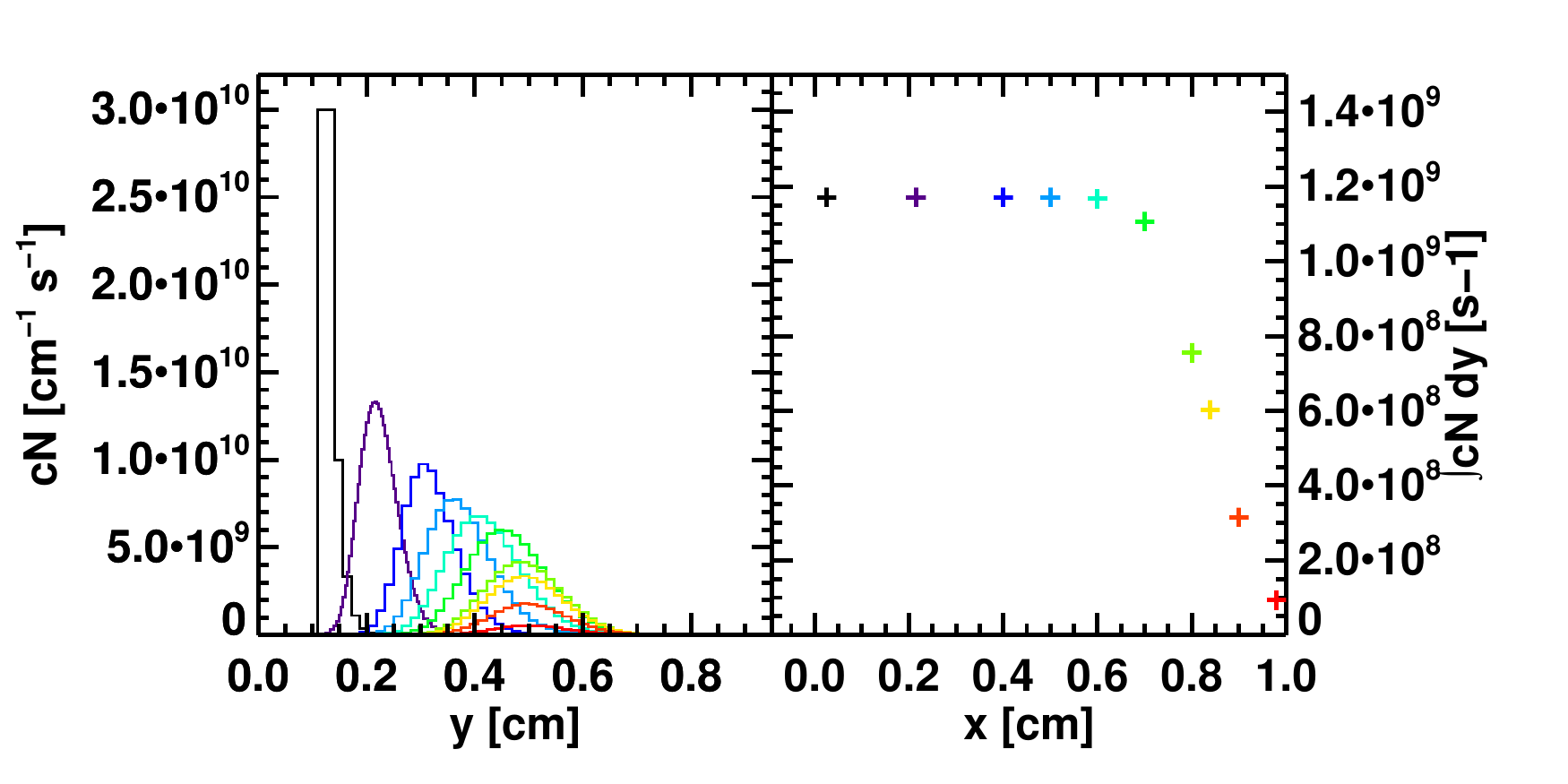}}
  \vspace{-6.mm}
  \subfloat{\includegraphics[width=0.3\textwidth]
    {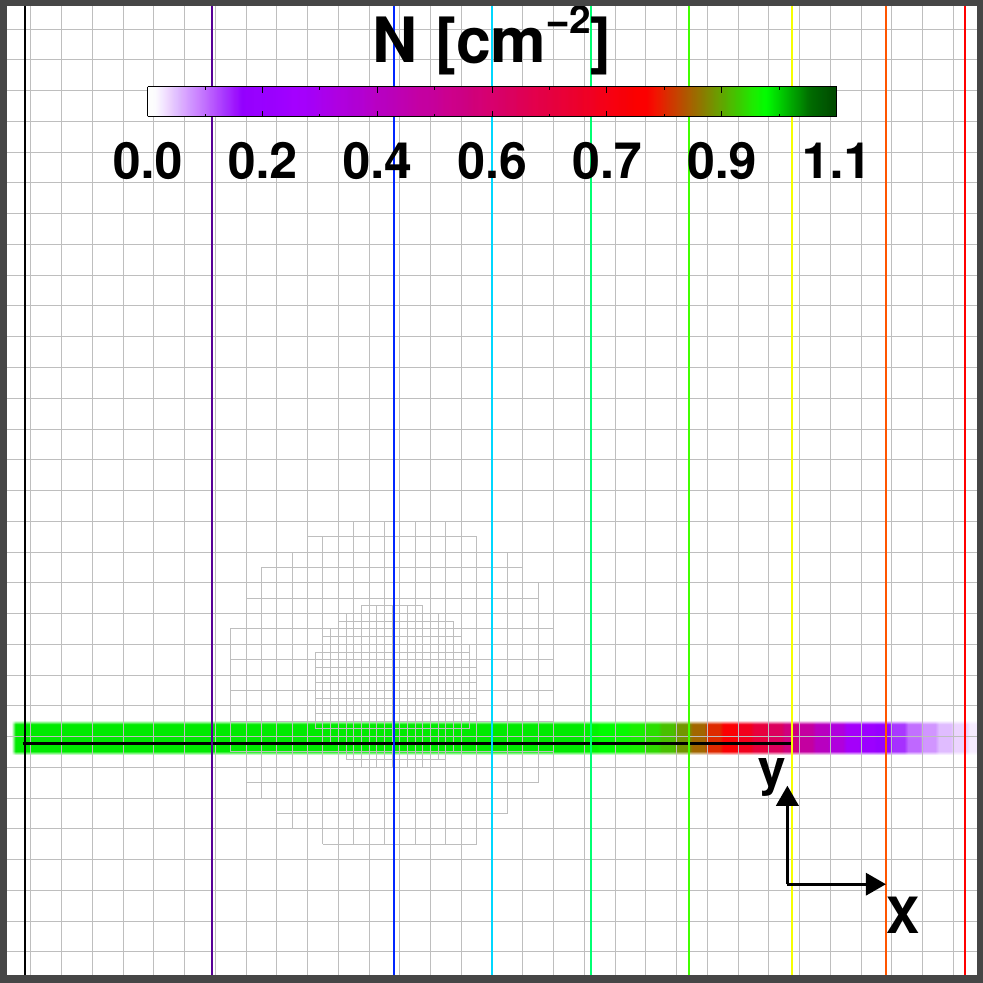}}
  \subfloat{\includegraphics[width=0.65\textwidth]
    {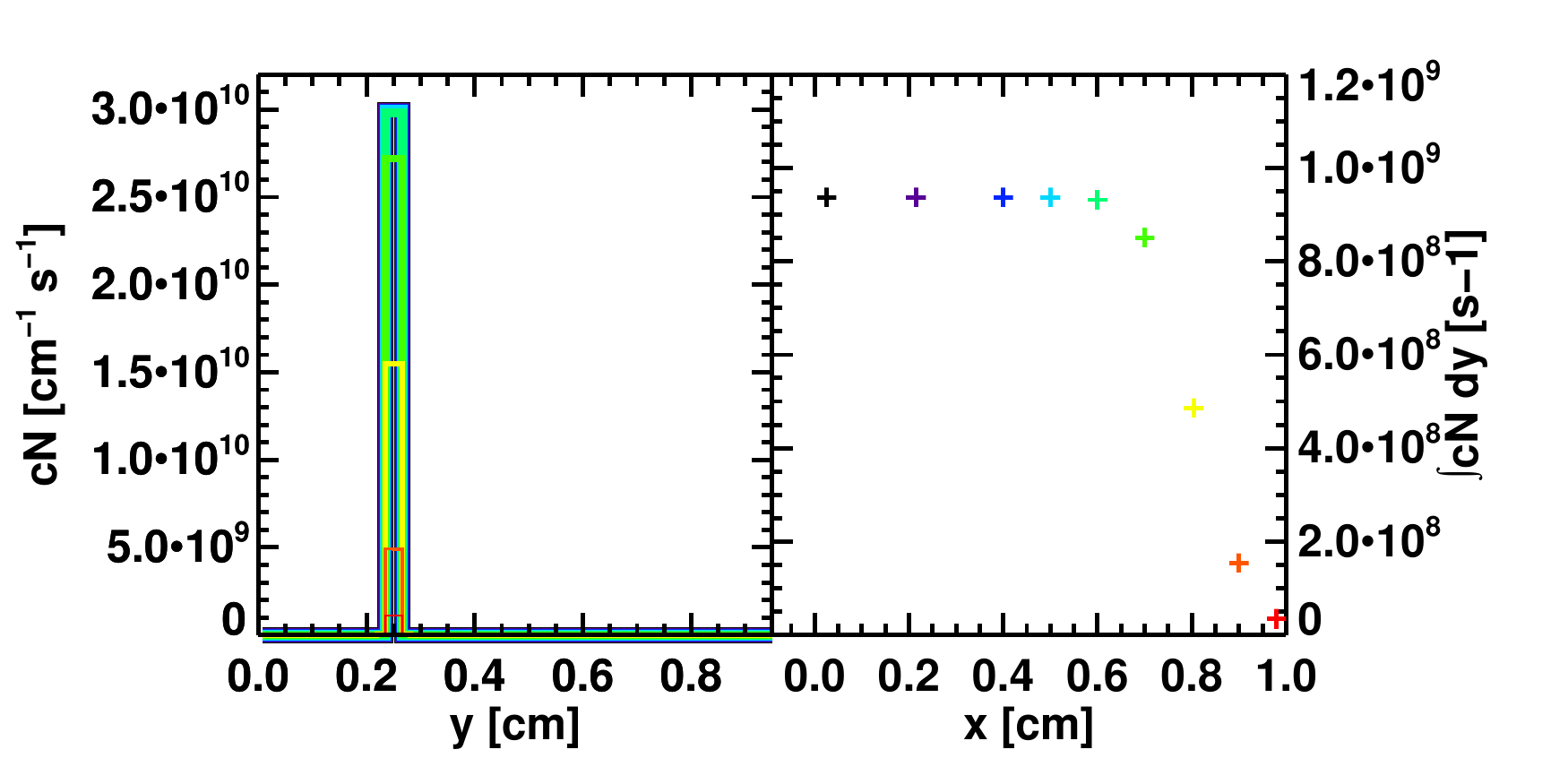}}
  \caption
  {\label{beam_amr.fig}2D beam experiments, demonstrating photon flux
    conservation across changing refinement levels. The upper panel
    shows an experiment with an off-axis beam, $26.5^{\circ}$ from the
    horizontal, and the lower panel shows an identical experiment,
    except the beam is horizontal. The maps on the left show photon
    number density, with the grid structure overplotted in grey (which
    is kept constant throughout the experiments). \joki{Black lines
      plotted over the beams mark the light-crossing distance at the
      time the snapshots are taken.} Coloured vertical lines mark
    x-positions at which photon flux profiles are plotted in the left
    plots. The right plots show integrals of each profile, i.e. the
    total photon flux across each x-coordinate.}
\end{figure*}
\begin{figure*}
  \centering
  \subfloat{\includegraphics[width=0.3\textwidth]
    {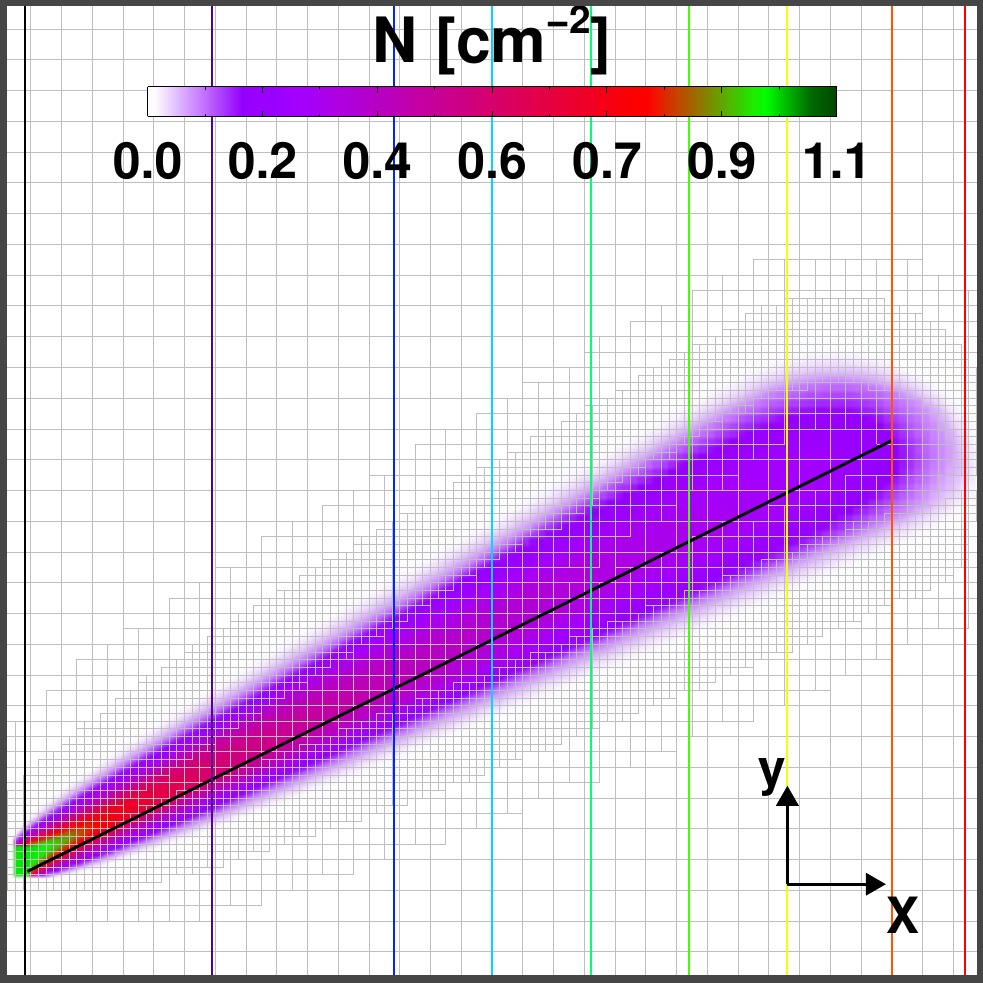}}
  \subfloat{\includegraphics[width=0.65\textwidth]
    {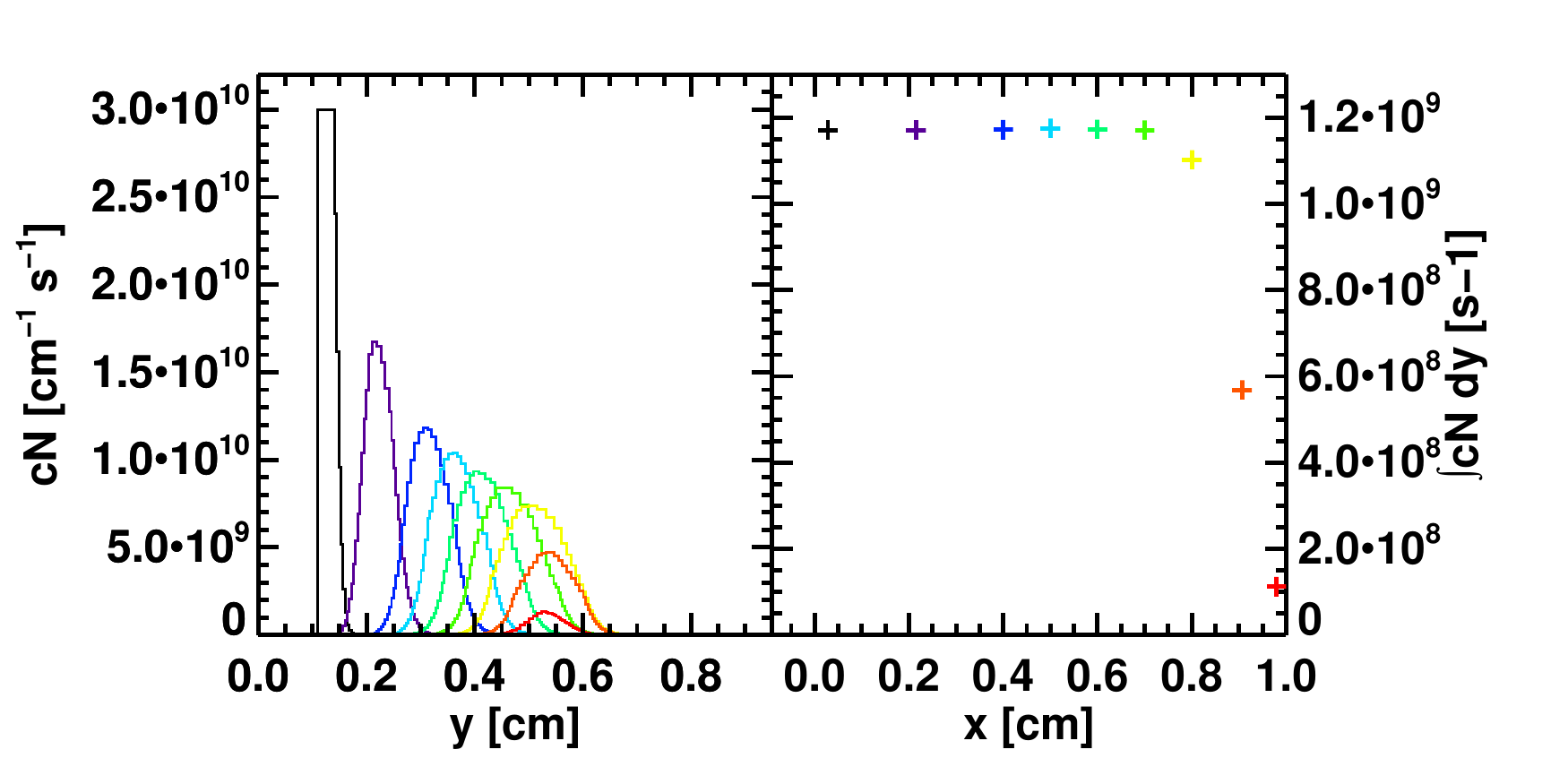}}
  \caption
  {\label{beam_amr_live.fig}2D beam experiment, same as
    \Fig{beam_amr.fig}, but with on-the-fly AMR refinement.}
\end{figure*}

Of the seven standard RT and RHD tests described in
Section~\ref{tests.sec}, five include active or inactive grid
refinement, demonstrating that the radiation hydrodynamics perform
robustly in conjunction with (on-the-fly) cell
refinements/de-refinements. In addition, we demonstrate in
\Fig{beam_amr.fig} how radiation flux is well retained across changes
in grid refinement. The upper left map of the figure shows a beam of
radiation in a 2D \ramsesrt{} experiment, where we use the HLL flux
function and deactivate radiation-gas interactions \joki{(with zero
  photoionization cross sections)}. The beam is injected into two
cells in the bottom left corner by imposing a unity reduced photon
flux of $3 \times 10^{10}$ photons $\sm \ \cmmone$, corresponding to a
photon density of $1 \ \ccitwod$, at an angle of $26.5^{\circ}$ from
the horizontal. The beam traverses a circular region of 2 successive
levels of increasing refinement, going from refinement level 6 to 8,
i.e. effective resolutions of $64^2$ to $256^2$ cells. We use here
straight injection (i.e. no interpolation) for inter-level cell
fluxes, but linear interpolation gives identical results. \joki{The
  snapshot is taken at $t=3.04 \times 10^{-11}$ s, just before the
  beam has had time to cross to the right edge of the $1$ cm wide
  box.}  To the right of the map we plot photon flux profiles, $cN$,
across the coloured vertical lines in the map. The beam experiences
diffusion, as can be seen by the widening of the flux profiles, but
this is exclusively due to the intercell flux function and independent
of the refinement changes. The far left plot shows the integrals
across each flux profile, i.e. the total photon flux across each
line. \joki{The values are consistent until around $x=0.6$, and then
  reduce to zero towards the edge of the beam. We've verified that if
  the test is let to run for double the time, i.e. about $6 \times
  10^{-11}$ s, the total flux is consistent throughout the whole box
  width to about $1$ in $10^4$, so photons are very well conserved
  across the changes in refinement.}

To further demonstrate flux conservation, the lower panel in the same
figure shows an identical experiment except that the beam is
horizontal, such that it can be perfectly maintained with the HLL flux
function. \joki{To stay just under a light crossing time, we consider
  a shapshot at $2.6 \times 10^{-11}$ s. Here again, the flux is well
  preserved towards the edge of the beam, and we have verified that in
  two crossing times, the total flux is retained perfectly to the
  number precision, which here is 7 decimals.}

We also consider another beam with the same setup, shown in
\Fig{beam_amr_live.fig}, where instead of a static refinement region,
the grid is actively refined on inter-cell gradients in photon density
$N$. According to the criterion, two adjacent cells at positions $i$
and $i+1$ are refined if
\begin{equation}\label{beam_refine.eq}
  2 \left|
    \frac{N^i-N^{i+1}}{N^i+N^{i+1}+10^{-3} \ {\rm{cm}}^{-2}}
  \right| > 0.4.
\end{equation}
Straight injection (no interpolation) is used here for inter-level
fluxes and cell refinements, but the results are identical when linear
interpolation is used for inter-level fluxes and cell refinements.
\joki{The snapshot here is taken at $3.3 \times 10^{-11}$ s ($\sim$ a
  crossing time). The plot on the far right shows the flux
  conservation across different x-coordinates. (Note the total flux is
  slightly different from that in \Fig{beam_amr.fig} because of the
  different geometry of the beam injection.) The total flux is again
  well maintained towards the beam edge. We verified that in two light
  crossing times, the discrepancy of the beam flux at different
  $x-$coordinates levels out to within $0.03\%$.}

These simple beam experiments demonstrate that the code accurately
transports radiation across (even dynamically) changing refinement
levels. The main errors are the artificial diffusion of radiation on
the grid, which is not caused by refinement, but rather by the
inter-cell flux function, and the dipole approximation inherent to the
M1 closure, which does not allow opposing streams of radiation to pass
through one another. Note though that while the diffusion is
artificial, the total flux is well maintained, i.e. energy is
conserved.

\joki{
\subsubsection{Speed of light}
The AMR transport tests also demonstrate that radiation in \ramsesrt{}
propagates at the correct speed, i.e. at the speed of light. In each
beam map (Figs. \ref{beam_amr.fig}-\ref{beam_amr_live.fig}), a black
line has been plotted over the beam, starting at the beam injection
and ending at the light-crossing distance, i.e. $t\times c$, where $t$
is the snapshot time. Qualitatively it can be seen that the beam ends
roughly at the same position as the black line, and in the flux plots
on the far right side of each beam map it can be seen that the beam
has roughly half the original flux at this end position. The far end
of the beam is smooth over a few cell widths rather than
discontinuous, because of numerical diffusion.}

\subsection{Cosmological settings}\label{cosmo.sec}
\ramses{} uses super-comoving variables to allow for the impact of the
cosmological expansion on the Poisson equation, the equations of
hydrodynamics (\ref{euler1.eq}-\ref{consE.eq}) and particle
propagation \citep{{Martel:1998gf}, {Teyssier:2002fj}}: a change is
made from the physical variables to super-comoving ones with
\begin{align*}
  d\tilde{t}   = \frac{H_0}{a^2} \ dt, \ \ \ \ \ \ \ \ \ \ 
  \tilde{x}    = \frac{1}{aL} \ x,   \ \ \ \ \ \ \ \ \ \ 
  \tilde{\rho} = \frac{a^3}{\Omega_m \rho_c H_0^2} \ \rho, \\
  \tilde{\vel} = \frac{a}{H_0 L} \ \vel,   \ \ \ \ \ \ \ 
  \tilde{\etherm} = \frac{a^5}{\Omega_m \rho_c H_0^4 L^2} \ \etherm, \ \ \
  \ \ \ \ \ \ \ \ \ 
\end{align*}
where $H_0$ is the Hubble constant, $\Omega_m$ the matter density
parameter, $L$ the comoving width of the simulation box (physical
width at $a=1$), and $\rho_c$ the critical density of the
Universe. When these variables are used instead of the physical ones,
the cosmological expansion is accounted for, while all relevant
equations remain unchanged, Euler equations included.

For consistency, and to partly account for the effect of cosmological
expansion on the radiative transfer, the additional change is made in
\ramsesrt{} to super-comoving RT variables for the photon transport:
\begin{align*}
  \tilde{N}   = a^3 \ N, \ \ \ \ \ \ \ \ 
  \tilde{\bF} = \frac{a^4}{H_0 L} \ \bF, \ \ \ \ \ \ \ \ 
  \tilde{c}   = \frac{a}{H_0 L} \ c.
\end{align*}
The dilution ($\propto a^{-3}$) of photon number density is thus
accounted for, while it can easily be verified that
Eqs. \eq{RTfin1.eq}-\eq{RTfin2.eq} remain unchanged with the new
variables -- including the reduced flux \eq{Udir.eq} used in the M1
tensor \eq{edd2.eq}.

Note that when reduced light speed is used, the photons will be
over-diluted in cosmological simulations, since the time taken for
them to get from source to destination will be overestimated. Note
also that wavelength stretching with redshift, which in reality adds a
fourth power of $a$ to the dilution of $N_{\gamma}$, is not accounted
for here. This is actually non-trivial to do: one could add one power
of $a$ to the definitions of $\tilde{N}$ and $\tilde{\bF}$, but it
would be a very crude approximation of the wavelength dilution, as the
wavelength shift that should feed photons from one group to the next
is neglected. In any case, this effect is likely to be important only
in the context of reionization, where the photons have a chance of
travelling cosmological distances before they are absorbed.
\joki{While cosmological diffusion and redshifting is difficult to
  account for in ray-tracing methods, where the radiation is typically
  traced as far as it can get in one moment in time, moment-based
  approach are more straightforwardly able to model these effects
  \citep[e.g.][]{Ricotti:2002dh,Petkova:2009fx, Finlator:2011ji}. }


\section{Radiative transfer tests}\label{tests.sec}
\input{RamsesRT_RTtests}

\section{Discussion}
\label{Discussion.sec}

In this paper, we have presented a new implementation of radiation
hydrodynamics in the \ramses{} code.  It is based on a moment
representation of the radiation field, where we have used the M1
closure relation to define a purely local variable Eddington
tensor. Because the resulting system is a set of hyperbolic
conservation laws, we have exploited the Godunov methodology to design
a time-explicit, strictly photon conserving radiation transport
scheme.  The resulting algorithm is first order accurate in space and
time, and uses various Riemann solvers (GLF and HLL) to compute
radiation fluxes. The main novelty compared to our previous
implementation (AT08) is the coupling between gas and radiation,
resulting in a fully consistent radiation hydrodynamics solver, and
the introduction of adaptive mesh techniques in the radiation
transport step, making use of both the AMR and parallel computing
capabilities of \ramses{}. Overall, the code was quite easy to
implement, owing to the explicit nature of the time integration
scheme. The price to pay is the need to resolve the propagation of
hyperbolic waves traveling at or close to the speed of light. Among
many different options available to overcome this constraint, we have
chosen to use the ``reduced speed of light" approximation. This
approximation is valid when the propagation speed of I-fronts is still
slower than the (reduced) light speed. We have developed a recipe to
assess the validity of this approximation, based on the light-crossing
time of Str{\"o}mgren spheres. We have verified that this framework
indeed allows us to estimate in advance the speed-of-light reduction
factor reliably. We have shown, for example, that in cosmological
problems, such as cosmic reionization, using the correct value for the
speed of light is crucial, and using either a reduced or an infinite
speed of light (like in some ray-tracing codes) might result in large
inaccuracies.

This new algorithm has already been used in galaxy formation studies,
exploiting the coupling between radiation and hydrodynamics offered by
\ramsesrt{}. In \cite{{Rosdahl:2012bt}}, we studied the impact of
ionizing radiation in determining the thermal state of cold filaments
streaming into high redshift galaxies, allowing us to make accurate
observational predictions and demonstrating a possible link between
cold streams and Lyman-alpha blobs. More recently, we have also
explored the role of ionizing radiation in the overall efficiency of
stellar feedback (Geen et al. 2013, Powell et al. 2013, both in
preparation). Beyond ionizing radiation, possible extensions of
\ramsesrt{} are the inclusion of photo-dissociating radiation and the
thermochemistry of molecules, as well as the effect of dust as an
additional source of opacity and thermal regulation inside star
forming galaxies. This would require introducing additional photon
groups (such as a far UV and IR photons) and the associated
microphysics, but the overall methodology would remain very similar.

In order to improve the current algorithm, we have many possibilities
ahead of us. One obvious development is to develop a second-order
sequel of our current first order Godunov solver. Second-order Godunov
schemes, both in time and space, are used routinely in hydrodynamics
codes (such as the MUSCL scheme in \ramses{}). This might reduce
significantly the rather large diffusivity of our current
implementation. However, since photo-ionization and photo-dissociation
problems are governed to a large extent by the thermo-chemistry, it is
not clear how much the accuracy of the results would depend on the
advection scheme. A second route we would like to explore in the
future is the optional introduction of radiation sub-cycles during
each adaptive hydro step. This is quite challenging since it would in
principle require decoupling in time of the various AMR levels,
resulting in the loss of strict photon conservation. In some cases,
however, it is advantageous to sacrifice the exact number conservation
of photons in favour of modelling the correct speed of light with many
radiation sub-cycles. In any case, this would offer us a new tool with
greater flexibility. Along the same lines, because of the
fundamentally different propagation properties of I-fronts in the IGM
on one hand and deep inside galaxies on the other, we could couple
\ramsesrt{} to \aton{}: use \aton{} to transport radiation on the
coarse grid with GPUs at the full speed of light, and use \ramsesrt{}
on the fine AMR levels at a reduced light speed. This would require us
to define two photon group populations that mirror each other: a large
scale, low density photon population that propagates at the correct
speed of light and makes use of GPU acceleration (if available), and a
small scale, high density photon population that makes use of the
``reduced speed of light" approximation. Coupling properly the two
photon group populations will of course be quite challenging and at
the heart of this new avenue of research. A last development we have
in mind is the introduction of radiation pressure as a new channel of
coupling radiation with hydrodynamics. This is highly relevant for
studies focusing on radiation pressure on dust, from both young star
clusters and supermassive black holes.

\section*{Acknowledgements}
We are grateful for the help and insight provided by Stephanie Courty,
Julien Devriendt, Yohan Dubois, and Leo Michel-Dansac. We thank the
anonymous referee whose effort and remarks clearly helped us improve
the paper. This work was funded in part by the Marie Curie Initial
Training Network ELIXIR of the European Commission under contract
PITN-GA-2008-214227, by the European Research Council under the
European Union’s Seventh Framework Programme (FP7/2007-2013) / ERC
Grant agreement 278594-GasAroundGalaxies, and the Marie Curie Training
Network CosmoComp (PITN-GA-2009-238356). The tests were performed
using the HPC resources of CINES under the allocation 2011-c2011046642
made by GENCI (Grand Equipement National de Calcul Intensif), and the
Cray XT-5 cluster at CSCS, Manno, Switzerland.  We also acknowledge
computing resources at the CC-IN2P3 Computing Center
(Lyon/Villeurbanne - France), a partnership between CNRS/IN2P3 and
CEA/DSM/Irfu. JB acknowledges support from the ANR BINGO project
(ANR-08-BLAN-0316-01).


\bibliography{ref,ref_main}

\appendix

\section{RamsesRT non-equilibrium
  thermochemistry}\label{neq_cooling.sec}
We describe here in detail the non-equilibrium thermochemistry we have
implemented for \ramsesrt{} to accommodate for the interactions
between photons and gas.  A thermochemistry step in \ramsesrt{}
considers a single cell of gas at a time with a given state
$\state=(\rho,\rho\vel,\edens,\rho\xhii,\rho\xheii,\rho\xheiii,
\Nphot_i,\bF_i)$ (respectively, mass density, momentum density, energy
density, hydrogen and helium ion abundances, photon densities and
fluxes\footnote{Here we ignore the metal mass density, which is
  optionally stored in every cell, but at this time is not used in the
  non-equilibrium thermochemistry.}) and evolves numerically over a
time-step $\Delta t$ the thermochemical state
\begin{equation}\label{state.eq}
   \state_T= 
   \left( \begin{matrix} 
    \etherm, \
    \xhii, \
    \xheii, \
    \xheiii, \
    \Nphot_1, \
   .. \
    \Nphot_M, \
    \Fphot_1, \
   .. \
    \Fphot_M
\end{matrix}\right)
\end{equation}
(where $\etherm=E-1/2\rho\vel^2$ is the thermal energy density),
i.e. solves the set of $4+2M$ coupled equations
\begin{equation}\label{U_int_1.eq}
  \frac{\partial \state_T}{\partial t} = \rate,
\end{equation}
where $\rate \equiv \dot{\state_T}$.

Due to the stiffness of the thermochemistry equations, it is feasible
to solve them implicitly, i.e. using $\rate(\state_T^{t+\Delta t})$ on
the right hand side (RHS), which guarantees stability and convergence
of the solver. However a fully implicit solver is complicated in
implementation, computationally expensive and not easily adaptable to
changes, e.g. a varying number of photon groups or additional
ion/chemical abundances. Instead we take an approach inspired by
\cite{{Anninos:1997in}}. The idea is to solve one equation at a time
in a specific order, and on the RHS use forward-in-time (FW) values,
i.e. evaluated at $t+\Delta t$, wherever available, but otherwise
backwards-in-time (BW) values, evaluated at $t$. So for the first
variable we choose to advance in time, there are no FW variables
available. For the next one, we can use the FW state of the first
variable, and so on. In that sense the method can be thought of as
being partially implicit.

The cell-thermochemistry is called once every RT time-step of length
$\Delta t_{RT}$, but in each cell it is split into local sub-steps of
length $\Delta t$ that adhere to the $10\%$ rule,
\begin{equation}\label{dt condition.eq}
  \rm{max}\left(\left| \frac{\Delta \state_T}{\state_T} 
    \right|\right) \le 0.1,
\end{equation}
where $\Delta \state_T$ is the change in $\state_T$ during the sub-step.
The RT step thus contains a loop for each cell, which calls the
$\tt{thermo\_step(\state_T,\Delta t)}$ routine once or more often: first
with $\Delta t = \Delta t_{RT}$, then possibly again a number of times
to fill in $\Delta t_{RT}$ if the first guess at $\Delta t$ proves too
long to meet the condition set by \eq{dt condition.eq}.

\vsk The $\tt{thermo\_step(\state_T,\Delta t)}$ routine performs the
following tasks:

\begin{enumerate}
\item $\Nphot$ and $\Fphot$ update       \label{N.upd}
\item $\edens$ update                    \label{Tmu.upd}
\item $\xhi$ update   \label{xh.upd}
\item $\xheii$ and $\xheiii$ update \label{xhe.upd}
\item  Check if we are safe to use a bigger time-step \label{time.upd}
\end{enumerate}

Tasks \ref{Tmu.upd} to \ref{xhe.upd} are in the same order as in
\cite{{Anninos:1997in}}, but they don't include radiative transfer in
their code, so there is no photon update. The argument we have for
putting it first rather than anywhere else is that the photon
densities appear to be the most dynamic variables and so are also most
likely to break the time-step condition (\ref{dt condition.eq}). This
we want to catch early on in the thermochemistry step so we avoid
doing calculations of tasks \ref{Tmu.upd} to \ref{time.upd} that turn
out to be useless because of the too-long time-step.

We now describe the individual tasks. Temperature dependent
interaction rates frequently appear in the tasks - their expressions
are given in \App{rates.sec}. The temperature can at any point be
extracted from the energy density and ionization state of the gas via 
\begin{equation} \label{T conv}
  T = \etherm \ \frac{(\gamma-1) m_H}{\rho \kb} \ \mu ,
\end{equation}
where $\gamma$ is the ratio of specific heats (usually given the value
of $5/3$ in \ramses{}, corresponding to monatomic gas), $m_H$ the
proton mass, $\kb$ the Boltzmann constant and $\mu$ is the average
mass per particle in the gas, in units of $m_H$. 

\subsection*{\ref{N.upd} Photon density and flux update}
The photon number densities and fluxes, $\Nphot_i$ and $\Fphot_i$, are
updated one photon group $i$ at a time. For the photon density the
equations to solve are
\begin{equation}\label{cool_N_update.eq}
  \frac{\partial \Nphot_i}{\pt} = \Npr_i + 
  C_i -  \Nphot_i \, D_i,
\end{equation}
where $\Npr_i$ represents the time derivative of $\Nphot_i$ given by
the RT transport solver (which is nonzero only if the \textit{smoothed
  RT} option is used), $C_i$ represents photon-creating
re-combinations, and $D_i$ represents photon-destroying
absorptions. The creation term is non-existent if the OTSA is used
(emitted photons are assumed to be immediately reabsorbed), but is
otherwise given by
\begin{equation}\label{cool_N_cr.eq}
  C_i = \sum_j^{\hii, \heii, \heiii}  
  b^{rec}_{ji} \ (  \recA_{j} - \recB_{j} ) \ n_j \ n_e,
\end{equation}
where the $b^{rec}_{ji}$ factor is a boolean (1/0) that states which
photon group $j$-species recombinations emit into and $\recA_{j}$
and $\recB_{j}$ are the temperature dependent case A and B
recombination rates for the recombining species.  The photon destruction factor is
given by
\begin{equation}\label{cool_N_de.eq}
  D_i = \sum_j^{\hi, \hei, \heii} \cred \ \csn_{ij} \ n_j,
\end{equation}
where $\cred$ is the (reduced) light speed and $\csn_{ij}$ is the
cross-section between species $j$ and photons in group $i$.

Photon emission from recombination is assumed to be spherically
symmetric, i.e. to go in all directions. It is therefore purely a
diffusive term, and the photon flux equation only includes the
photo-absorbtions:
\begin{equation}\label{cool_F_update.eq}
  \frac{\partial \Fphot_i}{\pt} = \Fpr_i - \Fphot_i \, D_i,
\end{equation}
where $\Fpr_i$ is the time derivative used only in smoothed RT and the
destruction factor remains as in \eq{cool_N_de.eq}.

Equations \eq{cool_N_update.eq} and \eq{cool_F_update.eq} are solved
numerically using a partly semi-implicit Euler (SIE) formulation, in
the sense that they are semi-implicit in the photon density and flux
but otherwise explicit (in temperature and the ion abundances). A tiny
bit of algebra gives:
\begin{align}\label{SIE photons.eq}
    \Nphot_i^{\, t+\dt} &=\frac{\Nphot_i^{\, t}+\dt(\Npr_i+C_i)}
    {1+\dt D_i}, \\
    \Fphot_i^{\, t+\dt} &=\frac{\Fphot_i^{\, t}+\dt \Fpr_i} {1+\dt
      D_i}, 
\end{align}
where all the variables at the RHS are evaluated at the beginning of
the time-step, i.e. at $t$.

For each photon group update, the $10\%$ rule is checked: if
\begin{equation}
  \frac{ \left| \Nphot^{t+\Delta t}_i-\Nphot^t_i \right|}
  {\Nphot_i^t } > 0.1,
\end{equation}
the $\tt{cool\_step}$ routine returns with an un-updated state but
instead a recommendation for a new time-step length $\Delta
t_{new}=0.5 \ \Delta t$, so the routine can be called again with a
better chance of completing. 

\subsection*{\ref{Tmu.upd} Thermal update}
Due to the dependency of $\mu$ on the ionization fractions it is
easiest to evolve the quantity
\begin{equation} \label{Tmu.eq} 
\Tmu \equiv \frac{T}{\mu},
\end{equation}
where $\mu$ can be extracted via
\begin{equation}\label{mu.eq}
  \mu = \left[\, X(1+\xhii) + 
    Y/4 (1+\xheii+2\xheiii) \, \right]^{-1},
\end{equation}
with $X$ and $Y=1-X$ the hydrogen and helium mass fractions,
respectively. Here we ignore the metal contribution to $\mu$, which in
most astrophysical contexts is negligible.

The temperature is updated by solving
\begin{equation}\label{ramsesrt_cool1_eq}
  \frac{\partial \Tmu}{\partial t} = 
  \eTconv \ \CH,
\end{equation}
where $\CH\equiv\dot{\etherm}=\Heat + \Cool$, $\Heat$ is the photoheating
rate and $\Cool$ the cooling rate. These rates are calculated as
follows:

\textit{The photoheating rate} $\Heat$ is a sum of the heating
contributions from all photoionization events:
\begin{equation}\label{PHeat1.eq}
  \mathcal{H} = \sum_j^{\rm{\hi{}, \hei{}, \heii{}}} n_j
  \int_{0}^{\infty} \sigma_j(\nu) \Fphot(\nu) \left[
    h \nu - \epsilon_j \right] d\nu,
\end{equation}
where $\nu$ is photon frequency, $\Fphot(\nu)$ local photon flux and
$\epsilon_j$ photoionization energies. With the discretization into
$M$ photon groups, \eq{PHeat1.eq} becomes
\begin{equation}
  \mathcal{H} = \sum_j^{\rm{\hi, \hei, \heii}} n_j
  \sum_{i=1}^{M} \cred \Nphot_{i} 
  \left( \egy_{i} \cse_{ij}  - \epsilon_j \csn_{ij}  \right) \label{PHeat2.eq}
\end{equation}
where $\egy_{i}$, $\csn_{ij}$, $\cse_{ij}$ and are the photon average
energies, average cross sections and energy weighted cross sections,
respectively, for ionization events between group $i$ and species $j$
(see Eqs. \ref{csn_multi.eq}-\ref{cse_multi.eq}).

\textit{The primordial cooling rate} $\Cool$ is given by
\begin{align}\label{rate_cooling.eq}
  \Cool &= \left[ \zeta_{\hi}(T) + \psi_{\hi}(T) \right] \ \nel \ \nhi
  \\
   &+ \zeta_{\hei}(T) \ \nel \ \nhei
  \nonumber \\
   &+ \left[ \zeta_{\heii}(T) + \psi_{\heii}(T) + 
     \eta^{\rm{A}}_{\heii}(T) + \omega_{\heii}(T) \right] \nel \nheii
  \nonumber \\
   &+ \eta^{\rm{A}}_{\hii}(T) \ \nel \ \nhii
  \nonumber \\
   &+ \eta^{\rm{A}}_{\heiii}(T) \ \nel \ \nheiii
  \nonumber \\
   &+ \theta(T) \ \nel \left( \nhii+\nheii+4\nheiii \right)
  \nonumber \\
   &+ \varpi(T) \ \nel, \nonumber
\end{align}
where the various cooling processes are collisional ionizations
$\zeta$, collisional excitations $\psi$, recombinations $\eta$,
dielectronic recombinations $\omega$, bremsstrahlung $\theta$ and
Compton cooling $\varpi$, all analytic (fitted) functions of
temperature taken from various sources. The complete expressions are
listed (with references) in \App{rates.sec}. If the OTSA is used, the
$\eta^{\rm{A}}$ coefficients are replaced with $\eta^{\rm{B}}$.

The temperature update \eq{ramsesrt_cool1_eq} is solved numerically
using semi-implicit formulation in $\Tmu$, using FW values of photon
densities and BW values of H and He species abundances. The
temperature is updated to
\begin{equation}\label{T integration num eq 2}
    \Tmu^{t+\dt}= \Tmu^{t} + \frac{\CH K \dt}{1-\CHp K \dt},
\end{equation}
where $K \equiv \eTconv$. The temperature-derivative, $\CHp \equiv
\frac{\partial \Cool}{\partial \Tmu}$,
is found by algebraically differentiating each of the primordial
cooling rate expressions in the case of $\Cool$ (and using
$\frac{\partial \Cool}{\partial \Tmu}=\mu\frac{\partial
  \Cool}{\partial T}$). The temperature derivative of the heating rate
is zero.

With $\Tmu^{t+\dt}$ in hand, the time-stepping condition is checked,
i.e if
\begin{equation}
  \frac{\left| \Tmu^{t+\dt}-\Tmu^t \right|}{\Tmu^t} > 0.1,
\end{equation}
$\tt{cool\_step}$ is re-started with half the time-step length. In
tests we've found that the usual time-step constraint given here is
not enough to ensure stability, as the temperature in some cases
oscillates, even in a divergent way. $\CH$ and $\CHp$ are both
evaluated backwards in time, i.e. at $t$, and the large difference
that can exist in these values from $t$ to $t+\dt$ appears to cause
these instabilities. To fix that we include also a first-order
time-step constraint on the temperature, i.e. if
\begin{equation}
\frac{\left|K\CH\dt\right|}{\Tmu^t} > 0.1,
\end{equation}
the time-step length is halved.  With this fix, we have not seen
further temperature oscillations, but there is no guarantee that
numerical instabilities are eliminated.

\subsection*{\ref{xh.upd} Hydrogen ionized fraction
  update} The \hii{} abundance is affected by collisional
ionizations, photoionizations, and recombinations, i.e.
\begin{equation}\label{nhii_diff.eq}
  \frac{\pa \nhii}{\pt} = 
  \nhi \left( \beta_{\hi} \nel  
  + \sum_{i=1}^M \csn_{i \hi} \cred \Nphot_i \right)
  - \nhii \recAH \nel, 
\end{equation}
where $\beta_{\hi}(T)$ is the rate of collisional ionizations by
electrons and $\recAH(T)$ the case A hydrogen recombination rate,
which is replaced here by $\recBH$ if the OTSA is used. In terms of
ionization fraction, \eq{nhii_diff.eq} becomes
\begin{align}
  \frac{\pa \xhii}{\pt} &=
  (1-\xhii) \left[ \beta_{\hi} \nel  
  + \sum_{i=1}^M \csn_{i \hi} \cred \Nphot_i 
\right]
  - \xhii \recAH \nel  \nonumber \\
  &=(1-\xhii)\ C - \xhii\  D \nonumber \\
  &=C - \xhii\  (C+D), \label{xhii_diff.eq}
\end{align}
where we have in the second line separated the rates into \hii{}
creation $C$ and destruction $D$, and in the third line collected
multiples of $\xhii$.

To prevent stiffness-induced instabilities, we have gone for an
approach which is semi-implicit in $\xhii$:
\begin{equation}\label{xhii_update_2.eq}
    \xhii^{t+\dt}=\xhii^t + \dt \frac{C-\xhii^t(C+D)}{1-J\dt} ,
\end{equation}
where 
\begin{align}
  J &\equiv \frac{\pa \dot{x}_{\hii}}{\pa \xhii} \nonumber \\
  &= \frac{\pa C}{\pa \xhii} - (C+D) - 
  \xhii\left(\frac{\pa C}{\pa \xhii} + \frac{\pa D}{\pa \xhii} \right),
\end{align}
and the creation and destruction derivatives are given by
\begin{align}
  \frac{\pa C}{\pa \xhii} &=\nh \beta_{\hi} - 
  \nel \Tmu \mu^2 X \frac{\pa \beta_{\hi}}{\pa T} \\
  \frac{\pa D}{\pa \xhii} &=\nh \recAH - 
  \nel \Tmu \mu^2 X \frac{\pa \recAH}{\pa T}.
\end{align}

We end with the usual check if the $10\%$ rule is broken, i.e. if
\begin{equation}
  \frac{\left|\xhii^{t+\dt}-\xhii^t \right|}{\xhii^t} > 0.1,
\end{equation} 
$\tt{cool\_step}$ is restarted with half the time-step length. Like
with the temperature a first-order check is also made, i.e.
\begin{equation}
 \frac{\left|C-\xhii^t(C+D)\right|}{\xhii^t}\dt > 0.1.
\end{equation}

\subsection*{\ref{xhe.upd} Helium ionized fractions update}
Though the \hei{} fraction is not a cell variable (it can be obtained
via $\xhei=1-\xheii-\xheiii$), it is evolved in order to make a
consistency check at the end of the helium updates. Before each of the
helium fraction updates, we recalculate $\nel$ and $\mu$ to reflect
the new FW abundances. 

The \hei{} fraction is set by
\begin{align}\label{xhei_diff.eq}
  \frac{\pa \xhei}{\pt} &=
  \xheii \recA_{\heii} \nel
  - \xhei \left( \beta_{\hei}\nel
  - \sum_{i=1}^M \csn_{i \hei} \cred \Nphot_i \right)
  \nonumber \\
  &= \ C \ -  \ \xhei \ D,
\end{align}
i.e. \heii{} recombinations and collisional- and photo-ionizations of
\hei{}. As usual $\recA$ is replaced by $\recB$ in the case of the
OTSA. In the second line of \eq{xhei_diff.eq} we've separated the RHS
into \hei{} creation $C$ and destruction $D$.

Here we follow \cite{{Anninos:1997in}} and do the \hei{} update with
\begin{equation}\label{xhei_update.eq}
  \xhei^{t+\dt}=\frac{\xhei^t+C\dt}{1+D\dt}.
\end{equation}
The update is partly implicit, since it uses updated values of
$N_i^{t+\dt}$, $\Tmu^{t+\dt}$, and $\xhii^{t+\dt}$ ($\rightarrow$
$\mu$ and $\nel$), but un-updated values of $\xheii^t$ and
$\xheiii^t$. 

We then evolve the \heii{} fraction. The differential equation to
solve is 
\begin{align}
  \frac{\pa \xheii}{\pt} &= 
  \xhei \left( \beta_{\hei} \nel
    + \sum_{i=1}^M \csn_{i \hei} \cred \Nphot_i \right)
    + \xheiii \recA_{\heiii}\nel
  \nonumber \\
  &- \xheii \left( \beta_{\heii} \nel
  + \recA_{\heii}\nel
  + \sum_{i=1}^M \csn_{i \heii} \cred \Nphot_i \right)
  \nonumber \\
  &= C \ -  \ \xheii \ D. \label{xheii_diff.eq}
\end{align}
The RHS terms are, in order of appearance, \hei{} collisional
ionizations, \heiii{} recombinations, \hei{} photoionizations (with an
optional homogeneous background in parentheses), \heii{} collisional
ionizations, \heii{} recombinations and \heii{} photo-ionizations. In
the third line we have grouped the terms into a creation term $C$ and
destruction terms $D$.

The discrete update is done with the same formulation as
\eq{xhei_update.eq}, i.e.
\begin{equation}\label{xheii_update.eq}
  \xheii^{t+\dt}=\frac{\xheii^t+C\dt}{1+D\dt},
\end{equation}
using updated values of $N_i^{t+\dt}$, $\Tmu^{t+\dt}$,
$\xhii^{t+\dt}$, and $\xhei^{t+\dt}$ ($\rightarrow$ $\mu$ and $\nel$),
and the un-updated value only of $\xheiii^t$.

The only variable left is the \heiii{} fraction. The differential
equation is
\begin{align}\label{xheiii_diff.eq}
  \frac{\pa \xheiii}{\pt} &=
  \xheii \left( \beta_{\heii} \nel +
  \sum_{i=1}^M \csn_{i \heii} \cred \Nphot_i \right) \nonumber \\
  &- \xheiii \ \recA_{\heiii}\nel \nonumber \\
  &= \ C \ -  \ \xheiii \ D.
\end{align}
In the third line we have as usual grouped the terms into creation and
destruction.

Again the update follows the same formulation,
\begin{equation}\label{xheiii_update.eq}
  \xheiii^{t+\dt}=\frac{\xheiii^t+C\dt}{1+D\dt},
\end{equation}
which is implicit in all variables.

Conservation of helium density is then enforced, i.e. that
\begin{equation}
  \xhei+\xheii+\xheiii=1,
\end{equation}
by lowering the largest of these fractions accordingly (in the case of
$\xhei$ being the largest there is no update).

The $10\%$ rule is not applied to the helium fractions. Instead, the
final $10\%$ check is done on the electron density, which is retrieved
from all the ionization fractions with 
\begin{equation} \nel =
  \xhii\nh + (\xheii+2\xheiii)\nhe.
\end{equation}
If 
\begin{equation}
  \frac{\left|\nel^{t+\dt}-\nel^t \right|}{\nel^t} > 0.1,
\end{equation}
$\tt{cool\_step}$ is restarted with half the time-step length.

\subsection*{\ref{time.upd} Time-step check} All the variables have
been updated, from $\state_T^t$ to $\state_T^{t+\dt}$, and the $10\%$
rule is not violated over the \joki{thermochemistry time-step just
  taken, $\dttc$. However, its length may have been unneccessarily
  short}, and if so, there is a large probability that it is also
unneccessarily short for the next call to $\tt{cool\_step}$\joki{,
  i.e. for the next thermochemistry time-step} (to fill the total
$\dt_{RT}$).

Therefore a final time-step check is made before finishing up, of how
close we were to breaking the $10\%$ rule \joki{over $\dttc$}. If the
maximally changed variable in $\state_T$ has changed by less than
$5\%$, i.e. if
\begin{equation}
  \rm{max}\left( \left| \frac{\state_T^{t+\dt}-\state_T^t}{\state_T^t}\right|
  \right) < 0.05,
\end{equation}
then the \emph{next} \joki{$\dttc$} in that cell is set to twice the
one just used. Note that this is on a cell-by-cell basis, and the next
\joki{$\dttc$} for each cell is only stored during the thermochemistry
subcycling and lost at the end of each $\dt_{RT}$ cycle. At the
beginning of each cell-cycle over $\dt_{RT}$, the first guess at a
timestep is always \joki{$\dttc = \dt_{RT}$}. If this is too large for
the $10\%$ rule to be obeyed, successive calls to $\tt{cool\_step}$
will quickly fix that by halving \joki{$\dttc$} until the rule is no
longer broken, and only then will $\tt{cool\_step}$ start to return
updated values of $\state_T$.

\section{Stellar UV emission and derived photon
  attributes}\label{SEDs.sec} 
\begin{figure*}\center{
  \subfloat[\textbf{BC03}]
  {\includegraphics[width=0.5\textwidth]
    {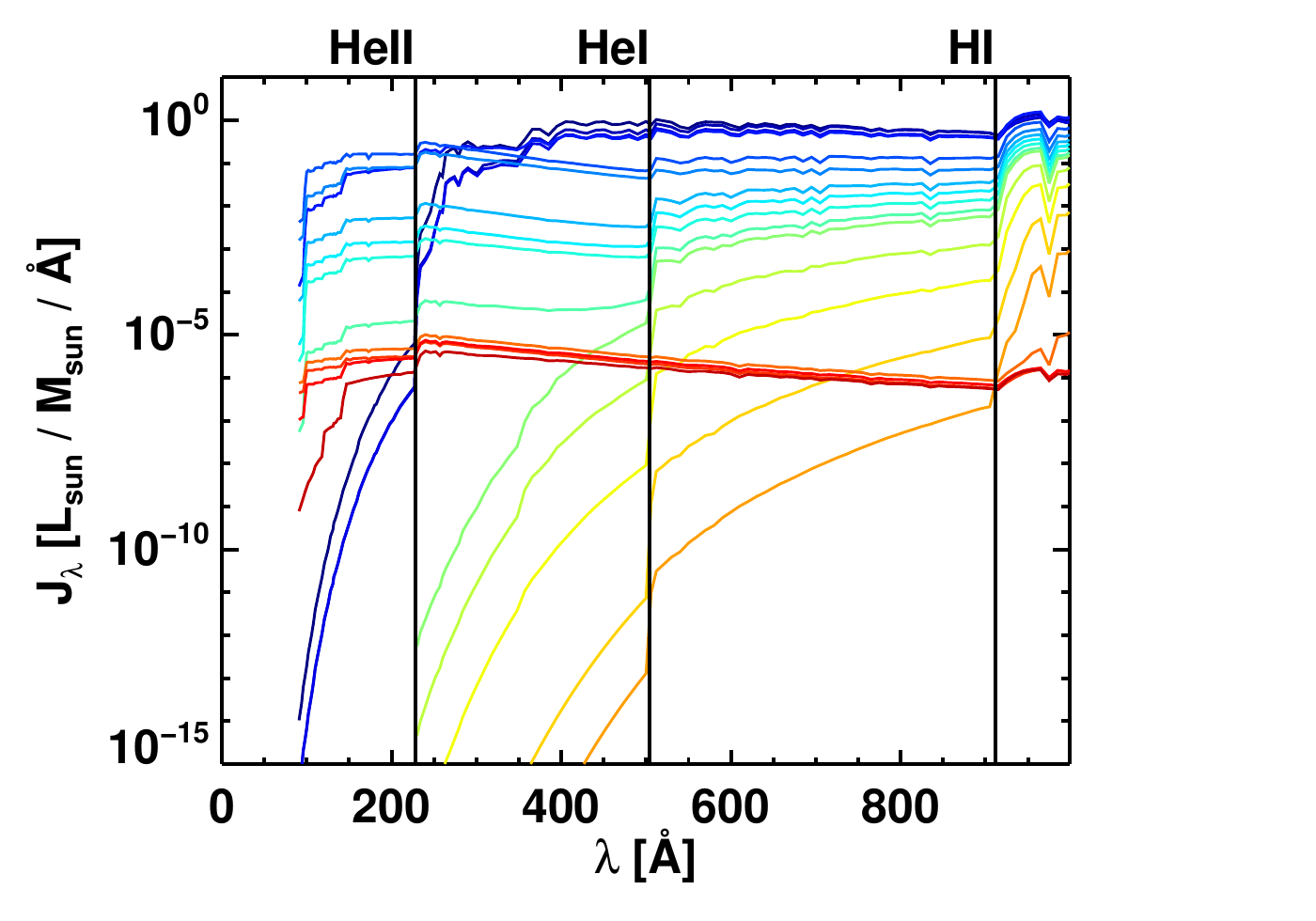}\label{sed_bc03.fig}}\hspace{-3.1cm}
  \subfloat[\textbf{Starburst99}]
  {\includegraphics[width=0.5\textwidth]
    {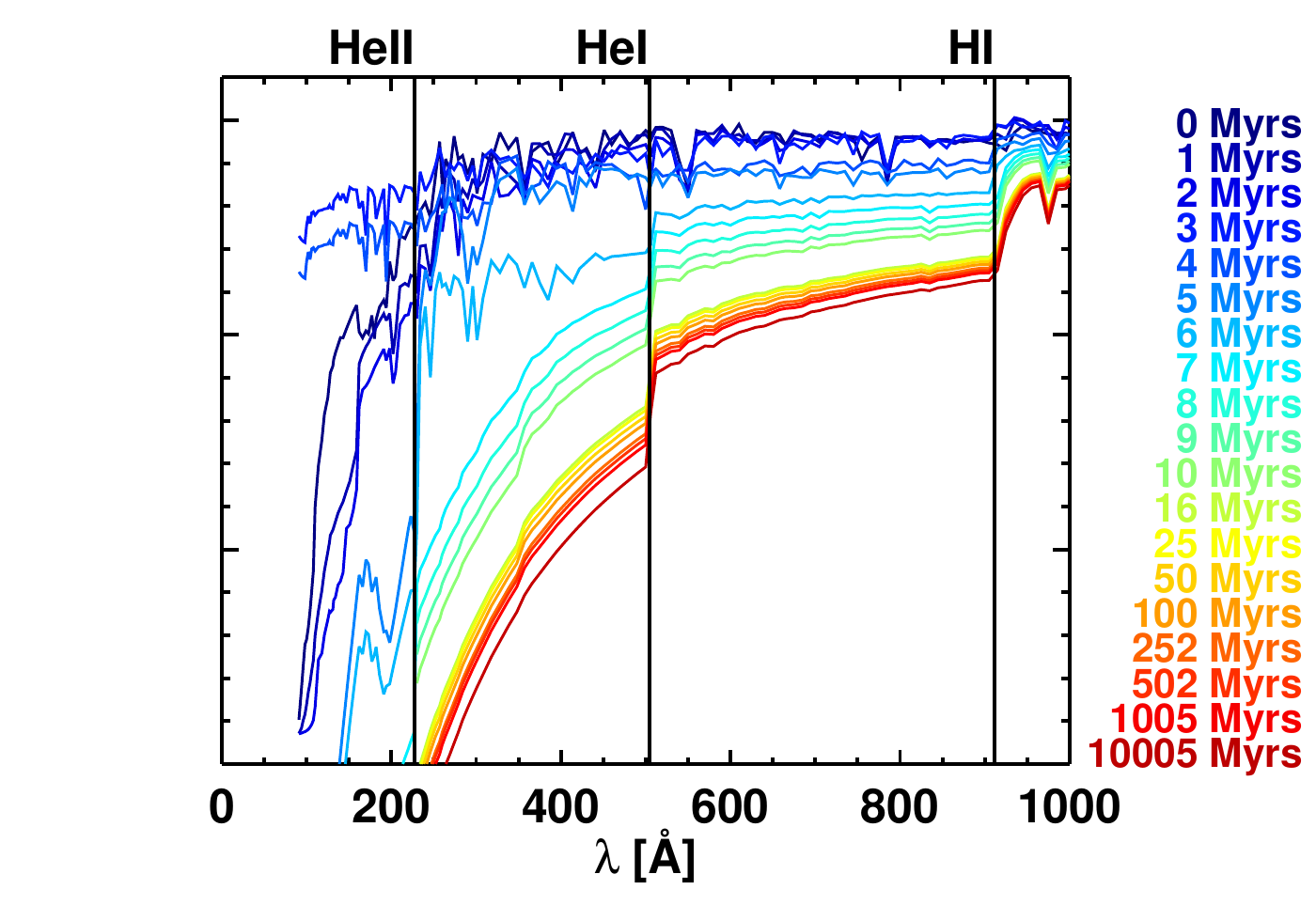}\label{sed_s99.fig}}}
  \caption[SED plots]
  {SED plots from (\textbf{a}) \cite{{Bruzual:2003ck}} and
    (\textbf{b}) Starburst99 \citep{{Leitherer:1999jt}} for solar
    metallicity at different stellar population ages. The spectral
    luminosity is given in solar luminosities ($3.8 \times 10^{33}\
    \ergs$) per solar mass ($2 \times 10^{33}$ g) per
    wavelength. Vertical lines mark the ionization wavelengths for
    $\hi{}$, $\hei{}$ and $\heii{}$, which correspond to the
    wavelengths marking the three photon groups we typically use in
    our simulations. The Starburst99 spectra are generated with the
    instantaneous formation of $10^6$ solar masses and a Salpeter
    initial mass function.}\label{sed_spectra.fig}
\end{figure*}

\begin{figure*}\center{
  \subfloat{\includegraphics[width=0.7\textwidth]
    {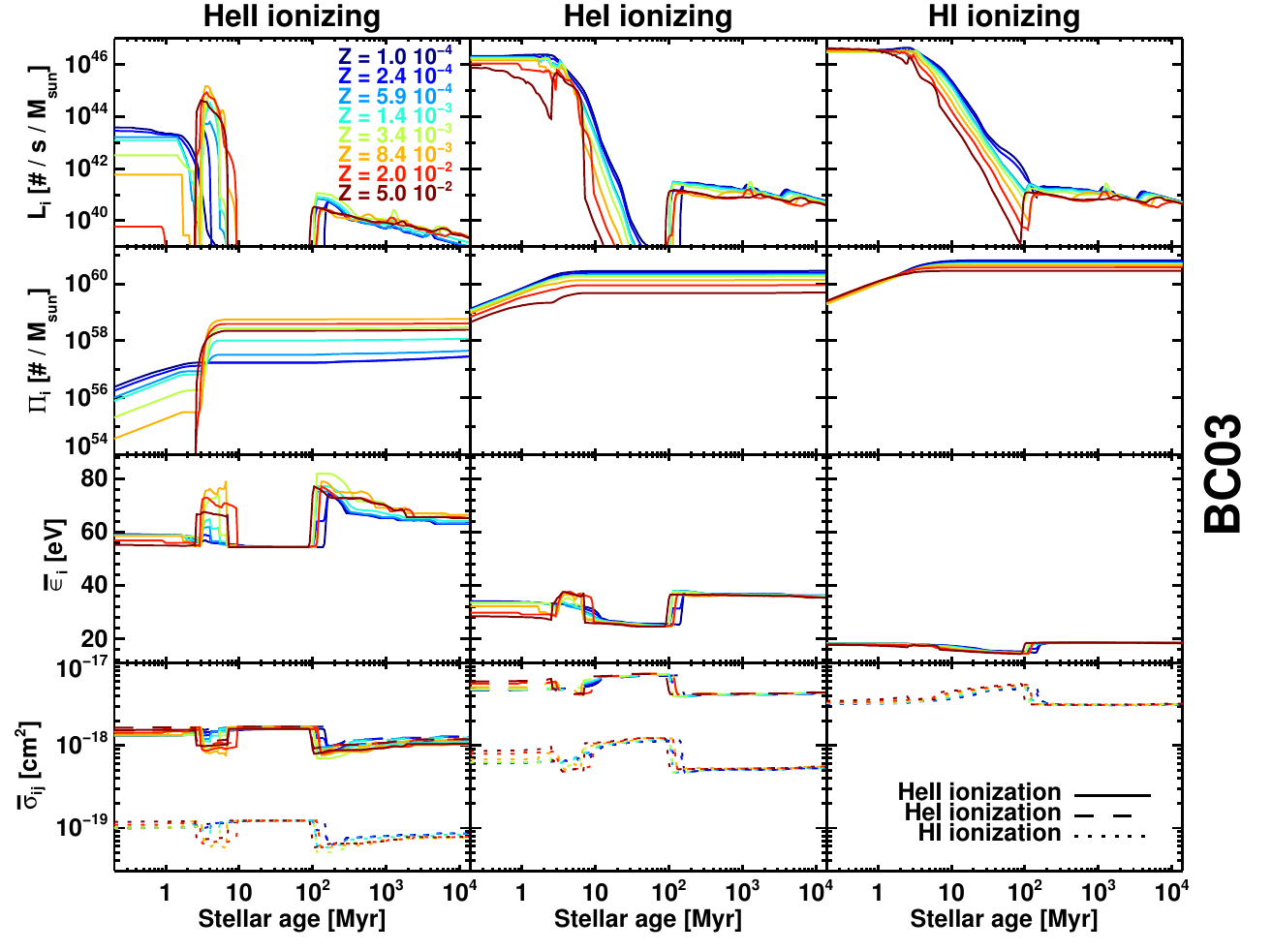}}

  \subfloat{\includegraphics[width=0.7\textwidth]
    {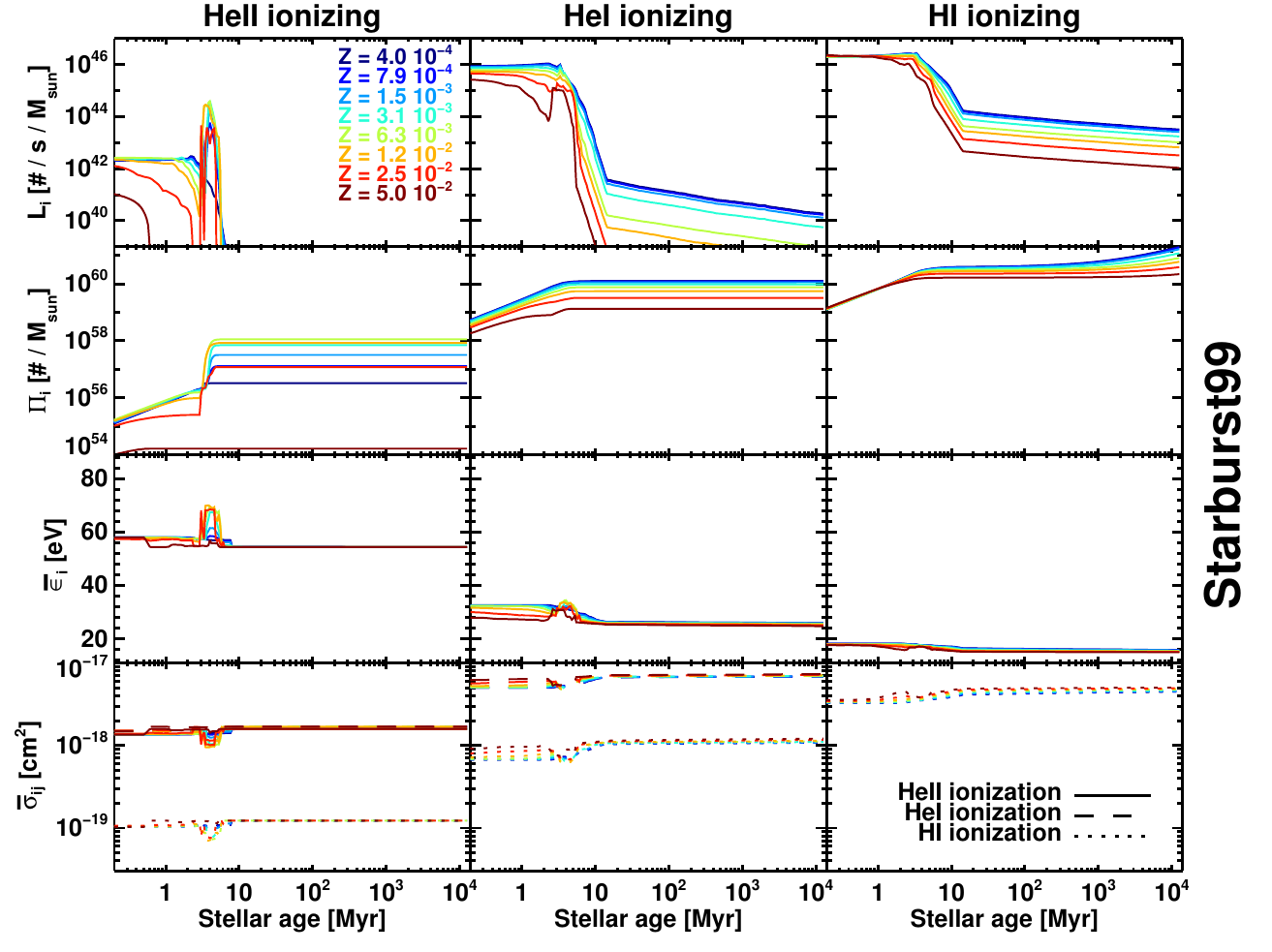}}
  \caption[Photon group attributes derived from two SED models]
  {\label{sed_pacs.fig}$\heii$, $\hei$, and $\hi$ ionizing
    luminosities and photon group attributes derived from the BC03
    (\textbf{top panel}) and Starburst99 SED models (\textbf{bottom
      panel}), as functions of age (x-axis) and metallicity
    (colors). The plot columns represent the three photon
    groups. \textbf{Top rows} show stellar luminosity, in the number
    of photons that goes into each group per second per solar
    mass. \textbf{Second rows} show accumulated number of photons
    emitted. \textbf{Third rows} show the average photon energies per
    interaction. \textbf{Bottom rows} show average cross sections per
    interaction.}}
\end{figure*}

In the photon injection step in
\ramsesrt{} (\Sec{Injection.sec}), the task is to inject photons into
each grid cell corresponding to the luminosities of stellar particles
that reside in it. Here we describe how we derive these luminosities
from stellar energy distribution (SED) models, along with the
photoionization cross sections and energies for each photon group.

\subsection{Stellar luminosities}
Stellar particles in \ramses{} represent stellar populations, so it
makes sense to use SED models to infer their luminosities. \ramsesrt{}
can read SED tables at startup and derive from them stellar
luminosities for photon injection, as well as photon group attributes
that can be updated to reflect the average emission from the stellar
particles populating the simulation.

We have hitherto used the SED model of \cite{{Bruzual:2003ck}} (BC03),
but it can be replaced with any other model, e.g. Starburst99
\citep{{Leitherer:1999jt}}, as long as the file format is adjusted to
match. The model should come in the form of spectra,
$J_{\lambda}(\age,Z)$, giving emitted energy in Solar units per Solar
mass per wavelength per second, binned by stellar population age and
metallicity. In \Fig{sed_spectra.fig} we show $J_{\lambda}$ from BC03
and Starburst99 at solar metallicity for various population ages.

Age and metallicity dependent population luminosity $\Lum$, given in
number of photons emitted per second into photon group $i$, is
calculated from the SED model by
\begin{equation}\label{SEDlum.eq}
  \Lum_i(\age,Z)=\int_{\nui}^{\nuf}{J_{\nu}(\age,Z)/h\nu \ d\nu},
\end{equation}
where $J_{\nu}=c/\nu^2J_{\lambda(\nu)}$. The cumulative population
luminosity is then
\begin{equation}\label{SEDcum.eq}
  \Pi_i(\age,Z)=\int_{0}^{\age}{\Lum_{i}(t,Z) \ dt}.
\end{equation}
Since both the photon injection and the calculation of photon group
attributes are done on the fly, $\Lum_i(\age,Z)$ and $\Pi_i(\age,Z)$
must be evaluated as quickly as possible for given stellar particle
ages and metallicities. Values of $\Lum_i$ and $\Pi_i$ are therefore
only calculated from the SED spectra via \eq{SEDlum.eq} and
\eq{SEDcum.eq} at simulation startup, and tabulated with
equally-spaced logarithmic bins of age and metallicity, so that they
can be retrieved with minimum computational effort via linear
interpolation, e.g. when injecting photons into cells via
\Eq{injnum.eq}.

\subsection{Photon group attributes}
There are three sets of global attributes for each photon group.
These are average photon energies $\egy_{i}$, average photoionization
cross sections $\csn_{ij}$ and energy weighted cross sections
$\cse_{ij}$, that are defined in \Sec{Method.sec}
(Eqs. \ref{csn_multi.eq}-\ref{cse_multi.eq}). For an age and
metallicity dependent reference spectrum $J_{\nu}(\age,Z)$, these are
\begin{align}
  \egy_{i}(\age,Z)=
  \frac{\int_{\nui}^{\nuf}{J_{\nu} \ d\nu}}
  {\int_{\nui}^{\nuf}{J_{\nu} /h\nu \ d\nu}}, \label{egy_multi2.eq} \\
  \csn_{ij}(\age,Z)=
  \frac{\int_{\nui}^{\nuf}{\sigma_{\nu j} J_{\nu} /h\nu \ d\nu}}
  {\int_{\nui}^{\nuf}{J_{\nu} /h\nu \ d\nu}}, \label{csn_multi2.eq} \\
  \cse_{ij}(\age,Z)=
  \frac{\int_{\nui}^{\nuf}{\sigma_{\nu j} J_{\nu} \ d\nu}}
  {\int_{\nui}^{\nuf}{J_{\nu} \ d\nu}}. \label{cse_multi2.eq}
\end{align}

Since there are three ionizeable species in the current implementation
of \ramsesrt{}, each photon group has three values of $\csn$ and three
of $\cse$. These attributes can be set as run parameters to reflect
some typical stellar spectra, e.g. a blackbody or a SED. It can also
be left to \ramsesrt{} to set them on the fly to reflect the
in-simulation stellar populations, using the expressions
\eq{egy_multi2.eq}-\eq{cse_multi2.eq}, with the loaded SED spectra
representing $J_{\nu}$ and the expressions from \cite{{Verner:1996dm}}
for $\sigma_{\nu j}$ (see \App{rates_cs.sec}).  Due to the averaged
nature of the photon groups, we must however suffice to set the group
attributes to reflect the \textit{average} stellar emission in the
simulation\joki{, weighted by the stellar luminosities}
\footnote{\joki{This infers that local variations in cross sections
    and energy, due to variations in stellar age and metallicity, are
    ignored. For example, it can be seen in \Fig{sed_pacs.fig} that
    stellar populations temporarily (at $\sim3-5$ Myr) become very
    luminous in high-energy photons: while this is reflected in the
    luminosities of the stellar particles, the energies and cross
    sections of the photons emitted from them are simply the
    luminosity weighted averages over all stellar populations, which
    are the same everywhere.} Note also that the on-the-fly update of
  photon attributes according to \eq{cse_avg.eq} and \eq{egy_avg.eq}
  infers that existing photons attributes are changed, i.e. the
  attributes of photons that have already been emitted change in
  mid-air.  }.  If this option is used, the photon group attributes
are updated every $n$ coarse time-steps (where $n$ is an adjustable
parameter) by polling all the stellar particles in the simulation and
setting for each group $i$ and species $j$,
\begin{align}
  \egy_{i}=
  \frac{\sum\limits^{\textrm{all stars}}_{\star}
    \egy_{i}(\sage,Z_{\star})\ m_{\star}\ \Lum_i(\sage,Z_{\star}) } 
  {\sum\limits^{\textrm{all stars}}_{\star}\ 
    m_{\star}\ \Lum_i(\sage,Z_{\star})}. \label{egy_avg.eq}
\end{align}
\begin{align}
  \csn_{ij}=
  \frac{\sum\limits^{\textrm{all stars}}_{\star}
    \csn_{ij}(\sage,Z_{\star})\ m_{\star}\ \Lum_i(\sage,Z_{\star}) } 
  {\sum\limits^{\textrm{all stars}}_{\star}\ 
    m_{\star}\ \Lum_i(\sage,Z_{\star})},  \label{cse_avg.eq}
\end{align}
\begin{align}
  \cse_{ij}=
  \frac{\sum\limits^{\textrm{all stars}}_{\star}
    \cse_{ij}(\sage,Z_{\star})\ m_{\star}\ \Lum_i(\sage,Z_{\star}) } 
  {\sum\limits^{\textrm{all stars}}_{\star}\ 
    m_{\star}\ \Lum_i(\sage,Z_{\star})}. \label{csn_avg.eq}
\end{align}
The values of each stellar particle's $\Lum_i(\sage,Z_{\star})$,
$\egy_{ij}(\sage,Z_{\star})$, $\csn_{ij}(\sage,Z_{\star})$ and
$\cse_{ij}(\sage,Z_{\star})$ are interpolated from tables that are
generated at startup via \eq{SEDlum.eq} and
\eq{egy_multi2.eq}-\eq{cse_multi2.eq}.

Although one is free to use many photon groups to resolve
frequencies, it is practical to only use a handful, due to limitations
in memory and computation. We typically use three photon groups in our
simulations, representing $\hi$, $\hei$ and $\heii$ ionizing photons,
as indicated by vertical lines in the plots of \Fig{sed_spectra.fig}.
The stellar luminosities, instantaneous and accumulated, average cross
sections and energies for those groups are plotted in
\Fig{sed_pacs.fig} for BC03 (top) and Starburst99 (bottom), as
calculated via \eq{SEDlum.eq}-\eq{cse_multi2.eq}.  From the luminosity
plots (top rows), it can be seen that the stellar populations emit
predominantly for the first $\sim3-6$ Myrs and the luminosity
drastically goes down as the most massive stars in the population
begin to expire.

\section{Non-equilibrium thermochemistry tests}
\label{CoolTests.sec}
\input{RamsesRT_chemtests}

\section{On multi-stepping in the AMR level hierarchy}
\label{App_interlevel}
\joki{As discussed in \Sec{ramsesrt.sec}}, solving hydrodynamics over
an AMR grid with a multi-stepping approach always leaves ill-defined
states at inter-level boundaries between the start and finish of the
coarse level timestep. \joki{This imposes severe constraints on how
  the RT can be coupled to the hydrodynamics, and essentially means
  that RT cannot be sub-cycled within multi-stepping
  hydrodynamics. Here we will clarify this point in detail.}

Hydrodynamic advection across the boundaries of a cell is performed in
an operator-split fashion, such that the advection is solved
separately across a discretized timestep for each boundary. In order
for the solver to be consistent, i.e. for the result at the end of
the timestep to be \emph{independent} of the order in which the
boundaries are accounted for, the solver must work from the same
initial cell state $\state$ for all the inter-cell updates. Thus, a
copy is first made of the original cell states involved, i.e.
\begin{equation}
\state \rightarrow \staten,
\end{equation}
where we can term $\state$ the \emph{source state} and
$\staten$ the \emph{destination state}.  Using $\state$
as source terms for the intercell fluxes, the advection can be solved
with some computational method (e.g. Godunov solver for the
hydrodynamics in \ramses{} and an HLL/GLF flux function for the RT
advection in \ramsesrt{}), which performs the update on
$\staten$. To take a concrete example, each RT advection
update, \Eq{dUdt_sol.eq}, uses $\staten$ for the update (the
LHS term and the first RHS term), but the intercell fluxes are derived
from $\state$, i.e. $\stateF(\state)$. Once all the updates (6 per
cell) have been collected, the cell update is made final by:
\begin{equation}
\staten \rightarrow \state.
\end{equation}

In the $\tt{amr\_step(\ell)}$ hierarchy in \ramses{}, such copies are
made of all $\ell$ cells before the AMR recursion, and the update is
made final after the recursion has returned and the
$\tt{hydro\_solver}$ has been called at the current level,
i.e. advection has been performed over the timestep over the current
level and all finer levels. 

\begin{figure}
  \centering
  \includegraphics[width=0.47\textwidth]
    {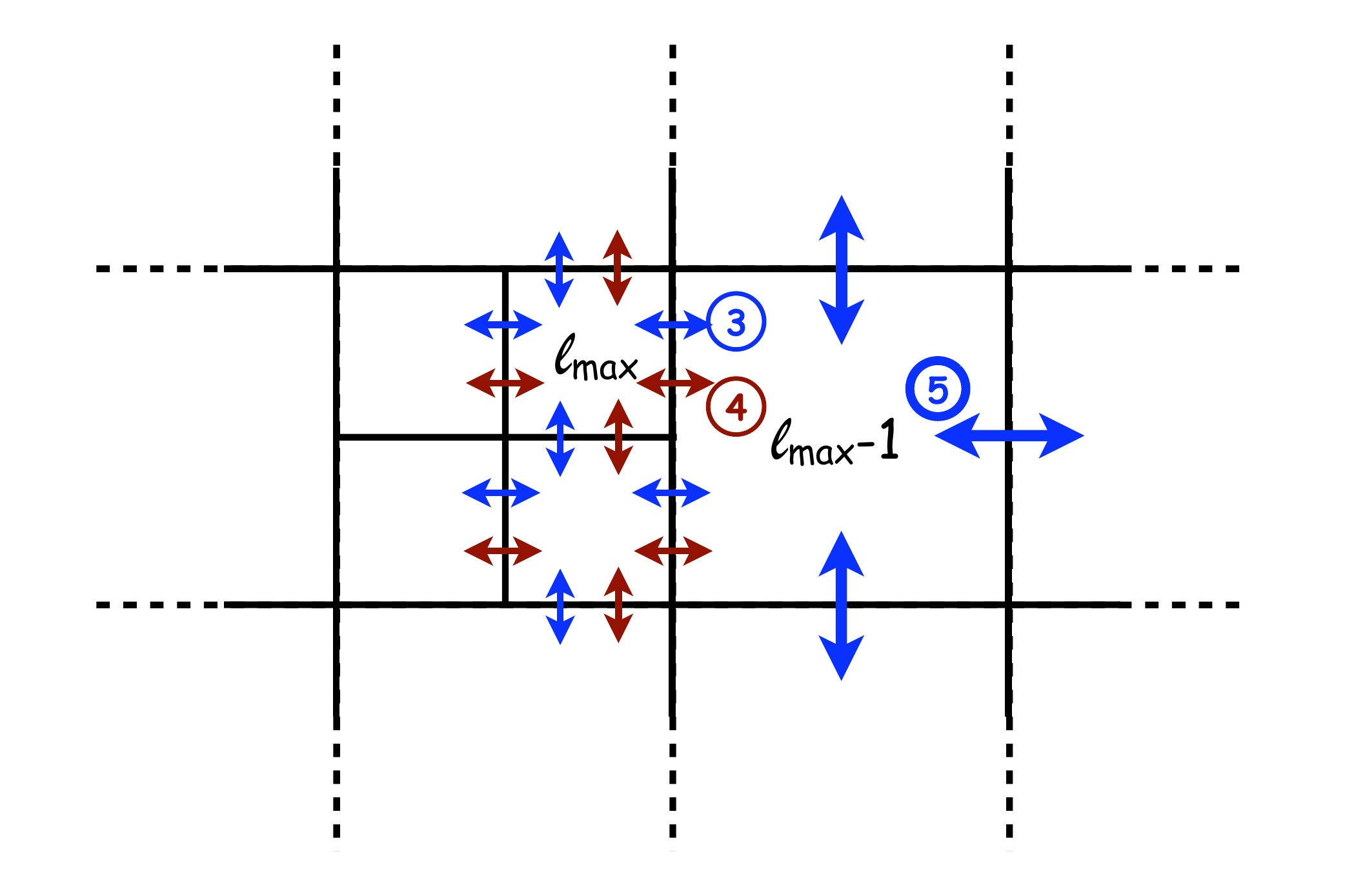}
  \caption[Down-level update of cell gas state]
  {\label{iflux.fig}Level $\ell$ gas state updates via
    intercell fluxes also perform partial gas updates in neighbouring
    cells at level $\ell-1$. The example shown corresponds to the
    hierarchy from \Fig{amr_step.fig}. Steps 3 and 4 at the finest
    level also include partial updates of neighbouring $\lmax-1$
    cells, but these neighbour cell states are not fully updated until
    all the intercell fluxes are taken into account, which is in step
    5 from \Fig{amr_step.fig}.}
\end{figure}

This allows cell states to be updated not only at the current level,
but also (twice) in all neighbouring cells at the next coarser level.
The coarser level update is only \textit{partial} though, because it
only reflects the intercell fluxes across inter-level boundaries, and
fluxes across other boundaries (same level or next coarser level) will
only be accounted for when the coarser level time-step is
advanced. Until then, these coarser level neighbour cells have
\emph{two} gas states, $\state$ and $\staten$.  This is shown
schematically in \Fig{iflux.fig}.

If RT subcycling is to be done at each AMR fine-level step, over the
whole grid, the question is, which cell state do we use for the
thermochemistry, i.e. the interaction between photons and gas, in
those inter-level boundary cells?

Choosing one but not the other leads to an obvious and severe
inconsistency between the source and destination states. If the
thermochemistry does the update on $\staten$, then a gas element which
is transported from one cell to a neighbour during the following hydro
transport is not thermochemically evolved over the time-step, because
it originates from $\state$. If instead the update is done on
$\state$, a gas element which stays still in any cell over the
following hydro transport step is not thermochemically evolved over
the time-step.  One might then just update both states via
thermochemistry, i.e. apply it on each cell twice. This does not
really make sense for these inter-level intercell boundary cells that
have $\state \ne \staten$, as $\staten$ doesn't represent a true state
but is rather an intermediate and temporary quantity that exists
between well-defined times. Also, it would be really non-trivial to
implement: applying thermochemistry on each of the states also implies
transporting the photons through two different states in each cell,
which creates alternative time-lines for the radiative transfer!

\joki{Thus, subcycling RT within multi-stepping hydrodynamics in a
  conservative way is not possible (or at least non-trivial), which
  has led us to disallow RT subcycling within the hydro timestep in
  our implementation.}

\section{Interaction rate coefficients adopted in RamsesRT}
\label{rates.sec}
Here we collect the rate coefficients used in \ramsesrt{} for hydrogen
and helium interactions, which are fitted functions taken from various
sources. These are, in order of appearance, collisional ionization
rates, recombination rates, cooling rates (collisional ionization,
recombination, collisional excitation, bremsstrahlung, Compton and
dielectric recombination), and photoionization cross sections.

\subsection{Collisional ionization rate coefficients}
\label{rates_coll.sec}
Those are in units of [$\ccs$] and are taken from \cite{{Cen:1992dq}},
with temperature everywhere assumed in Kelvin:
\begin{align*}
  \beta_{\hi}(T)   &= 5.85 \times 10^{-11} \ \sqrt{T}  
  \ \left( 1 + \sqrt{\frac{T}{10^5}} \right)^{-1}
  \ e^{-157 \ 809.1/T} \\ 
  \beta_{\hei}(T)   &= 2.38 \times 10^{-11} \ \sqrt{T}  
  \ \left( 1 + \sqrt{\frac{T}{10^5}}  \right)^{-1}
  \ e^{-285 \ 335.4/T} \\
  \beta_{\heii}(T)   &= 5.68 \times 10^{-12} \ \sqrt{T}  
  \ \left( 1 + \sqrt{\frac{T}{10^5}}  \right)^{-1}
  \ e^{-631 \ 515/T}
\end{align*}

\subsection{Recombination rate coefficients}
\label{rates_rec.sec}
These are all taken from \cite{{Hui:1997vj}}. For readability, we use
the following unitless functions:
\begin{align*}
  \lambda_{\hi}(T) &=   \frac{315 \ 614 \ K}{T}  \\
  \lambda_{\hei}(T) &=  \frac{570 \ 670 \ K}{T} \\
  \lambda_{\heii}(T) &= \frac{1 \ 263 \ 030 \ K}{T} 
\end{align*}
The coefficients are as follows, all in units of [$\ccs$]: 
\begin{align*}
  \recA_{\hii}(T)   &= 1.269 \times 10^{-13}  \ \frac{\lambda^{1.503}_{\hi}}
  {[1+(\lambda_{\hi}/0.522)^{0.47}]^{1.923}} \\ 
  \recA_{\heii}(T)  &= 3 \times 10^{-14}      \ \lambda^{0.654}_{\hei} \\
  \recA_{\heiii}(T) &= 2.538 \times 10^{-13}  \ \frac{\lambda^{1.503}_{\heii}}
  {[1+(\lambda_{\heii}/0.522)^{0.47}]^{1.923}}                 \\
  \recB_{\hii}(T)   &= 2.753 \times 10^{-14}  \ \frac{\lambda^{1.5}_{\hi}}
  {[1+(\lambda_{\hi}/2.74)^{0.407}]^{2.242}}                   \\
  \recB_{\heii}(T)  &= 1.26 \times 10^{-14}   \ \lambda^{0.75}_{\hei}   \\
  \recB_{\heiii}(T) &= 5.506 \times 10^{-14}  \ \frac{\lambda^{1.5}_{\heii}}
  {[1+(\lambda_{\heii}/2.74)^{0.407}]^{2.242}}  
\end{align*}

\begin{table*}
  \centering
  \caption
  {Photoionization cross section parameters -- see \Eq{PhI.eq}}
  \label{sigma.tbl}
  \begin{tabular}{c|ccccccc}
    \toprule
    Ion species & $\epsilon_0$ [eV] & $\sigma_0$ [$\cs$] & $P$   & 
    $y_a$       & $y_w$             & $y_0$              & $y_1$ \\
    \midrule
    \hi     & $0.4298$ & $5.475 \ 10^{-14}$ & $2.963$ & 
    $32.88$ & $0$      & $0$               & $0$ \\
    \hei    & $0.1361$ & $9.492 \ 10^{-16}$ & $3.188$ & 
    $1.469$ & $2.039$  & $0.4434$          & $2.136$ \\
    \heii   & $1.720$  & $1.369 \ 10^{-14}$ & $2.963$ & 
    $32.88$ & $0$      & $0$               & $0$ \\
    \bottomrule
  \end{tabular}
\end{table*}
\subsection{Cooling rate coefficients}\label{rates_cool.sec}
The temperature used in these coefficients is assumed everywhere in
Kelvin. Collisional ionization cooling rate coefficients [$\coolU$]
\citep{{Cen:1992dq}}:
\begin{align*}
  \zeta_{\hi}(T)   &= 1.27 \times 10^{-21} \ \sqrt{T}  
  \ \left( 1 + \sqrt{\frac{T}{10^5}} \right)^{-1}
  \ e^{-157 \ 809.1/T} \\ 
  \zeta_{\hei}(T)   &= 9.38 \times 10^{-22} \ \sqrt{T}  
  \ \left( 1 + \sqrt{\frac{T}{10^5}}  \right)^{-1}
  \ e^{-285 \ 335.4/T} \\
  \zeta_{\heii}(T)   &= 4.95 \times 10^{-22} \ \sqrt{T}  
  \ \left( 1 + \sqrt{\frac{T}{10^5}}  \right)^{-1}
  \ e^{-631 \ 515/T}
\end{align*}
Case A and B recombination cooling rate coefficients [$\coolU$]
\citep{{Hui:1997vj}}:
\begin{align*}
  \eta^{A}_{\hii}(T)   &= 1.778 \times 10^{-29}  \ T \
  \frac{\lambda^{1.965}_{\hi}}
  {[1+(\lambda_{\hi}/0.541)^{0.502}]^{2.697}} \\ 
  \eta^{A}_{\heii}(T)   &= \kb \ T  \recA_{\heii}
  \ = \ \kb \ T  \ 3 \times 10^{-14}      \ \lambda^{0.654}_{\hei} \\ 
  \eta^{A}_{\heiii}(T)   &= 8 \times 1.778 \times 10^{-29}  \ T \
  \frac{\lambda^{1.965}_{\heii}}
  {[1+(\lambda_{\heii}/0.541)^{0.502}]^{2.697}} \\ 
  \eta^{B}_{\hii} (T)  &= 3.435 \times 10^{-30}  \ T \
  \frac{\lambda^{1.97}_{\hi}}
  {[1+(\lambda_{\hi}/2.25)^{0.376}]^{3.72}} \\ 
  \eta^{B}_{\heii}(T)   &= \kb \ T  \recB_{\heii}
  \ = \ \kb \ T  \  1.26 \times 10^{-14}   \ \lambda^{0.75}_{\hei} \\ 
  \eta^{B}_{\heiii}(T)   &= 8 \times 3.435 \times 10^{-30}  \ T \
  \frac{\lambda^{1.97}_{\heii}}
  {[1+(\lambda_{\heii}/2.25)^{0.376}]^{3.72}}
\end{align*}
Collisional excitation cooling rate coefficients [$\coolU$]
\citep{{Cen:1992dq}}:
\begin{align*}
  \psi_{\hi}(T)   &= 7.5 \times 10^{-19}   
  \ \left( 1 + \sqrt{\frac{T}{10^5}} \right)^{-1}
  \ e^{-118 \ 348/T} \\
  \psi_{\heii}(T)   &= 5.54 \times 10^{-17} \ T^{-0.397}  
  \left( 1 + \sqrt{\frac{T}{10^5}} \right)^{-1}
  e^{-473 \ 638/T} 
\end{align*}
Bremsstrahlung cooling rate coefficients [$\coolU$]
\citep{{Osterbrock:2006ul}}:
\begin{align*}
  \theta_{\hii}(T)   &= 1.42 \times 10^{-27} \sqrt{T} & \\
  \theta_{\heii}(T)   &= 1.42 \times 10^{-27} \sqrt{T} & \\
  \theta_{\heiii}(T)   &= 4\times 1.42 \times 10^{-27} \sqrt{T} &
\end{align*}
Compton cooling/heating rate coefficient [$\ergs$]
\citep{{Haiman:1996iz}}, with $a$ the cosmological expansion factor
and $T_{\gamma}\equiv 2.727/a$ K the temperature of the cosmic
background radiation.:
\begin{align*}
  \varpi(T,a)   &= 1.017 \times 10^{-37} \ \left( \frac{2.727}{a}\right)^4
    \ \left( T- \frac{2.727}{a}\right)
\end{align*}
Dielectronic recombination cooling rate coefficient [$\coolU$]
\citep{{Black:1981uk}}:
\begin{align*}
  \omega_{\heii}(T)   &= 1.24 \times 10^{-13} T^{-1.5}
  e^{-470 \ 000/T} 
  \left( 1 +0.3 e^{-94 \ 000/T} \right) 
\end{align*}

\subsection{Cross sections}\label{rates_cs.sec}
\begin{table}
  \centering
  \caption[Photoionization energies and corresponding frequencies]
  {Photoionization energies and corresponding frequencies}
  \label{sigma_thresh.tbl}
  \begin{tabular}{c|cc}
    \toprule
    Ion species & $\epsilon_{ion}$  & $\nu_{ion}$\\
    \midrule
    \hi     & $\epsilon_{\hi}=13.60$ \ \ eV 
            & $\nu_{\hi}=3.288 \ 10^{15}$ \ \    $\sm$ \\
    \hei    & $\epsilon_{\hei}=24.59$ \ \ eV 
            & $\nu_{\hei}=5.946 \ 10^{15}$ \ \    $\sm$ \\
    \heii   & $\epsilon_{\heii}=54.42$ \ \ eV 
            & $\nu_{\heii}=1.316 \ 10^{16}$ \ \    $\sm$ \\
    \bottomrule
  \end{tabular}
\end{table}

Expressions for frequency dependent photoionization $\hi-$, $\heii-$
and $\heiii-$ cross sections are used in \ramsesrt{} to derive photon
group attributes from stellar populations (\App{SEDs.sec}). These
expressions are taken from \cite{{Verner:1996dm}}
\citep[via][]{{Hui:1997vj}} and are given in [$\rm{cm}^2$] as a
function of photon energy $\epsilon$ by
\begin{align}\label{PhI.eq}
  \sigma(\epsilon)=
      \sigma_0
      \left[ (x-1)^2+y_w^2  \right]
      \frac{y^{0.5P-5.5}}{(1+\sqrt{y/y_a})^P}
      &\mbox{, if $\epsilon \ge \epsilon_{ion}$},
\end{align}
(and $0 \ \rm{cm}^2$ otherwise), where
\begin{displaymath} 
  x\equiv \frac{\epsilon}{\epsilon_0}-y_0,
\end{displaymath} 
and
\begin{displaymath} 
  y\equiv \sqrt{x^2+y_1^2},
\end{displaymath} 
and the fitting parameters $\sigma_0$, $\epsilon_0$, $y_w$, $P$,
$y_a$, $y_0$, and $y_1$ are given in \Tab{sigma.tbl}. The ionization
energies $\epsilon_{ion}$ and corresponding frequencies $\nu_{ion}$
are given in \Tab{sigma_thresh.tbl}.


\end{document}


%% file: RamsesRT_RTtests.tex
The tests described in this section come from two papers that were
born out of a series of workshops on radiative transfer. Tests with
simple analytic results to compare to are hard to engineer in
radiative transfer, so the solution was to instead make simple tests
where the correct result is not necessarily well known but the results
of many different codes can instead be compared. Thus it is likeliest
that the correct results are usually where most of the codes agree,
and if a code stands out from all or most of the others in some way,
this would most likely be a problem with that particular code. These
tests have become sort of benchmark tests for RT codes, and most
publications that present new implementations use some or all of these
tests for validation.

The first paper is \cite{Iliev:2006jz}, hereafter known as \Ila{}
-- it describes four RT post-processing tests, i.e. with the
hydrodynamic advection turned off, and shows the results for 11 RT
codes. The second paper is \cite{{Iliev:2009kn}}, hereafter known as
\Ilb{} -- it describes three additional tests, and results for 9
codes, where the RT is coupled to the hydrodynamics. 

The tests results from \Ila{} and \Ilb{} are normally downloadable on
the web, but at the time of this writing the links have been down for
some time. However, Ilian Iliev has been kind enough to provide all
test results for one of the codes, the grid based short
characteristics code \C2R{}, which is described in detail in
\cite{{Mellema:2006cl}}. We thus present here \ramsesrt{} results with
comparisons to those of \C2R{}. The inclusion of the \C2R{} results in
the plots shown here should be useful to guide the eye if one then
wants to compare with the other codes in \Ila{} and \Ilb{}.

As prescribed by the test papers, all tests use hydrogen only gas. We
use smooth RT in the \ramsesrt{} runs for all tests, but remark that
turning off the smoothing has no discernible effect on the results
(only calculation speed). \joki{Unless noted otherwise in the
  following tests, the GLF intercell flux function is used
  (\Sec{Transport.sec}), and the the on-the-spot approximation is
  applied (\Sec{otsa.sec}). In all except test 1, where the radiation
  is monochromatic, the radiation energy distribution is assumed to be
  a $T_{\rm{eff}}=10^5$ K blackbody, which is approximated with three
  photon groups bordered by the hydrogen and helium ionization
  energies:}
\begin{equation}\label{groups.eq}
  ]13.6, \ 24.59], \ \ \ ]24.59, \ 54.42], \ \ \
  ]54.42, \ \infty[ \ \ \ \rm{eV}.
\end{equation}
\joki{A reduced speed of light fraction of $\fc = 1/100$ is used
  unless otherwise noted.  \AT{} contain an analysis of the effect of
  different light speeds in the first three tests from \Ila{}, and
  find the results start diverging non-negligibly somewhere between
  $\fc=10^{-2}$ and $10^{-3}$, which matches well with our analysis in
  \Sec{reduced_c.sec}. The prescribed resolution in the tests is
  $128^3$ cells, but in most tests we use adaptive refimenent for
  demonstrative purposes, with a coarse resolution of $64^3$ cells,
  and an \emph{effective} resolution of $128^3$ cells. We use a
  Courant factor of 0.8, so the RT timestep is set by}
\begin{equation}\label{dt_courant.eq}
  \dtrt =  0.8  \ \frac{\Delta x}{3\cred},
\end{equation}
\joki{where $\Delta x$ is the cell width and $\cred$ the reduced light
  speed. Taking as an example the test 1 setup, which has a box width
  of $6.6$ kpc, a simulation time of $500$ Myr, and a reduced light
  speed fraction $\fc=10^{-2}$, this translates into a (fine level)
  timestep length of $\sim4,500$ yr, so $\sim 10^5$ fine-level steps need
  to be computed to run the test.}

\subsection{\Ila{} test 0: \ \   The basic thermochemistry physics}
\label{Iliev_tests0.sec}
\begin{figure}
  \centering
  \includegraphics[width=.35\textwidth]{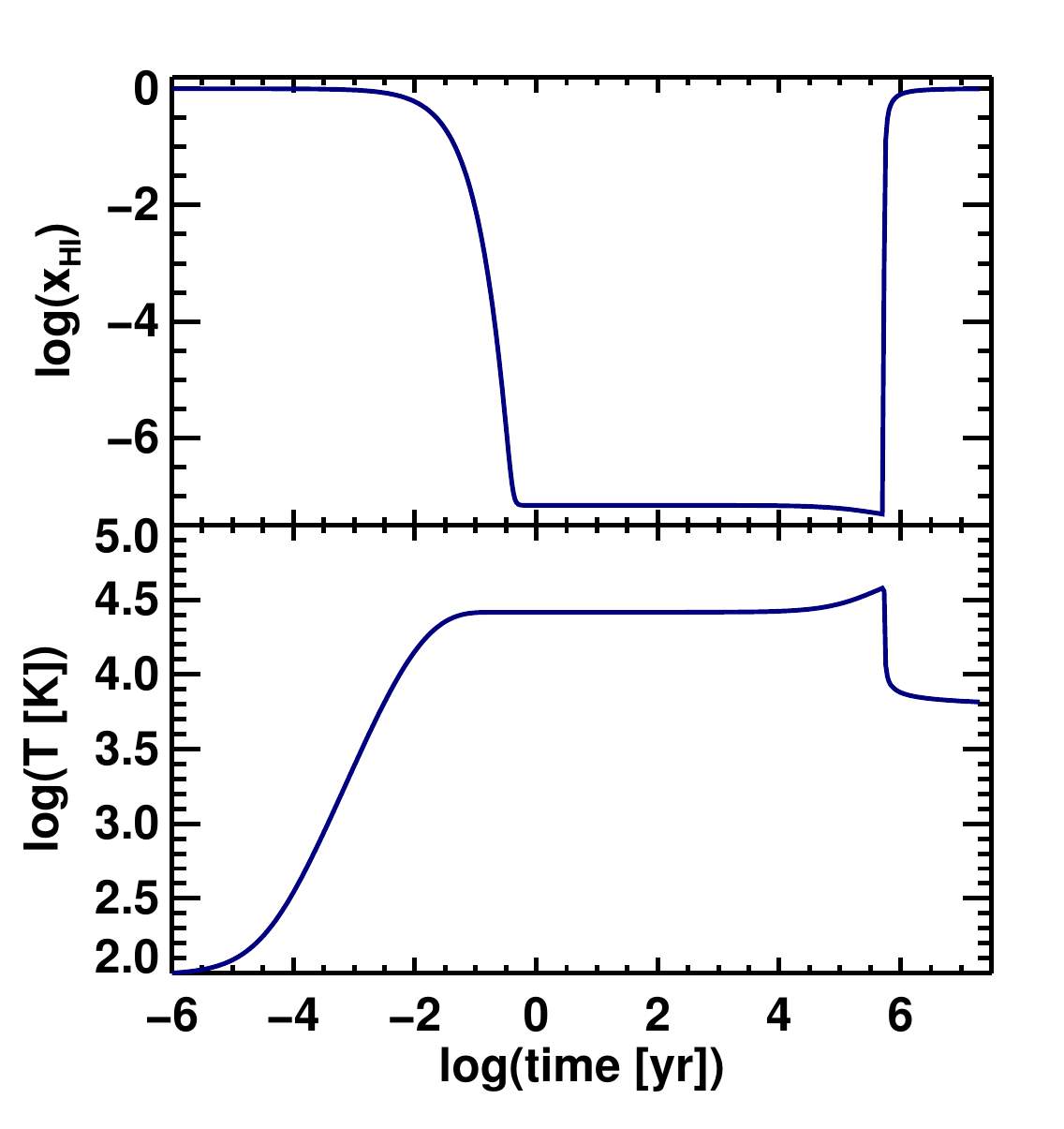}
  \caption[\Ila{} test 0]
  {\label{test0.fig}\Ila{} test 0. Single-zone photoheating
    and ionization with subsequent cooling and recombinations.}
\end{figure}

This is essentially a one-cell test of the non-equilibrium
thermochemistry and not radiative transfer per se, so it doesn't
really count with the rest of the comparison project tests (hence test
\textit{zero}). It is important nontheless since thermochemistry is a
major new component in \ramsesrt{}.

We start with completely neutral hydrogen gas with density $n_H=1 \;
\cci$ and temperature $T=100 \; \mathrm{K}$ at $t=0$. A photo-ionizing
flux of $F=10^{12} \ \flux$ with a $10^5$ K blackbody spectrum is
applied to the gas and maintained until $t=0.5$ Myr at which point it
is switched off. The run is continued for a further $5$ Myr, allowing
the gas to cool down and recombine. \joki{The run-time is separated
  into 500 logarithmically equally spaced timesteps, and the
  thermochemistry solver sub-cycles these timesteps adaptively (see
  \Sec{10prule.sec}). The photon flux is not evolved, i.e. it is kept
  fixed (until $0.5$ Myr) thoughout the integration.} The resulting
evolution of the neutral fraction and temperature of the gas is shown
in \Fig{test0.fig}. The evolution closely follows that of the codes
described in \Ila{}, with the exception of \simplex{} and \ffte{}
which stand out somewhat, and we don't see any sign of the
stiffness-induced oscillations that can be seen in the \crash{} code
test.

\subsection[\Ila{} test 1: \ \  Pure hydrogen isothermal 
  \hii{} region expansion]
{\Ila{} test 1: \ \  Pure hydrogen isothermal 
  HII region expansion}\label{iliev1.sec}
A steady monochromatic ($h\nu=13.6$ eV) source of radiation is turned
on in a homogeneous neutral gas medium, and we follow the resulting
expansion of a so-called Str\"omgren sphere of ionized gas. Heating and
cooling is turned off and the temperature is set to stay fixed at
$T=10^4$ K.

The box is a cube of width $L_{box}=6.6$ kpc. The gas density is
$\nh=10^{-3} \, \cci$ and the initial ionization fraction is $\xhi=1.2
\times 10^{-3}$, corresponding to collisional ionization
equilibrium. The radiative source is in the corner of the box and the
emission rate is $\dot{N}_{\gamma}=5 \times 10^{48} \, \emrate$.  The
simulation time is $t_{sim}=500$ Myr. To demonstrate on-the-fly AMR at
work (and speed up the runtime), we use a base resolution of $64^3$
cells, but allow for one level of further refinement, i.e. to the
effective prescribed resolution of $128^3$ cells. \joki{Typically, AMR
  refinement is applied on mass-related criteria, since massive
  structures are usually the objects of interest in
  simulations. However, since the density field is homogeneous in this
  test, we apply refinement on gradients in $\xhi$ and $\xhii$: two
  adjacent cells at positions $i$ and $i+1$ are refined if }
\begin{equation}\label{Il1_refine.eq}
  2 \left|
    \frac{x^i-x^{i+1}}{x^i+x^{i+1}}
  \right| > 0.8,
\end{equation}
where $x$ is either $x_{\hisub}$ or $x_{\hiisub}$.

\begin{figure}
  \centering
  \includegraphics[width=0.3\textwidth]
    {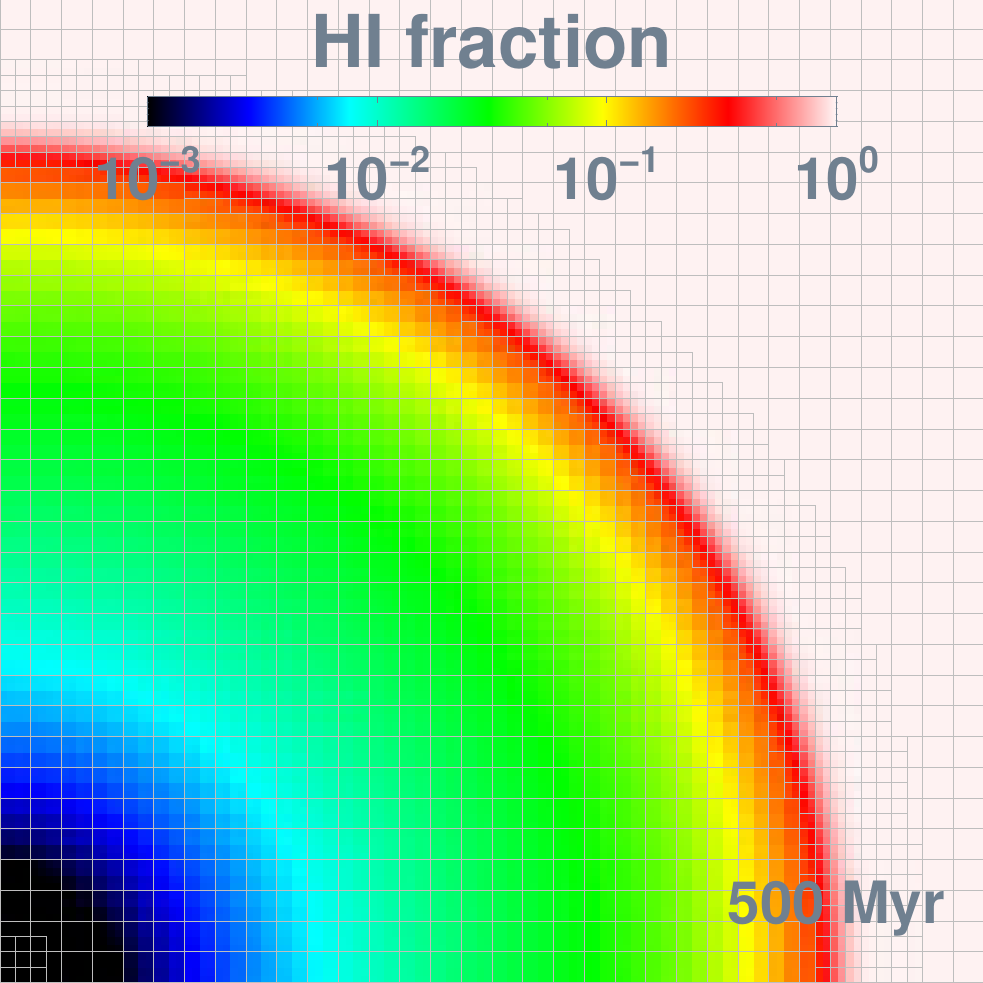}
  \caption[\Ila{} test 1]
  {\label{Il1Maps.fig}\Ila{} test 1. Map of the neutral fraction
    in a box slice at $z=0$, at $500$ Myr. Overplotted is the AMR
    grid, which is refined on the fly during the experiment from
    $64^3$ to $128^3$ cells effective resolution. Maximum refinement
    stays on the corner source throughout the run, and it adaptively
    follows the expansion of the I-front.}
\end{figure}

\begin{figure*}
  \centering
  \subfloat[]{\includegraphics[width=0.34\textwidth]
    {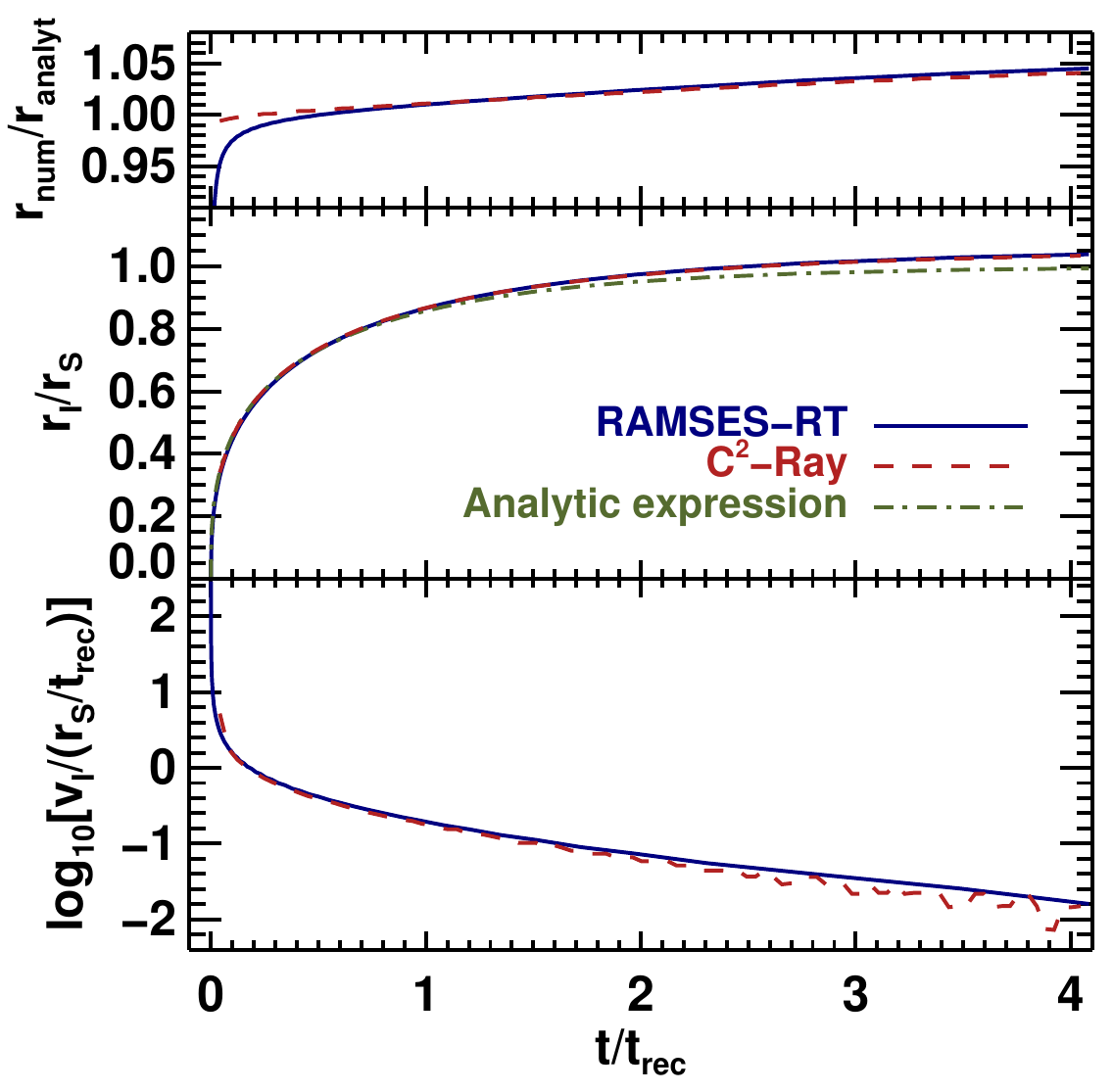}\label{Il1_Ifront.fig}}\hspace{-2mm}
  \subfloat[]{\includegraphics[width=0.34\textwidth]
    {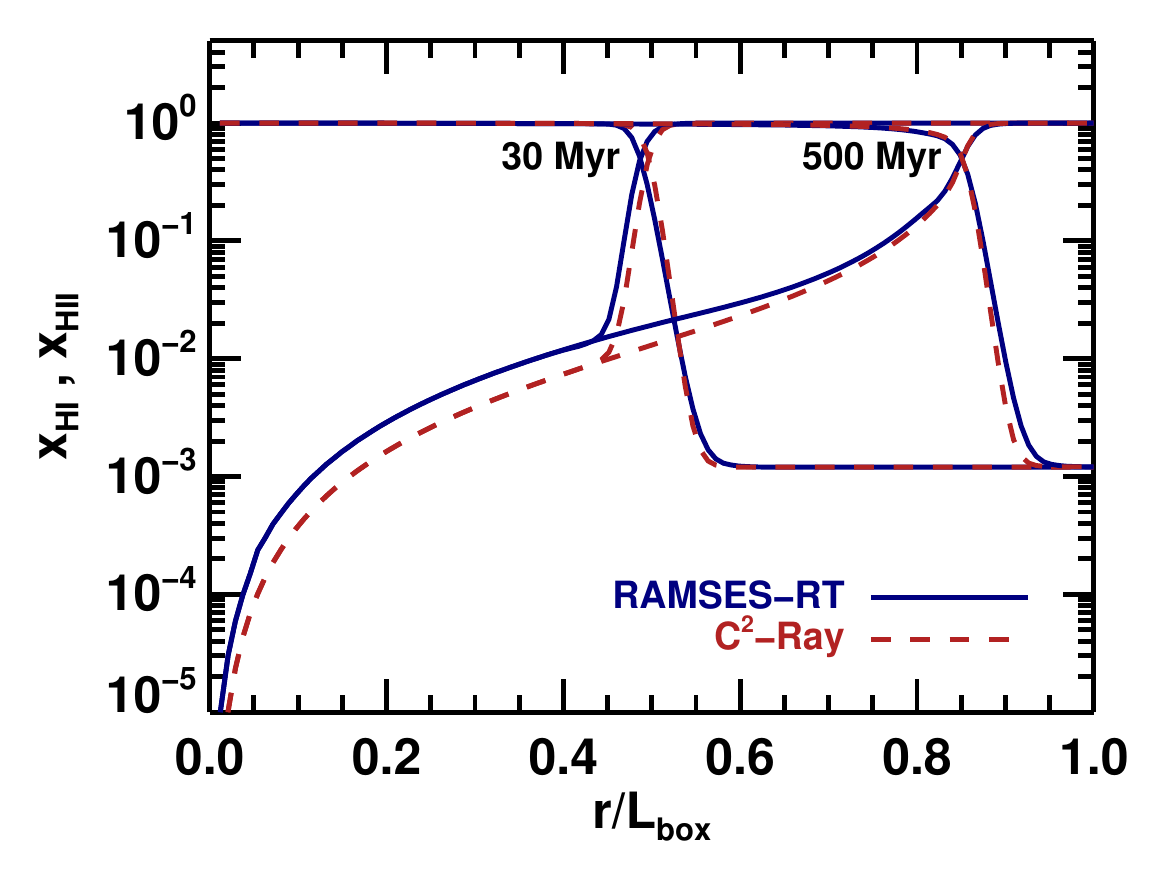}\label{Il1_Iprof.fig}}\hspace{-4mm}
  \subfloat[]{\includegraphics[width=0.34\textwidth]
    {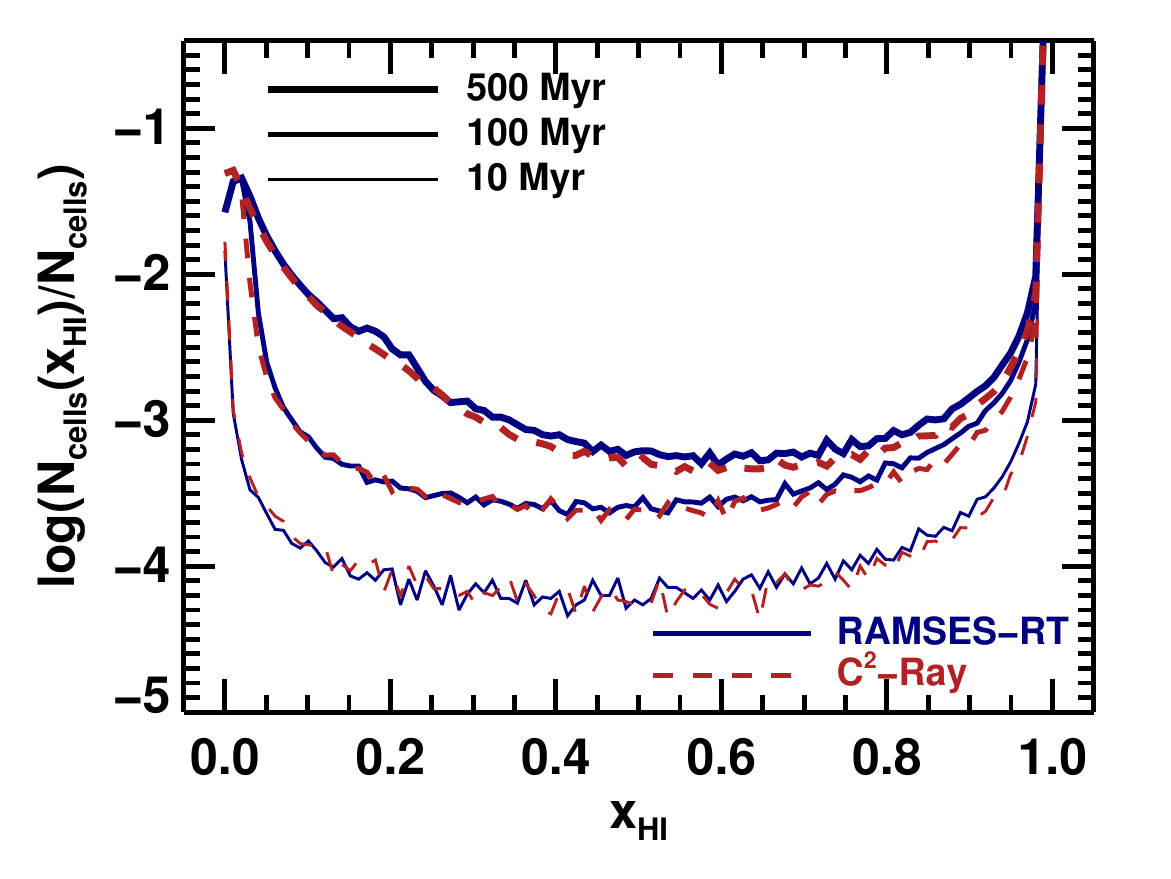}\label{Il1_hists.fig}}\hspace{-1.3mm}
  \caption{\Ila{} test 1. \textbf{(a)} Evolution of the I-front
    position and velocity. Blue solid lines show our result, red
    dashed lines show the \C2R{} result and green dot-dashed lines
    show the analytic expression. \textbf{(b)} Spherically averaged
    profiles for neutral fractions $\xhi$ and ionized fractions
    $\xhii$ at $30$ and $500$ Myr versus radius (in units of the box
    width $\Lbox$). \textbf{(c)} Histogram showing fractions of cells
    within bins of $\xhi$ at three simulation times.}
\end{figure*}




The Str\"omgren radius, $\rs$, is the radius of the ionization front
(I-front) from the center when steady state has been reached, and in
the case of fixed density and temperature it has the simple analytical
result shown in \Eq{Stromgren_radius.eq}.
In this result the I-front evolves in time according to
\begin{equation}\label{r_analyt.eq}
  \ri=\rs \left[ 1-e^{-t/\trec} \right]^{1/3},
\end{equation}
where $\trec=(\nh \recBH)^{-1}$ is the recombination time. For the
parameters of this experiment, $\trec=122.4$ Myr and $\rs=5.4$ kpc.

\Fig{Il1Maps.fig} shows maps at $500$ Myr of the neutral fraction,
with the grid refinement overplotted, in a box slice at $z=0$. The
Str\"omgren sphere is nicely symmetric and qualitatively it can be
seen to agree well with results from the RT codes described in \Ila{}
(their Fig. 6).

\Fig{Il1_Ifront.fig} shows the evolution of the I-front position and
velocity with \ramsesrt{} (solid blue), compared with the analytic
expression (green dot-dashed) and the result for the \C2R{} code (red
dashed), which is typical for the RT code results presented in \Ila{}
and does not stand out particularly in this test. Our result can be
seen to match the \C2R{} one, though we have an initial lag due to the
reduced speed of light that can best be seen in the top plot showing
the fraction of the numerical result's I-front radius versus
$\rs$. The analytic $\ri$ is typically ahead of $\rs$ by $\la 5\%$,
which is simply because the analytic result is step-like with complete
ionization within $\rs$ and none outside, whereas the real result has
a gradually evolving ionization profile with radius. Indeed,
\cite{{Pawlik:2008kk}} computed the exact analytic result to this
problem, accounting for an equilibrium neutral fraction inside the
Str\"omgren sphere, and found an equilibrium I-front radius which is
exactly $1.05 \ \rs$.

\Fig{Il1_Iprof.fig} shows spherically averaged radial profiles of the
gas ionization state at $30$ and $500$ Myr. Again we see a good match
with the \C2R{} result. There is still a little lag in the I-front
position at $30$ Myr due to the RSLA and $\xhi$ is somewhat lower
inside the Str\"omgren sphere in \ramsesrt{}. However, the \C2R{}
result stands out a little in this test in \Ila{} as being most
effective at ionizing the gas within the Str\"omgren sphere (i.e. has
the lowest values of $\xhi$), and the \ramsesrt{} result is
typical of the \Ila{} codes' results in this plot.

\begin{figure}
  \centering
  \includegraphics[width=0.37\textwidth]
    {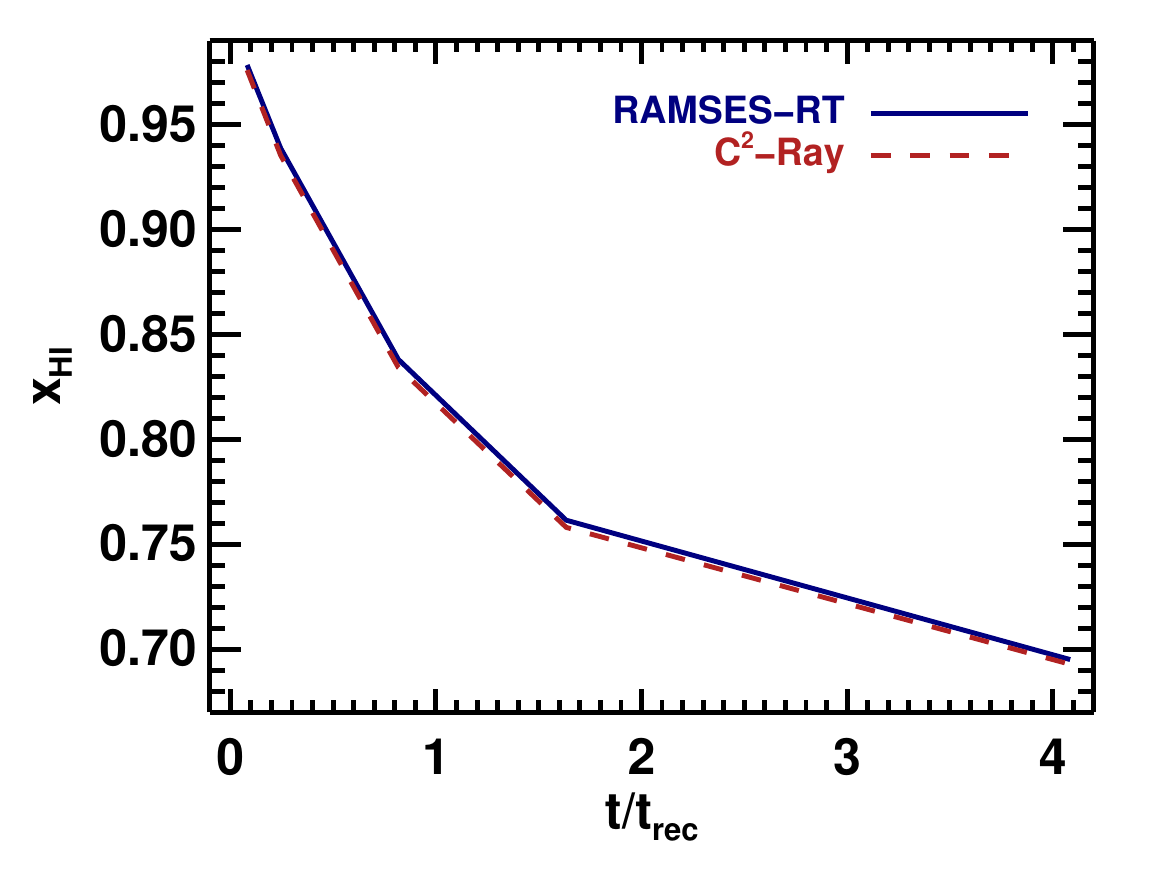}\hspace{-1mm}
    \caption{\label{Il1_xEvol.fig}\Ila{} test 1. Evolution of the
      globally averaged neutral fraction.}
\end{figure}

\begin{figure*}
  \centering
  \subfloat{\includegraphics[width=0.23\textwidth]
    {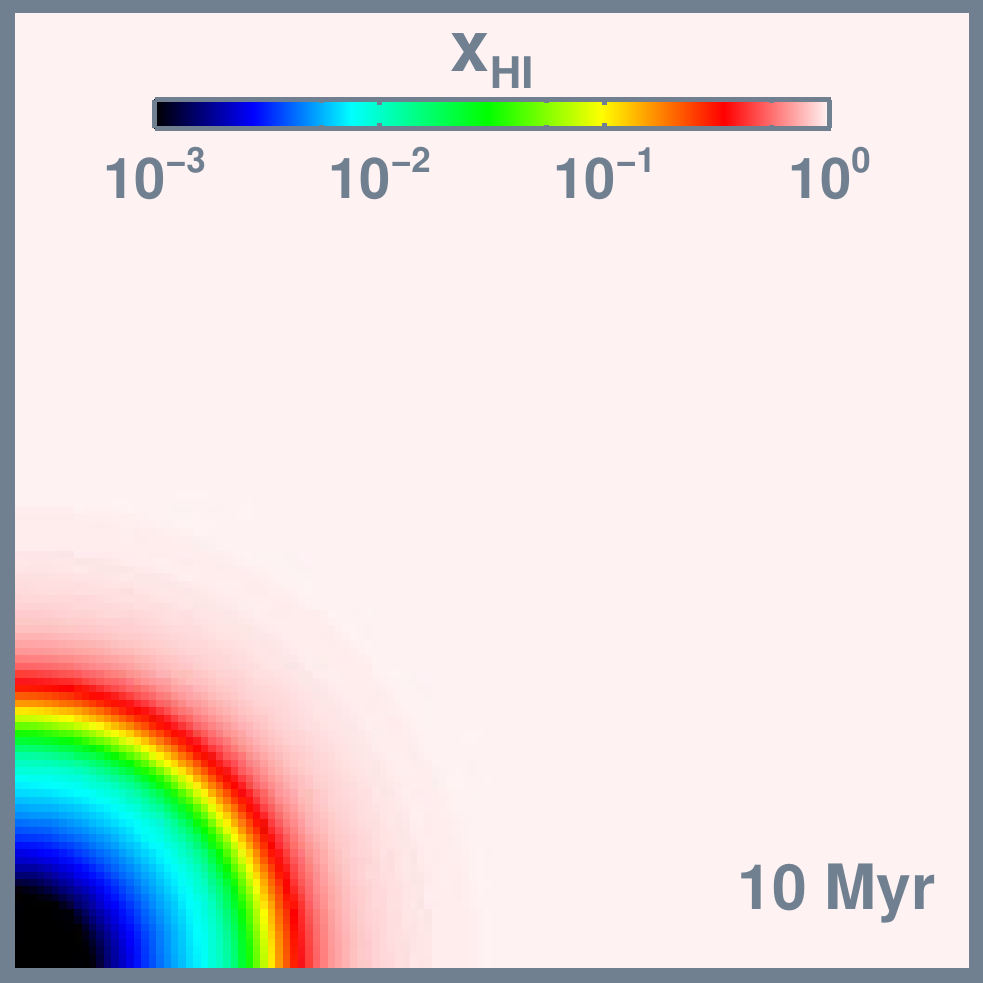}}\hspace{-1.3mm}
  \subfloat{\includegraphics[width=0.23\textwidth]
    {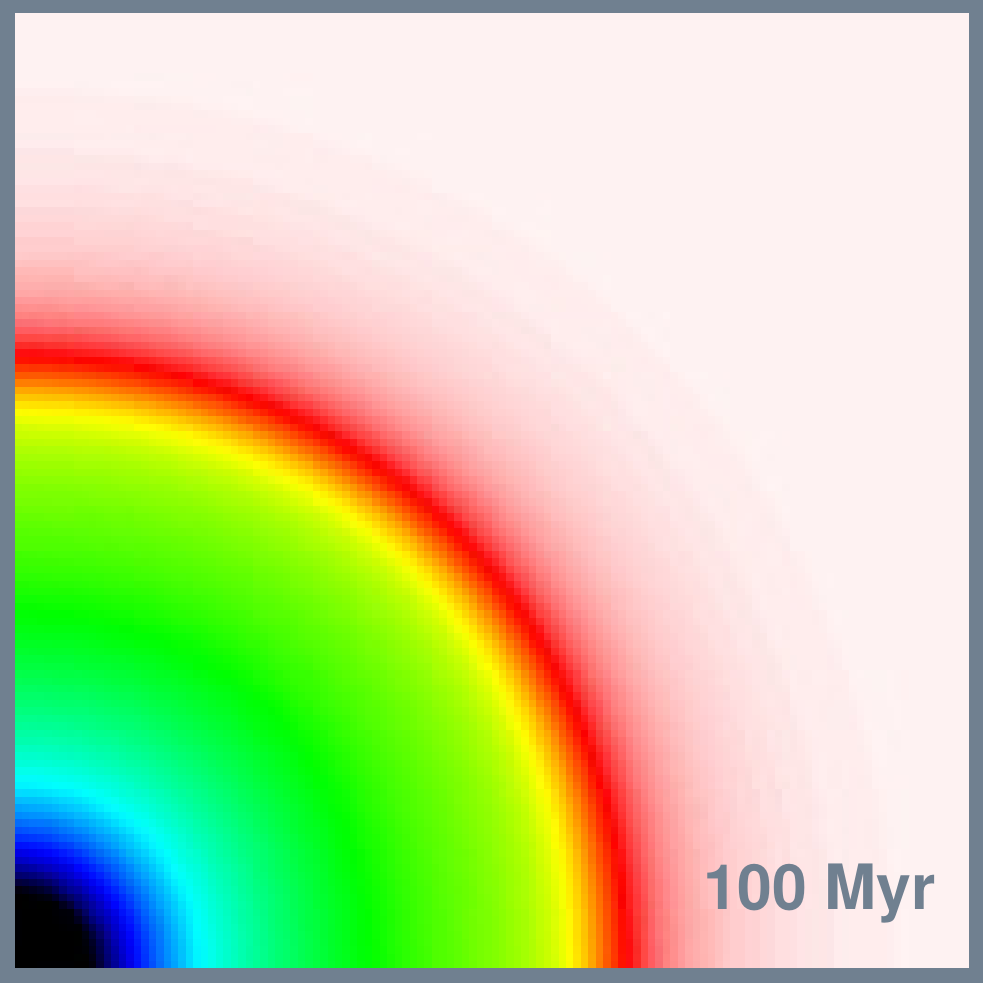}}\hspace{ 0.mm}
  \subfloat{\includegraphics[width=0.23\textwidth]
    {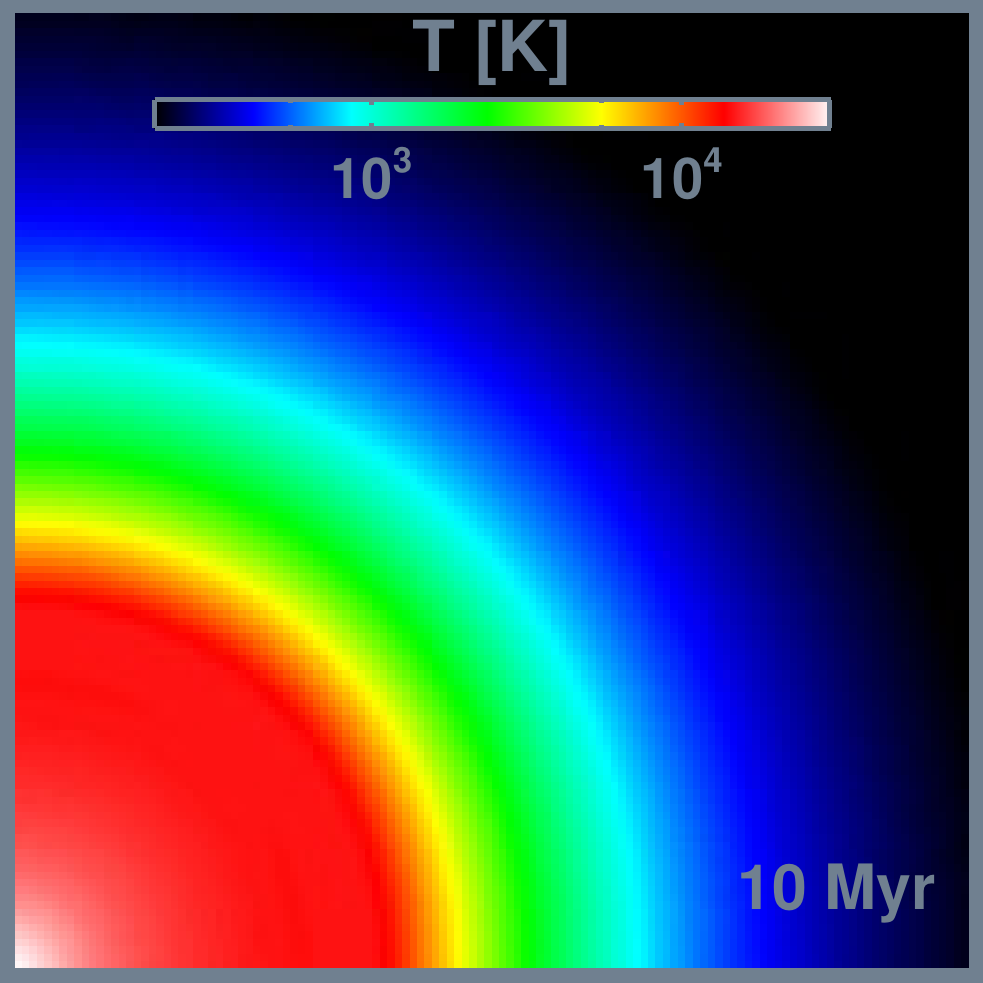}}\hspace{-1.3mm}
  \subfloat{\includegraphics[width=0.23\textwidth]
    {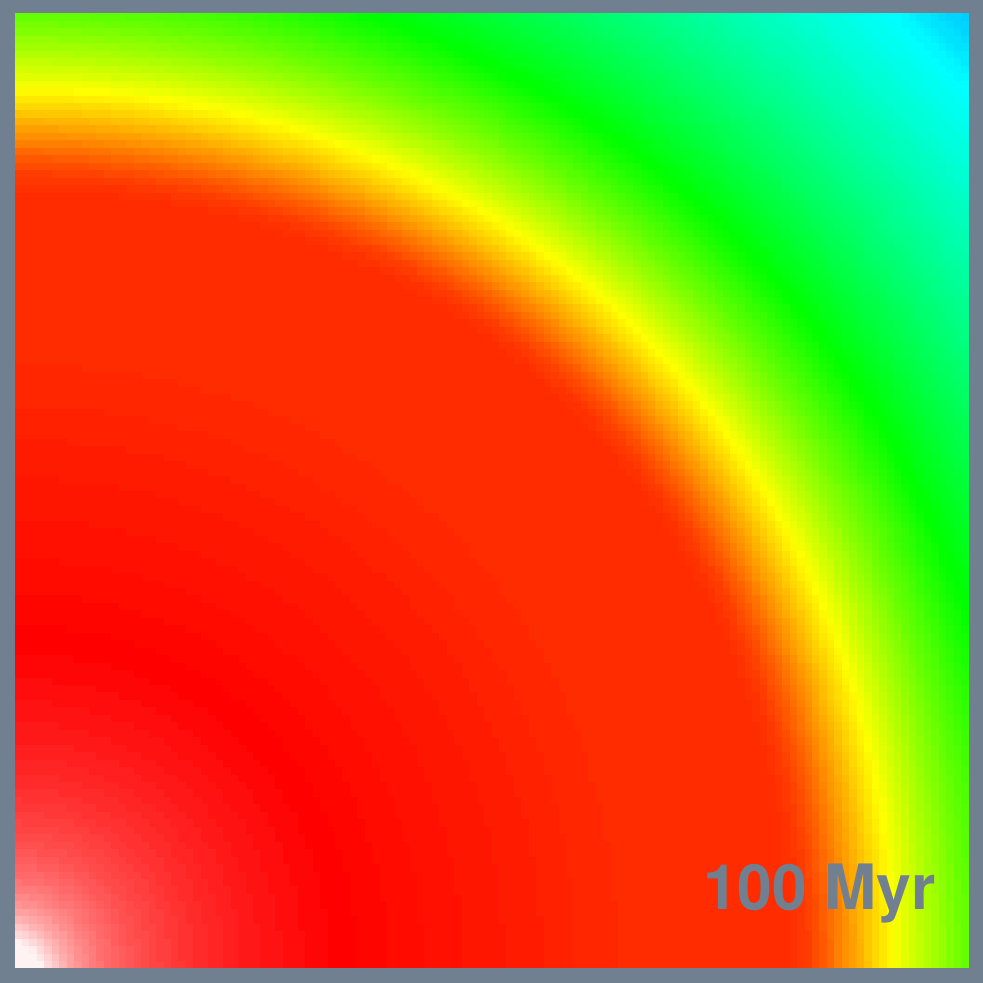}}\hspace{-1mm}
  \vspace{0.5mm}
  \caption[\Ila{} test 2 - maps]
  {\label{Il2maps.fig}\Ila{} test 2. Maps showing slices at $z=0$ of
    the neutral fraction and temperature at 10 Myr and 100 Myr.}
\end{figure*}

A further comparison is made in \Fig{Il1_hists.fig}, here comparing
ionization fraction histograms at three simulation times. Again the
\ramsesrt{} result closely matches the \C2R{} one, whose histograms
fall into a group with the codes \ift{}, \flashhc{} and \ffte{} that
stand out a little in \Ila{} (Fig. 9) as having less frequent
intermediate neutral fractions than the other codes.

Finally for this test, \Fig{Il1_xEvol.fig} shows a comparison with
\C2R{} of the globally averaged neutral fraction as a function of
time. It is a close match, and the \C2R{} result is here typical
for the \Ila{} codes.

All in all, there is nothing out of the ordinary in the \ramsesrt{}
result for \Ila{} test 1, except for a slight initial delay of the
I-front which is to be expected due to the RSLA. 

\joki{We note that performing this test with the full prescribed
  $128^3$ resolution, rather than using AMR like we've done here, has
  no discernible effect on the results. In the AMR run, the number of
  fine level cells is maximally (at the end of the run) $15\%$ of the
  number of fine level cells in the non-AMR run, and the computation
  time is $~30\%$ of that in the non-AMR run. The cost of the
  experiment (with AMR) is on the order of 50 cpu
  hours\footnote{defined as the wall-clock hours of the run, times the
    number of cpus used.}, which is a lot for a simple test in which
  little actually happens: for much of the run, the I-front is moving
  towards a stand-still at speeds which are much slower than our
  reduced speed of light ($\fc=0.01$), so barring the RT Courant
  condition, the timesteps taken could have dramatically increasing
  length towards the end of the test. Implicit transport solvers can
  take advantage of this (almost) static situation by on-the-fly
  adapting the timestep length (which is in the case of implicit
  solvers not constrained by the Courant condition), so presumably an
  implicit solver can run this test (and most of the tests described
  in this work) with considerably less computation than we
  do. However, in more realistic cosmological scenarios, such steady
  regimes simply do not happen over times longer than the typical age
  of stellar populations, which is on the order of $10$ Myr ($50$
  times shorter than the run-time for this test). Furthermore, stellar
  populations typically are turning on and off on even shorter
  timescales than that thoughout the simulation volume, which limits
  the dynamical time of ionization fronts even further. This
  presumably constrains the main advantage (possible long time steps)
  of implicit solvers severely, since even though they are not
  constrained by Courant-like conditions, they still need to resolve
  dynamical timescales.}

\begin{figure*}
  \centering
  \subfloat[]{\includegraphics[width=0.33\textwidth]
    {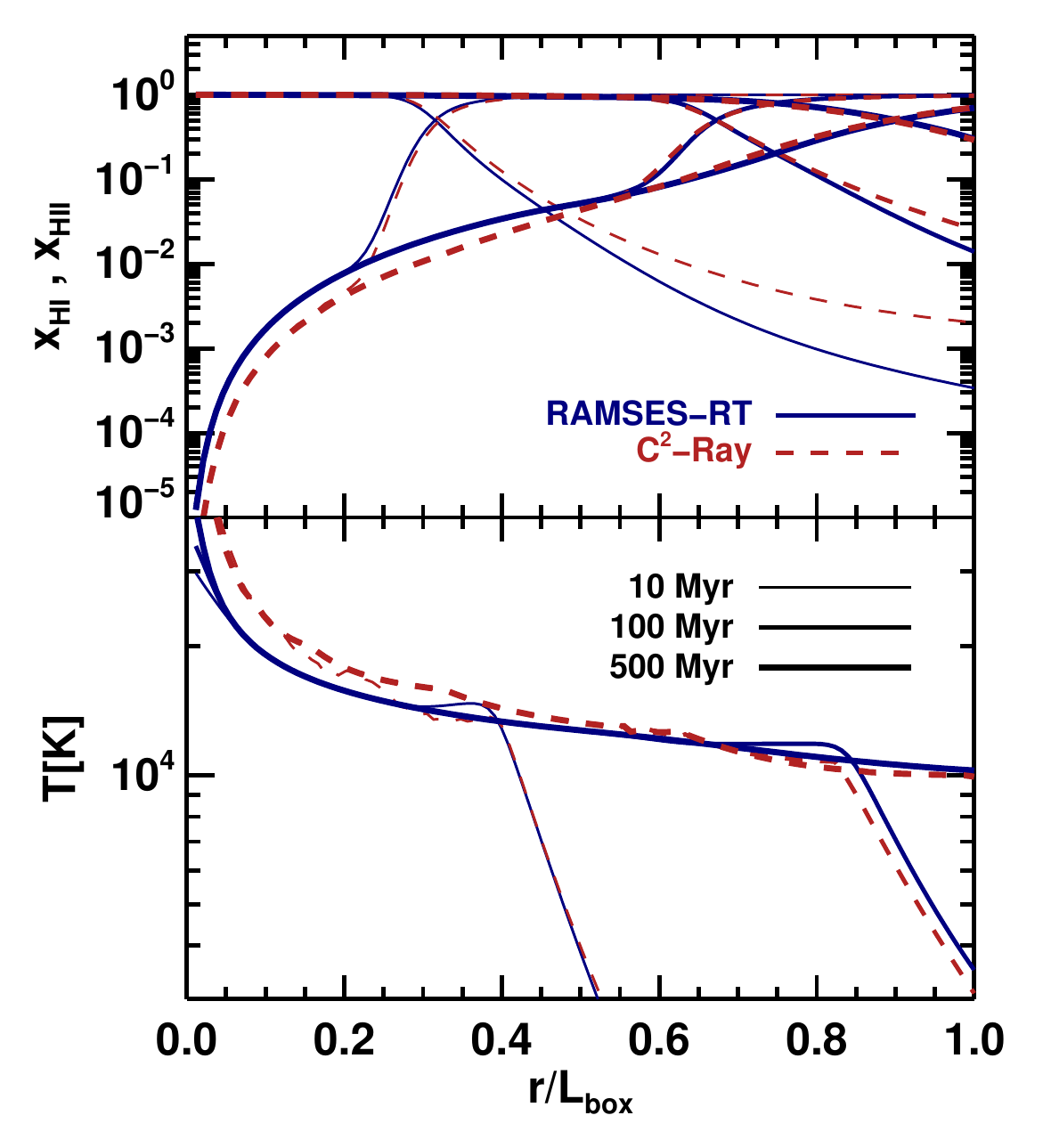}\label{Il2_profs.fig}}\hspace{-1.3mm}
  \subfloat[]{\includegraphics[width=0.33\textwidth]
    {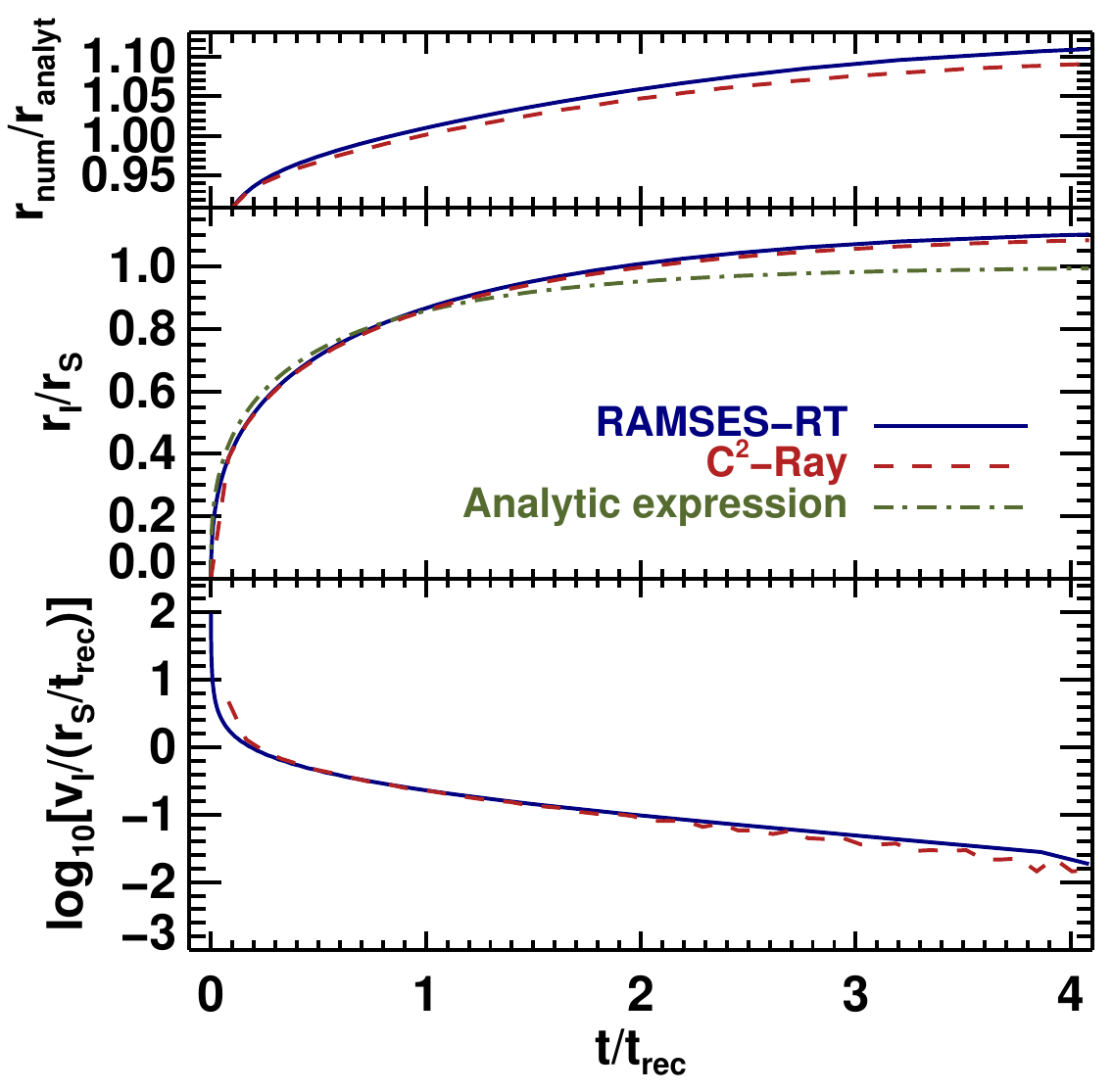}\label{Il2_ifront.fig}}\hspace{-1mm}
  \subfloat[]{\includegraphics[width=0.33\textwidth]
    {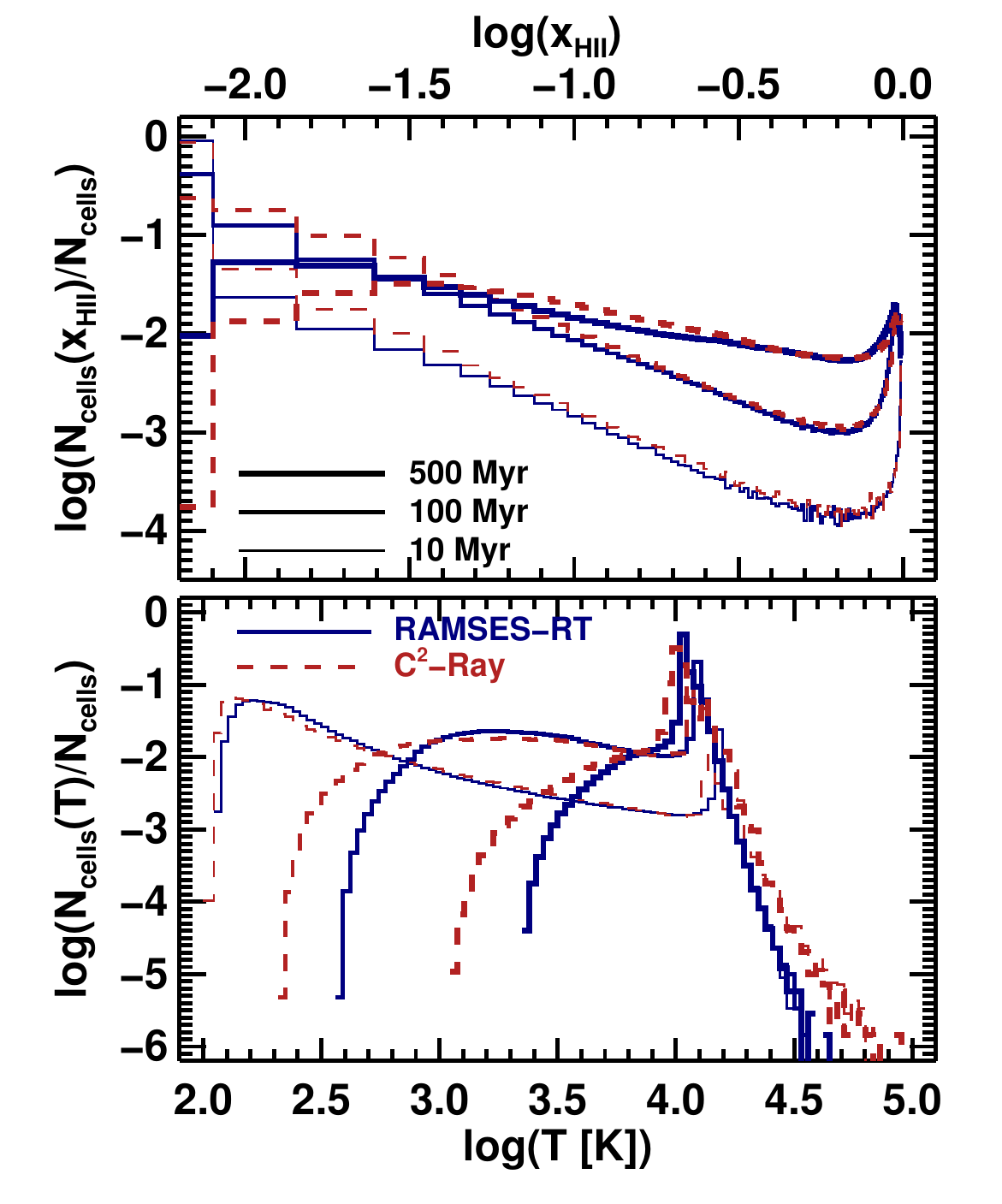}\label{Il2_hists.fig}}\hspace{-1.3mm}
  \caption{\Ila{} test 2. \textbf{(a)} Evolution of the temperature
    and ionization state profiles. \textbf{(b)} Evolution of the
    ionization front. The top plot shows the ratio of the radius of
    the I-front in the tests, $r_{\rm{num}}$ versus the time-evolving
    radius $r_{\rm{analyt}}$ in the analytic result from test 1
    (\Eq{r_analyt.eq}). The middle plot shows the ratio of the test
    I-front radius versus the steady-state radius in the same analytic
    result (\Eq{Stromgren_radius.eq}). The bottom plot shows the speed
    of the I-front, $\vi$ in units of a `characteristic' speed, given
    by $\rs/\trec$.  \textbf{(c)} Histograms of temperature and
    ionized fraction.}
\end{figure*}

\subsection[\Ila{} test 2: \ \ \hii{} region expansion and the
temperature state] {\Ila{} test 2: \ \ HII region expansion and the
  temperature state}
The setup here is the same as in \Ila{} test 1, except for the
following points:
\begin{itemize}
\item We allow for cooling and photo-heating of the gas, i.e. the
  temperature is no longer constant, and the analytic result,
  \Eq{Stromgren_radius.eq} no longer applies (because of the
  non-constant recombination rate).
\item The initial temperature is $100$ K.
\item The initial ionization fraction of the gas is
  $\xhii=10^{-6}$. It should be fully neutral according to the test
  recipe \joki{in \Ila{}}, but this is (the default) minimum value for
  $\xhii$ in \ramsesrt{}, that exists in order to keep bounds on the
  subcycling of the thermochemistry. \joki{In any case, the specific
    value is not critical to the test results, as long as it is low.}
\item The radiation source is a $T=10^5$ K blackbody, modeled with the
  three photon groups defined by \eq{groups.eq}. The emission rate is
  the same as before, $\dot{N}_{\gamma}=5 \times 10^{48} \, \emrate$.
\item We don't use grid refinement in this test. The grid is
  homogeneous and the resolution is $128^3$ grid cells, as prescribed
  in \Ila{}.
\end{itemize}

Slice maps at $z=0$ of the neutral fraction and temperature are shown
in \Fig{Il2maps.fig}. Both the ionization and heating fronts are
smooth and symmetric, and the maps agree qualitatively with other
codes in \Ila{} (Figs. 11-14). In comparison with the same test with
\aton{} (\AT, Fig. 3), both fronts are clearly much thicker here,
which is due to our multi-frequency implementation (whereas \aton{}
used one photon group). More detailed comparison with the \Ila{} codes
can be made through the ionization state and temperature plots in
\Fig{Il2_profs.fig}, where we include the \C2R{} result. The
ionization state profile develops very similarly to that of \C2R{},
though we have less ionization on both sides of the front, especially
on the outer side where the difference in $\xhii$ is as high as a
factor of ten. Presumably this is due to the different implementations
of multi-frequency photo-heating and cooling. The thermal profiles are
also similar to \C2R{}, though we have considerably lower (up to a
factor of two) temperatures on the inside of the I-front, and
conversely higher temperatures on the outside. As can be seen in
Fig. 17 in \Ila{}, \C2R{} has the strongest heating of any code on the
inside of the I-front in this test and most codes have stronger
heating on the outside, so our thermal profiles (as the ionization
state profiles) are fairly typical of the ones presented in \Ila{} for
this test.

\Fig{Il2_ifront.fig} shows the evolution with time of the ionization
front, compared with \C2R{} and the analytic result from test 1. The
front moves more slowly here than in test 1 due to the lower initial
temperature, so we no longer lag behind in the initial front
propagation. Our front propagates slightly further than in \C2R{}, and
ends at almost exactly the same radius as the \ffte{} code, which has
the furthest expanding I-front of any code in this test in
\Ila{}. Still the difference between the codes is small, with the
ratio between the numerical and analytic results
($r_{\rm{num}}/r_{\rm{analyt}}$) ranging between $1.01$ and $1.11$.

\Fig{Il2_hists.fig} shows histograms of the ionized fraction and
temperature at different times in the test for \ramsesrt{} and
\C2R{}. The ionized fraction histograms are quite similar, the biggest
difference being a higher fraction of almost completely neutral gas
$\xhii \la 10^{-2}$ in \ramsesrt{}, which we already saw in
\Fig{Il2_profs.fig} (top) beyond the I-front. The temperature
histogram for \ramsesrt{} differs a bit from \C2R{} in having less
extreme temperatures (\C2R{} has both hotter gas and colder gas) but
are very similar to those for the codes \art{}, \rsph{} and \crash{}
in \Ila{}.

Finally, \Fig{Il2_xEvol.fig} shows the time evolution of the volume
averaged neutral fraction in \ramsesrt{} and \C2R{}, and here we see a
close match. There is quite a lot of discrepancy between the different
codes in the analogue plot in \Ila{} (Fig. 20), with 3 groups of
results, and our result closely follows those of \C2R{}, \crash{} and
\rsph{}.

As with test 1, there is nothing out of the ordinary in the \ramsesrt{}
result for \Ila{} test 2, except perhaps for an ever so slightly further
advanced I-front than most codes in \Ila{} have.

\begin{figure}
  \centering
  \includegraphics[width=0.35\textwidth]
    {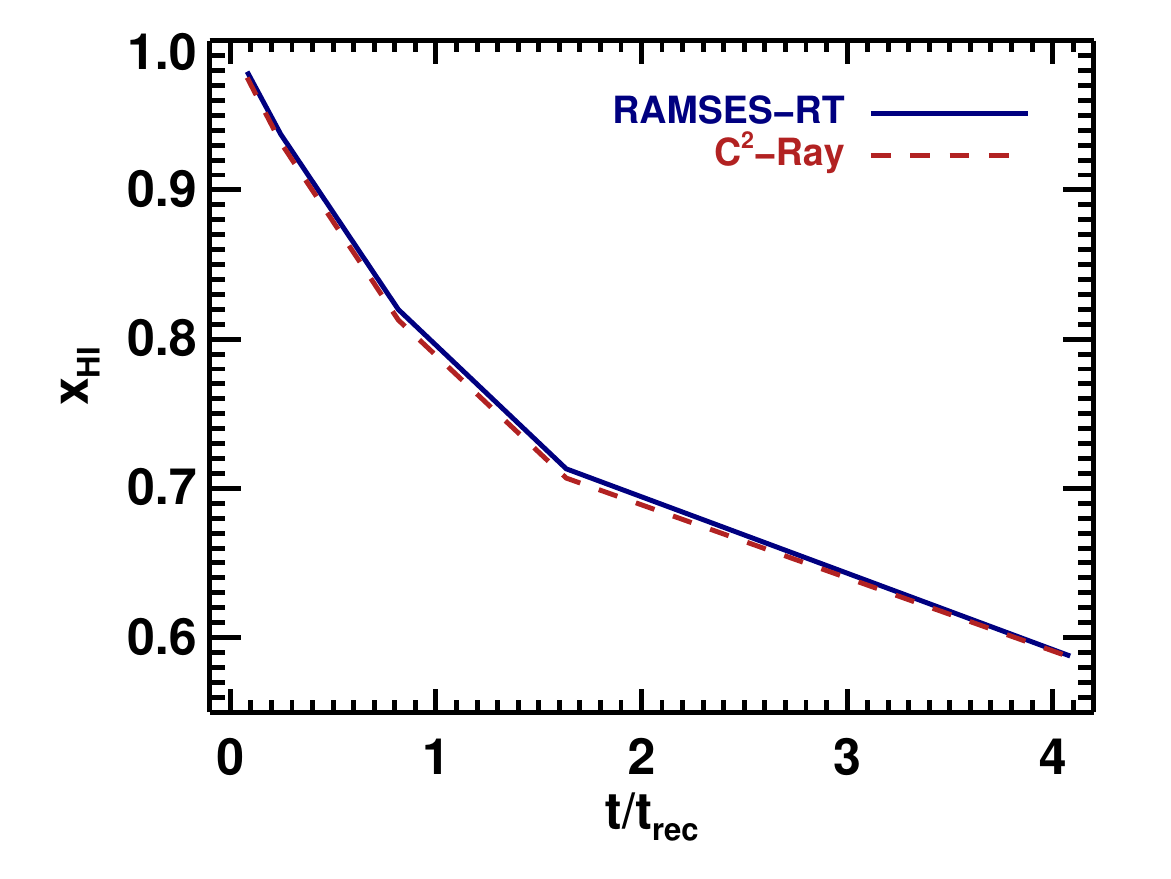}\hspace{-1mm}
    \caption{\label{Il2_xEvol.fig}\Ila{} test 2. Time evolution of the
      volume average neutral fraction.}
\end{figure}

\subsection{\Ila{} test 3: \ \ I-front trapping in a dense clump and the
  formation of a shadow}\label{iliev3.sec}
\begin{figure*}
  \centering
  \vspace{-3mm}
  \subfloat{\includegraphics[width=0.25\textwidth]
    {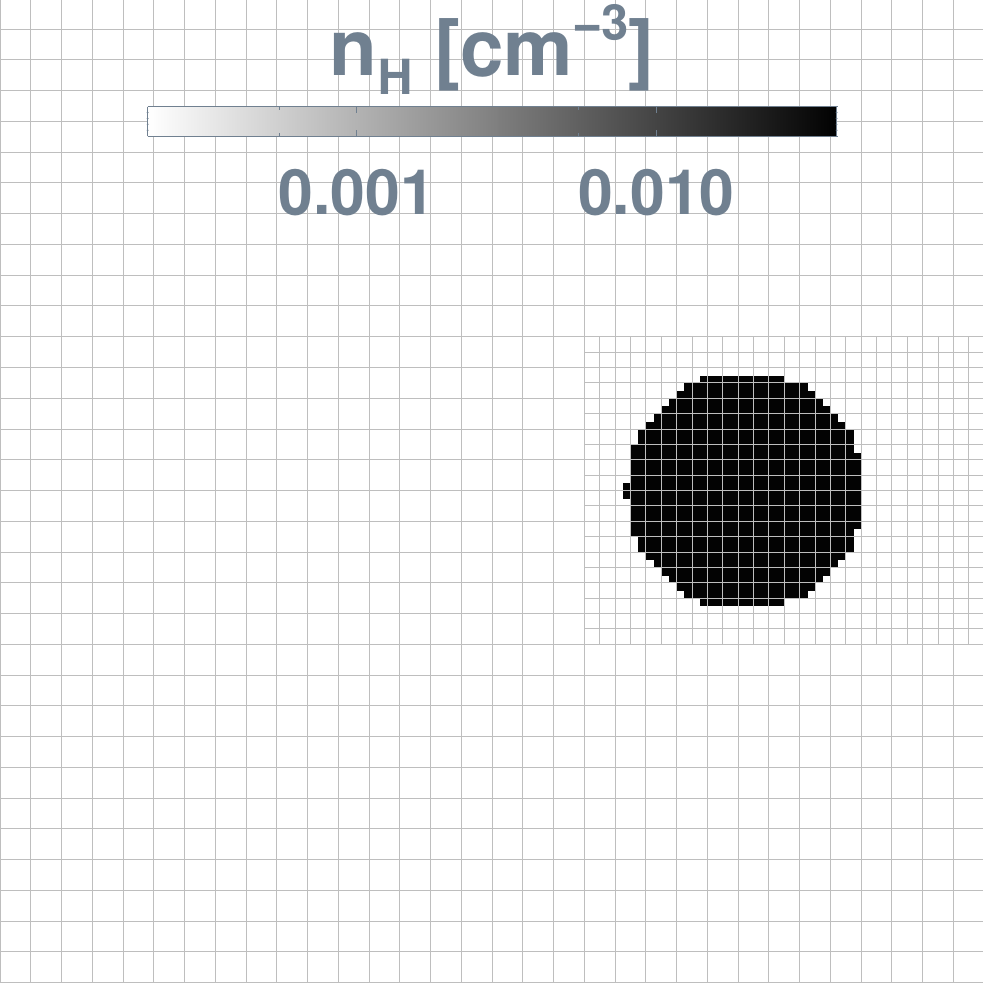}}  \vspace{-3mm}\hspace{0.mm}
  \hspace{100.mm} 
  \vspace{-4mm}
  \subfloat{\includegraphics[width=0.25\textwidth]
    {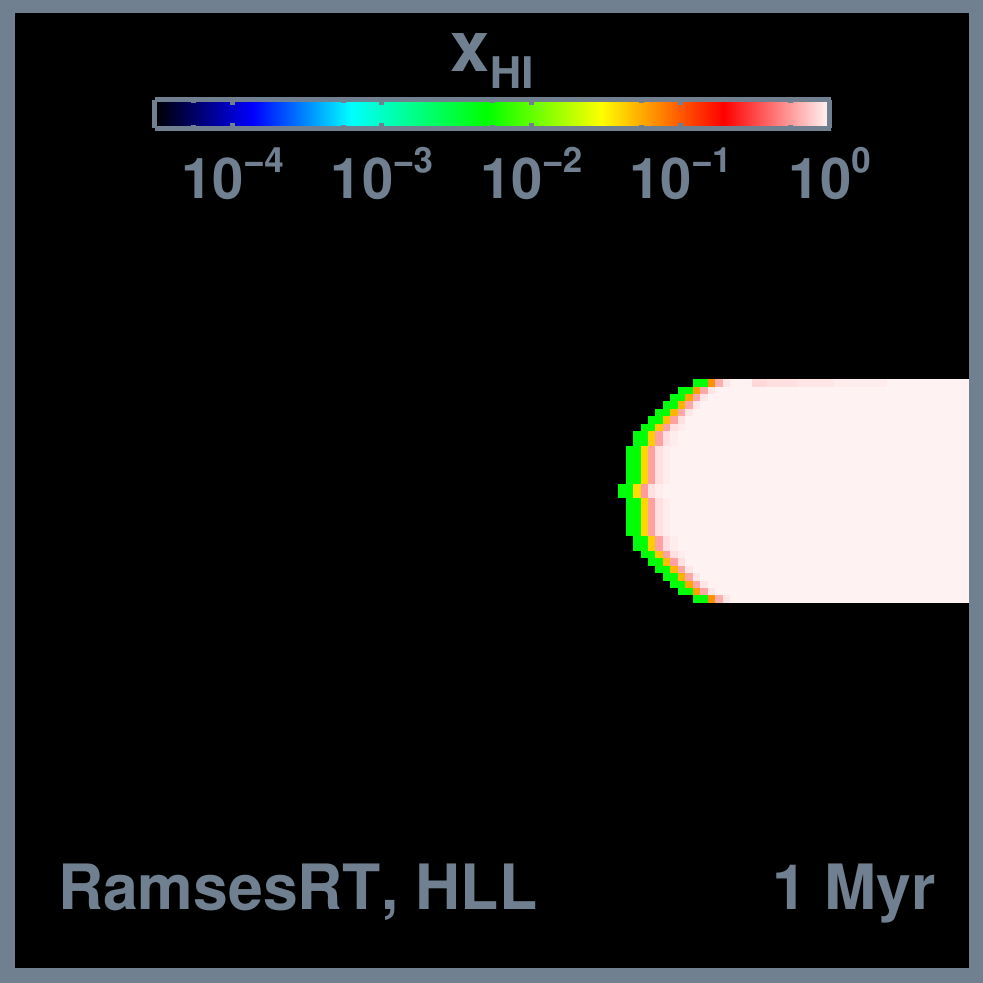}}\hspace{-1.3mm}
  \subfloat{\includegraphics[width=0.25\textwidth]
    {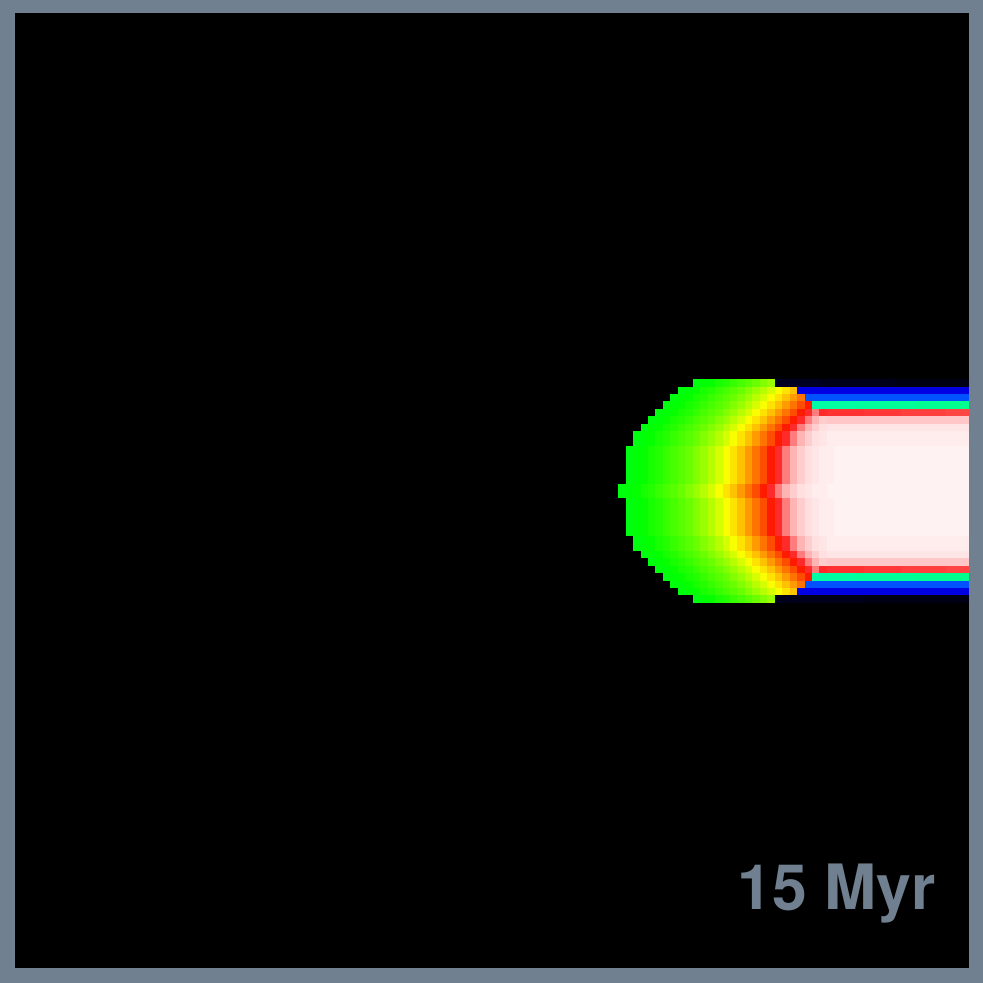}}\hspace{-1.3mm}
  \subfloat{\includegraphics[width=0.25\textwidth]
    {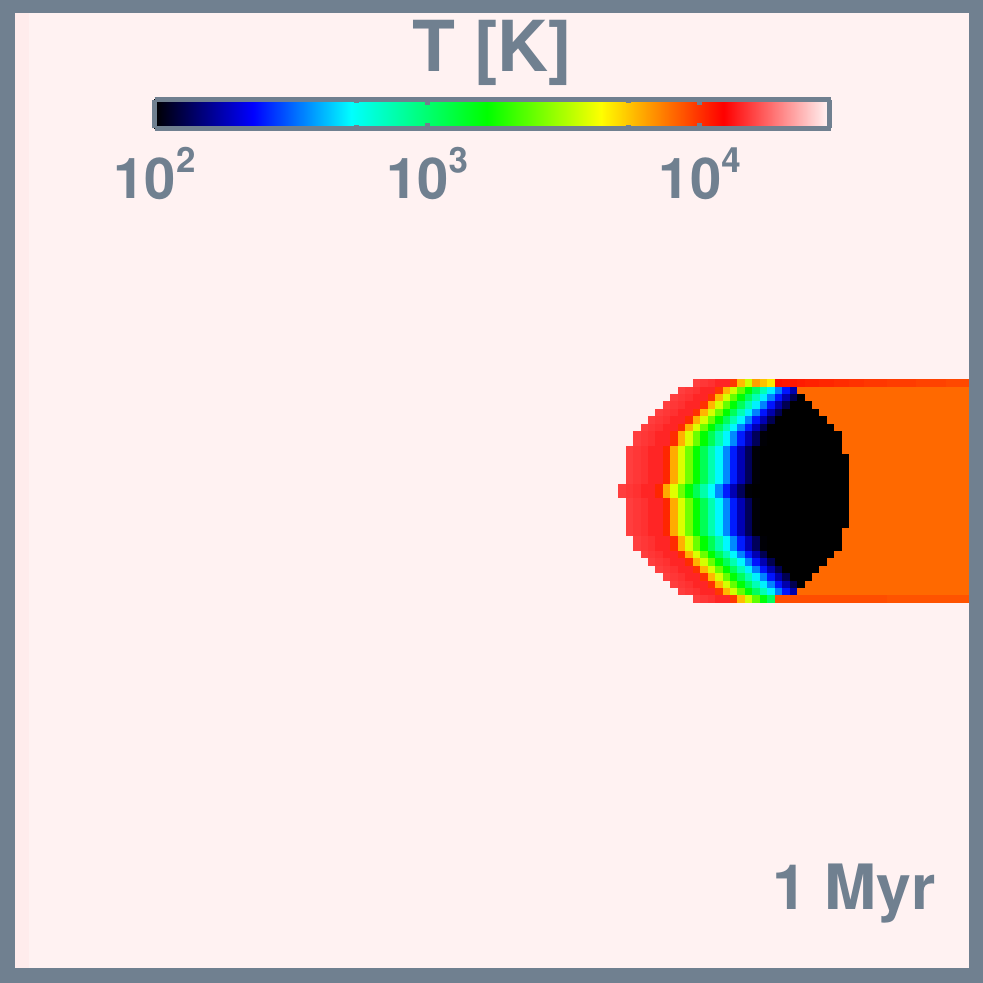}}\hspace{-1.3mm}
  \subfloat{\includegraphics[width=0.25\textwidth]
    {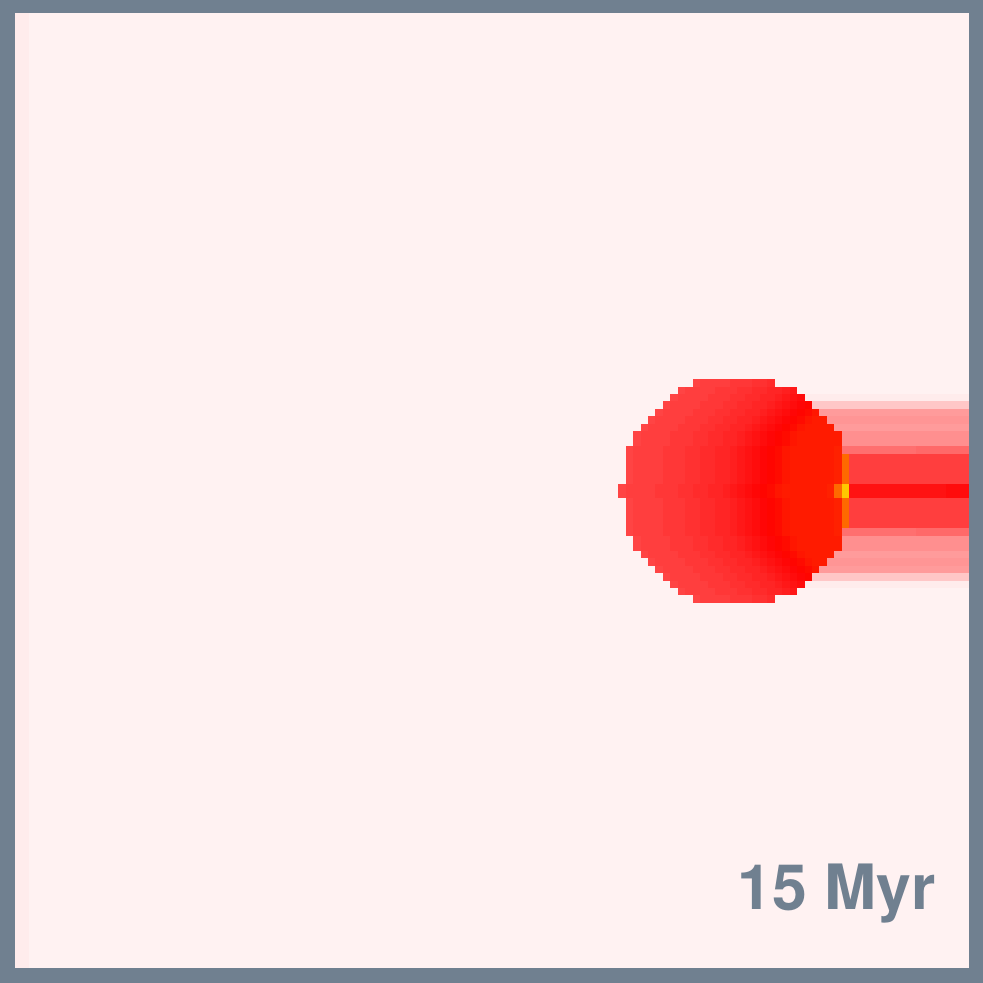}}\hspace{-1.3mm}
  \hspace{1mm}\vspace{-4mm}
  \subfloat{\includegraphics[width=0.25\textwidth]
    {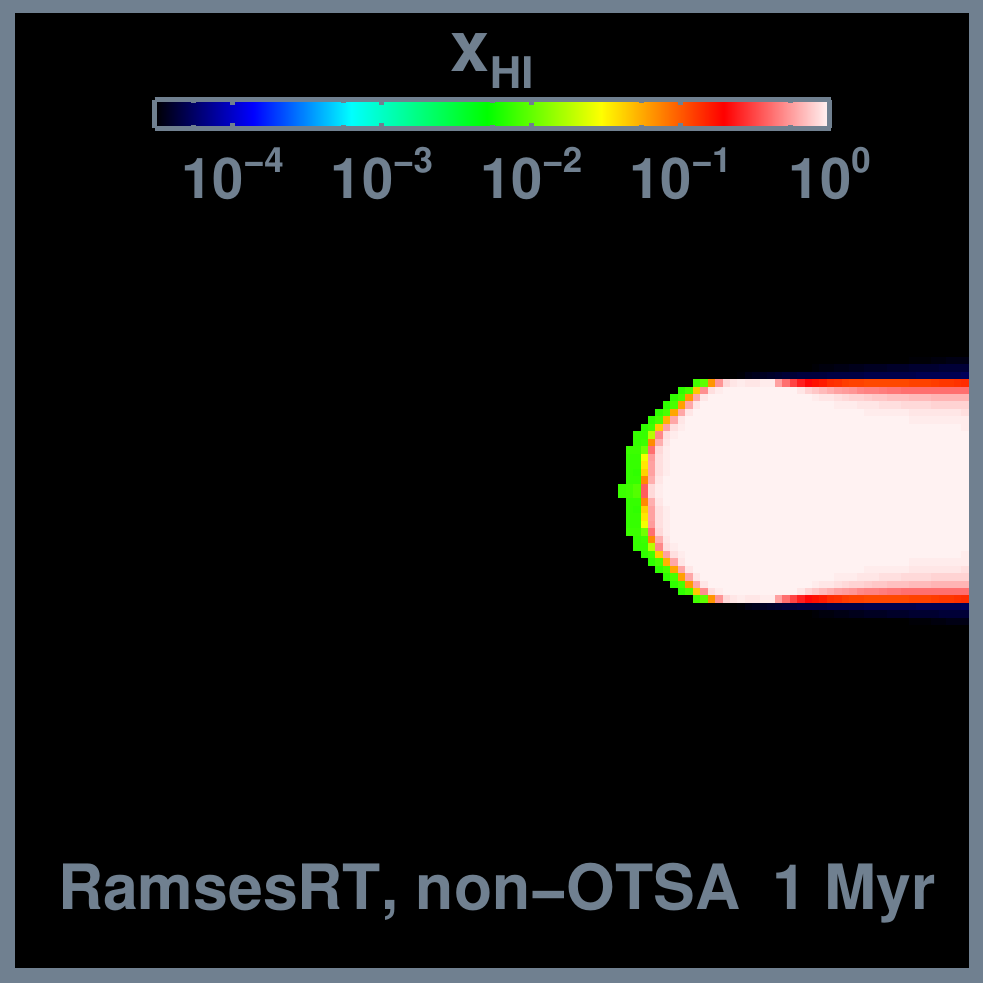}}\hspace{-1.3mm}
  \subfloat{\includegraphics[width=0.25\textwidth]
    {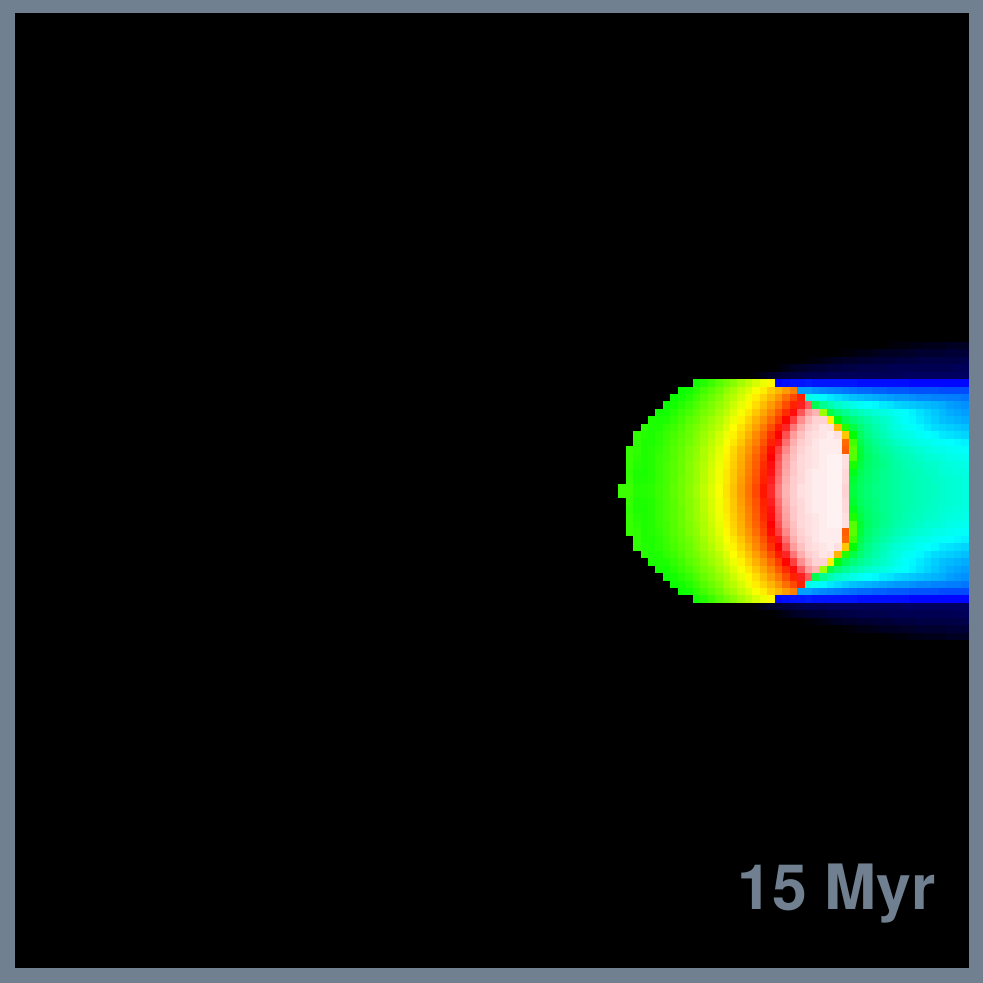}}\hspace{-1.3mm}
  \subfloat{\includegraphics[width=0.25\textwidth]
    {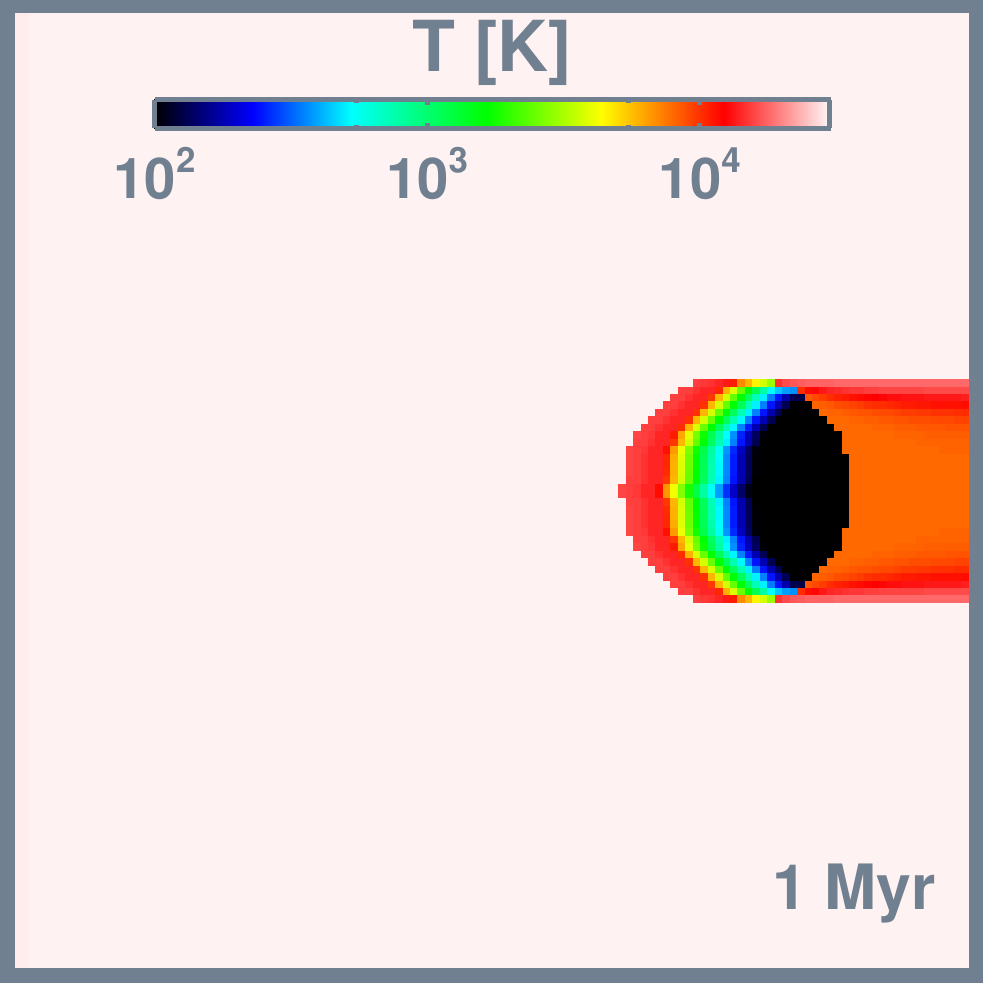}}\hspace{-1.3mm}
  \subfloat{\includegraphics[width=0.25\textwidth]
    {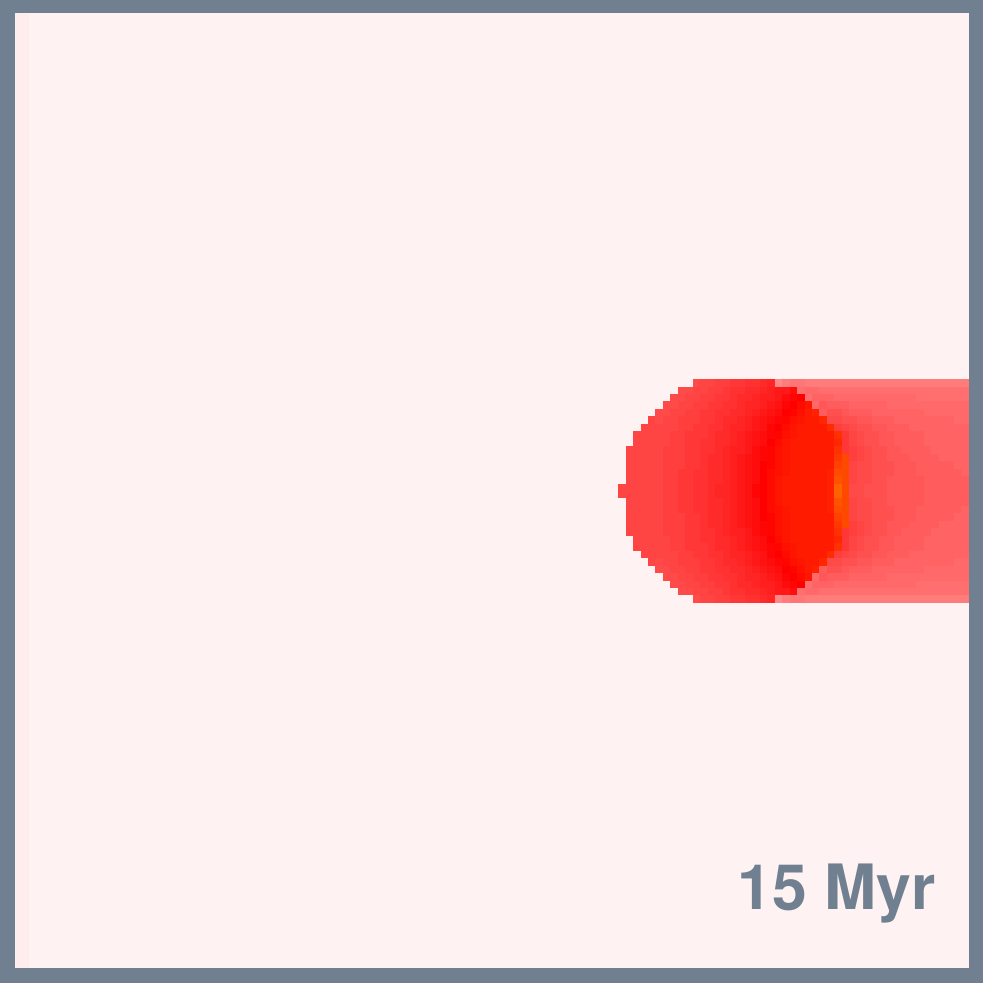}}\hspace{-1.3mm}
  \hspace{1mm}\vspace{-4mm}
  \subfloat{\includegraphics[width=0.25\textwidth]
    {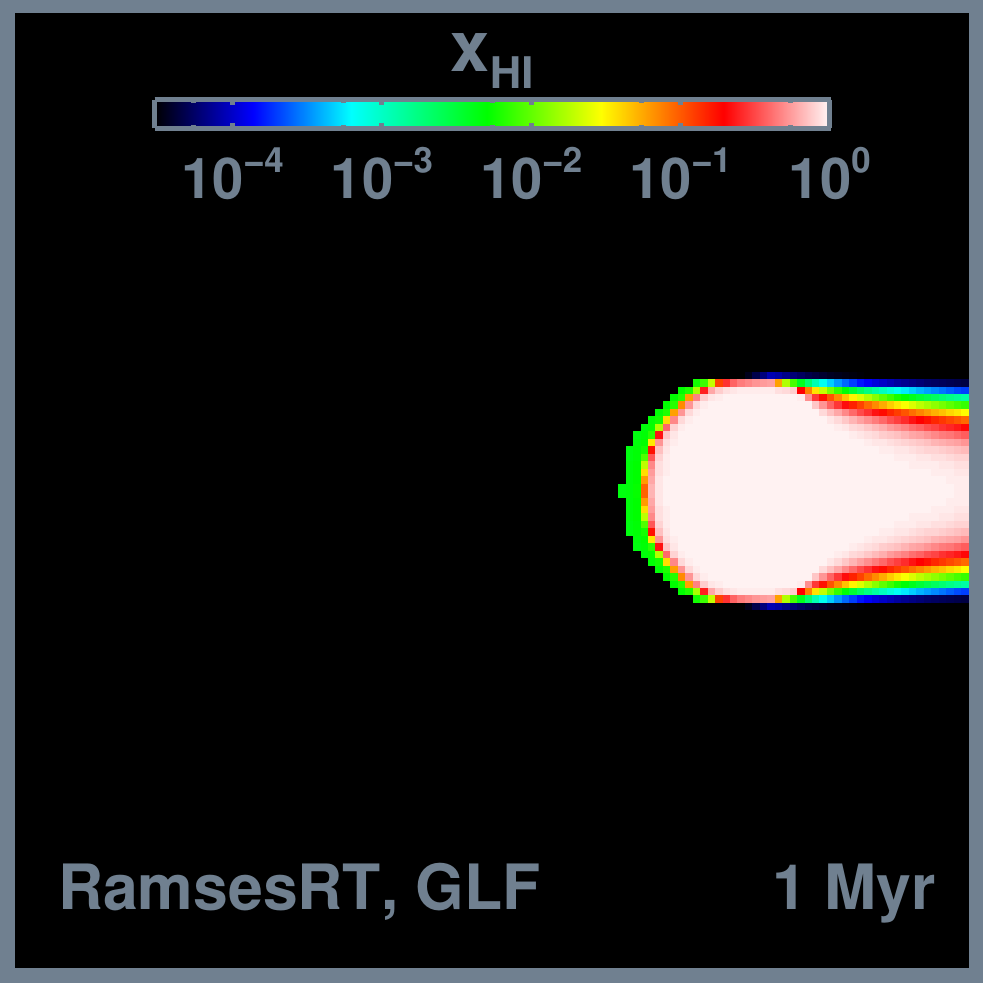}}\hspace{-1.3mm}
  \subfloat{\includegraphics[width=0.25\textwidth]
    {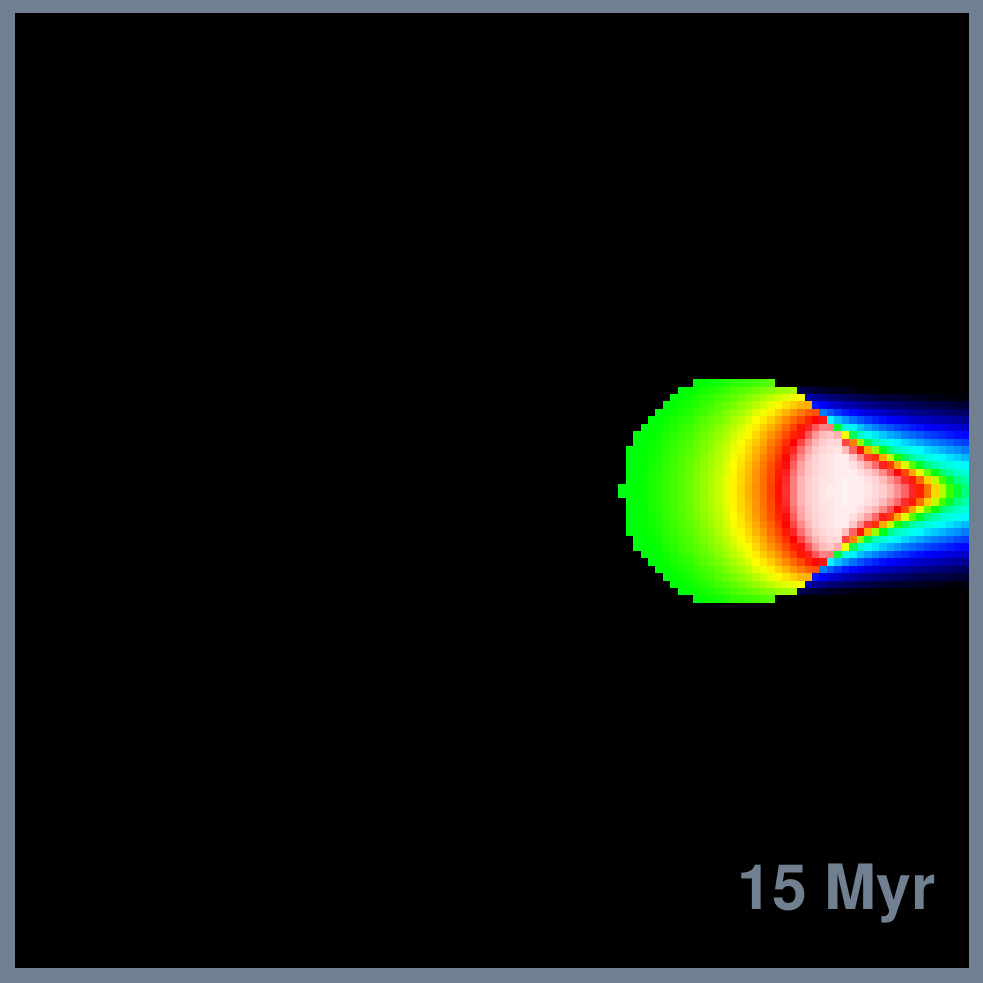}}\hspace{-1.3mm}
  \subfloat{\includegraphics[width=0.25\textwidth]
    {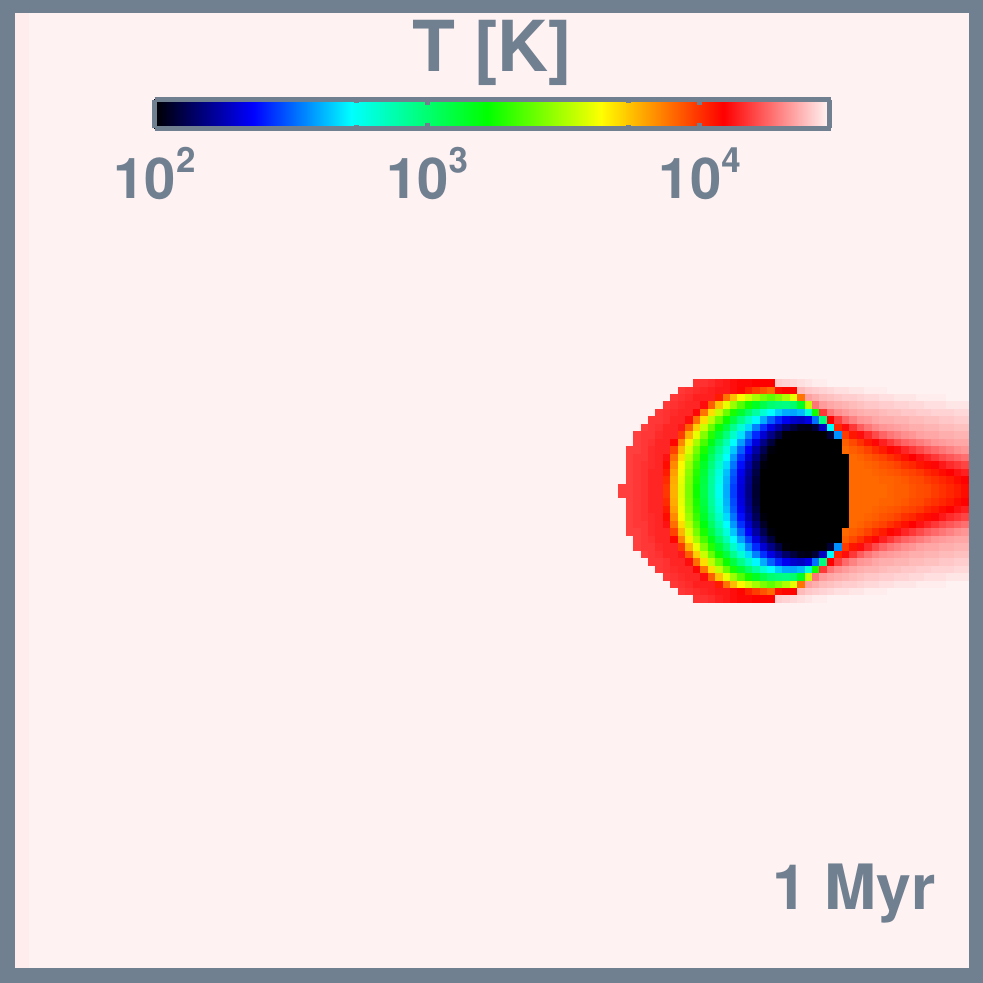}}\hspace{-1.3mm}
  \subfloat{\includegraphics[width=0.25\textwidth]
    {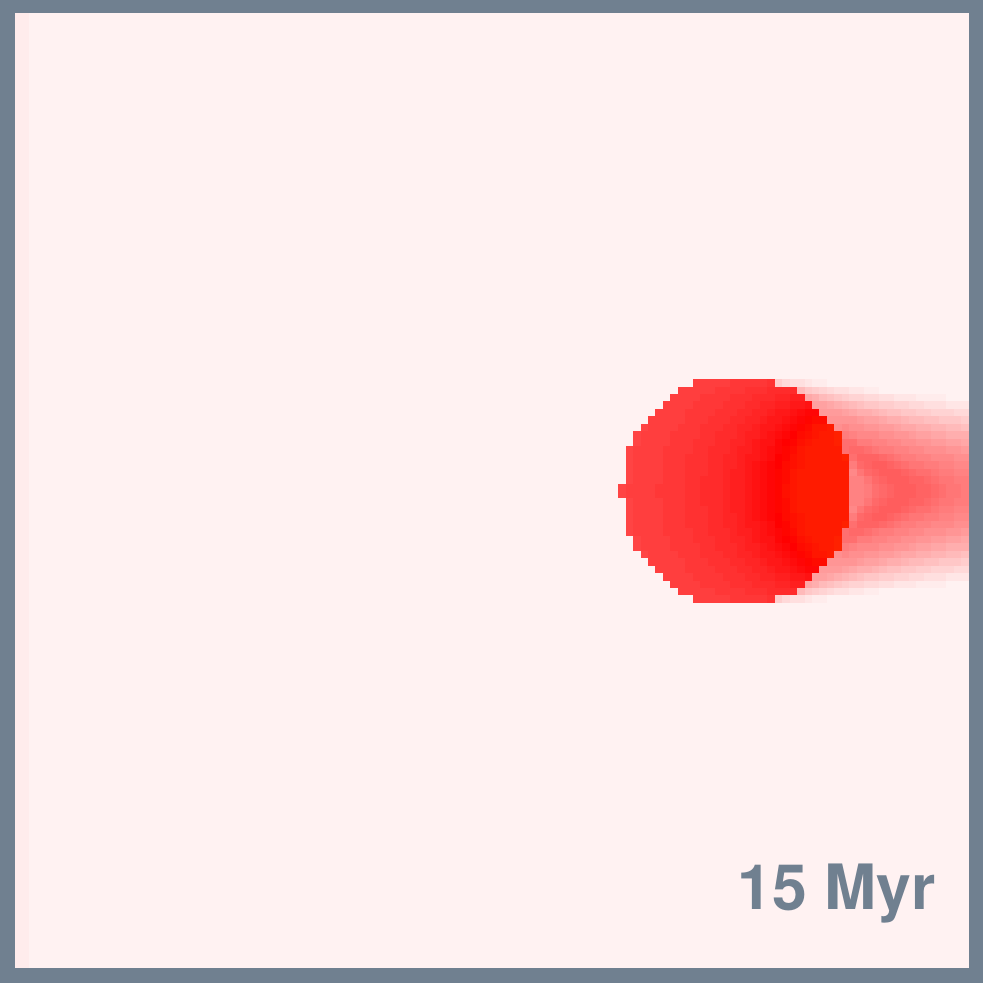}}\hspace{-1.3mm}
  \hspace{1mm}\vspace{-4mm}
  \subfloat{\includegraphics[width=0.25\textwidth]
    {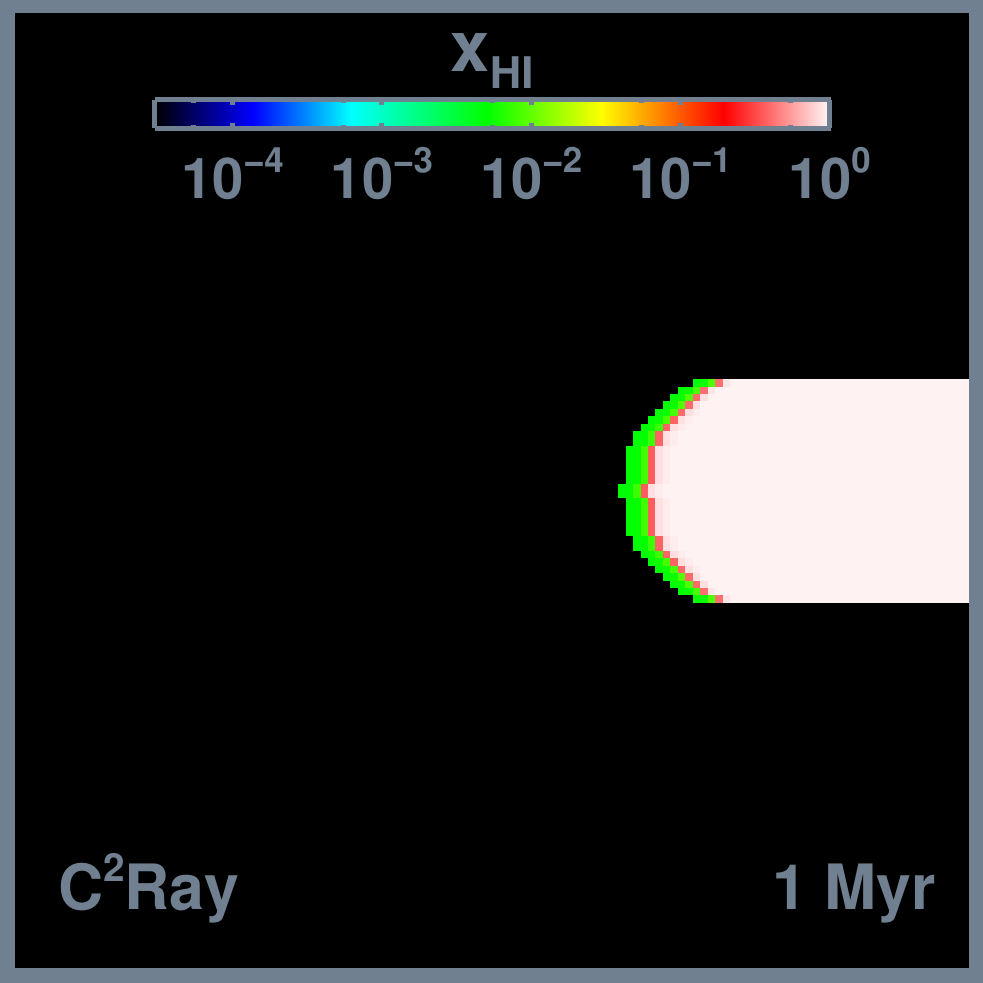}}\hspace{-1.3mm}
  \subfloat{\includegraphics[width=0.25\textwidth]
    {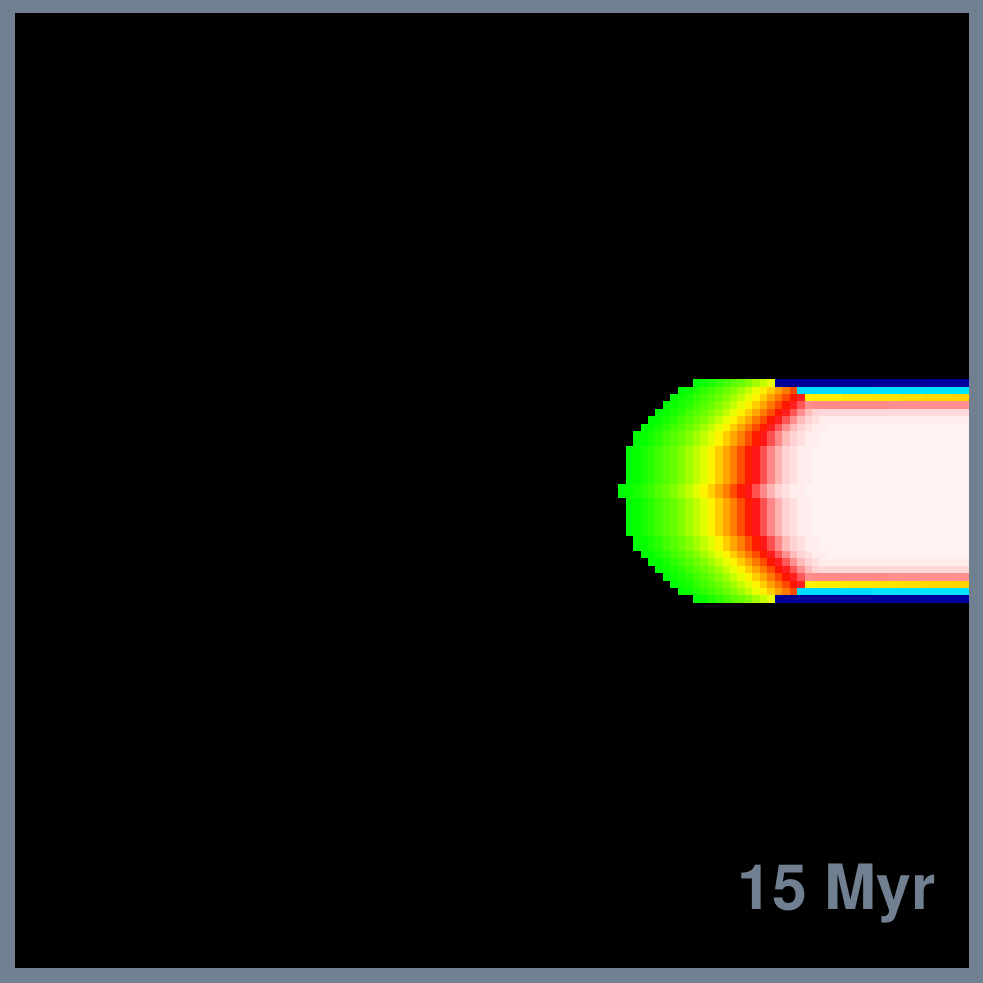}}\hspace{-1.3mm}
  \subfloat{\includegraphics[width=0.25\textwidth]
    {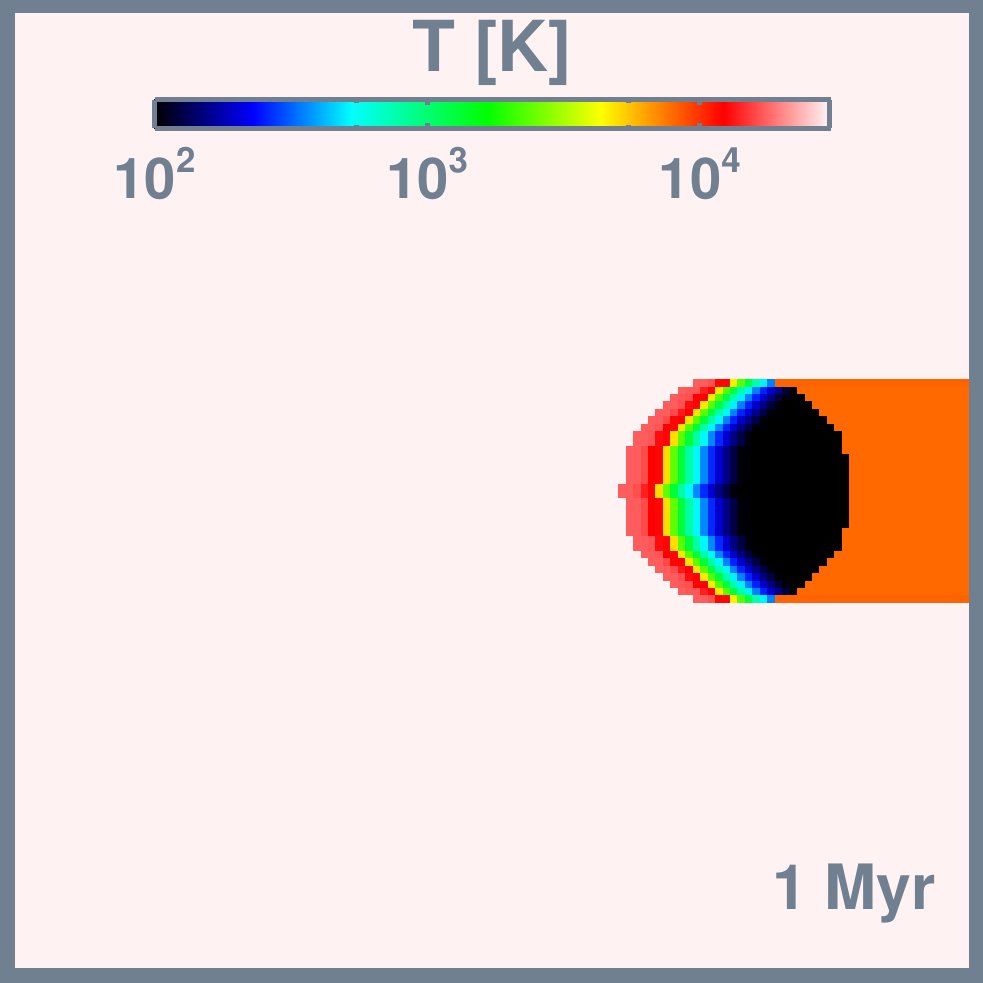}}\hspace{-1.3mm}
  \subfloat{\includegraphics[width=0.25\textwidth]
    {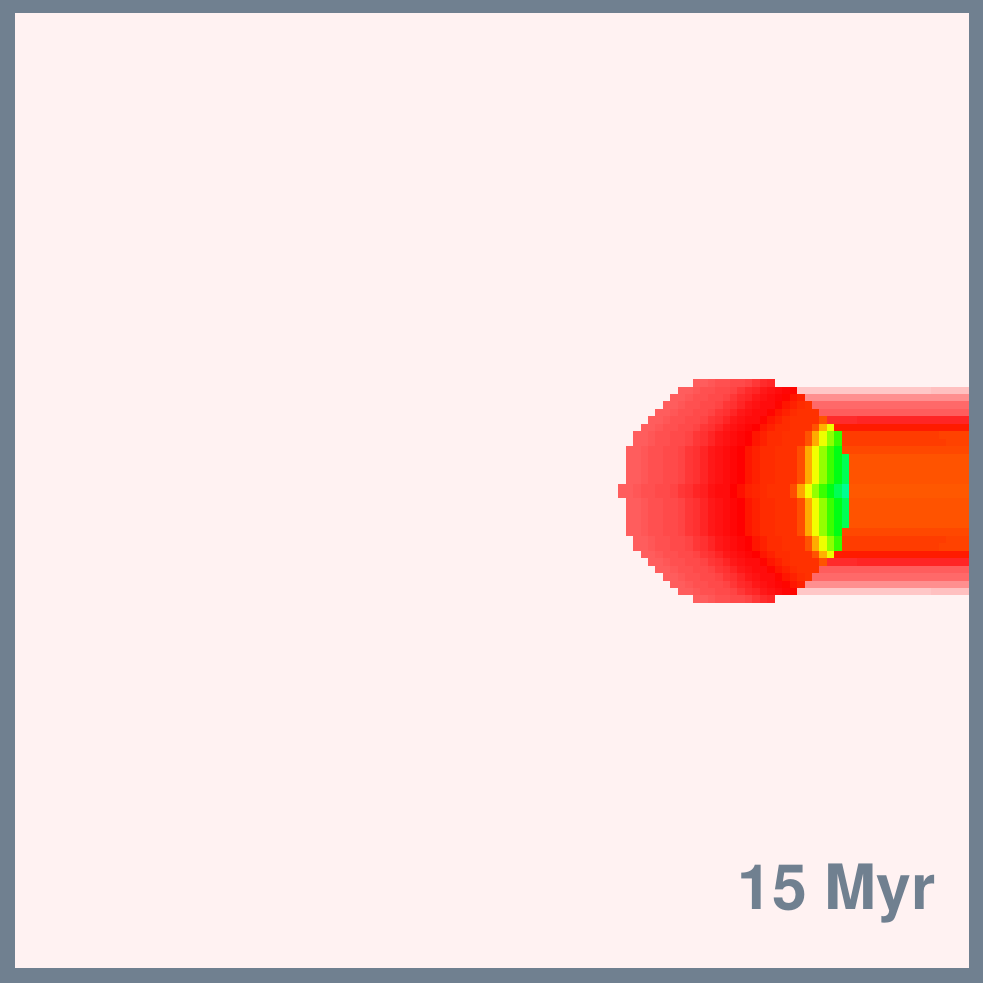}}\hspace{-1.3mm}
  \vspace{2mm}
  \caption[\Ila{} test 3 - maps]
  {\label{Il3maps.fig}\Ila{} test 3. Maps showing slices at $z=0.5 \
    \Lbox$. The \textbf{top} map shows the (constant) density field,
    with the static refinement overplotted. The \textbf{second row}
    shows the \ramsesrt{}+HLL results in terms of neutral fraction
    (left) and temperature (right) at 1 and 15 Myr. The \textbf{third
      row} Shows the \ramsesrt{}+HLL results without the on-the-spot
    approximation. The \textbf{fourth row} shows the \ramsesrt{}+GLF
    results. The \textbf{bottom row} shows the \C2R{} results for
    comparison.}
\end{figure*}

This test considers self-shielding within a dense gas cloud bombarded
on one side by UV radiation, and the shadow trailing on the `dark'
side - something which may find place with clouds close to sites of
star-formation.

The setup is as follows: the simulation box has width $\Lbox=6.6$
kpc. We place a spherical cloud of gas in the center of the
($y,z$)-plane, with radius $r_{\rm{cloud}}=0.8$ kpc, and it's center
at $(x_c,y_c,z_c)=(5,3.3,3.3)$, as seen in \Fig{Il3maps.fig}, top
left, showing an $(x,y)-$slice of gas density through the middle of
the box.  Outside the gas cloud we have $\nh^{\rm{out}}=2 \times
10^{-4}\, \cci$, $T^{\rm{out}}=8000$ K and $\xhii^{\rm{out}}=0$, and
inside we have $\nh^{\rm{cloud}}=200 \, \nh^{\rm{out}}=4 \times
10^{-2} \, \cci$, $T^{\rm{cloud}}=40$ K and
$\xhii^{\rm{cloud}}=10^{-6}$. We apply a constant ionizing photon flux
$F=10^6 \; \flux$ from the $x=0$ boundary of the box (left in the
\Fig{Il3maps.fig} maps), and run for $15$ Myr. \joki{We use a light
  speed fraction of $\fc=10^{-1}$. This is ten times higher than the
  ``norm'' in the RT tests, but it is needed for the light to have
  reached the cloud in the first snapshot under consideration, at 1
  Myr.} In order to best capture the formation of a shadow behind the
cloud, we apply the HLL flux function in this test rather than the
usual GLF function, and we use the OTSA. We have run identical tests
though, one with the GLF flux function, and one where we use the HLL
flux function but don't assume the OTSA, and we show maps of those
experiments for a qualitative comparison. As usual, the resolution
prescribed by \Ila{} is $128^3$ cells, but here we apply static AMR
refinement such that the coarse resolution is $64^3$ cells, but a
rectangular region that encompasses the gas cloud and the shadow
behind it has one level of additional refinement, making the effective
resolution in the cloud and its shadow $128^3$ cells. The refinement
region is shown in the top panel of \Fig{Il3maps.fig}, plotted over a
density map that shows the spherical gas cloud.  \joki{The fraction of
  volume at the fine resolution is $4\%$, and the computation time for
  the test is roughly a quarter of a an analogous uniform grid run
  (about $32/130$ cpu hours for the AMR/non-AMR runs).}

\begin{figure*}
  \centering
    \hspace{-6mm}
  \subfloat[]{\includegraphics[width=0.355\textwidth]
    {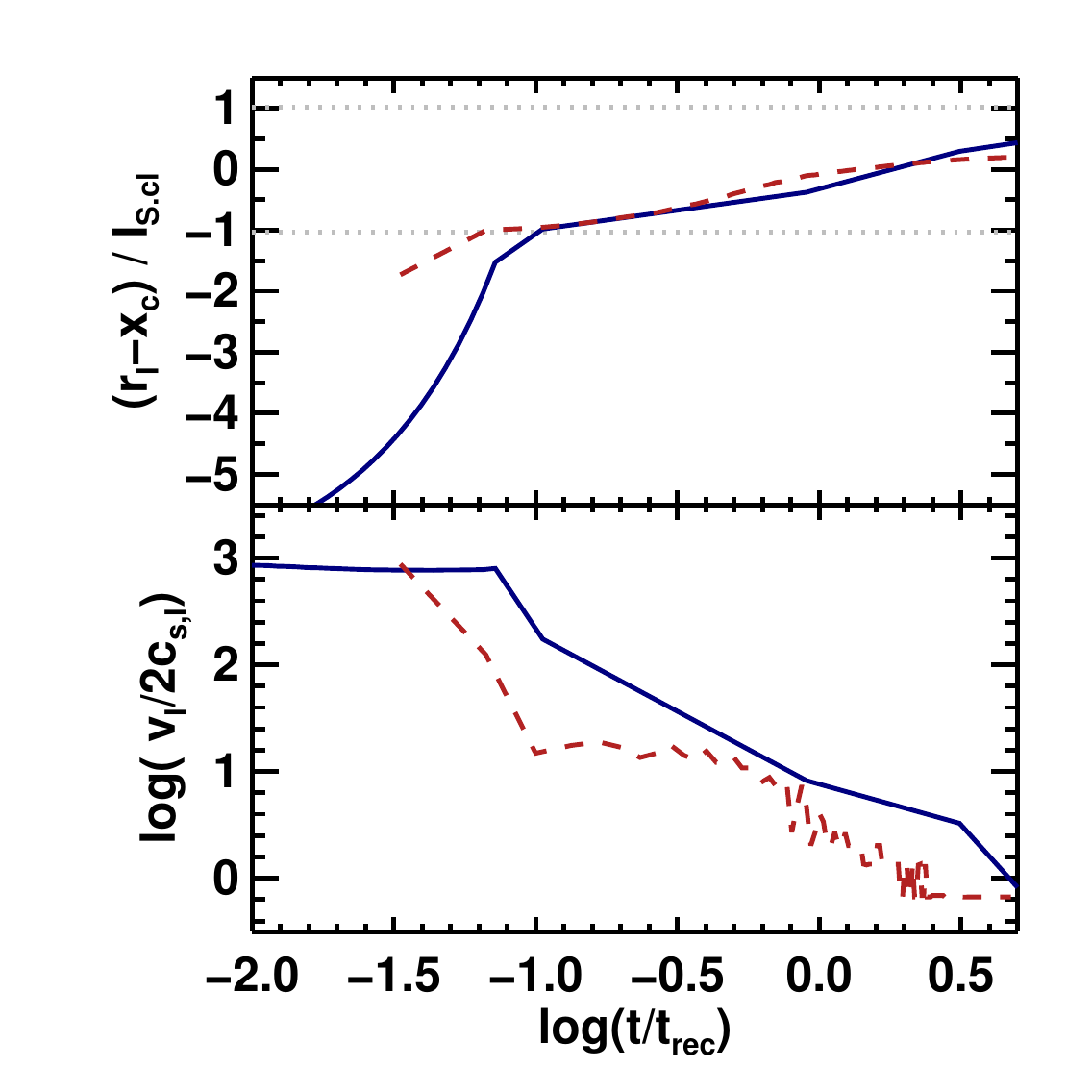}\label{Il3_ifront.fig}}\hspace{-3mm}
  \vspace{-0.3mm}
  \subfloat[]{\includegraphics[width=0.355\textwidth]
    {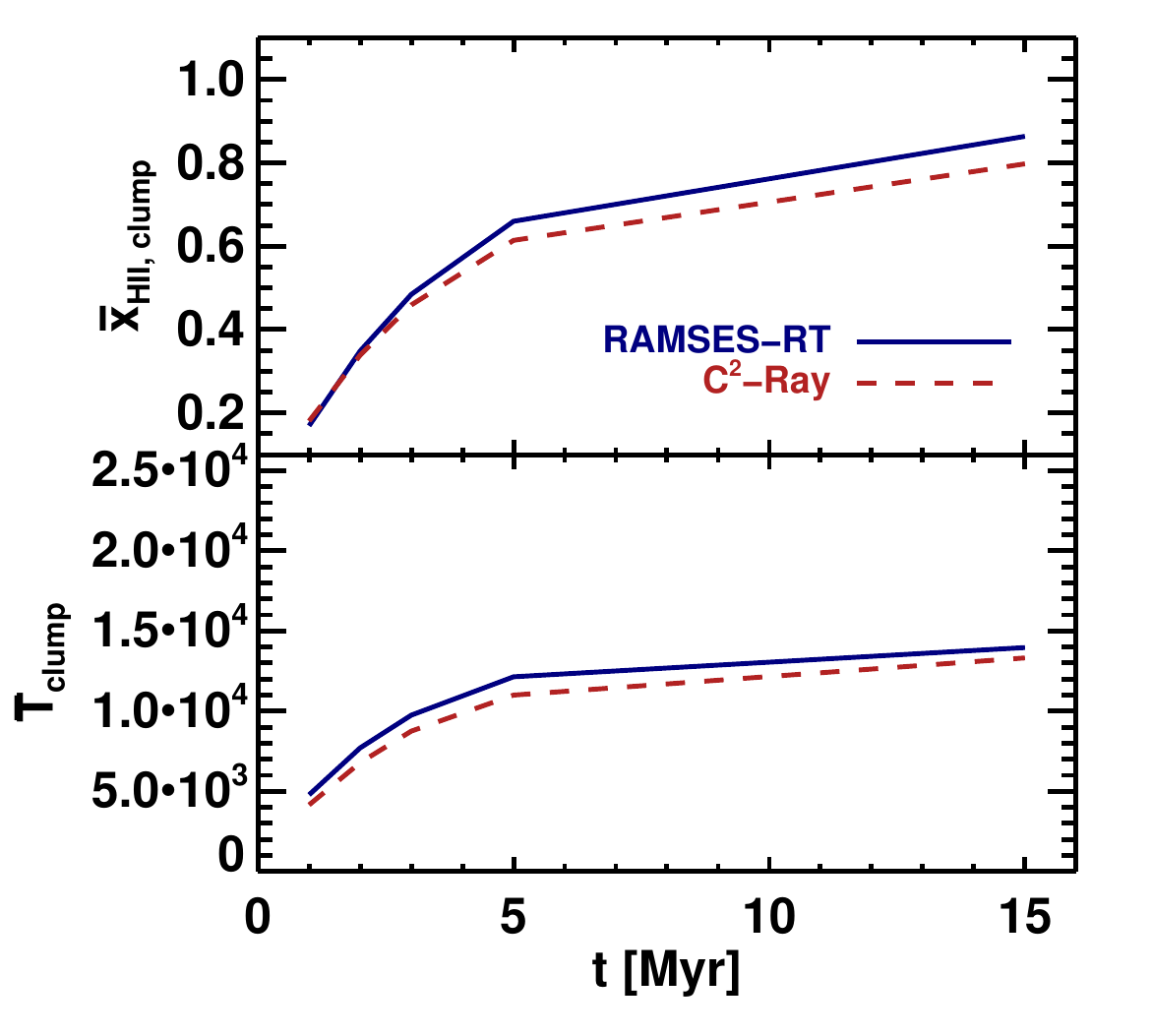}\label{Il3_xT.fig}}\hspace{-5mm}
  \vspace{-0.3mm}
  \subfloat[]{\includegraphics[width=0.355\textwidth]
    {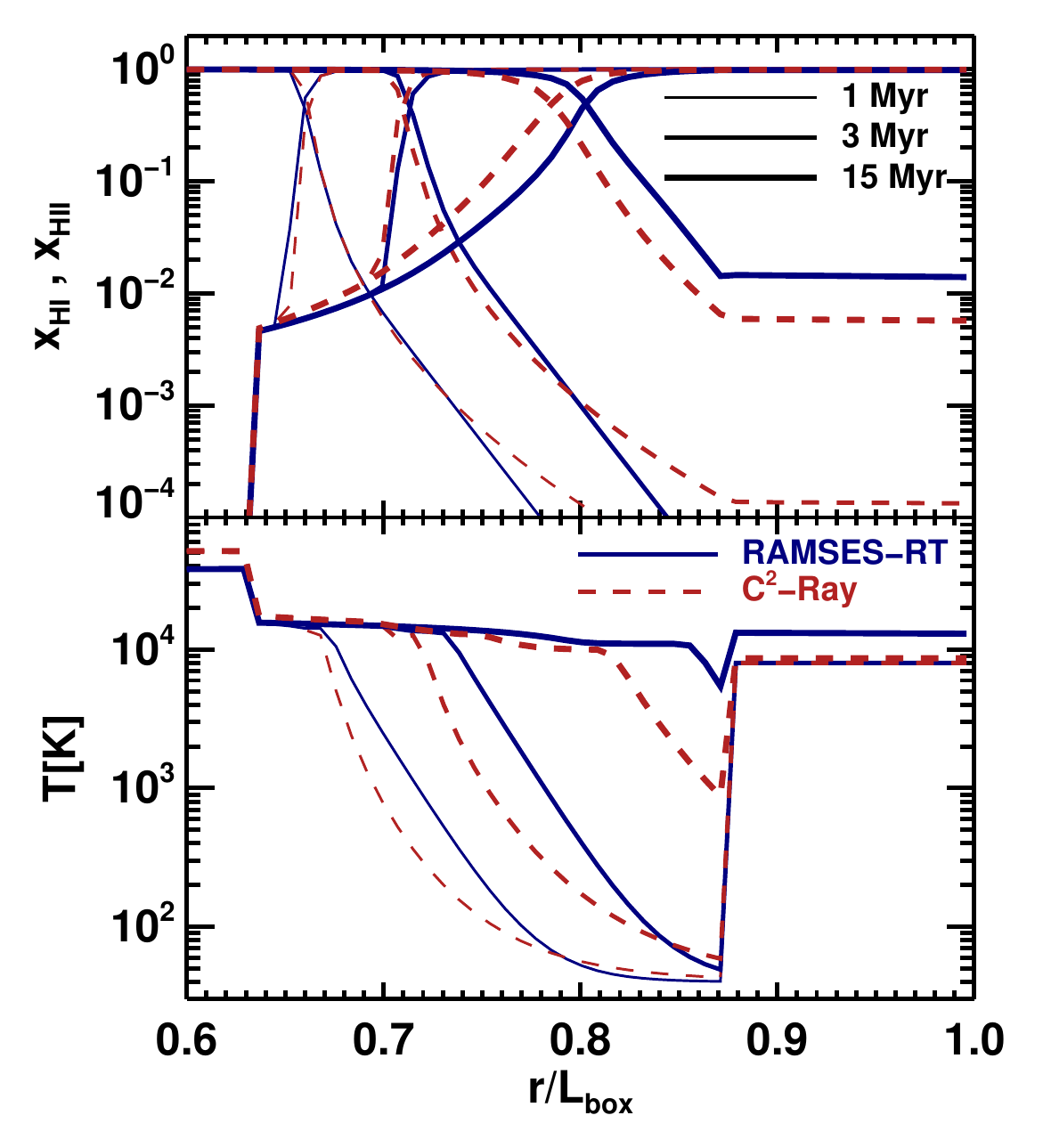}\label{Il3_profiles.fig}}\hspace{-5mm}
  \vspace{-0.3mm}
  \caption[\Ila{} test 3 - I-front and averages]
  {\Ila{} test 3: \ramsesrt{}+HLL versus \C2R{} comparinson.
    \textbf{(a)} Evolution of the position and speed of the I-front
    along the x-axis through the center of the box. The position plot
    (top) shows the x-position where $\xhii=0.5$, with respect to the
    center of the cloud, $x_C=5$ kpc, in units of the Str\"omgren
    length inside the cloud, $\ell_{S,cl}=0.78$ kpc. The dotted
    horizontal lines mark the edges of the cloud. The speed (bottom)
    is plotted in units of twice the isothermal sound speed in the
    cloud at $T=10^4$ K, $2c_{s,l}(10^4 \, \mathrm{K})=2.35 \times
    10^6$ cm/s. \textbf{(b)} Evolution of the average ionized fraction
    (top) and temperature (bottom) inside the dense
    cloud. \textbf{(c)} Profiles along the x-axis through the box
    center of the ionization state (top) and temperature (bottom), at
    1, 3 and 15 Myr.}
\end{figure*}

\Fig{Il3maps.fig} shows slices at $z=0.5 \ \Lbox$ of the neutral
fraction and temperature at $1$ and $15$ Myr.  From second top to
bottom row are shown \ramsesrt{}+HLL, \ramsesrt{}+HLL without the
OTSA, \ramsesrt{}+GLF (with the OTSA) and \C2R{}. The I-front travels
fast through the diffuse medium outside the cloud, but moves much more
slowly inside it, and a shadow is cast behind it. As the UV radiation
eats its way into the cloud, ionizing and heating it, the shadow also
very slowly diminishes in width because some photons manage to cross
through the edges of the cloud. The \ramsesrt{}+HLL maps compare very
well with \C2R{}, though the shadow is slightly thinner at $15$ Myr
and there is stronger heating inside the shadow; this could be due to
differences in the multifrequency approach and/or photoheating.
Without the OTSA, the shadow is diminished from the sides due to
photons being cast from the surrounding gas.  Using the GLF flux
function has much the same effect as not assuming the OTSA, though the
shadow is considerably more diminished here. The result with HLL but
without the OTSA is the most physical of the \ramsesrt{} results, as
one should expect recombination photons to be cast into the shadow.

\Fig{Il3_ifront.fig} shows the evolution of the position and speed of
the I-front through the center of the $(y,z)-$plane. In solid blue we
plot the \ramsesrt{} result and in dashed red is the \C2R{} result for
comparison. Horizontal dotted lines mark the edges of the cloud. There
is a large initial delay in the I-front compared to \C2R{}, which is
because in the diffuse gas outside the cloud, the I-front speed is
limited by the reduced speed of light. After the I-front gets into the
cloud (lower dotted line) it quickly catches up and then evolves in a
similar fashion in the two codes. If compared to the rest of the codes
in \Ila{}, it turns out that the evolution of the I-front in \C2R{}
slightly stands out from the rest of the codes (e.g. a small upwards
`bump' in the front position at $log(t/\trec)\sim0$, and a slightly
shorter distance of the I-front from the origin at the end of the
simulations), and most of the others in fact evolve very similarly to
that of \ramsesrt{}. The comparison appears best with \rsph{}, which
has the furthest extended I-front at the end-time of $15$ Myr. The
same can be said for the speed of the front. If we look away from the
initial $\sim 0.2$ Myr, when our I-front has to catch up, the speed
compares reasonably to \C2R{}, and quite well to the other codes in
\Ila{}.

\Fig{Il3_xT.fig} shows the evolution of the mean ionized fraction and
temperature inside the cloud, compared between \ramsesrt{} and \C2R{}. The
evolution is similar between the two codes in both cases. Compared
with the other codes in \Ila{}, the evolution of the ionized fraction is
most similar to \rsph{}, \ift{} and \coral{}, while the temperature in
\ramsesrt{} is consistently a little higher than in most codes (all
except \coral{} and \flash{} which stand out quite a lot in mean
temperature).

\begin{figure*}\begin{center}
  \includegraphics[width=.6\textwidth]
  {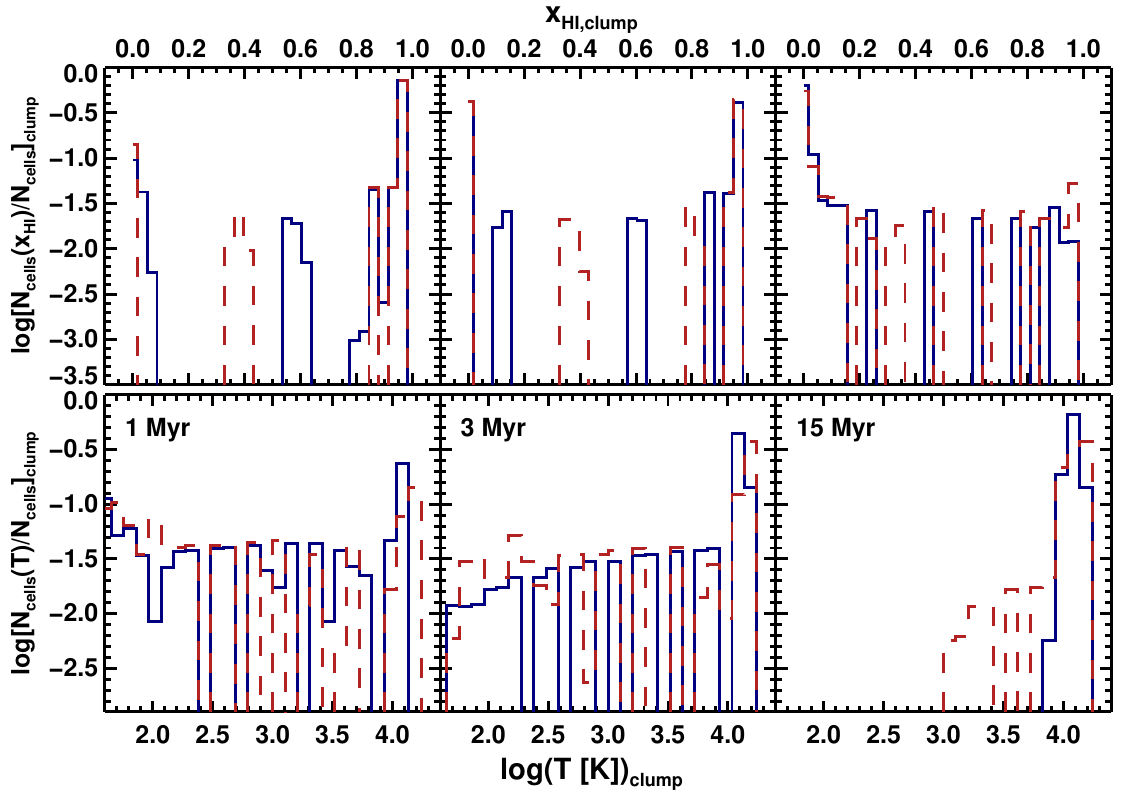}
  \caption[\Ila{} test 3 - histograms]
  {\label{Il3_hists.fig}\Ila{} test 3: \ramsesrt{}+HLL
    versus \C2R{} comparison.  Histograms of neutral fraction (top
    row) and temperature (bottom) inside the dense cloud at 1, 3 and
    15 Myr (from left to right).}
\end{center}\end{figure*}

\Fig{Il3_profiles.fig} shows profiles of the ionization state and
temperature along the x-axis at the center of the $(y,z)-$plane at
$1$, $3$ and $15$ Myr. The ionization state profile in \ramsesrt{} is
similar in most respects to that of \C2R{}, though it extends a bit
further at the end of the run-time. There is initially less ionization
on the far side of the front in \ramsesrt{}, but at the end of the run
this is reversed and we have slightly more ionization on the far side
in \ramsesrt{}. This `shift` can be explained by the temperature
profiles: at early times the cloud is efficiently shielding the far
side from even the high-energy photons in both codes, but at the end
of the \ramsesrt{} run the shielding buffer in the cloud is thin
enough that the high-energy photons can get through, hence efficiently
heating the gas inside the buffer as well as in the shadow, and the
gas in the shadow becomes slightly ionized as a consequence. The
analogue ionization state profiles for the other codes in \Ila{} are
mostly similar to ours. Most of them are actually closer to the
\ramsesrt{} than the \C2R{} profile, with the exception of \crash{}
which has a much more underdeveloped I-front and less ionization, and
\ffte{} and \ift{} which have an almost step-wise $\xhii$-profile on
the far side of the I-front. The temperature profiles differ pretty
widely between the codes. \ramsesrt{} doesn't particularly stand out,
though, and is most similar to that of \coral{} at $15$ Myr. The
temperature profile for \ramsesrt{} also differs notably from that of
\aton{}, where the shielded region inside the cloud is thicker and
more step-like both in the ionized fraction and temperature, due to
the monochromatic radiation.

Finally, \Fig{Il3_hists.fig} shows histograms of the neutral fraction
and temperature at $1$, $3$ and $15$ Myr for \ramsesrt{} and
\C2R{}. The comparison (also with the other codes in \Ila{}) is
qualitatively similar, though there is quite a difference between the
individual codes in these plots.

As with the previous tests, \ramsesrt{} performs well here and we
don't really have anything out of the ordinary in our results. One
should keep in note though that here we've used the non-diffusive HLL
flux function, whereas in most cosmological simulations it would be
more natural to use the more diffusive GLF function to have better
spherical symmetry around radiative stellar sources, which comes with
the price of less pronounced and shorter lived shadows than HLL. 
\joki{The survival of shadows in more realistic scenarios remains an
  open question, but considering the effects of recombination
  radiation, and the likelihood of any transparent region to have
  ionizing sources shining from different directions, it seems
  unlikely to us that shadowing is an efficient way of shielding
  gas from ionizing radiation.  }



\subsection{\Ila{} test 4: \ \ Multiple sources in a cosmological density
  field}\label{Ila4.sec}

This test involves the propagation of ionization fronts in a static
hydrogen-only density field taken from a cosmological simulation
snapshot at redshift 9. The density cube is $128^3$ cells and its
width is $500 \, h^{-1}$ co-moving kpc (corresponding to $50 \,
h^{-1}$ physical kpc). The Hubble factor is $h=0.7$. The initial
temperature is fixed at $100$ K everywhere. 16 radiative sources are
picked out corresponding to the most massive halos in the box and
these are set to radiate continuously for $0.4$ Myrs. The
mass-dependent radiation intensity for each halo is given in a
downloadable table (from the RT comparison project
website). \joki{Unlike in \Ila{}, we don't apply the OTSA in this
  test, i.e. we include the radiative transfer of recombination
  radiation, but we've verified that this has no discernible effect on
  the results. Our analysis from \Sec{reduced_c.sec} indicates that a
  reduced light speed gives incorrect results in this test. Thus we
  use a full light speed here (i.e. $f_{c}=1$), and for comparison
  with the codes from \Ila{}, which implicitly assume infinite light
  speed, we make an analogue run with a hundred-fold light speed
  ($f_{c}=100$).}

\begin{figure*}
  \centering
  \subfloat{\includegraphics[width=0.23\textwidth]
    {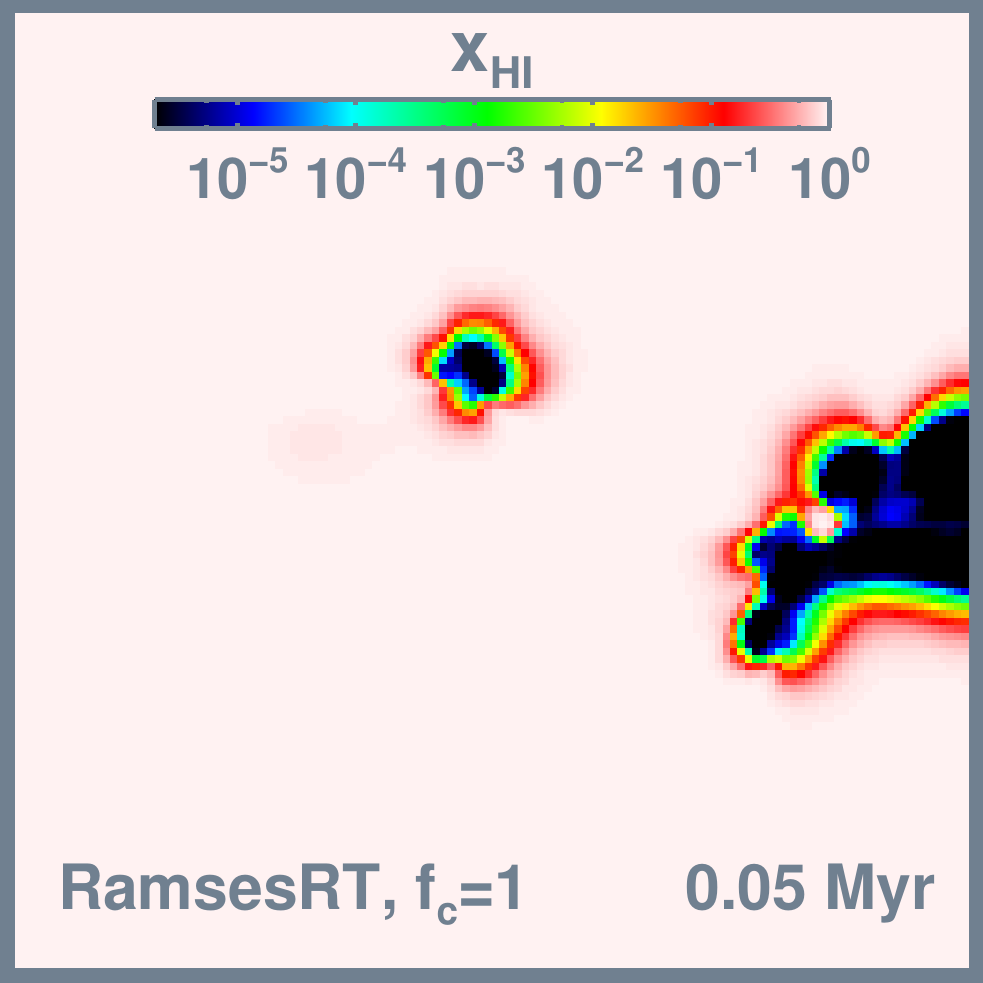}}\hspace{-1.2mm}
  \subfloat{\includegraphics[width=0.23\textwidth]
    {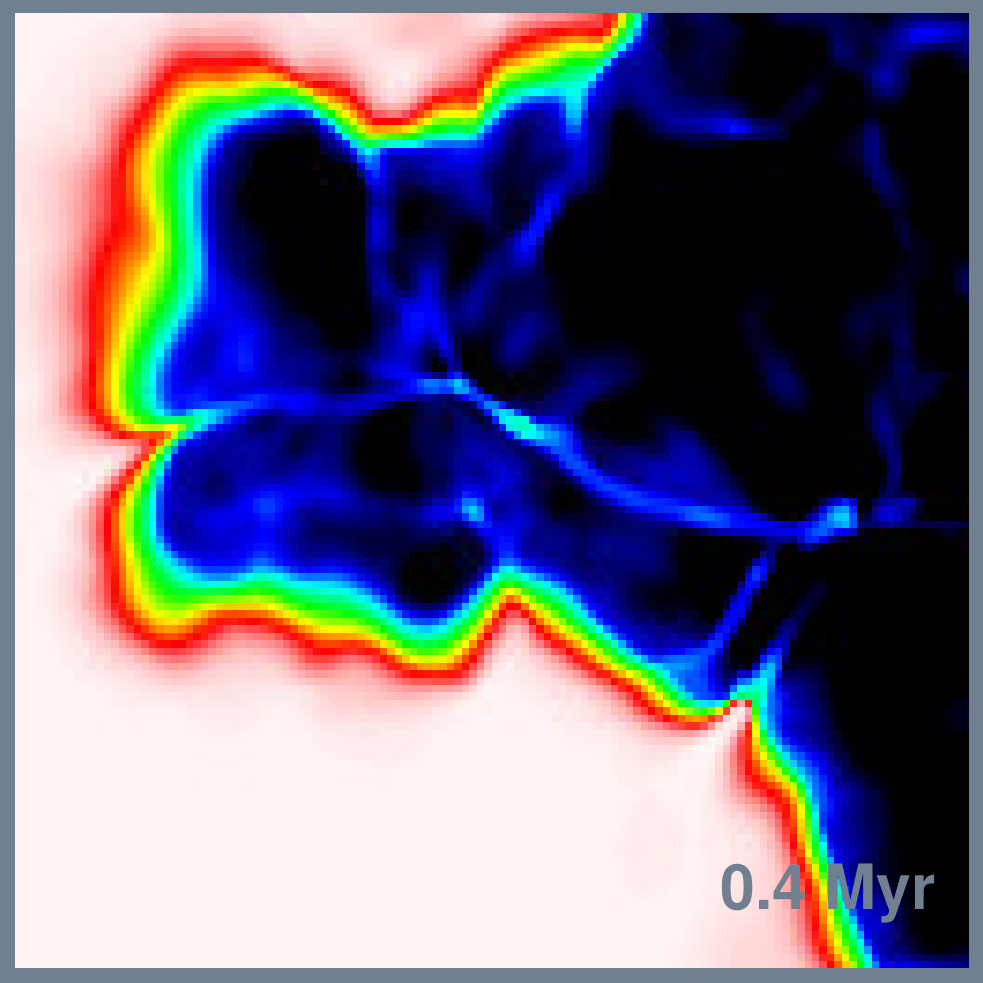}}\hspace{ 0.mm}
  \subfloat{\includegraphics[width=0.23\textwidth]
    {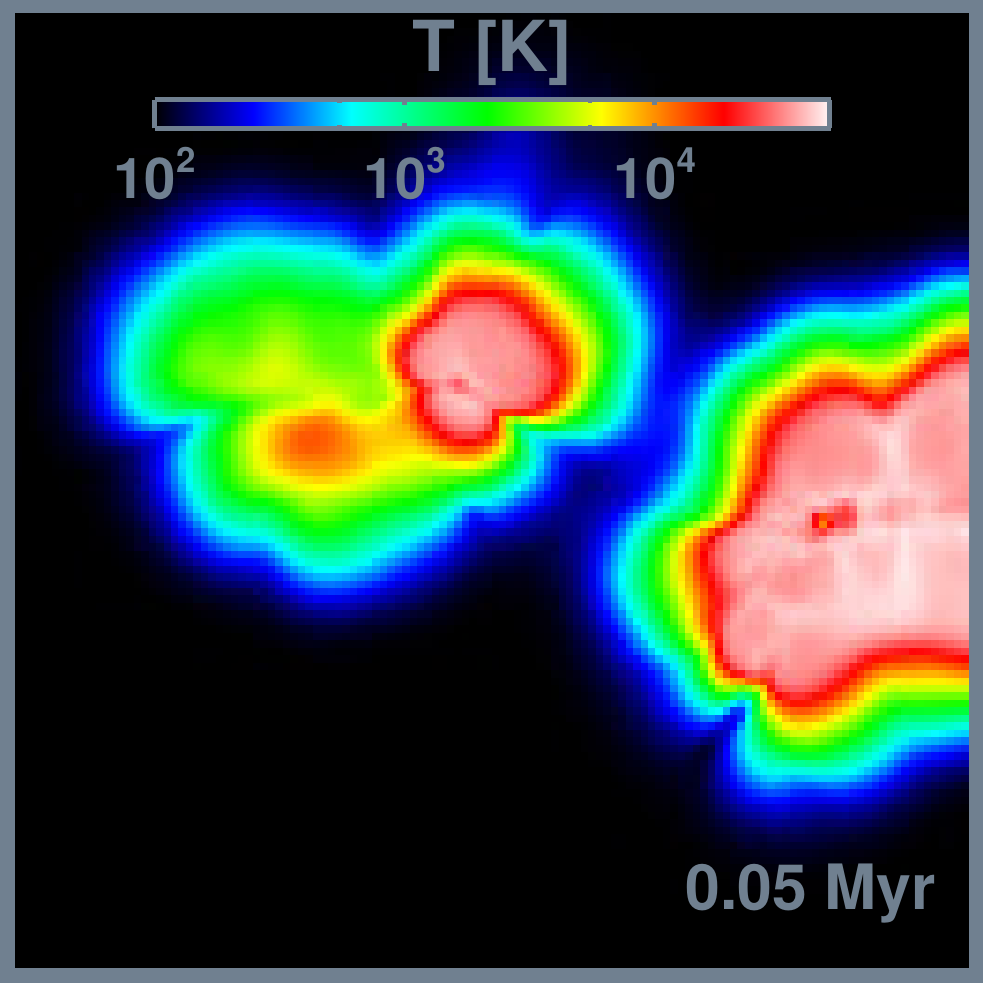}}\hspace{-1.2mm}
  \subfloat{\includegraphics[width=0.23\textwidth]
    {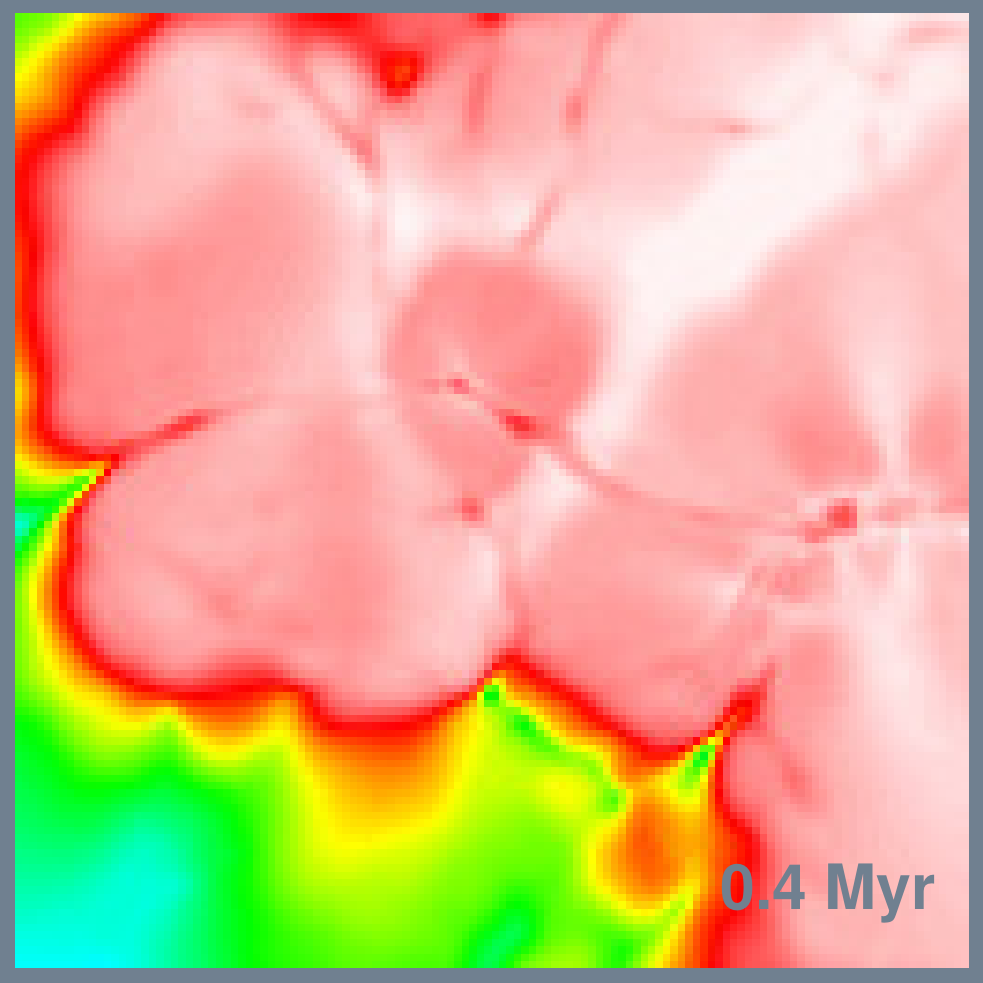}}\hspace{-1mm}
  \vspace{-2.5mm}

  \subfloat{\includegraphics[width=0.23\textwidth]
    {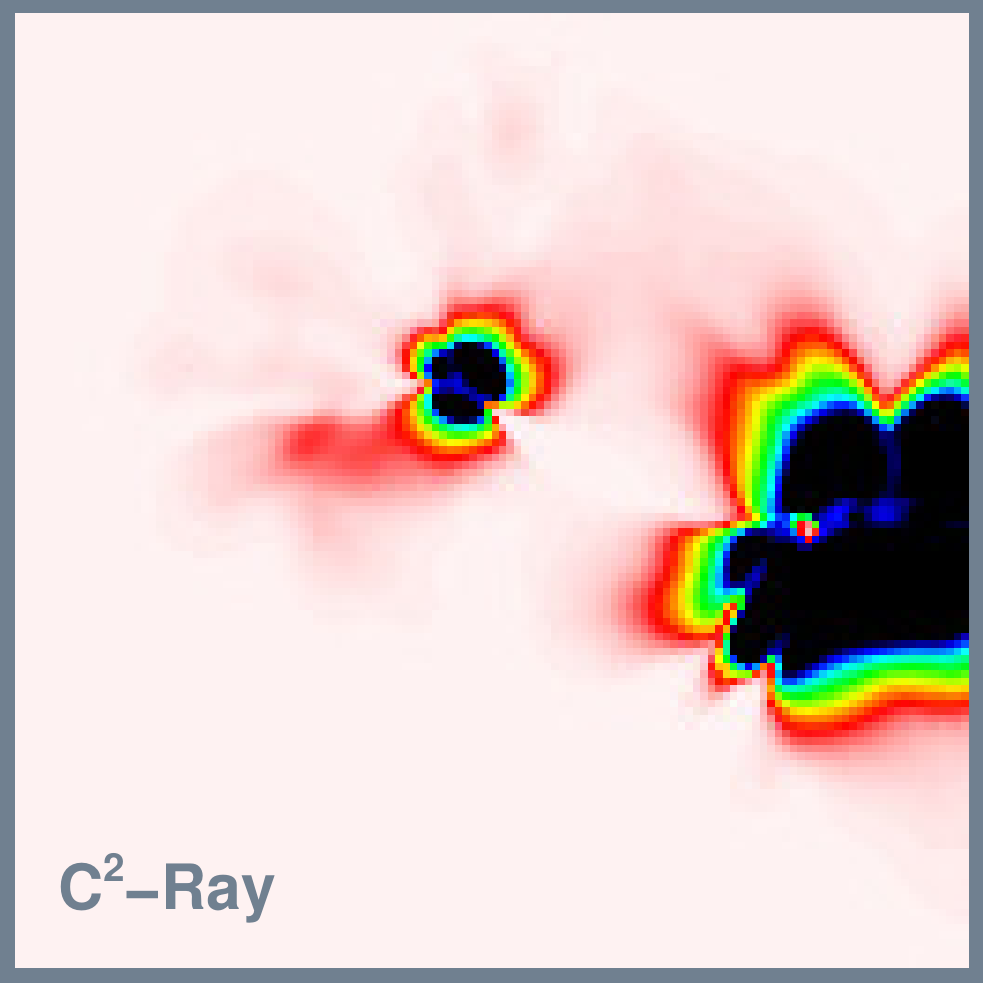}}\hspace{-1.2mm}
  \subfloat{\includegraphics[width=0.23\textwidth]
    {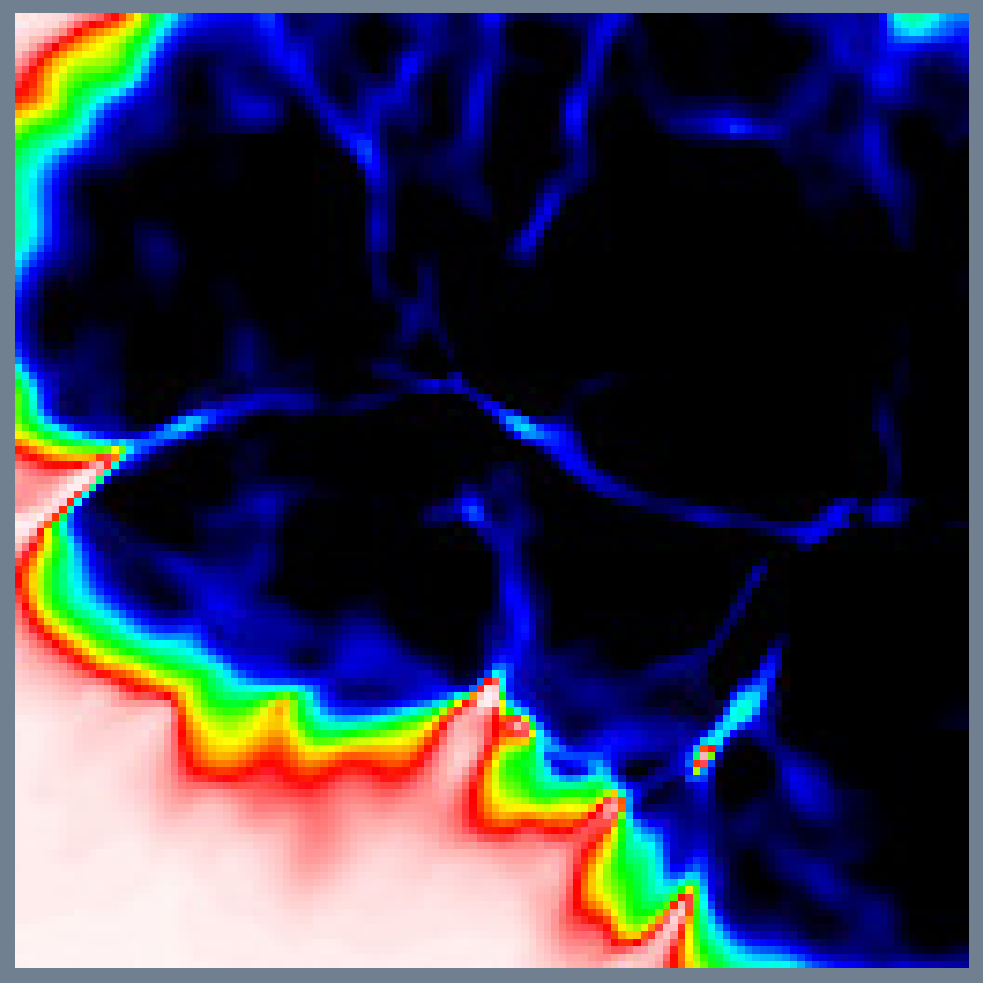}}\hspace{0.mm}
  \subfloat{\includegraphics[width=0.23\textwidth]
    {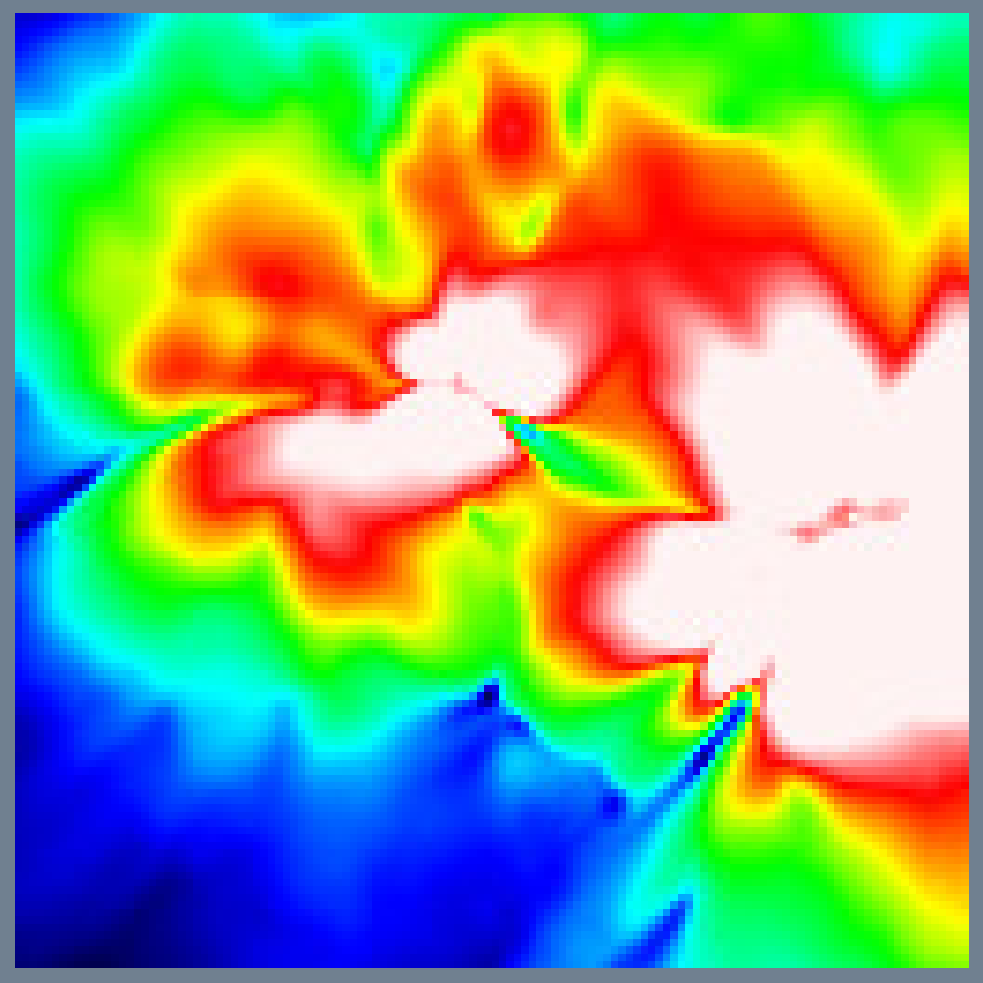}}\hspace{-1.2mm}
  \subfloat{\includegraphics[width=0.23\textwidth]
    {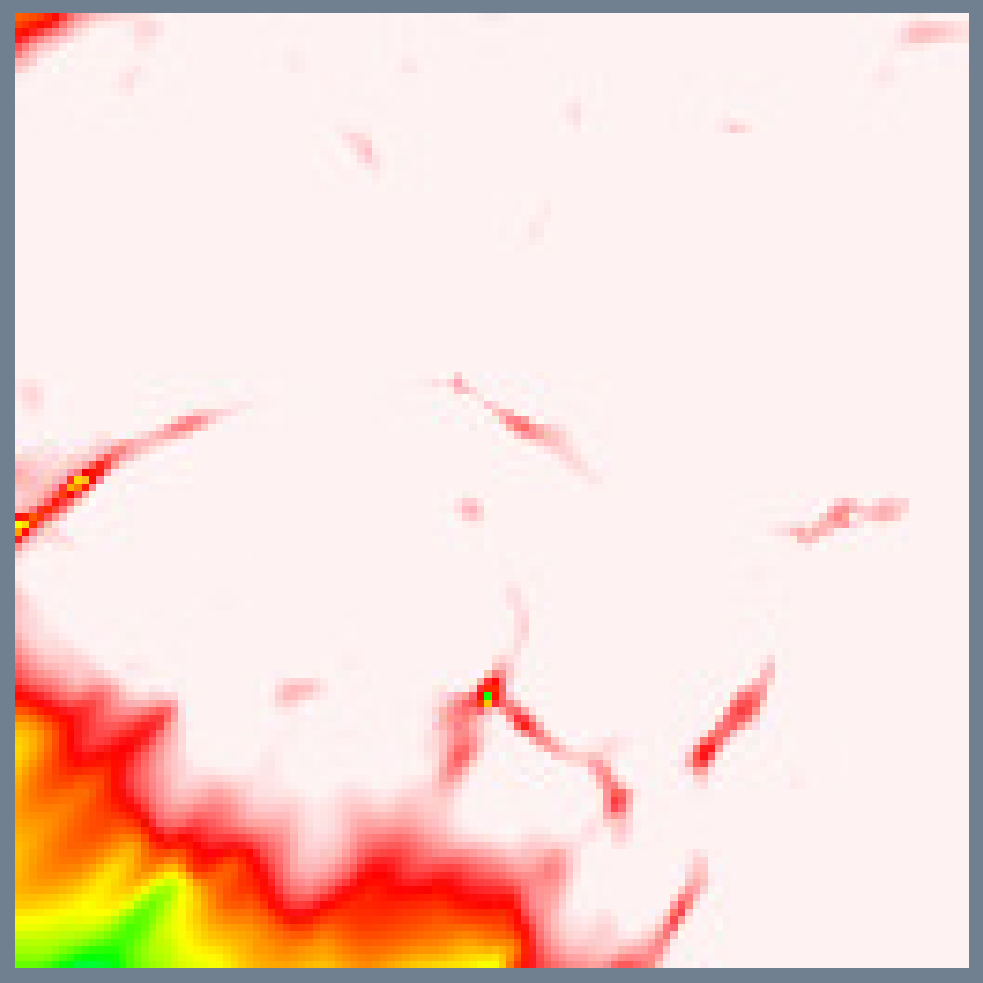}}\hspace{-1mm}
  \vspace{-2.5mm}

  \subfloat{\includegraphics[width=0.23\textwidth]
    {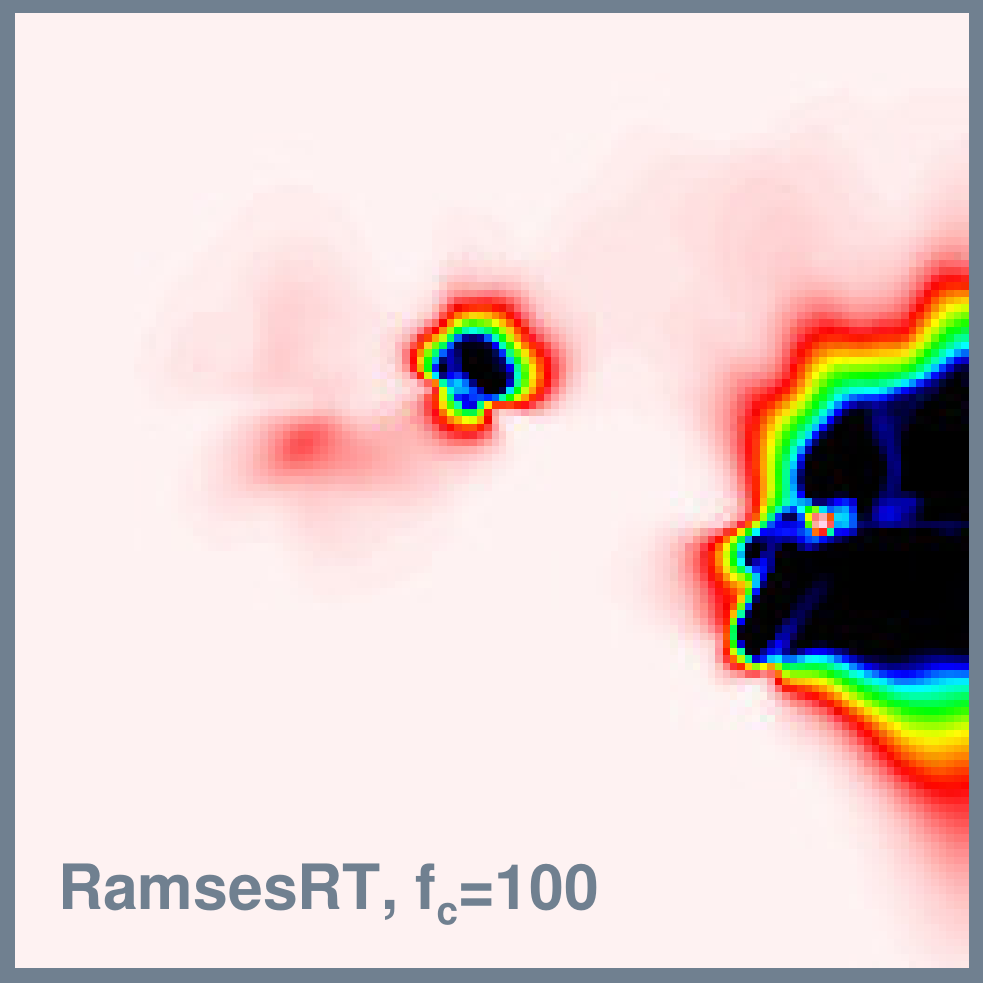}}\hspace{-1.2mm}
  \subfloat{\includegraphics[width=0.23\textwidth]
    {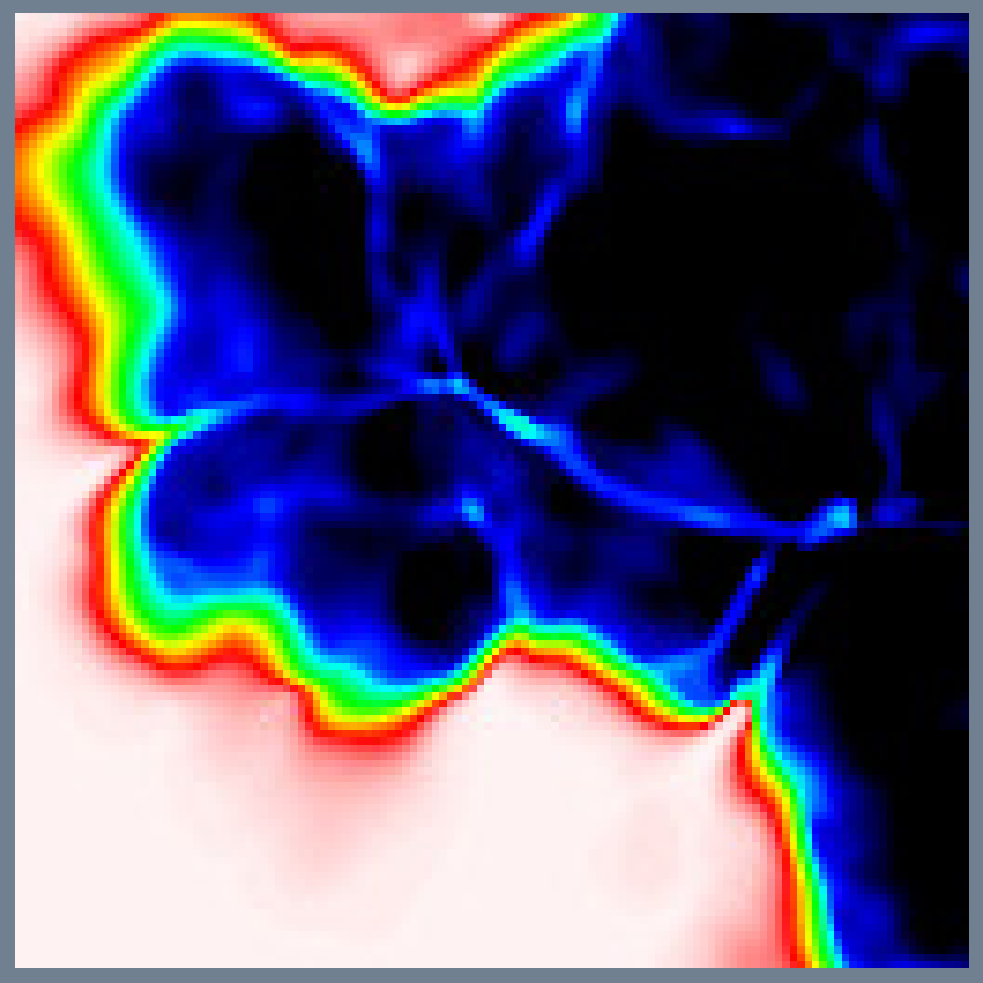}}\hspace{0.mm}
  \subfloat{\includegraphics[width=0.23\textwidth]
    {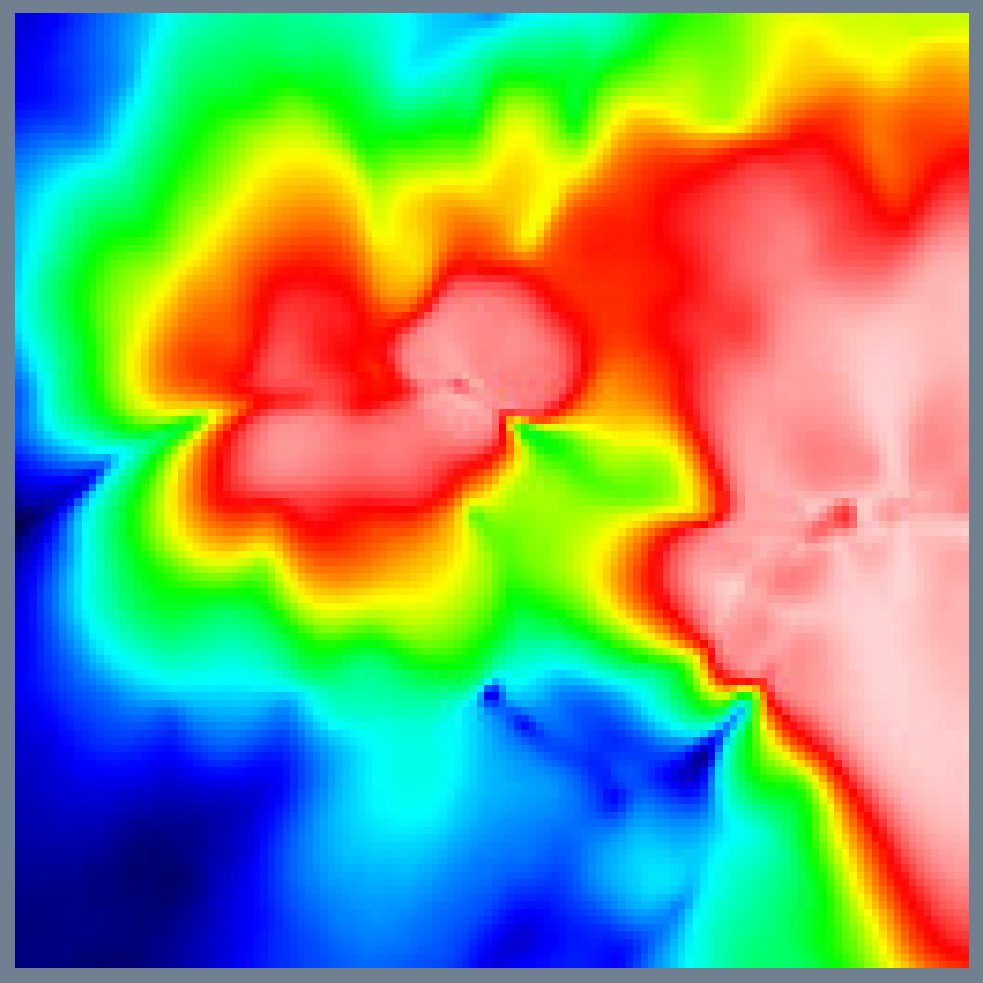}}\hspace{-1.2mm}
  \subfloat{\includegraphics[width=0.23\textwidth]
    {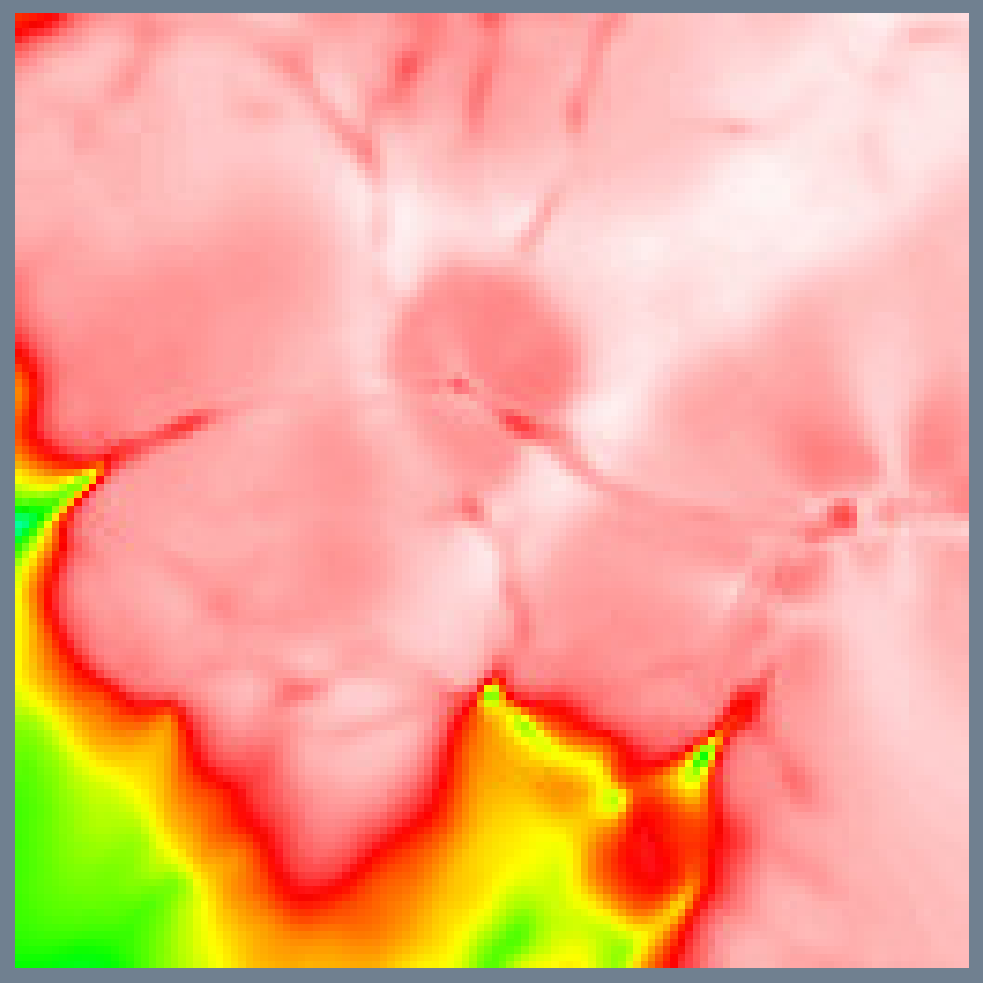}}\hspace{-1mm}
  \vspace{-1.5mm}
  \caption[\Ila{} test 4 - maps]
  {\label{Il4maps.fig}\Ila{} test 4. Maps showing slices at $z=0.5 \
    \Lbox$ of the neutral fraction and temperature at times 0.05 Myr
    and 0.4 Myr. Top row shows \ramsesrt{} results with physical light
    speed. The middle row shows the \C2R{} results (infinite light
    speed). The bottom row shows the \ramsesrt{} results with one
    hundred times the physical light speed.}
\end{figure*}

\begin{figure*}
  \centering
  \subfloat[]{\includegraphics[width=0.37\textwidth]
    {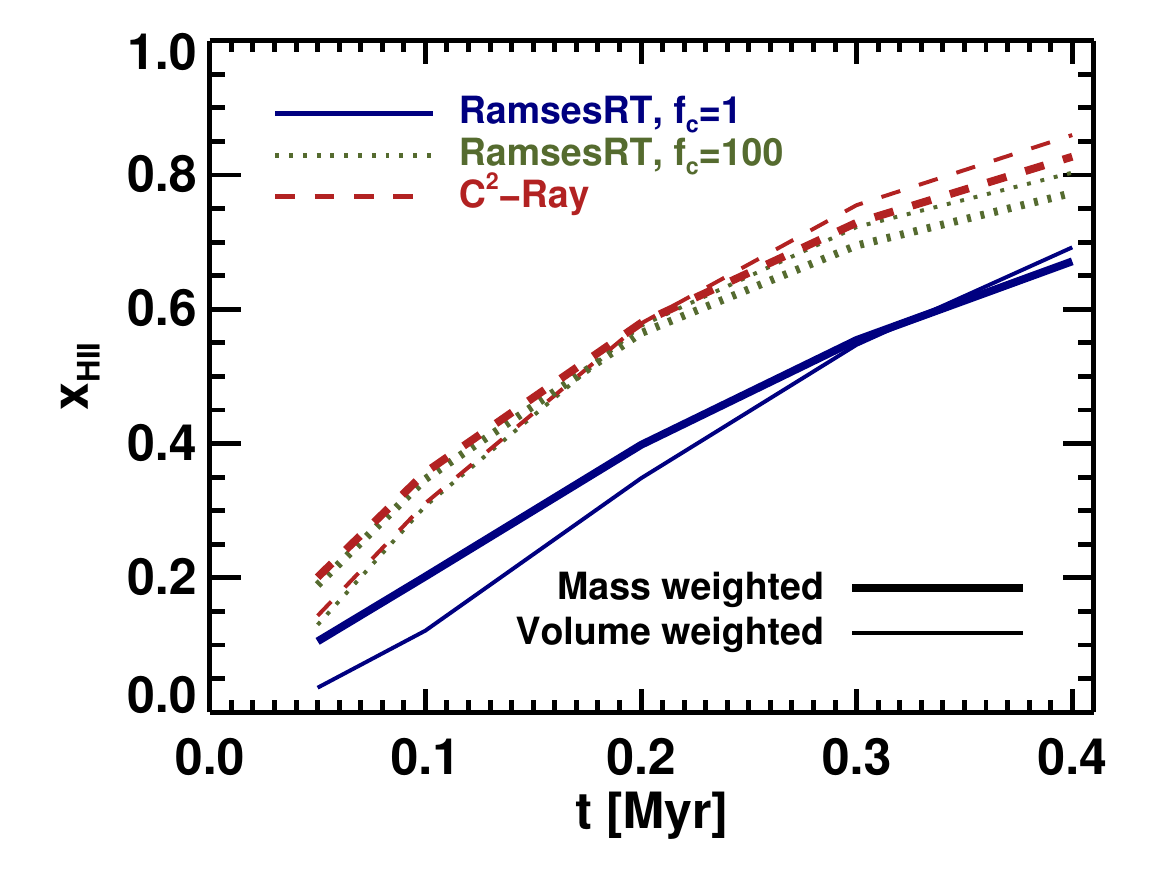}\label{Il4_xEvol.fig}}\hspace{-1.3mm}
  \subfloat[]{\includegraphics[width=0.37\textwidth]
    {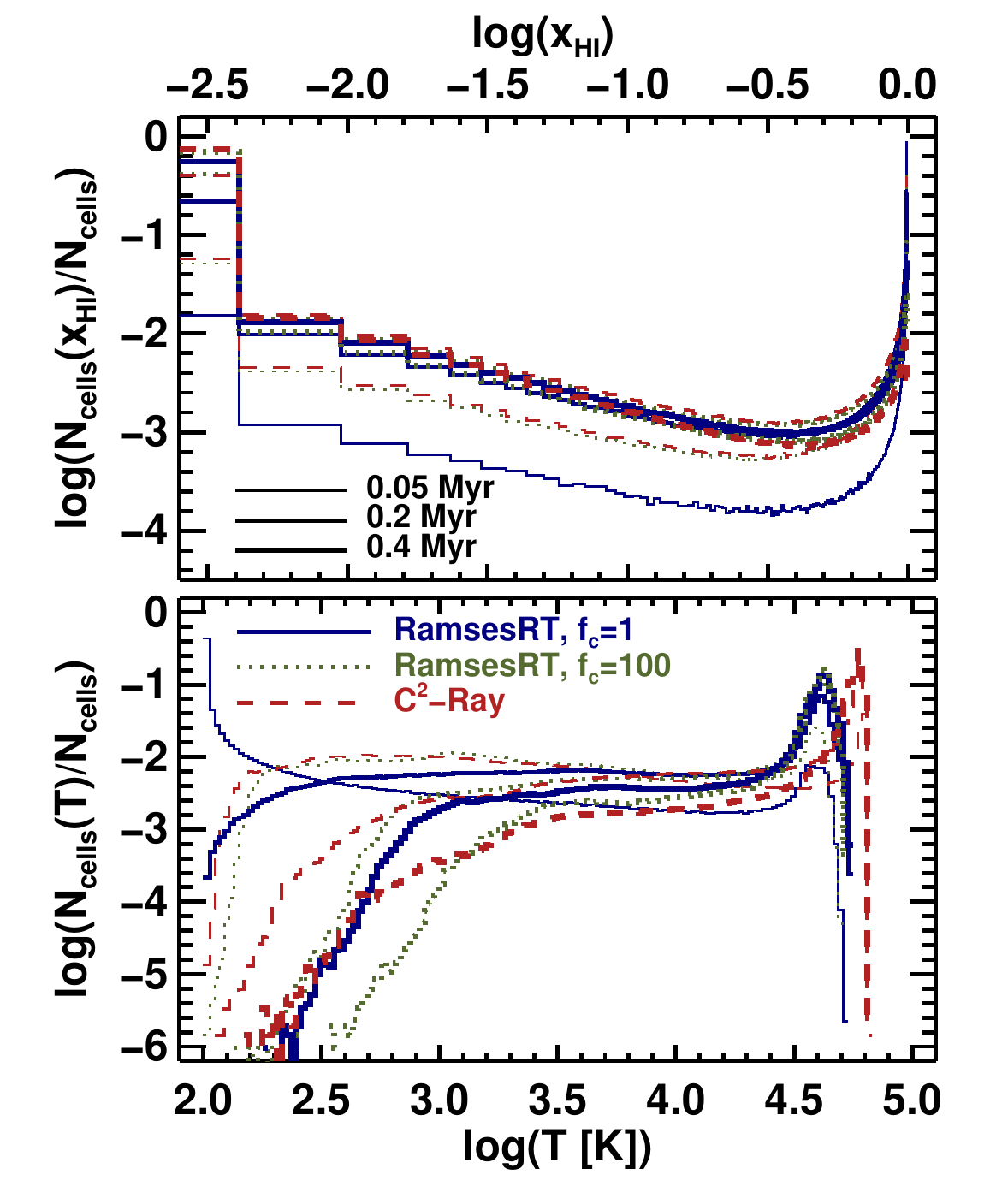}\label{Il4_hists.fig}}\hspace{-1mm}
  \vspace{-0.3mm}
  \caption[\Ila{} test 4 - time evolution and histograms]
  {\Ila{} test 4. \textbf{(a)} Time evolution of the mass
    weighted and volume weighted average ionized
    fractions. \textbf{(b)} Histograms of neutral fraction (top) and
    temperature (bottom).}
\end{figure*}

\Fig{Il4maps.fig} shows box slices, at $z=0.5 \ \Lbox$, of the neutral
fraction and temperature at times $0.05$ and $0.4$ Myr. Shown are our
two runs with different light speed fractions (top and bottom row),
and for comparison we show the result for the \C2R{} code, from
\Ila{}\footnote{Note that \Ila{} have likely mislabeled the maps
  showing the results from this test; their text and captions indicate
  the maps to be at $0.2$ Myr, but judging from the downloadable data
  they are at $0.4$ Myr.}: the I-fronts and photo-heating in our
$f_c=1$ run clearly lag behind the \C2R{} result, and there is also
less heating of the ionized gas. This is in accordance with the
\aton{} results described in \AT, where a similar delay was
found. They prescribed this delay to the fact that \aton{} is
monochromatic, but since our multi-frequency approximation (three
photon groups) gives results that are still much more similar to the
\aton{} results than those of \C2R{}, especially in terms of the
neutral fraction maps, we are inclined to blame the delay on another
factor, which is the speed of light. Our results with the speed of
light set to one-hundred times the physical value are shown in the
bottom row of \Fig{Il4maps.fig} and here the results are considerably
closer to those of \C2R{} in terms of the propagation of heating- and
I-fronts, although the maximum temperature in the ionized gas is still
colder in comparison. All four codes considered in the \Ila{} 4 test
use an infinite effective speed of light and this may give premature
fronts in the immediate vicinity of the sources and also further away
in under-dense regions. Thus we are perhaps not really dealing with a
delay in \ramsesrt{}, but rather premature fronts in the \Ila{}
codes. As \AT{} note, we are far from reaching a static state in the
fronts in this experiment in the run-time of $0.4$ Myr and we should
expect the different light speed runs to converge to similar results
when static state is reached. This is further corroborated by our
I-front light crossing time analysis from Section~\ref{reduced_c.sec}.

The smaller degree of photo-heating in the ionized gas compared to the
\C2R{} results is in line with the temperature profiles from the
previous tests (e.g. \Fig{Il2_profs.fig}), and presumably is a
consequence of the different ways multi-frequency is approximated.
Another notable difference in the maps in \Fig{Il4maps.fig} is that
our fronts are smoother and less jagged than those in \C2R{}. This is
an effect of the photon diffusion inherent in the GLF flux function
used here. Like \AT{} we find that using HLL instead gives more jagged
fronts.

\Fig{Il4_xEvol.fig} shows the evolution of the mass- and
volume-weighted ionized fractions, compared for the different
runs. The \ramsesrt{} run with the physical light speed gives ionized
fractions which are close (both mass- and volume-weighted) to the
\aton{} ones, whereas increasing the light speed by a factor of
hundred from the physical value gives results closer to \C2R{} (as
well as the three other codes that ran this test in
\Ila{}). Presumably we would converge further towards \C2R{} in the
limit of infinite light speed, but computational time constraints do
not allow to pursue that investigation. This is a further hint that
the correct speed of light is important in the non-steady regime of
ionization fronts.

Finally, \Fig{Il4_hists.fig} shows neutral fraction and temperature
histograms at three times in the test. Again there is a strong
discrepancy between the \ramsesrt{} run with $f_c=1$ and \C2R{},
especially at early times, and the gap all but closes when $f_c=100$
is used instead with \ramsesrt{}. There remains some difference though
in the minimum/maximum temperature, being smaller/larger for \C2R{}
than for our $f_c=100$ run, presumably because of our rather crude
multi-frequency approximation.

To summarize, there is notable discrepancy between the \ramsesrt{}
results and those presented in \Ila{}, in that the \ramsesrt{}
ionization front lags behind, which appears to be due to a finite
speed of light. This is corroborated to some degree by other papers in
the literature: \cite{{Wise:2011iw}} use a finite light speed and seem
to get results which are slightly lagging as well.
\cite{{Pawlik:2008kk}} specifically do a comparison between finite and
infinite light speed, with the finite one resulting in a delay which
is substantial, though maybe a bit less than ours, and they do comment
on the ionization bubbles in this test being unphysically large with
infinite light speed methods.  Other sources appear to be in conflict
with our conclusions, though: \cite{{Petkova:2011dz}} use a finite
light speed and get results which seem to compare well with those of
\C2R.

\begin{figure*}
  \subfloat{\includegraphics[width=0.2\textwidth]
    {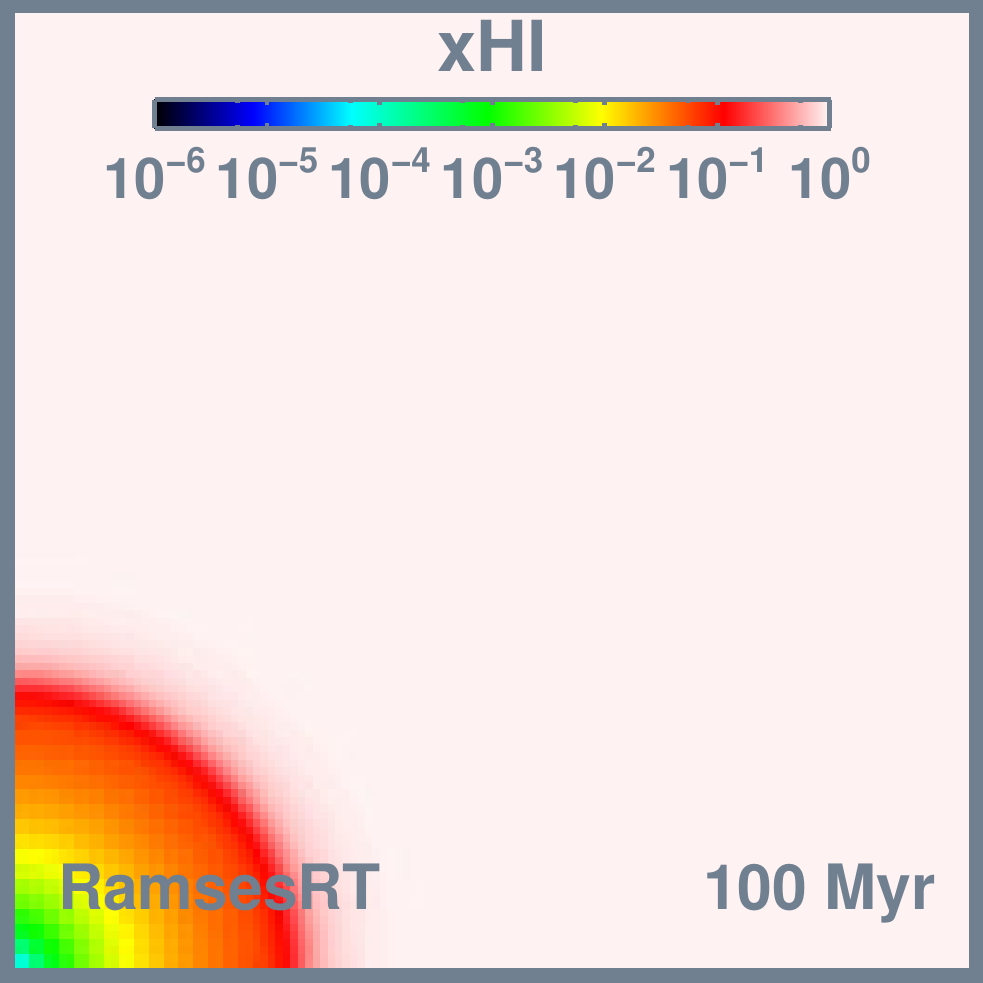}}\hspace{-1.2mm}
  \subfloat{\includegraphics[width=0.2\textwidth]
    {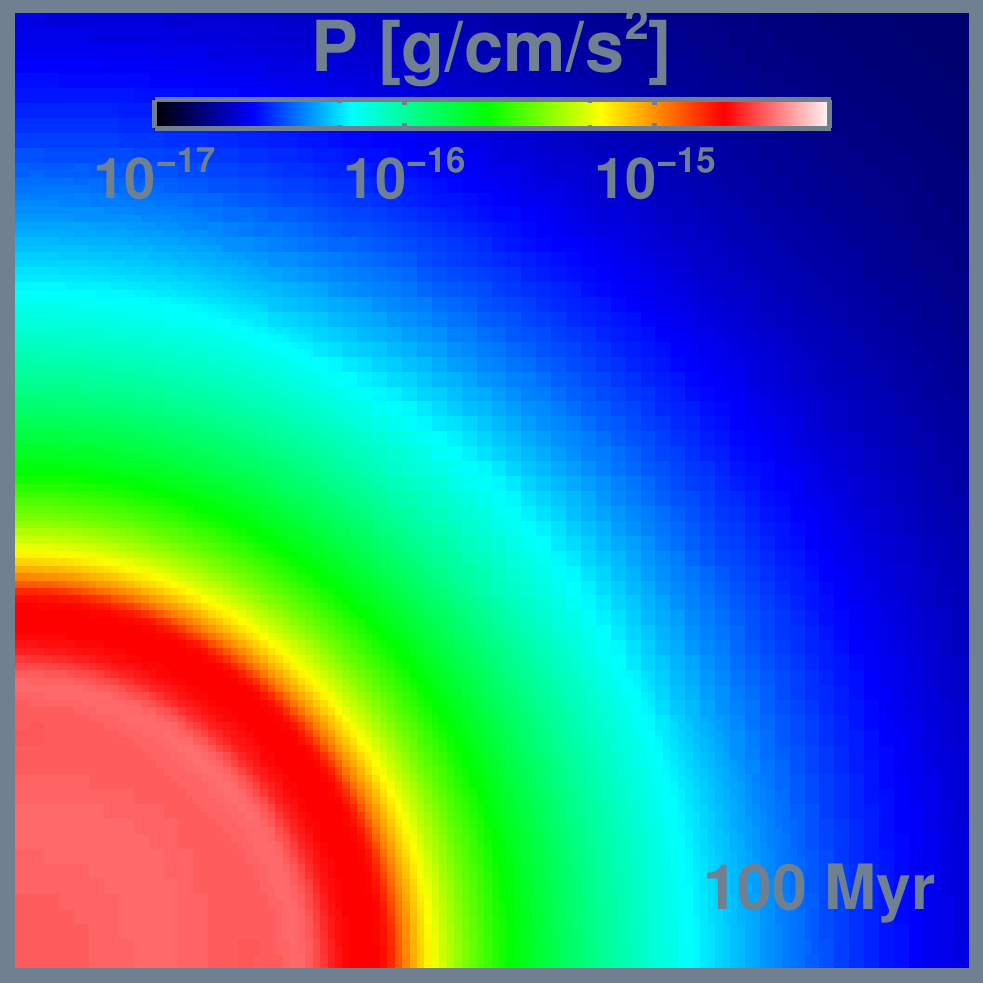}}\hspace{ -1.2mm}
  \subfloat{\includegraphics[width=0.2\textwidth]
    {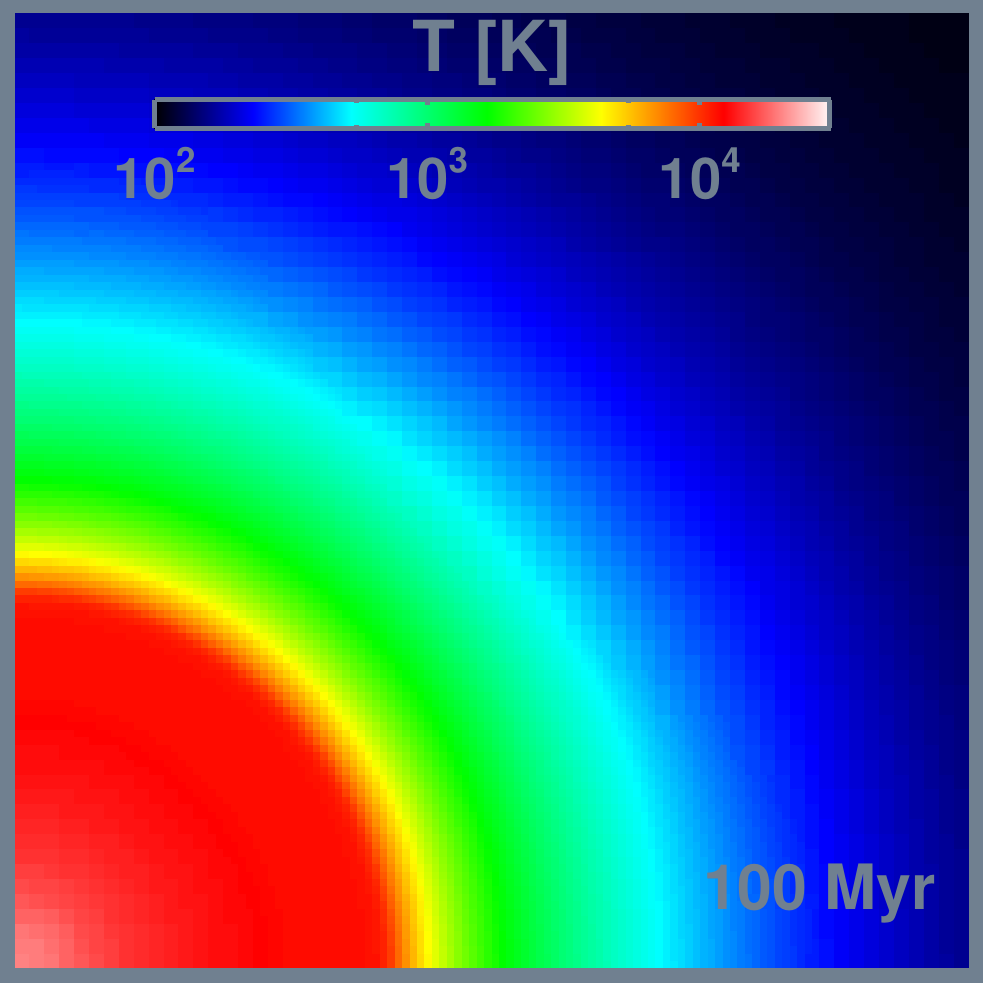}}\hspace{-1.2mm}
  \subfloat{\includegraphics[width=0.2\textwidth]
    {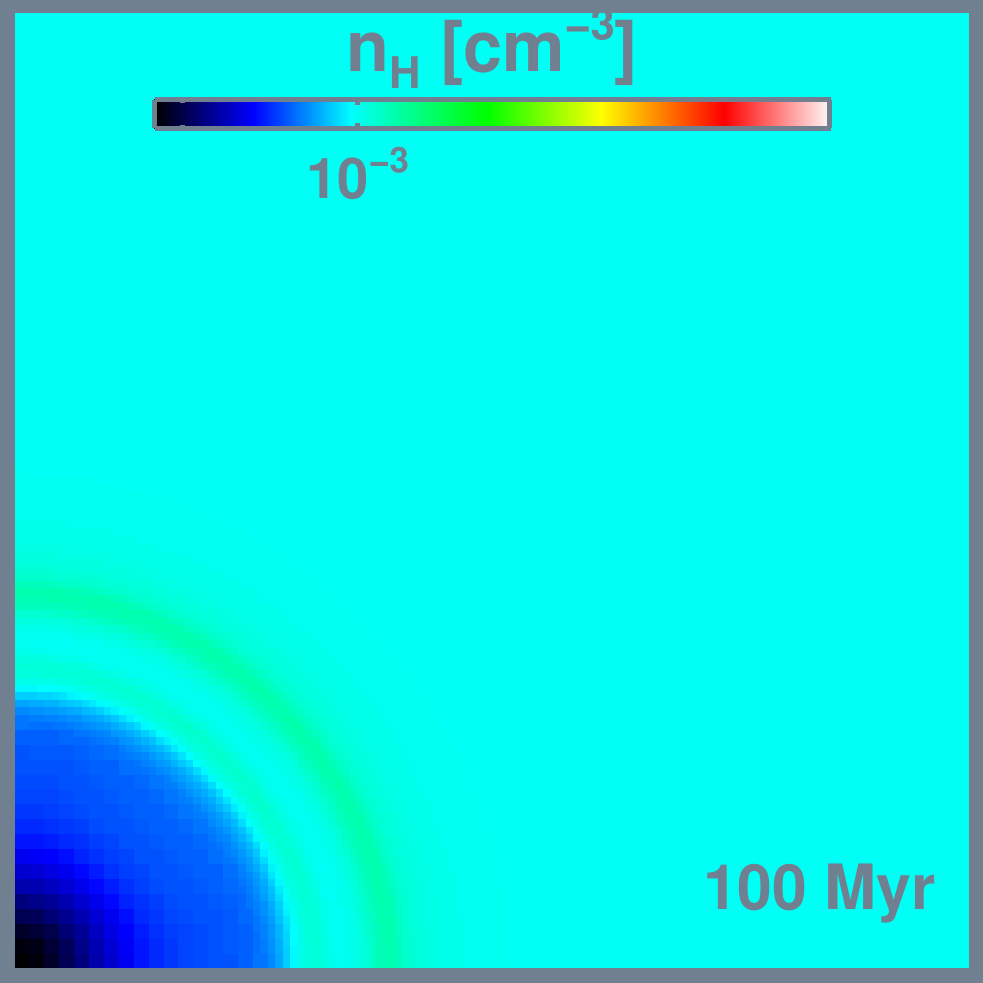}}\hspace{-1.2mm}
  \subfloat{\includegraphics[width=0.2\textwidth]
    {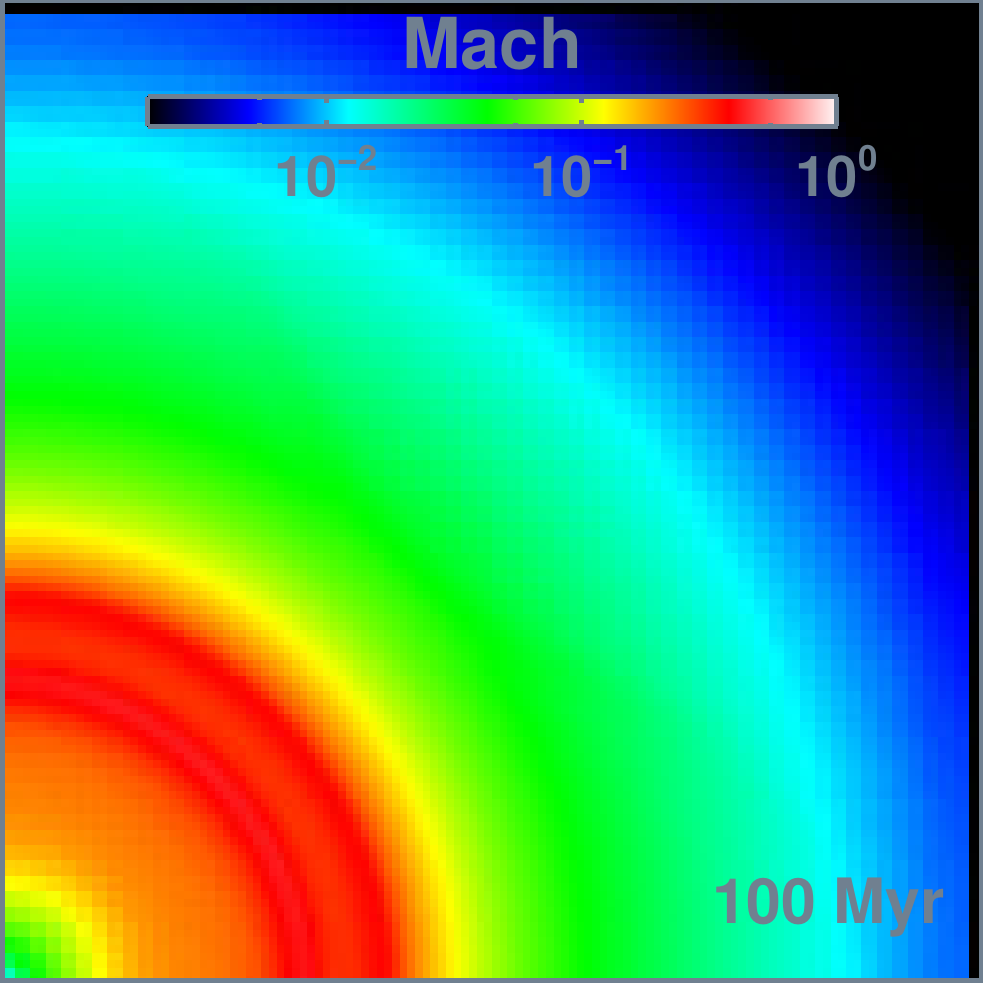}}\hspace{-1.2mm}
  \vspace{-4.mm}

  \subfloat{\includegraphics[width=0.2\textwidth]
    {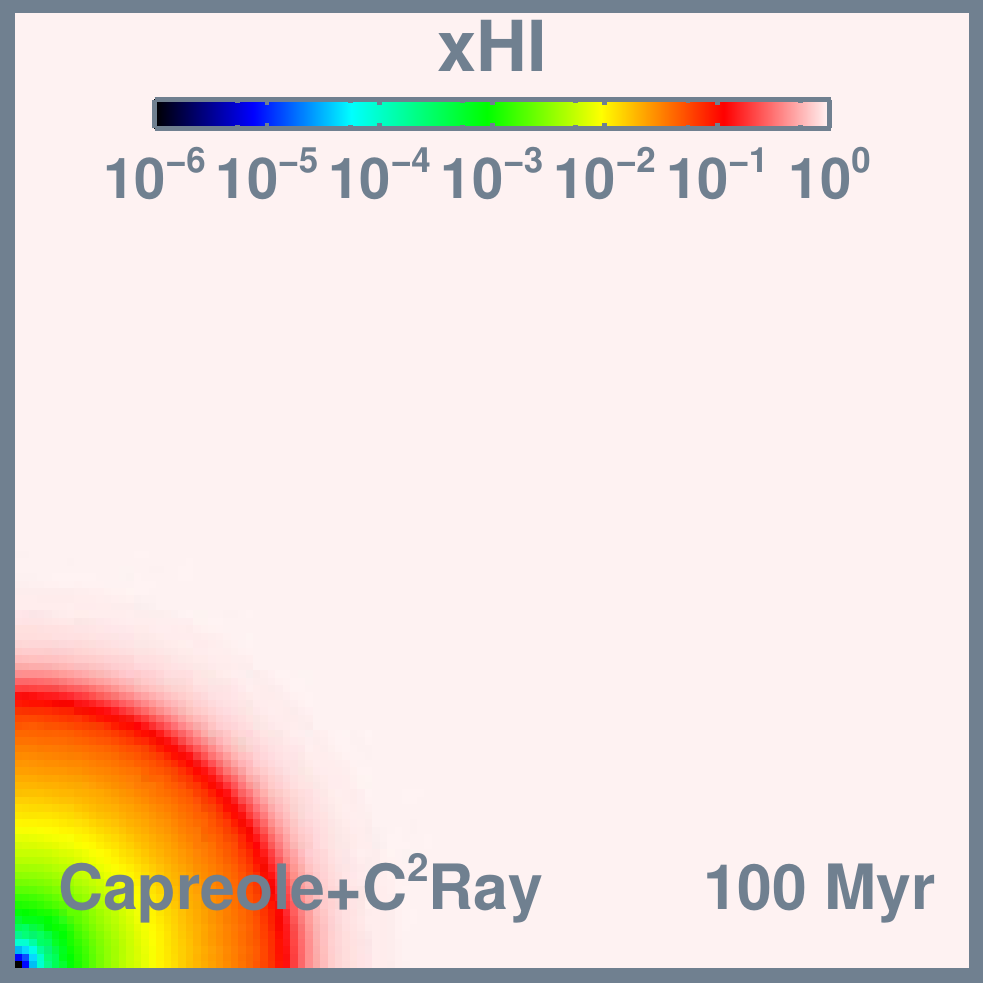}}\hspace{-1.2mm}
  \subfloat{\includegraphics[width=0.2\textwidth]
    {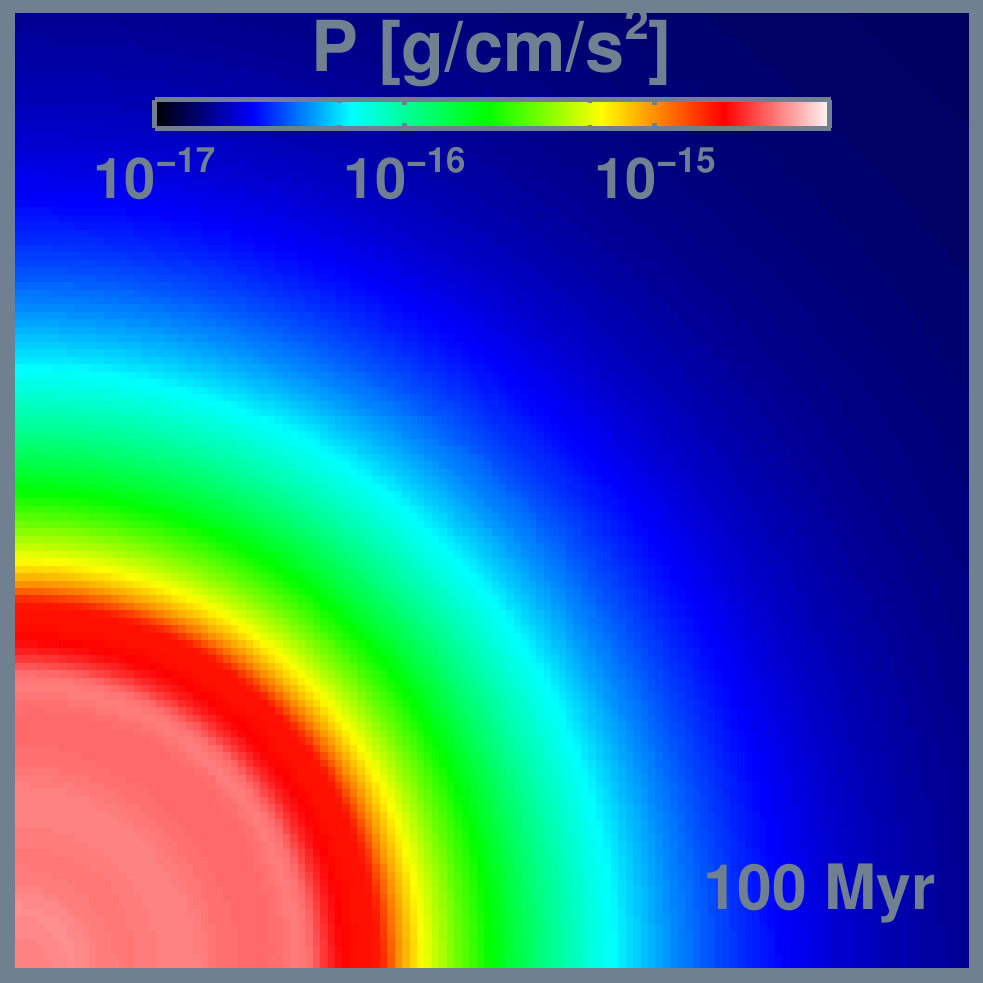}}\hspace{-1.2mm}
  \subfloat{\includegraphics[width=0.2\textwidth]
    {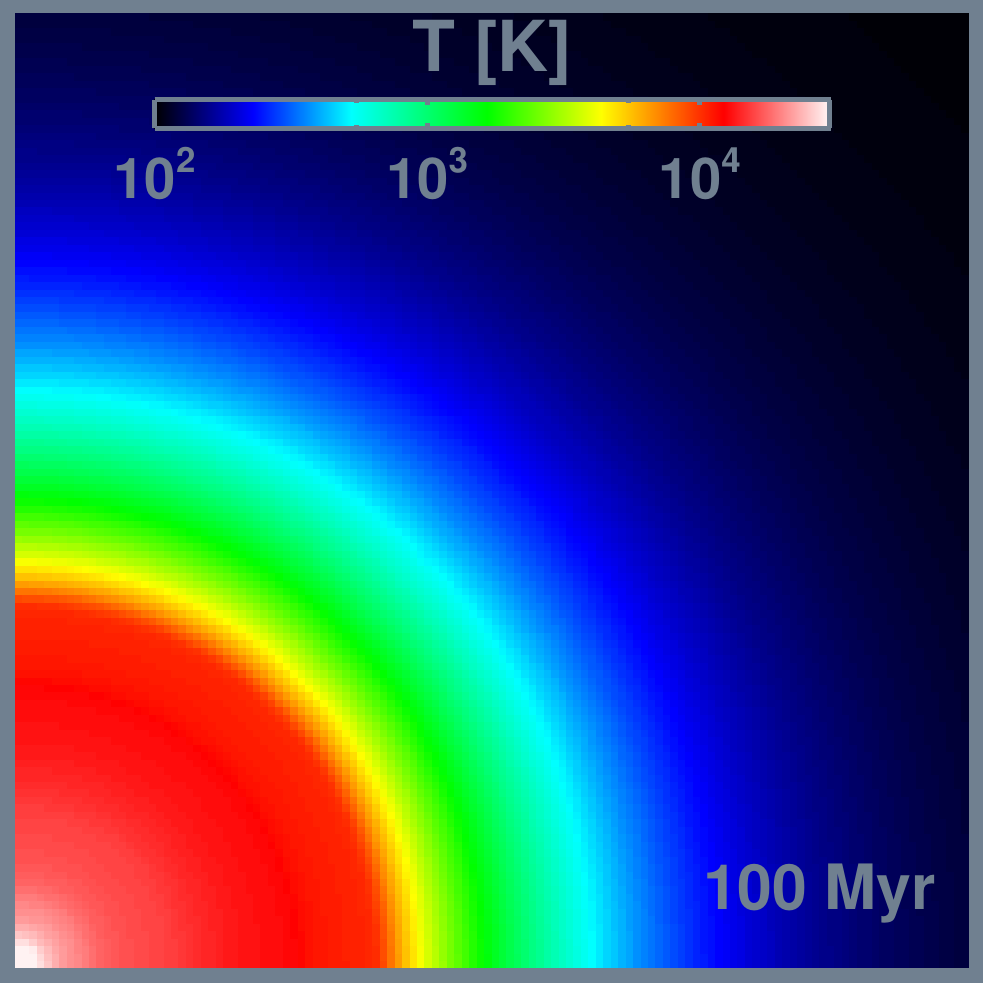}}\hspace{-1.2mm}
  \subfloat{\includegraphics[width=0.2\textwidth]
    {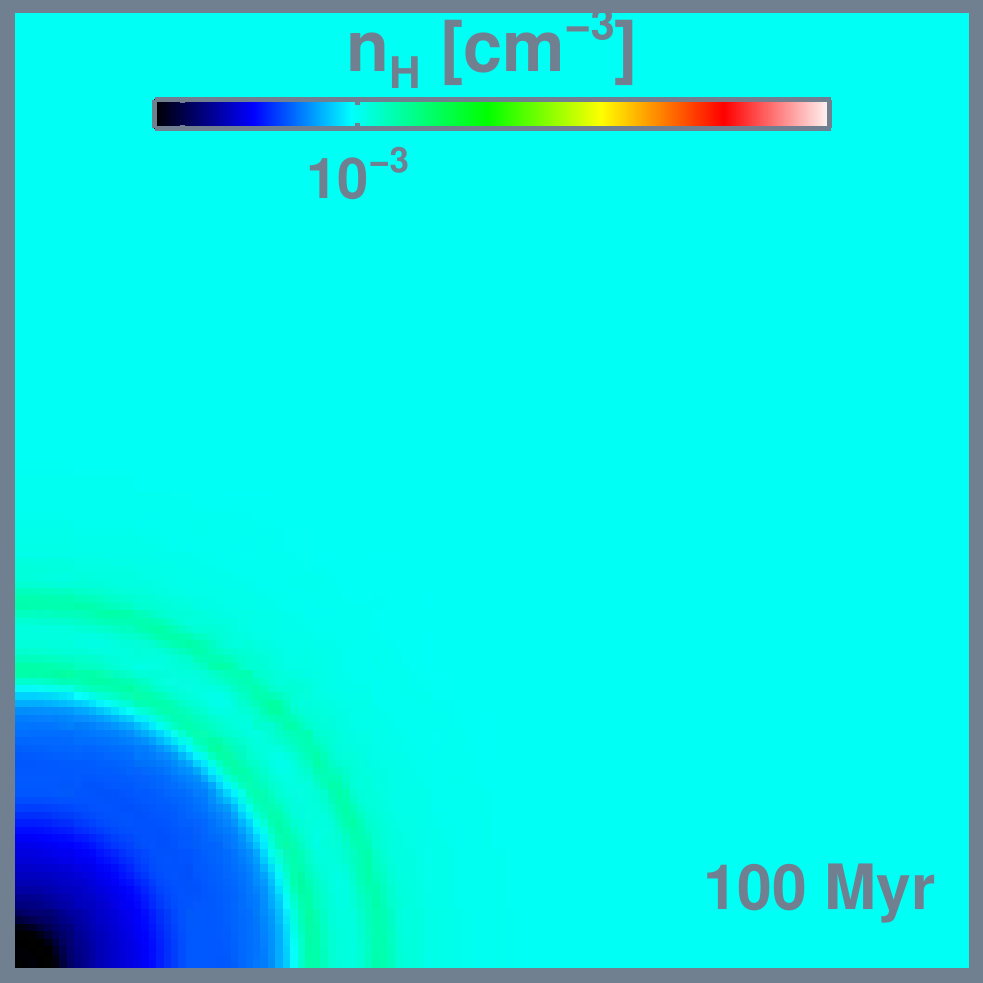}}\hspace{-1.2mm}
  \hspace{0.2\textwidth}\hspace{-0.2mm}

  \vspace{-2.5mm}

  \subfloat{\includegraphics[width=0.2\textwidth]
    {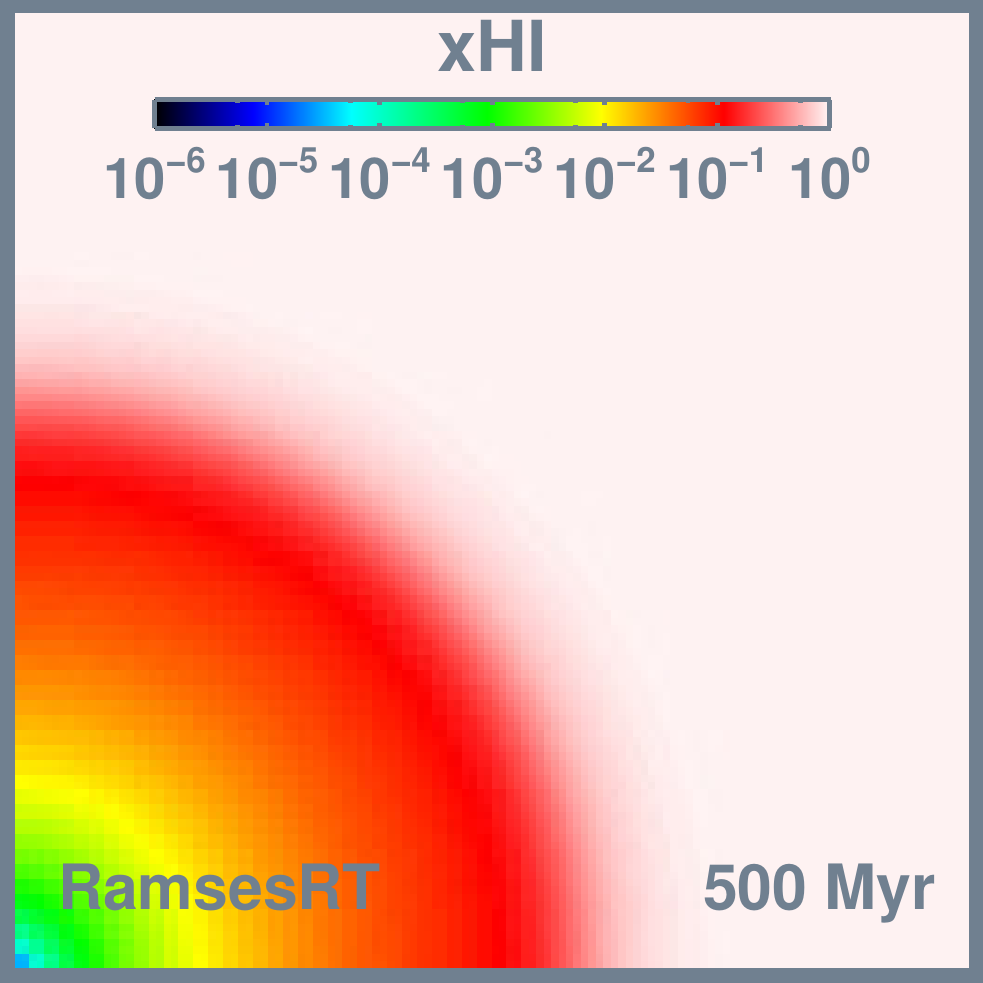}}\hspace{-1.2mm}
  \subfloat{\includegraphics[width=0.2\textwidth]
    {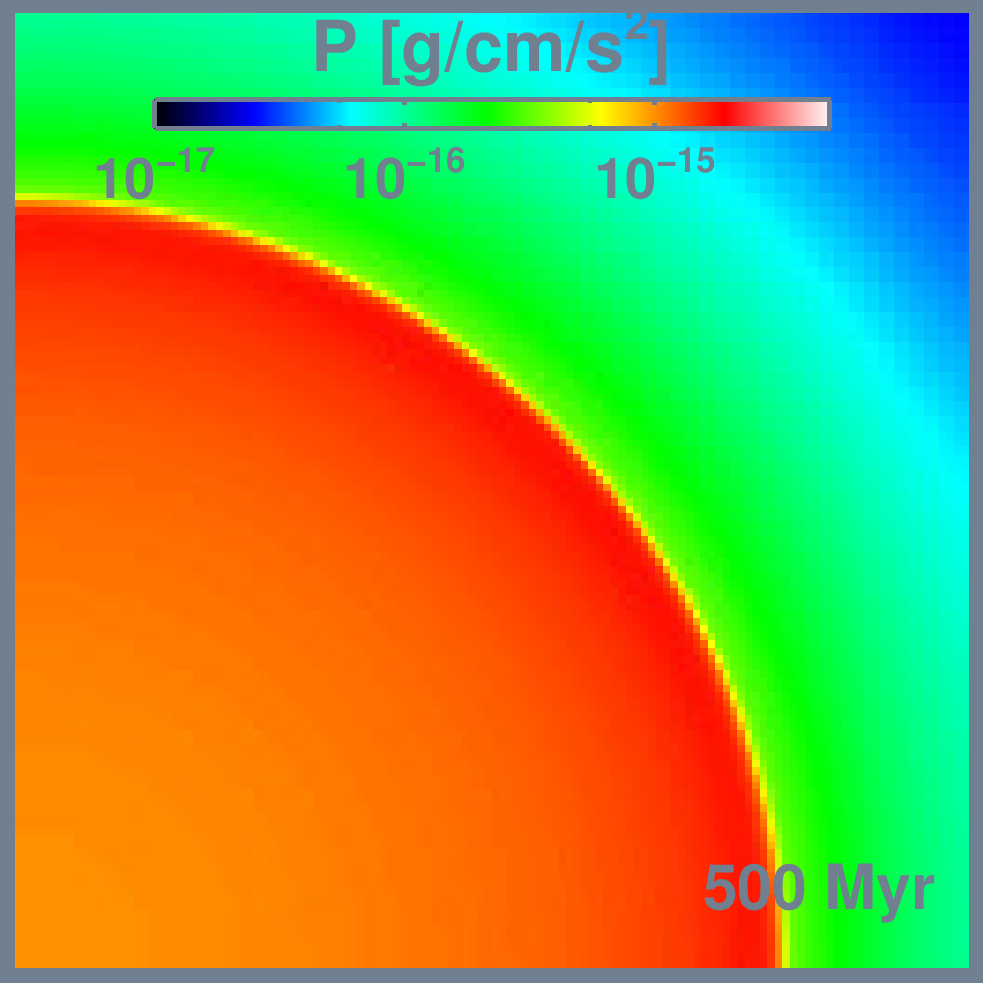}}\hspace{ -1.2mm}
  \subfloat{\includegraphics[width=0.2\textwidth]
    {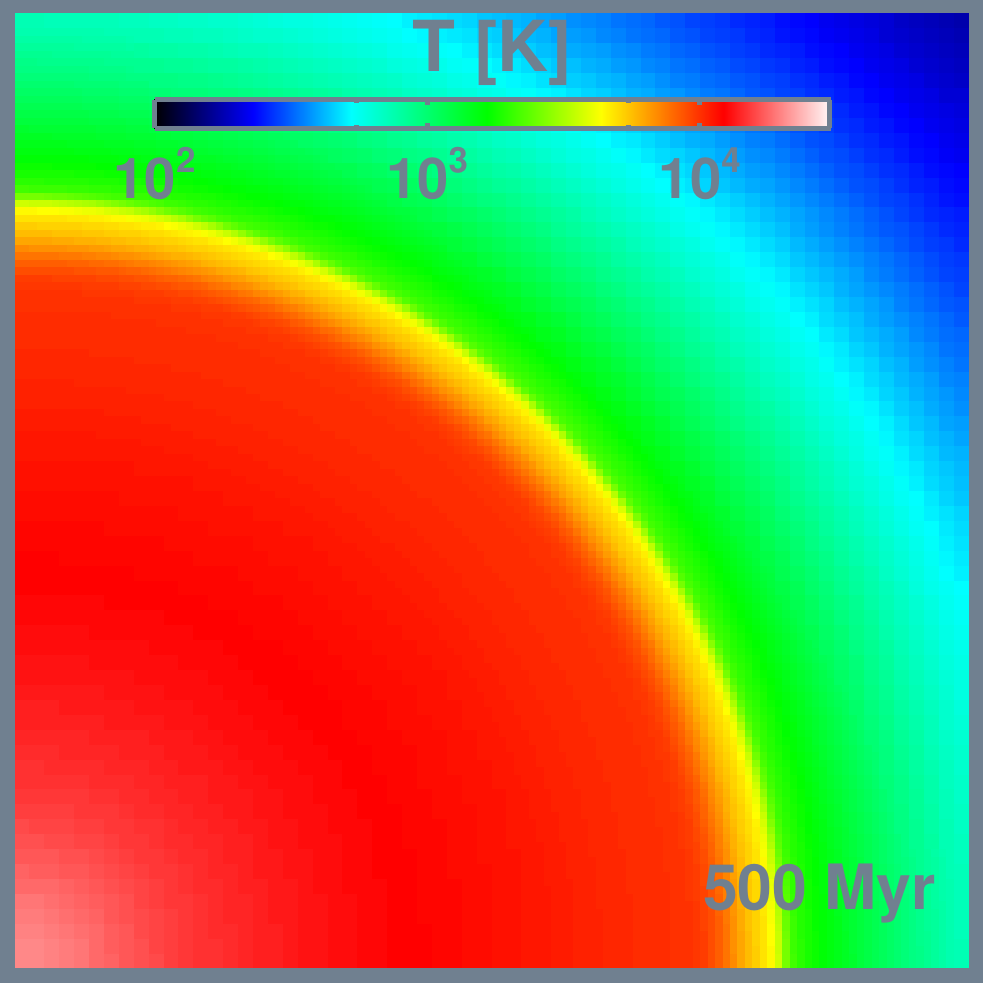}}\hspace{-1.2mm}
  \subfloat{\includegraphics[width=0.2\textwidth]
    {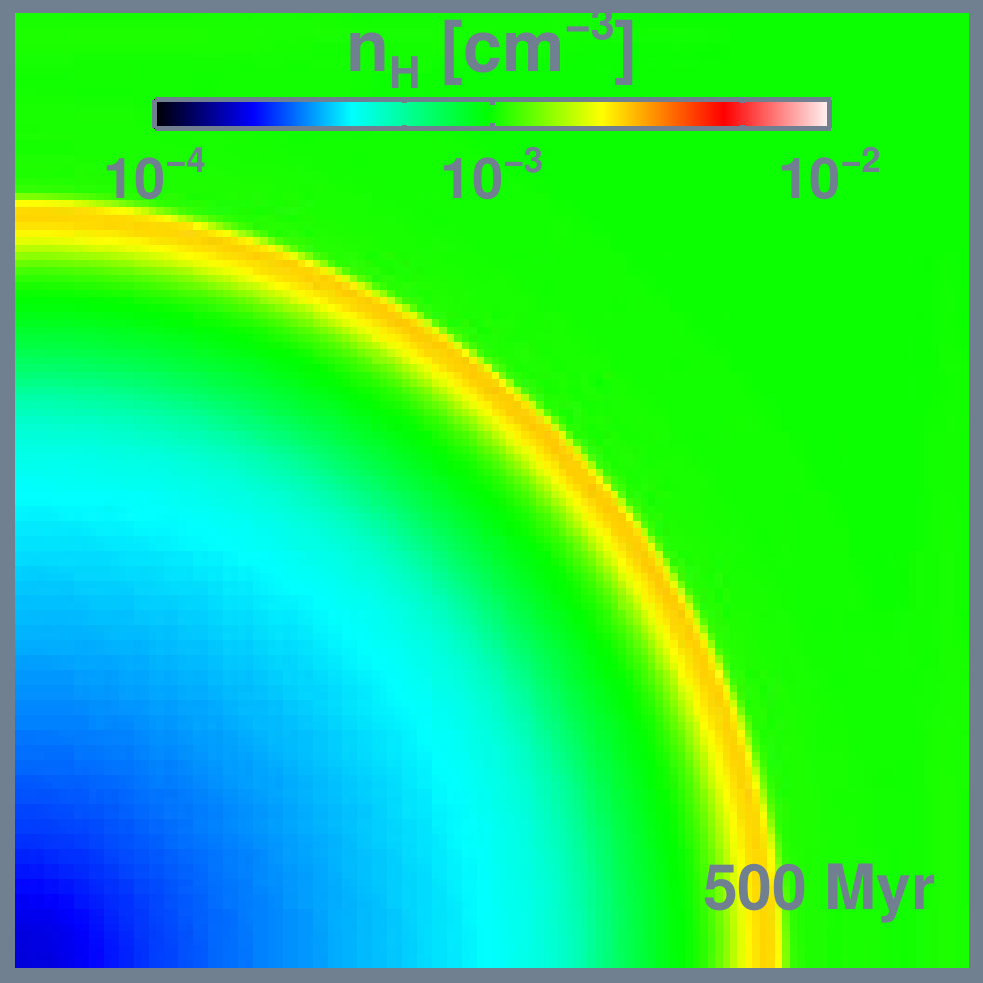}}\hspace{-1.2mm}
  \subfloat{\includegraphics[width=0.2\textwidth]
    {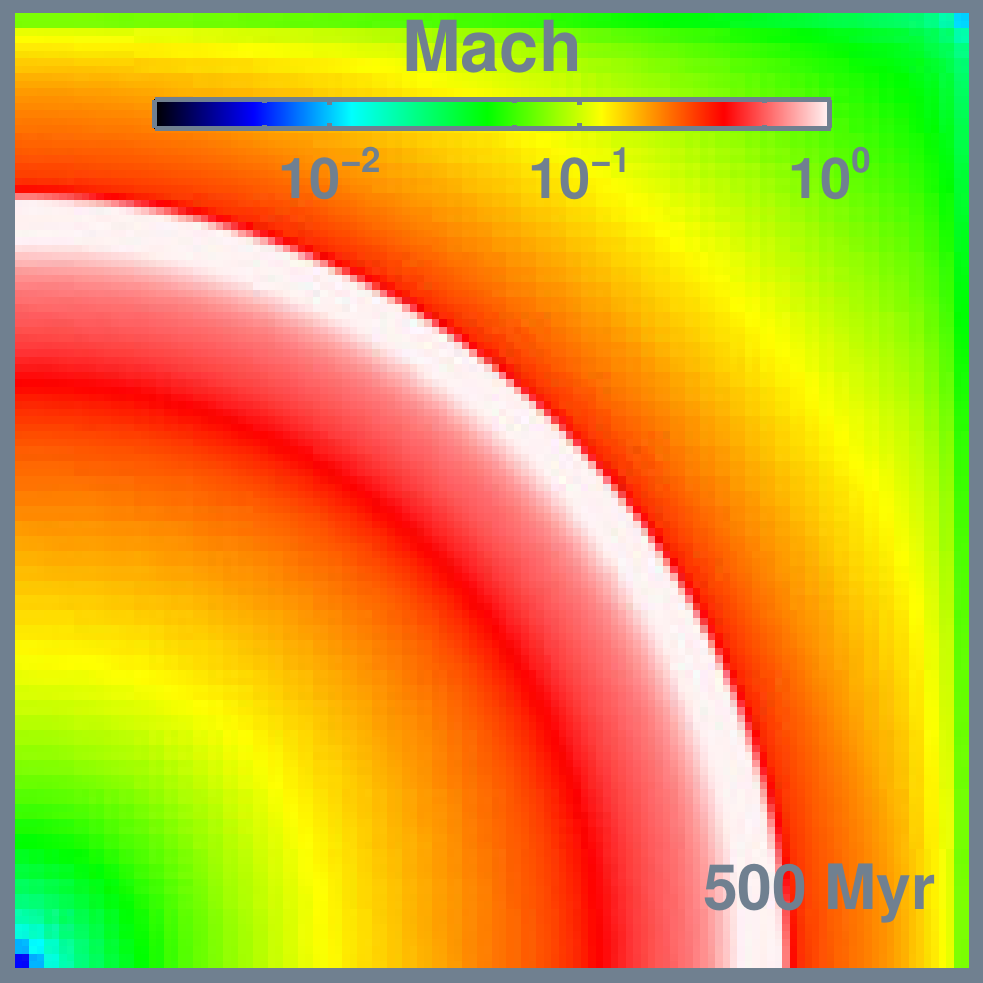}}\hspace{-1.2mm}
  \vspace{-4mm}

  \subfloat{\includegraphics[width=0.2\textwidth]
    {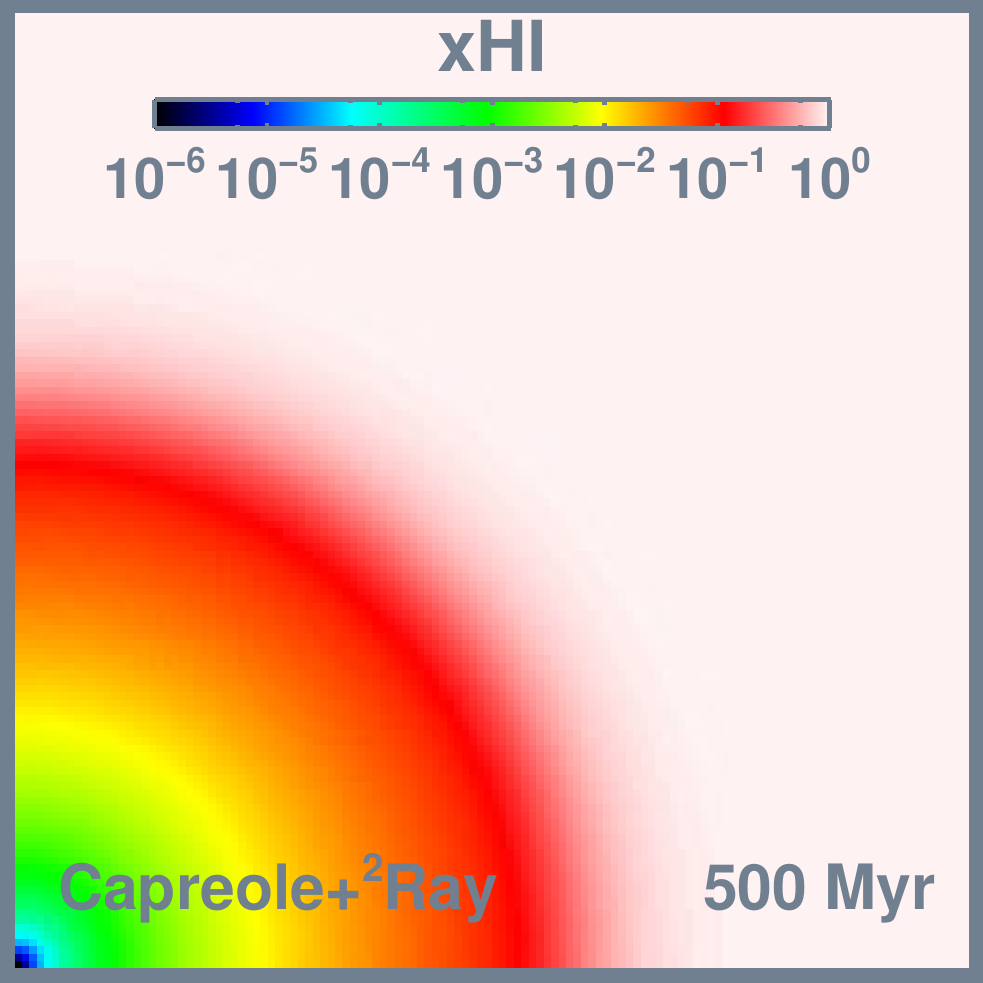}}\hspace{-1.2mm}
  \subfloat{\includegraphics[width=0.2\textwidth]
    {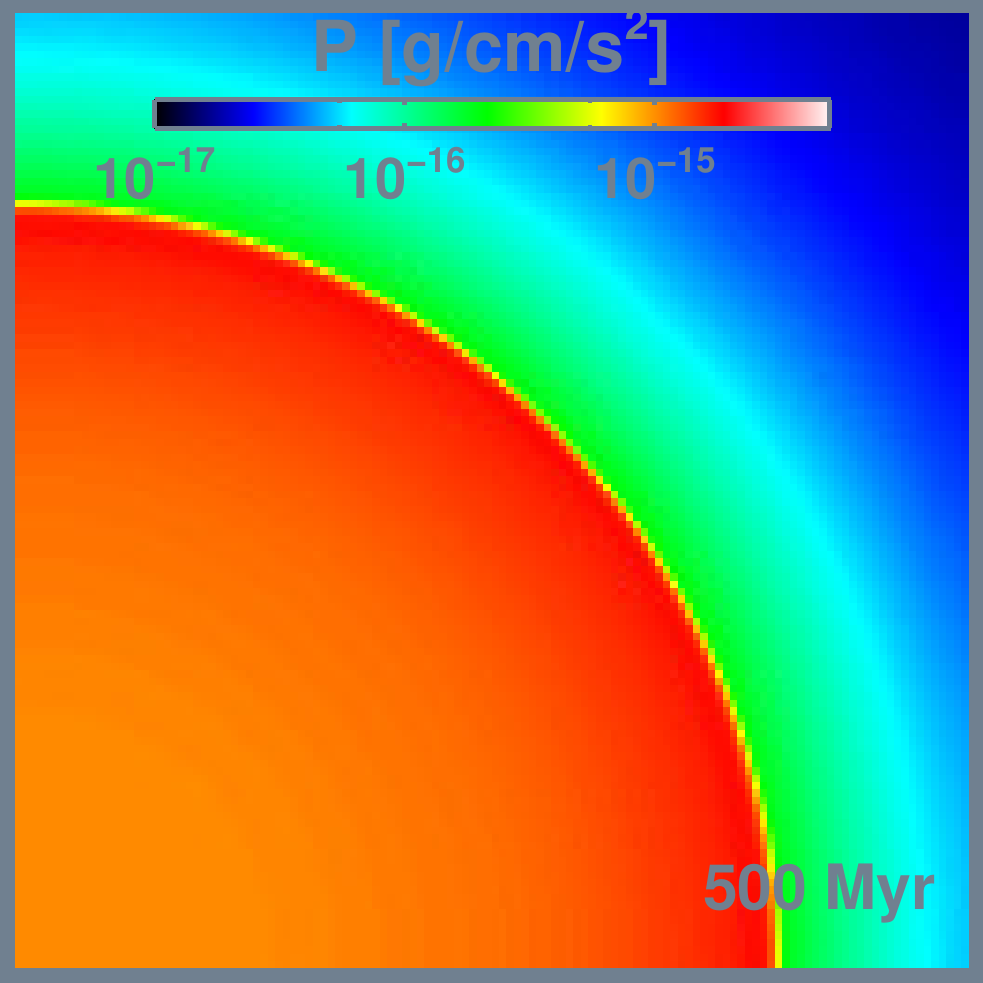}}\hspace{-1.2mm}
  \subfloat{\includegraphics[width=0.2\textwidth]
    {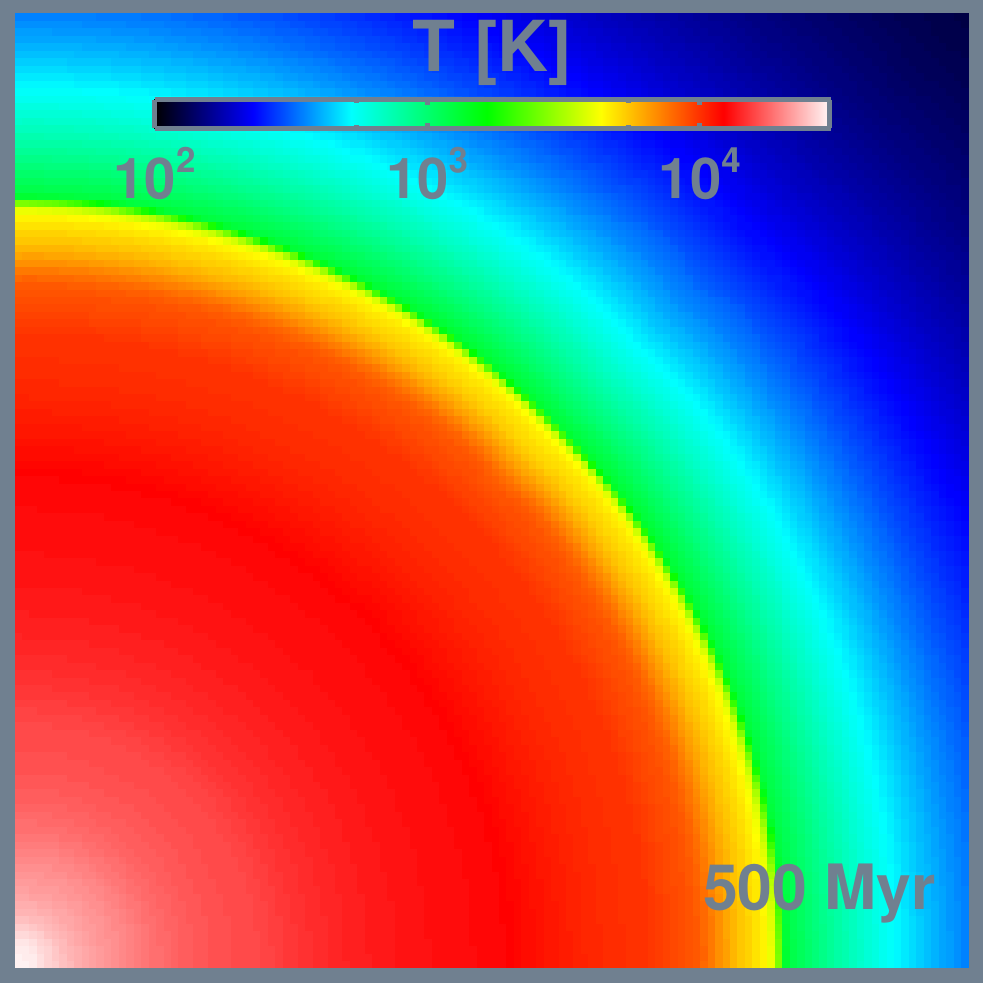}}\hspace{-1.2mm}
  \subfloat{\includegraphics[width=0.2\textwidth]
    {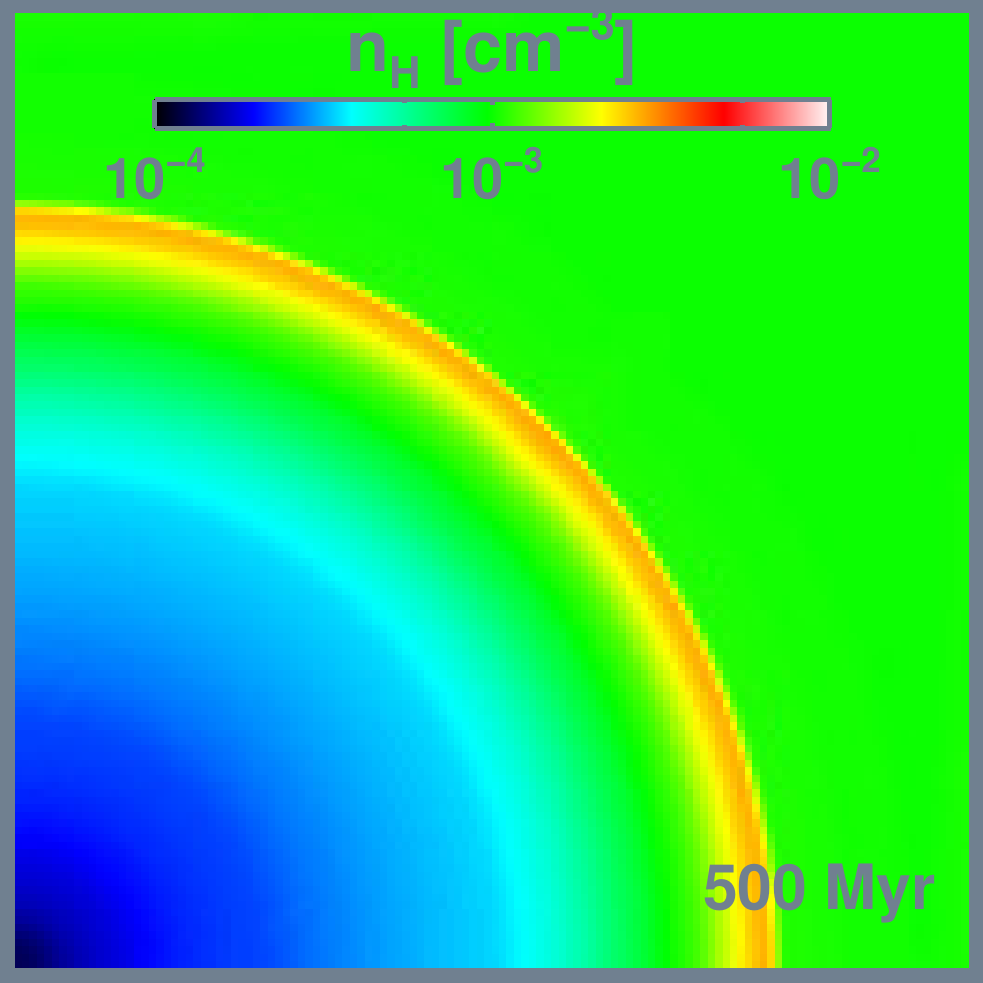}}\hspace{-1.2mm}
  \hspace{0.2\textwidth}\hspace{-0.2mm}

  \vspace{-2.5mm}
  \caption[\Ilb{} test 5 - maps]
  {\label{Il5maps.fig}\Ilb{} test 5. Maps showing slices at $z=0$ of
    various quantities at 100 Myrs (top panel) and 500 Myrs (lower
    panel). In each panel, the top row shows the \ramsesrt{} results
    and the lower row shows the \CC2R{} results for comparison.}
\end{figure*}

\subsection {\Ilb{} test 5: \ \ Classical HII region
  expansion}
We now come to the tests described in second radiative transfer codes
comparison paper by \cite{{Iliev:2009kn}}, which we denote as
\Ilb{}. This paper provides 3 code comparison tests to add to those in
\Ila{}, but with the important difference that whereas the \Ila{}
tests are pure radiative transfer post-processing tests with fixed
density fields, the tests in \Ilb{} are RHD tests, i.e. with the
radiative transfer directly coupled to the gas-dynamics. Thus we now
switch from the context of post-processing RT to hydro-coupled RHD.
Here, the pressure buildup in photo-heated gas causes it to expand.
Typically, the I-front is initially \textit{R-type}, where it expands
much faster than the gas response to it, which means RT-postprocessing
is a fairly good approximation. The I-front then begins to slow down
when it approaches the Str\"omgren radius, but gets moving again when
the gas catches up to it, and then the front is \textit{D-type},
i.e. moves along with the expanding gas.

\begin{figure*}\begin{center}
  \includegraphics[width=0.65\textwidth]
  {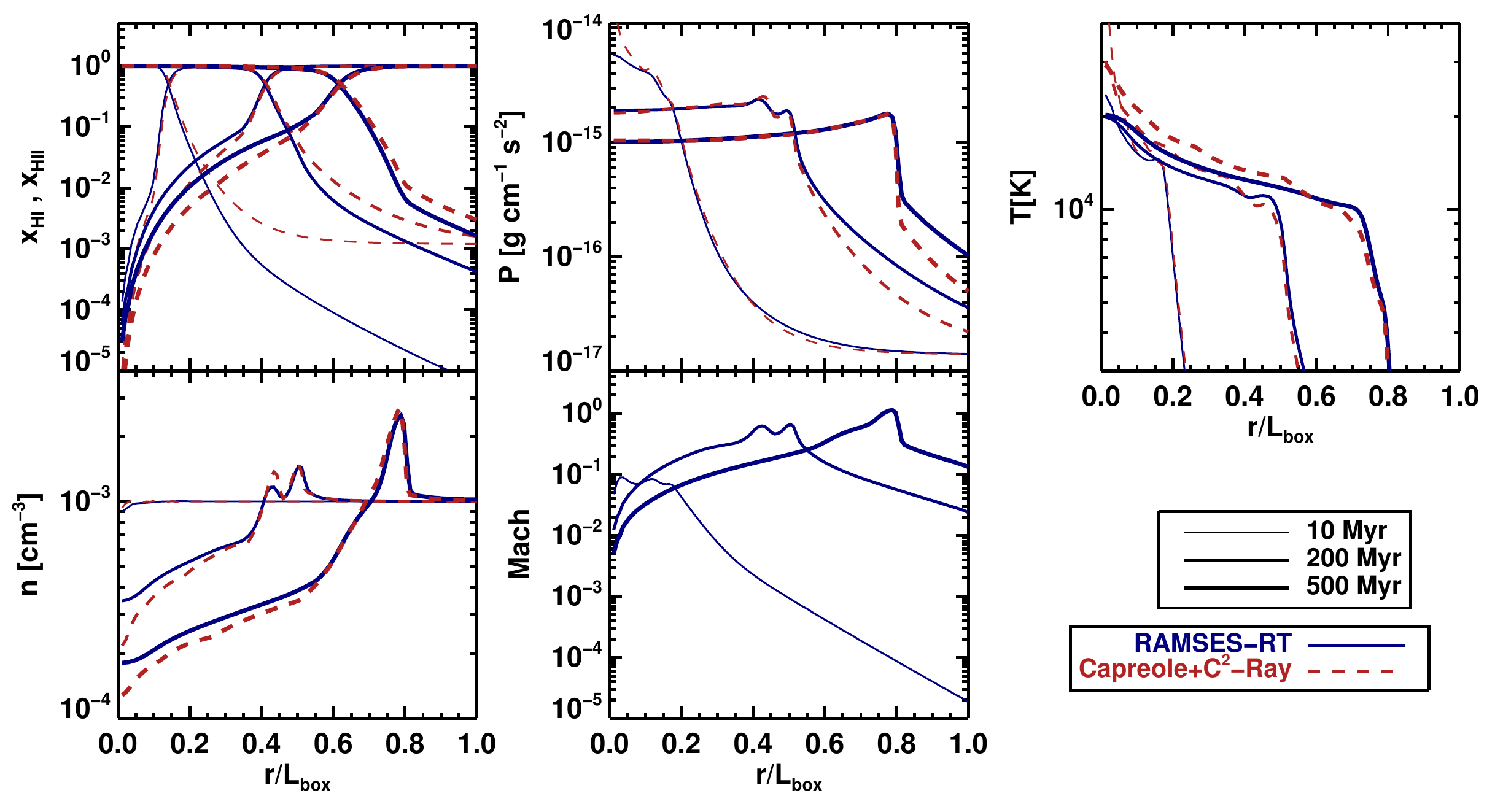}
  \caption[\Ilb{} test 5 - radial profiles]
  {\label{Il5_Profs.fig}\Ilb{} test 5. Radial profiles at
    10, 200 and 500 Myrs, compared to the \CC2R{}
    results. Clockwise from top left: ionization fractions, pressure,
    temperature, Mach number, atom number density.}
\end{center}\end{figure*}


As before we compare our \ramsesrt{} tests results with those of the
grid-based short characteristics ray-tracing code \C2R{}
\citep{{Mellema:2006cl}}, here coupled to the \Capreole{} code, which
employs a Riemann solver for the hydrodynamics. \joki{As the \CC2R{}
  combination is sensitive to numerical instabilities appearing in
  \Ilb{} test 6, we compare also in that particular test to \C2R{}
  coupled to the Eulerian TVD solver of \cite{{Trac:2004hp}} (that
  combination was not used in any other tests).} The test numbers
continue from the \Ila{} paper, thus we now come to \Ilb{} test 5,
which concerns the expansion of an ionization front due to a point
source in an \textit{initially} uniform-density medium.  The initial
setup, much like that of \Ila{} test 2, is as follows.

The box cube is $\Lbox=15$ kpc in width. The gas is hydrogen only as
usual, initially homogeneous with density $\nh=10^{-3} \, \cci$,
temperature $100$ K, and ionization fraction $\xhi=10^{-6}$ (\Ilb{}
prescribes $\xhi=0$). The radiative source is in the corner of the box
and the emission rate is $\dot{N}_{\gamma}=5 \times 10^{48} \,
\emrate$. We don't apply the OTSA in this test, i.e photons are
emitted from gas recombinations. The simulation time is $500$ Myr. The
base resolution of the box is $64^3$ cells and we apply on-the-fly
refinement on $\nh$ and $\xhii$ gradients (see \Eq{Il1_refine.eq}), so
that the ionization front has the prescribed effective resolution of
$128^3$ cells.

We first compare volume dissections at $z=0$ in the simulation
cubes at $100$ and $500$ Myr, for the \ramsesrt{} and \C2R{} results,
shown in \Fig{Il5maps.fig}. The maps show, from left to right, the
neutral fraction, pressure, temperature, density and mach number,
$M\equiv v/c_S$, where $c_s=\sqrt{1.4\ P/\rho}$ is the sound
speed. (Unfortunately the $M$ output is missing from the \C2R{}
results we've downloaded.) In these maps, the \ramsesrt{} results look
very similar to those of \C2R{}. The $\xhi$-maps show stronger
ionization immediately around the corner source in the \C2R{} result,
and correspondingly the temperature and density maps show this corner
gas is also hotter and more diffuse in the \C2R{} result than in
\ramsesrt{}. Conversely, the photo-heating region is somewhat
further-reaching in the \ramsesrt{} result than in \C2R{}, as can be
seen in the pressure and temperature maps. These small differences are
likely due to the different approaches in approximating
multi-frequency. Notably, the \C2R{} maps stand out in a very similar
way when compared to most of the corresponding maps from other codes
in \Ilb{}, i.e. a stronger effect close to the radiative source but
shorter-reaching photo-heating.

To paint a more quantitative picture, \Fig{Il5_Profs.fig} compares
radial profiles of the same quantities ($\xhi$, $P$, $T$, $\nh$ and
$M$) for \ramsesrt{} and \C2R{} at $10$, $200$ and $500$ Myr. The
ionization state profiles (top left) indeed show \C2R{} to ionize the
gas more strongly close to the radiative source, but \ramsesrt{} to
ionize more strongly beyond the I-front. The I-front itself is however
at very similar positions at all times. The pressure and temperature
plots show the same thing, but apart from these minor differences at
the extreme ends the shapes are very similar. The density plots show
that \C2R{} has more has more diffuse gas close to the source as a
result of the stronger photoheating, and also it appears to have a
more pronounced backflow peak around $200$ Myr (this double peak is a
temporary effect of photo-heating by high-energy photons beyond the
I-front). The smaller backflow peak in \ramsesrt{} is perhaps in part
a relic of on-the-fly refinement, though most of the codes in \Ilb{}
actually have backflow peaks similarly smaller than that of
\C2R{}. Unfortunately we can't compare the Mach profiles directly, but
the \ramsesrt{} profiles do look very similar in shape to those
presented in \Ilb{} (see their Fig. 15).

\begin{figure}\begin{center}
  \includegraphics[width=0.3\textwidth]
  {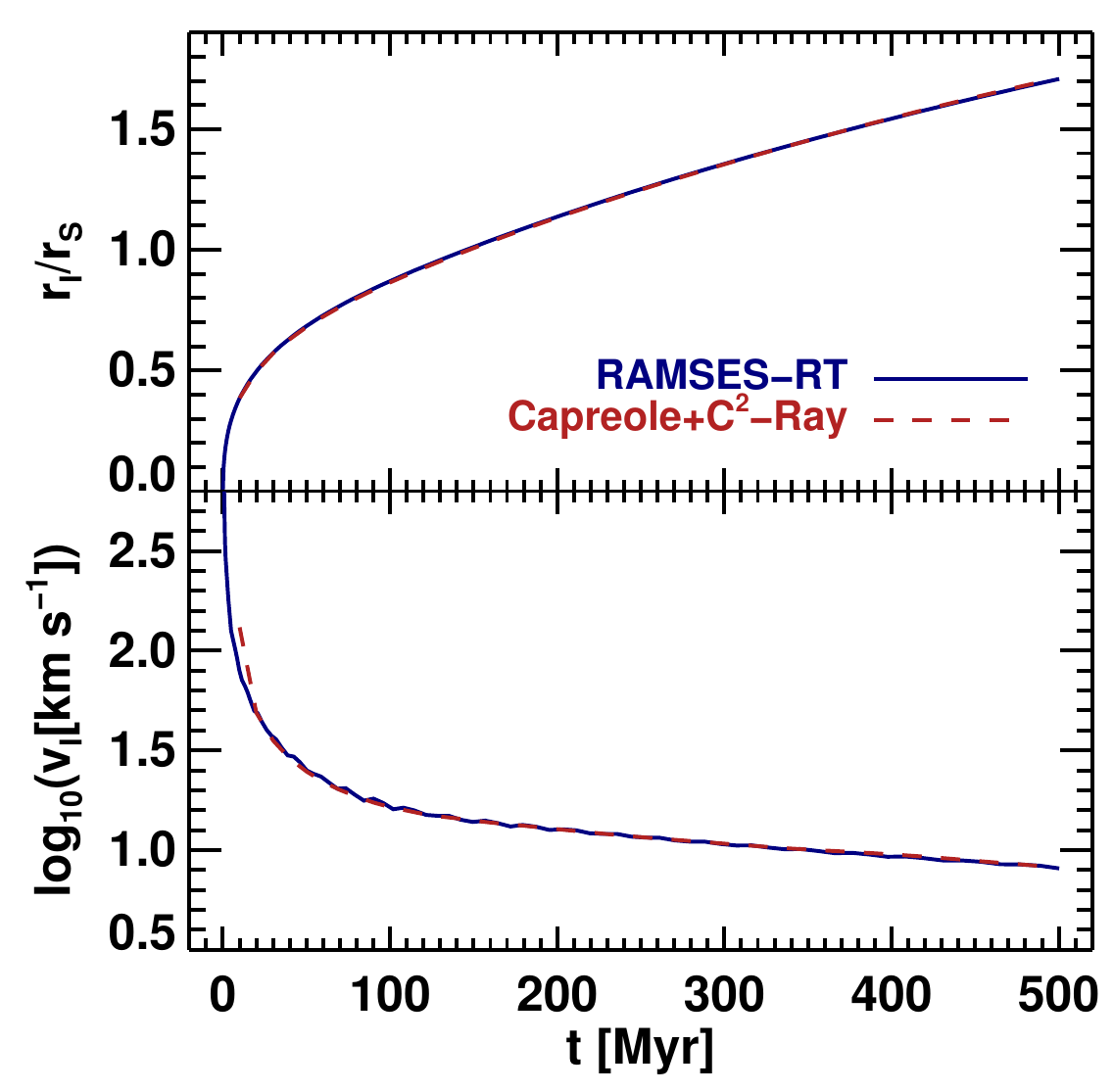}
  \caption{\label{Il5_Ifront.fig} \Ilb{} test 5: Time evolution of the
    ionization front, compared to the results from the \CC2R{}
    combination. Upper plot shows the radius of the Str\"omgren sphere
    in units of 5.4 kpc. The lower plot shows the speed of the front
    propagation.}
\end{center}\end{figure}

Finally, \Fig{Il5_Ifront.fig} shows how the position and velocity of
the I-front (defined as where the radial average of $\xhii$ is equal
to 0.5), for \ramsesrt{} and \C2R{}. The plots for the two codes are
virtually identical, the only noticeable difference being a slight
initial lag in the front speed. One might attribute this to the
reduced speed of light in the \ramsesrt{} run, but actually most other
codes described in \Ilb{} have a very similar lag in the initial front
speed compared to \C2R{}.

\joki{The fraction of the volume refined to the effective resolution
  of $128^3$ cells is $28\%$ at the end of the run, and the
  computational time is roughly half that of an analogous uniform grid
  run. The runs clock in at about double the cpu hours of test 1, even
  though test 1 had roughly twice the number of timesteps to perform,
  due to a smaller box width. This gives a qualitative idea of the
  added cost of adding two more photon groups (test 1 had one group)
  and coupling with the hydrodynamics, which totals to about four
  times the computational load.}

All in all, the \ramsesrt{} results for this test compare very well
with most of the codes presented in \Ilb{}. The \ramsesrt{} result
differs slightly from that of \C2R{} in some aspects, most notably in
the form of weaker photo-heating and ionization close to the radiative
source and wider I-fronts. However, these are precisely the aspects
where \C2R{} stands out from the other codes presented in \Ilb{}.

\subsection{\Ilb{} test 6: \ \ HII region expansion in a
  $\mathbf{r^{-2}}$  density profile}\label{iliev6.sec}
This test mimics a radiative source going off in a dense cloud, e.g. a
stellar nursery. The setup is much like that of the preceding test 5,
the main difference being that the gas is here inhomogeneous, the box
is much smaller, $\Lbox=0.8$ kpc in width, and the radiative corner
source is a hundred times more luminous, i.e. it radiates at
$\dot{N}_{\gamma}=5 \times 10^{50} \, \emrate$. As in the previous
test we don't apply the OTSA. The base resolution is $64^3$ cells, but
on-the-fly refinement on $\nh$ and $\xhii$ gradients ensures the
prescribed effective resolution of $128^3$ cells at ionization and
shock fronts. The initial temperature is $100$ K everywhere and the
running time is $75$ Myr.  The dense cloud is centered on the corner
source and is set up with a spherically symmetric, steeply decreasing
power-law density profile with a small flat central core of gas number
density $n_0=3.2$ $\cci$ and radius $r_0=91.5$ pc:
\begin{equation}
  \nh(r) = 
  \begin{cases} 
  \; n_0 &  \mathrm{if} \, r\le r_0 
  \\ 
  \; n_0(r_0/r)^{2} &  \mathrm{if} \, r\ge r_0. 
\end{cases}
\end{equation}
 
\begin{figure}\begin{center}
    \includegraphics[width=0.3\textwidth]
  {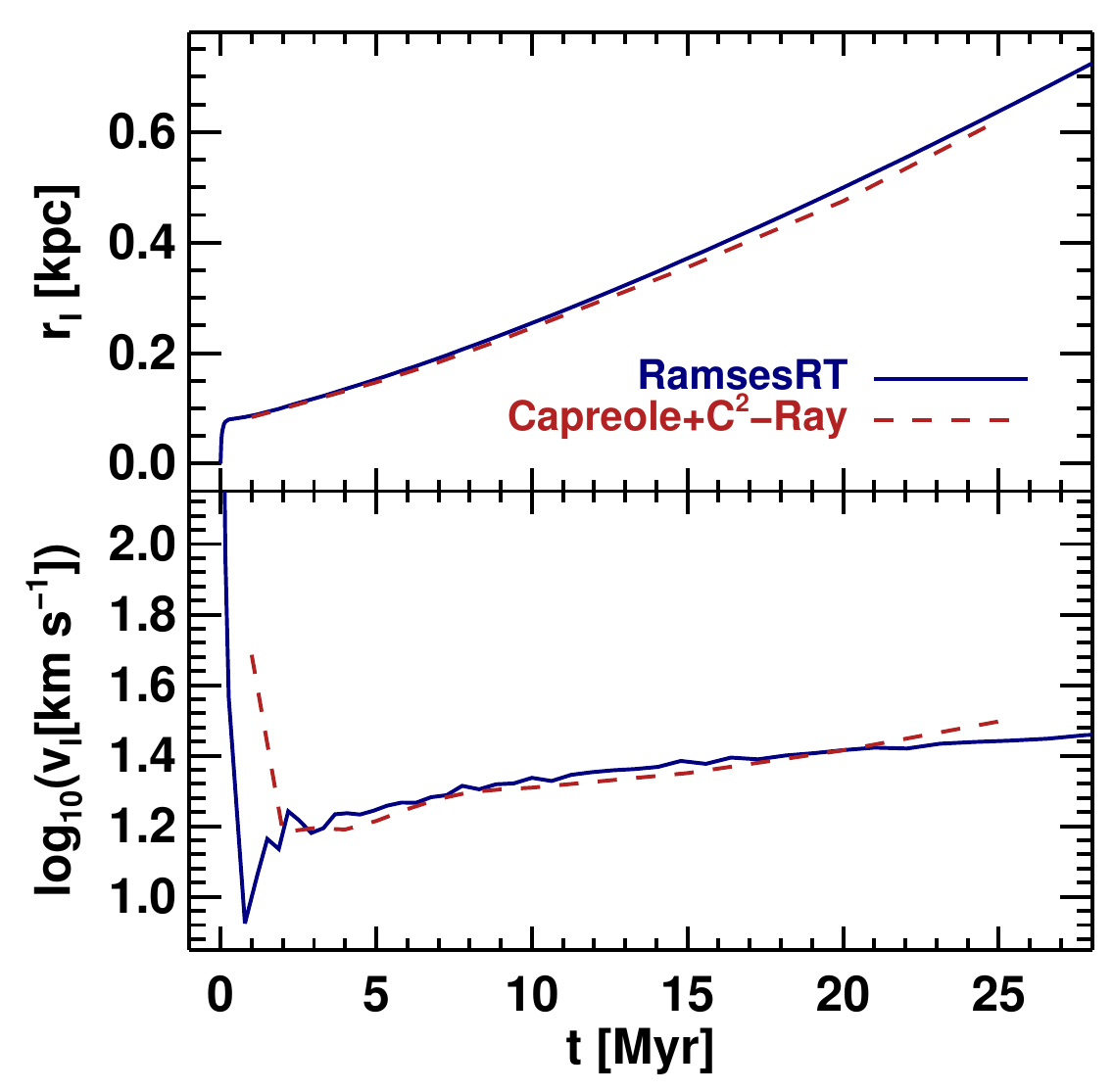}
  \caption{\label{Il6_Ifront.fig} \Ilb{} test 6: time evolution of the
    ionization front, compared to the \CC2R{} combination.}
\end{center}\end{figure}

\begin{figure*}
  \centering
  \subfloat{\includegraphics[width=0.2\textwidth]
    {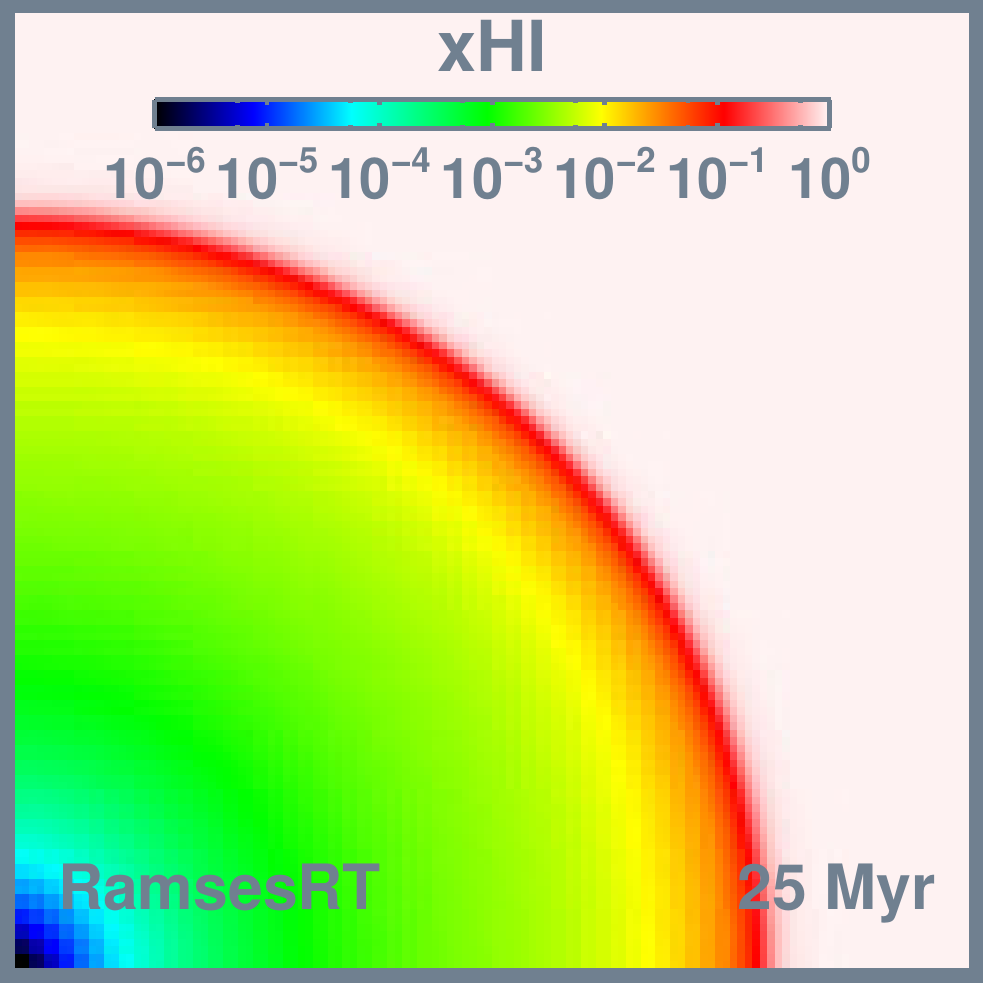}}\hspace{-1.2mm}
  \subfloat{\includegraphics[width=0.2\textwidth]
    {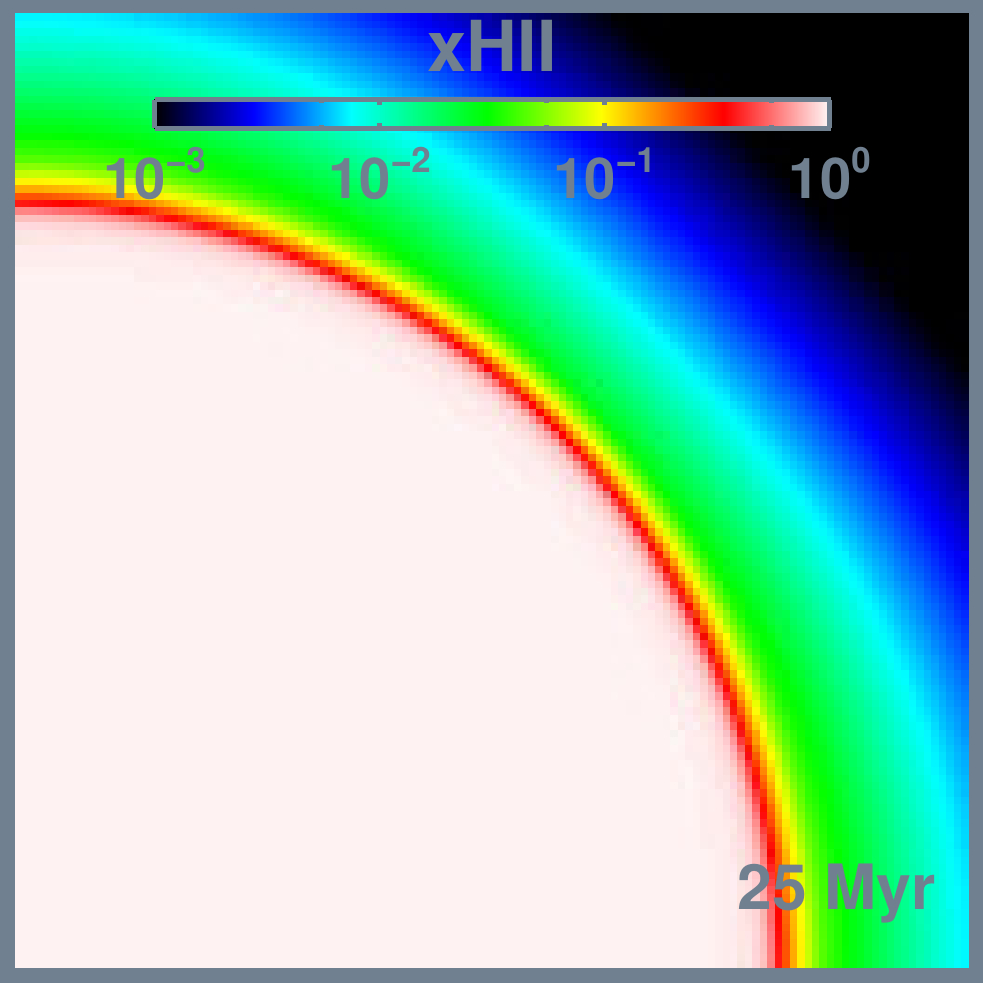}}\hspace{ -1.2mm}
  \subfloat{\includegraphics[width=0.2\textwidth]
    {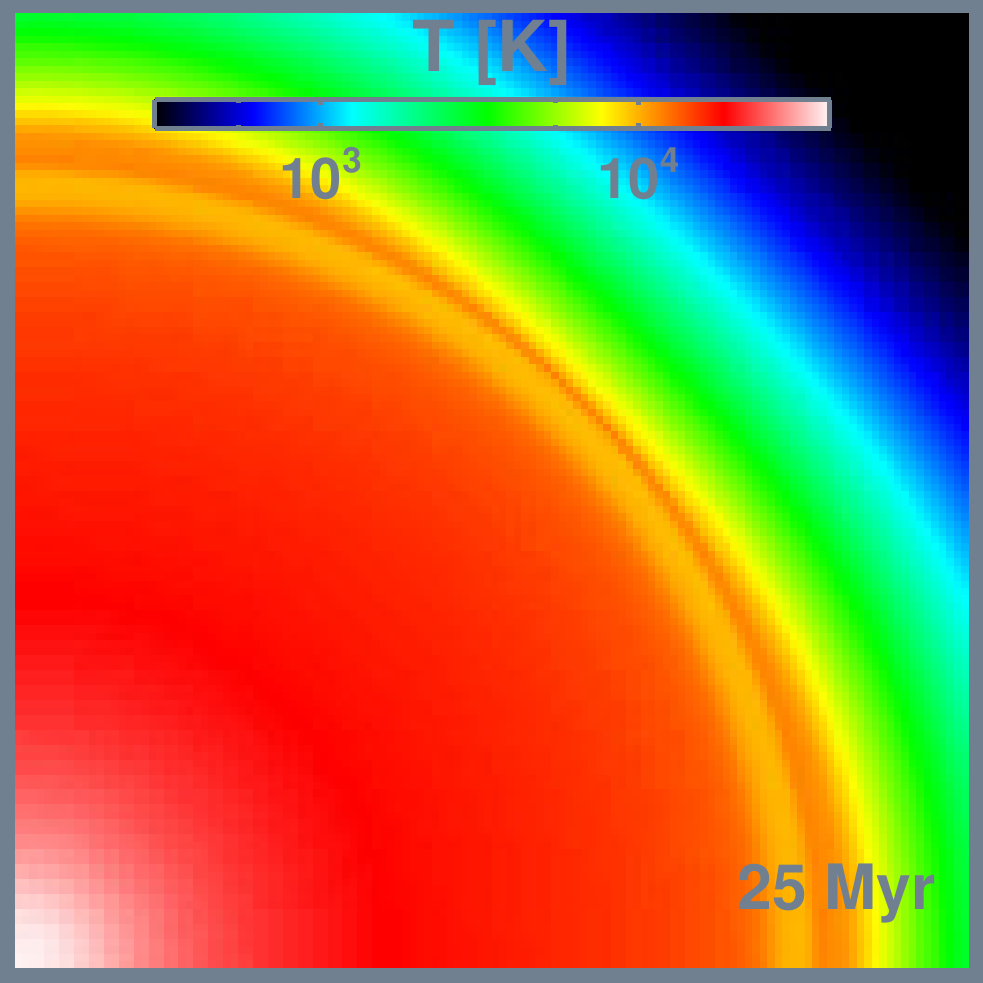}}\hspace{-1.2mm}
  \subfloat{\includegraphics[width=0.2\textwidth]
    {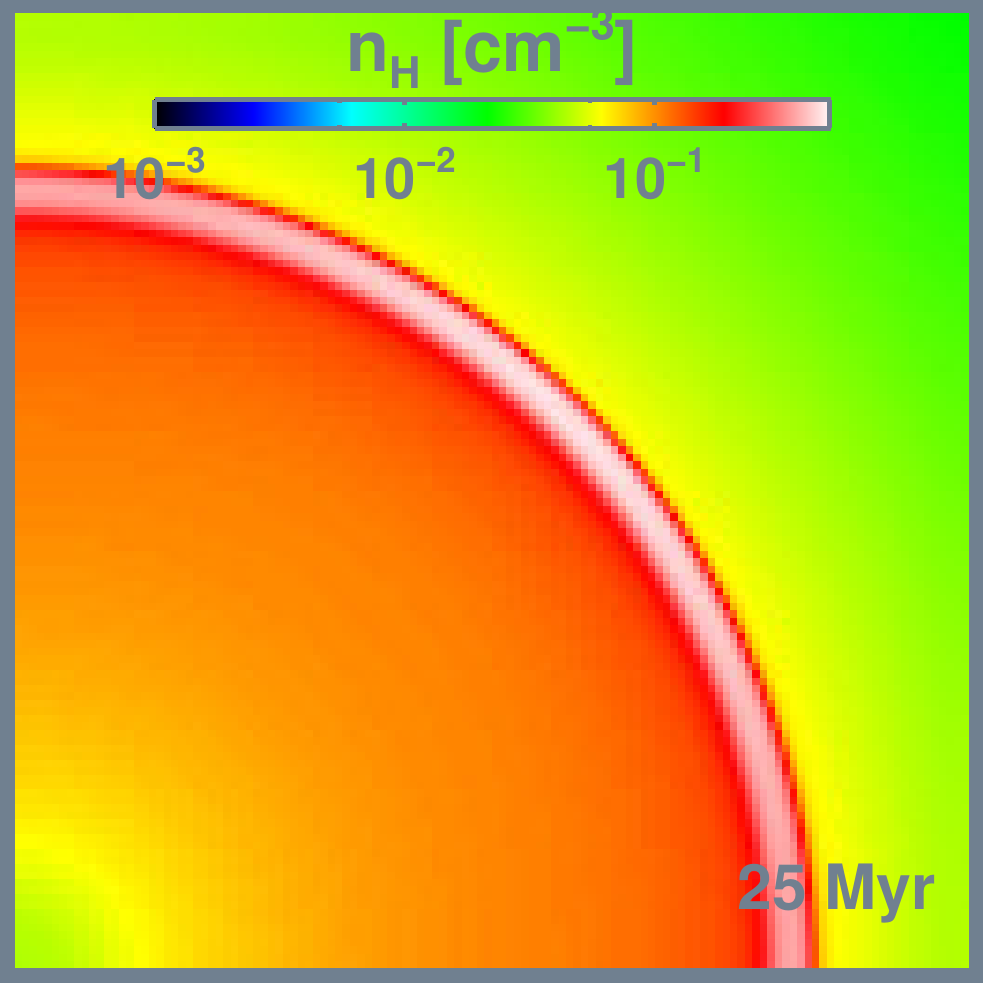}}\hspace{-1.2mm}
  \subfloat{\includegraphics[width=0.2\textwidth]
    {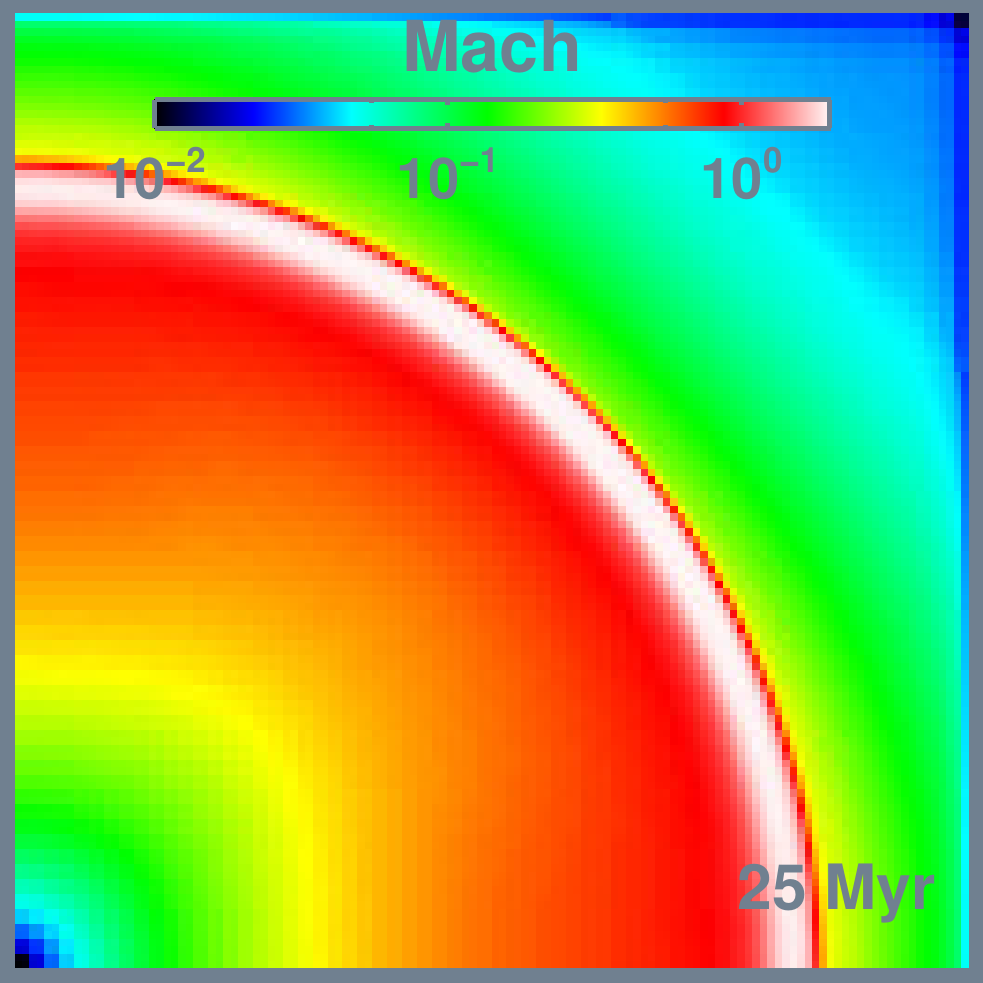}}\hspace{-1.2mm}
  \vspace{-4.mm}

  \subfloat{\includegraphics[width=0.2\textwidth]
    {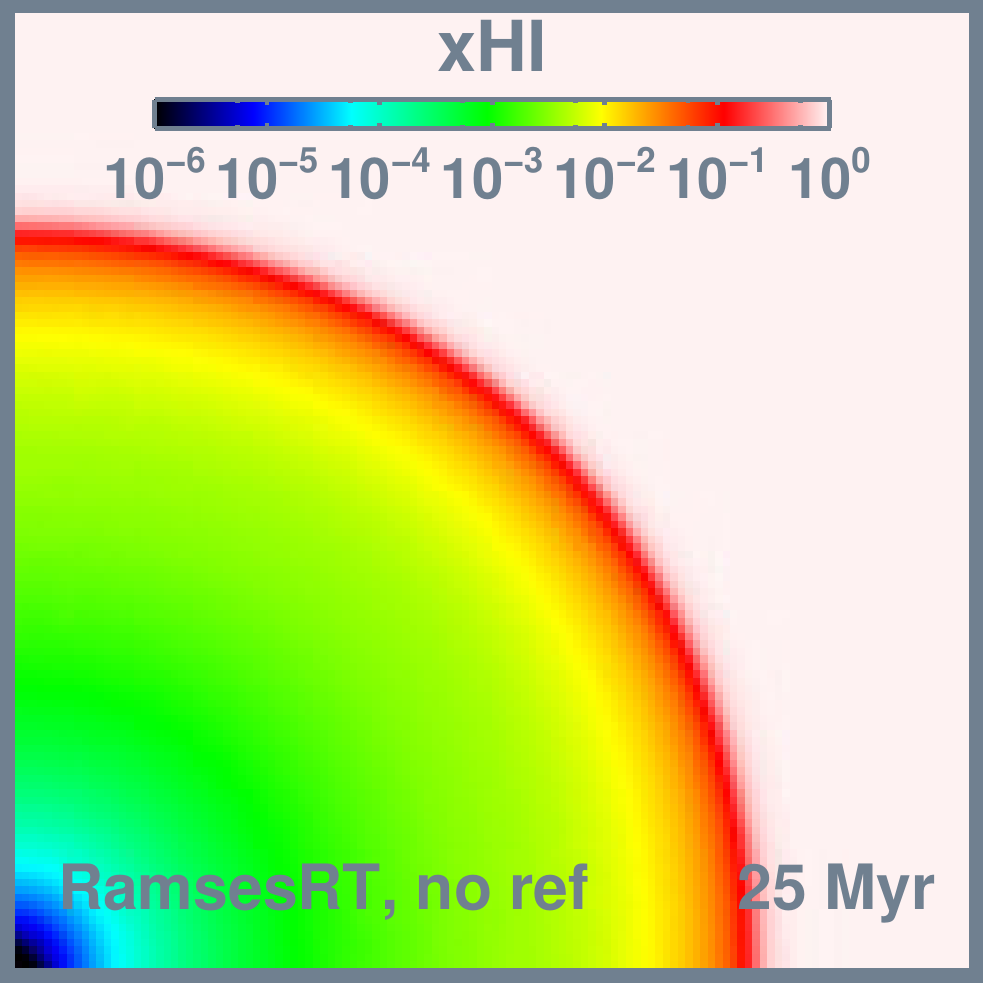}}\hspace{-1.2mm}
  \subfloat{\includegraphics[width=0.2\textwidth]
    {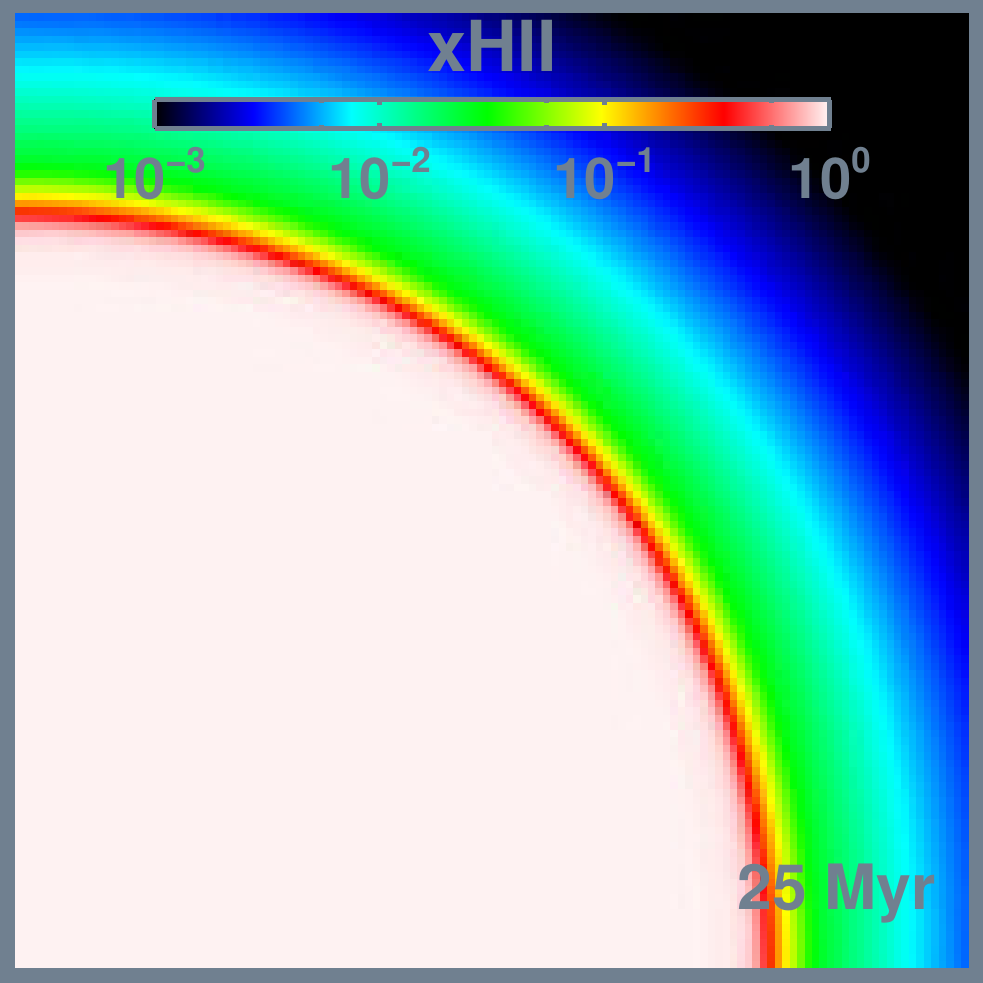}}\hspace{ -1.2mm}
  \subfloat{\includegraphics[width=0.2\textwidth]
    {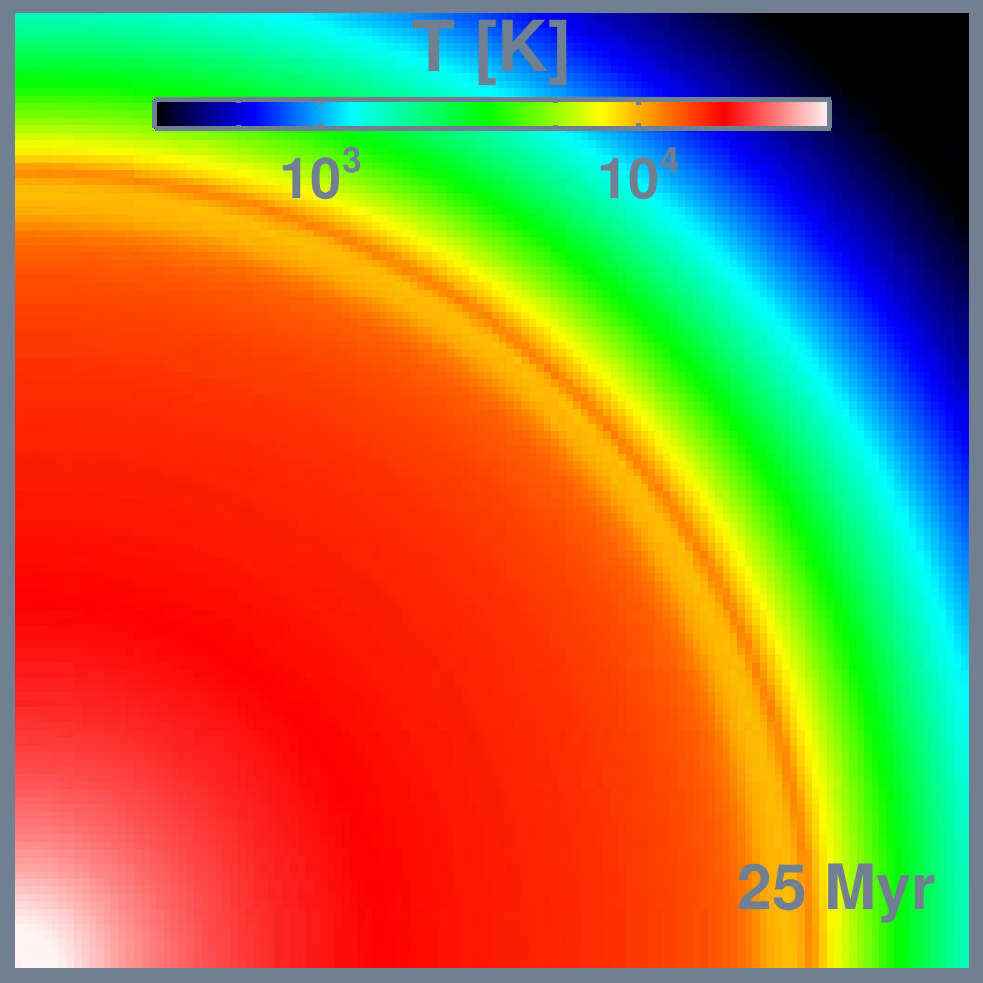}}\hspace{-1.2mm}
  \subfloat{\includegraphics[width=0.2\textwidth]
    {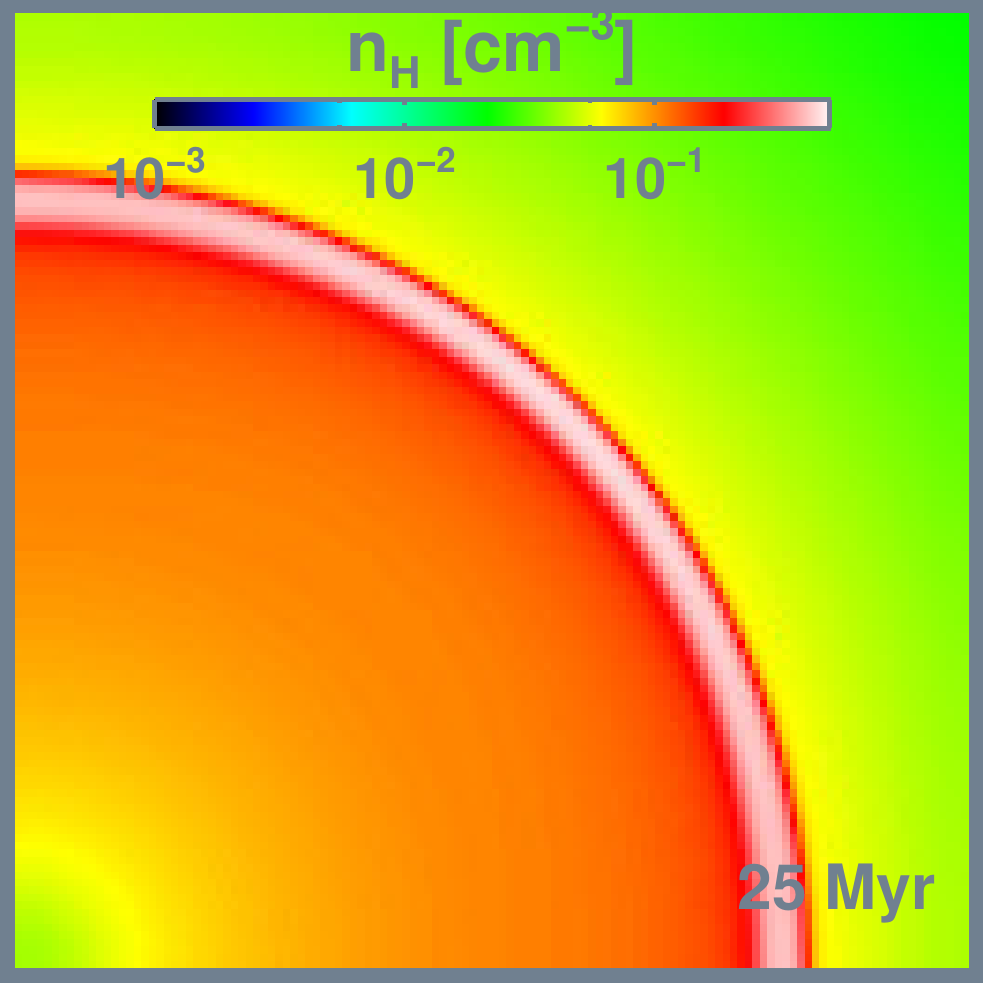}}\hspace{-1.2mm}
  \subfloat{\includegraphics[width=0.2\textwidth]
    {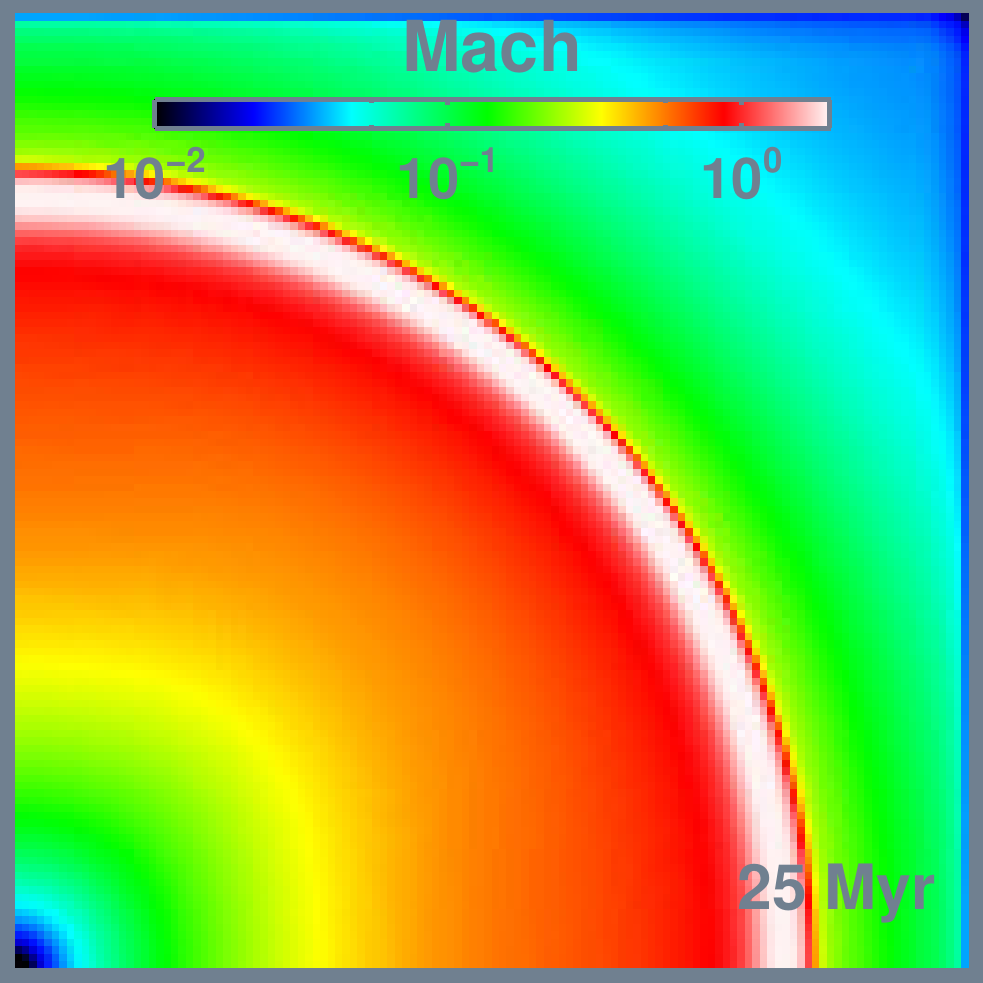}}\hspace{-1.2mm}
  \vspace{-4.mm}

  \subfloat{\includegraphics[width=0.2\textwidth]
    {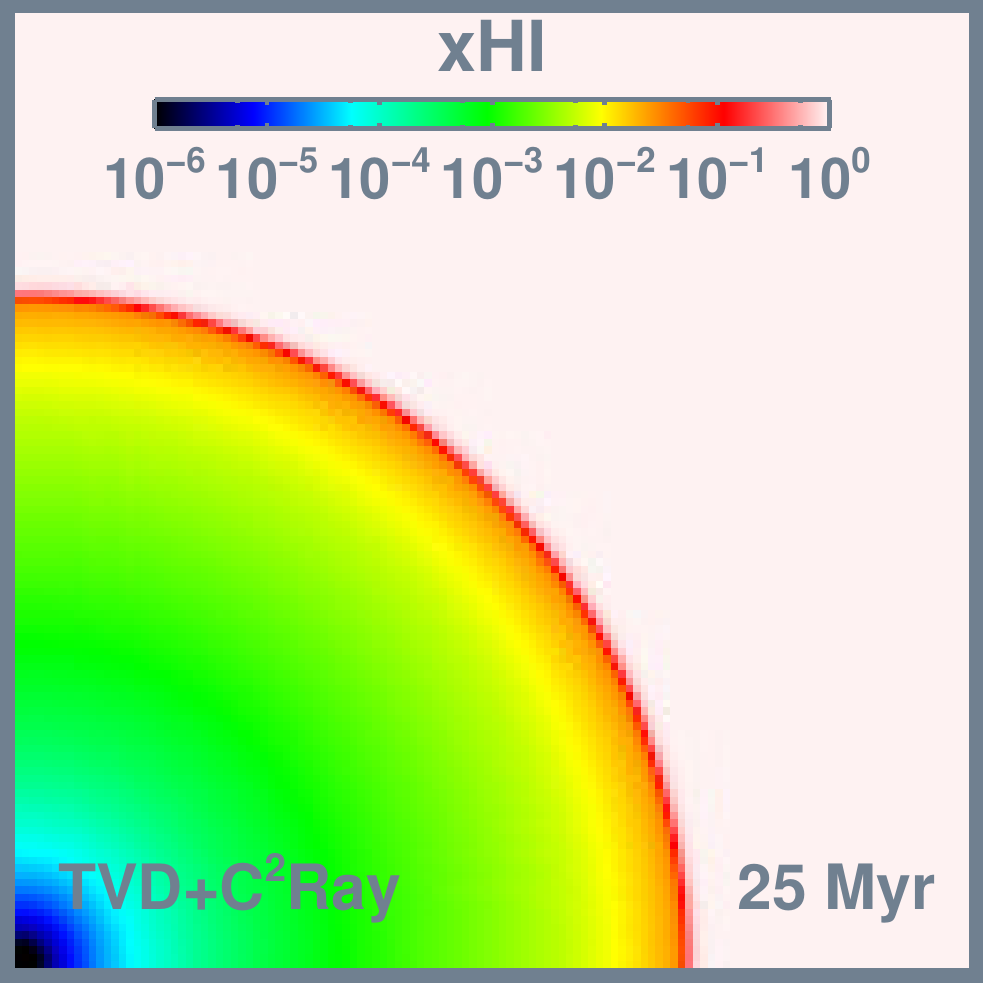}}\hspace{-1.2mm}
  \subfloat{\includegraphics[width=0.2\textwidth]
    {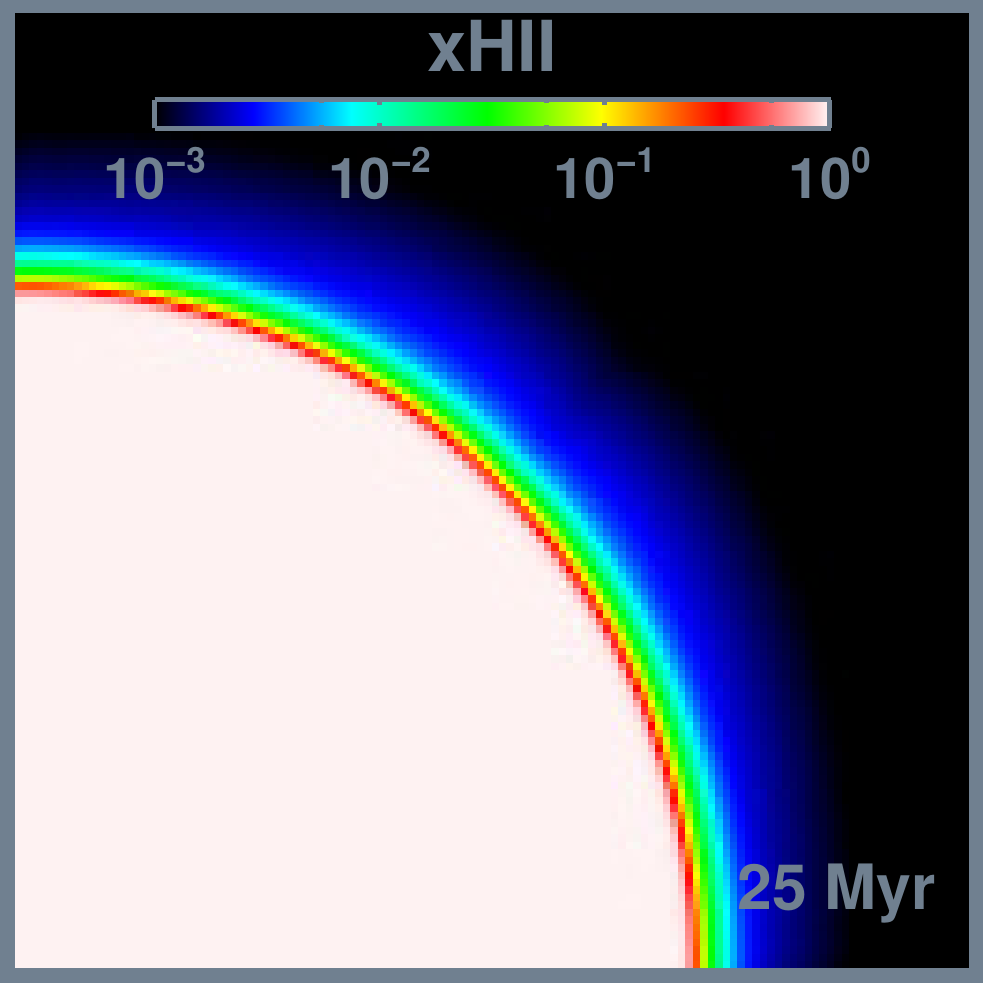}}\hspace{-1.2mm}
  \subfloat{\includegraphics[width=0.2\textwidth]
    {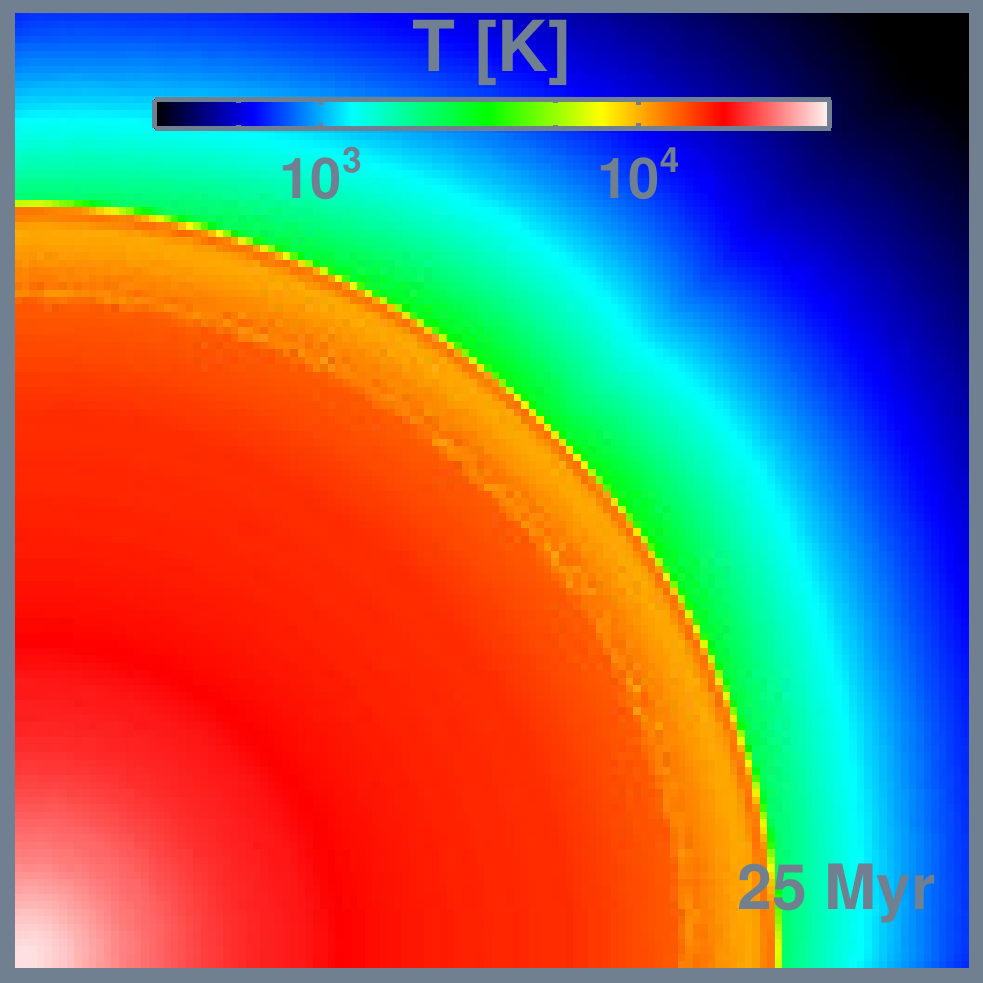}}\hspace{-1.2mm}
  \subfloat{\includegraphics[width=0.2\textwidth]
    {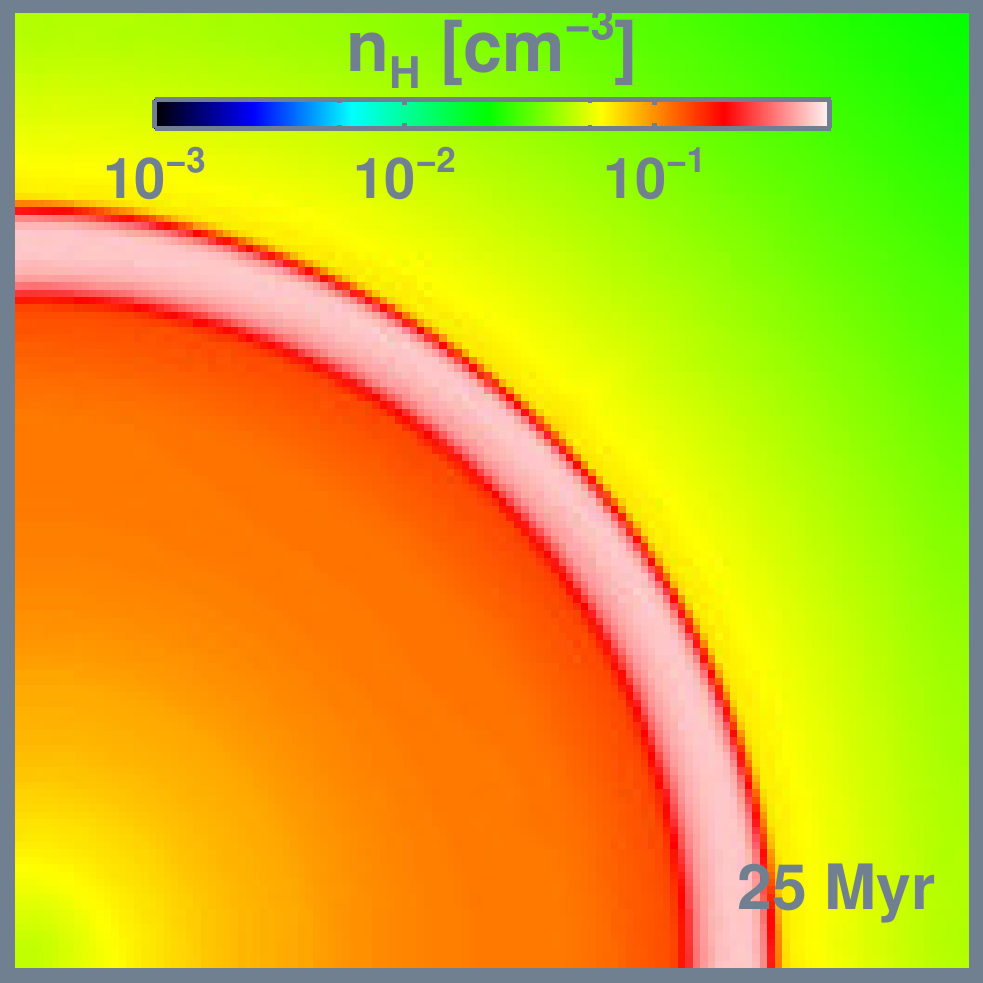}}\hspace{-1.2mm}
  \subfloat{\includegraphics[width=0.2\textwidth]
    {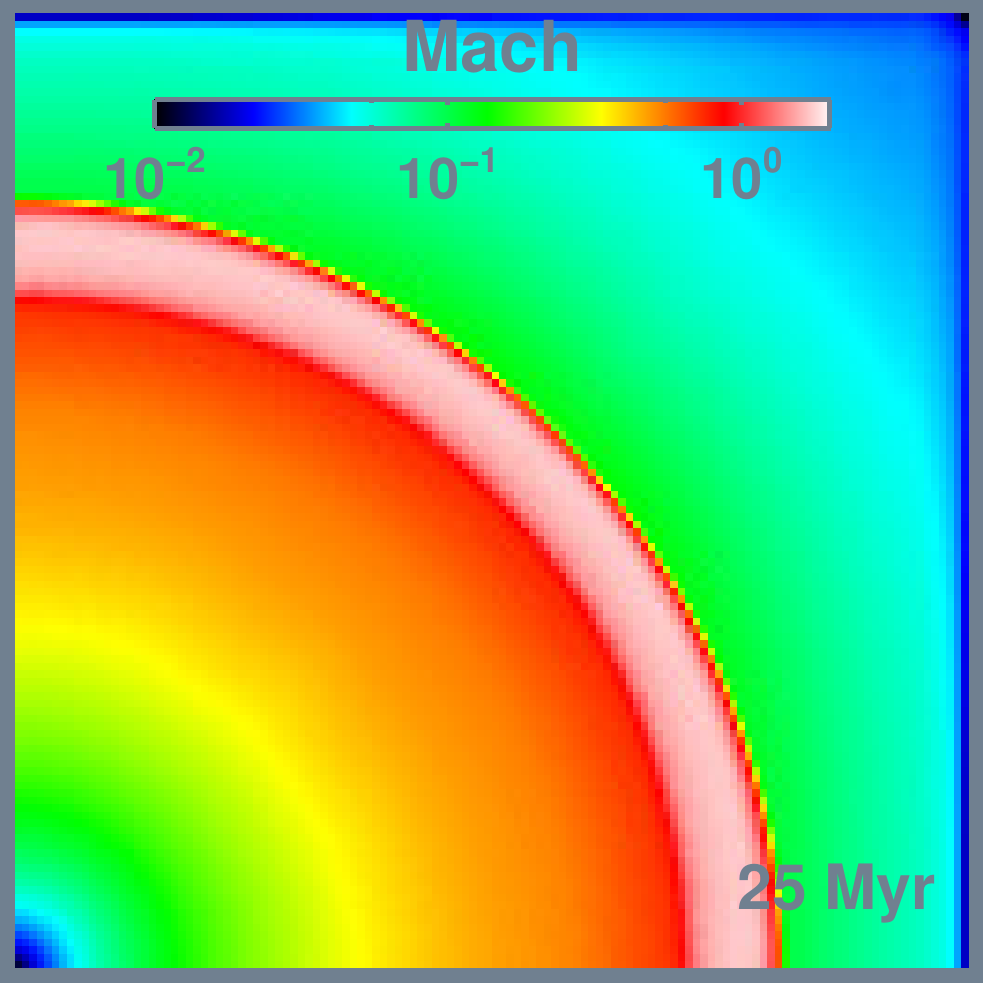}}\hspace{-1.2mm}
  \vspace{-0.7mm}

  \caption[\Ilb{} test 6 - maps]
  {\label{Il6maps.fig}\Ilb{} test 6. Maps showing slices at $z=0$ of
    various quantities at 25 Myrs. The top row shows the \ramsesrt{}
    results with adaptive refinement. The middle row shows results
    also from \ramsesrt{}, but with a fully refined box and adaptive
    refinement turned off.  The bottom row shows the \TC2R{} results
    for comparison.}
\end{figure*}

\begin{figure*}\begin{center}
  \includegraphics[width=.8\textwidth]
  {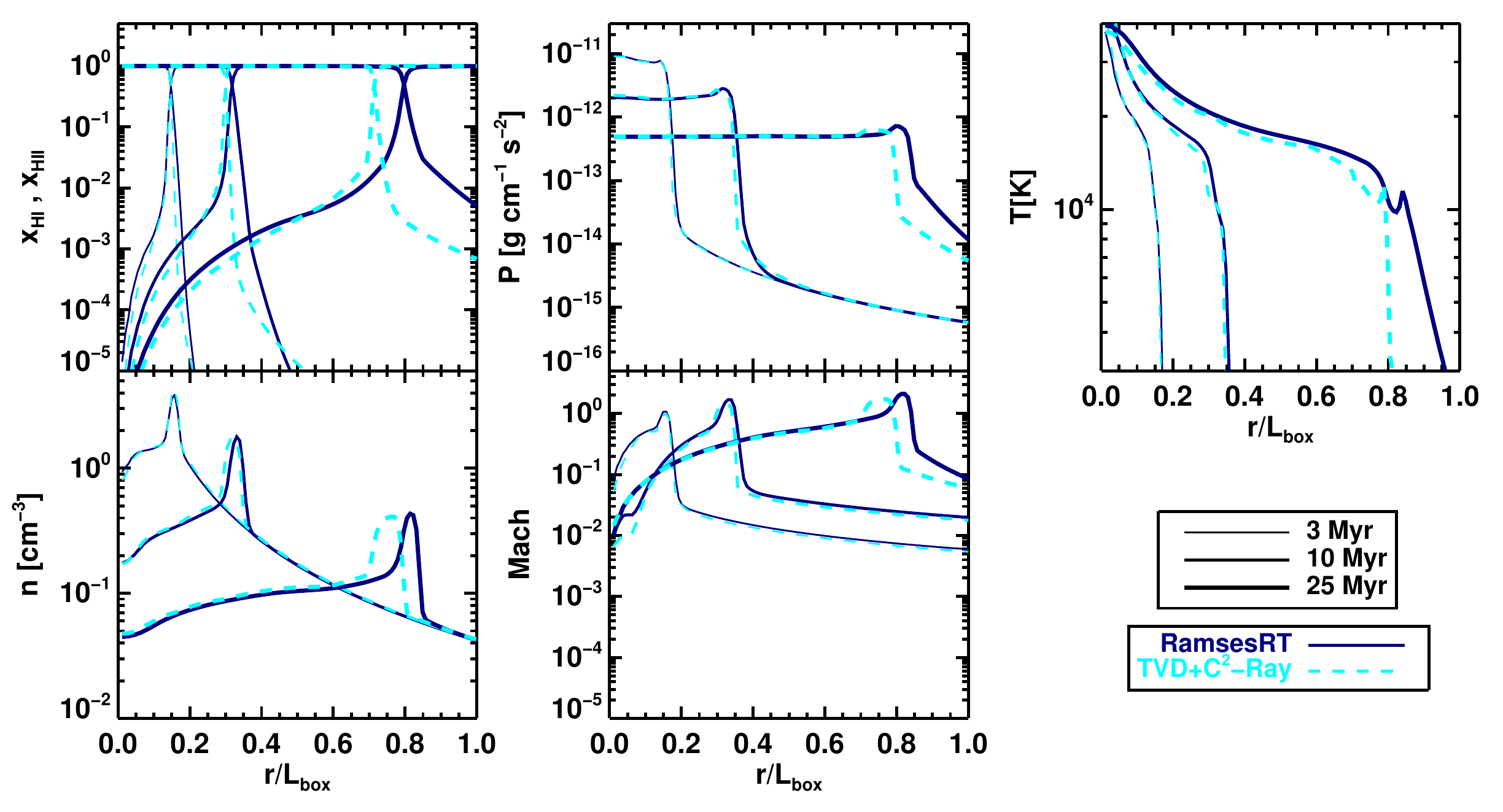}
  \caption[\Ilb{} test 6 - radial profiles]
  {\label{Il6_Profs.fig}\Ilb{} test 6. Radial profiles at
    3, 10 and 25 Myrs, compared to the \TC2R{} results. Clockwise from
    top left: ionization fractions, pressure, temperature, Mach
    number, atom number density.}
\end{center}\end{figure*}

The Str\"omgren radius for the core density, given by \Eq{r_analyt.eq},
is $r_S \approx 70$ pc, which lies within the flat core. Thus, the
I-front makes an initial transition from R-type to D-type within the
core, and then may accelerate back to R-type as it expands into
decreasingly dense gas outside the core.

We first compare the evolution of the position and speed of the
I-front, which is plotted in \Fig{Il6_Ifront.fig} for \ramsesrt{} and
the \CC2R{} combination. The I-front moves very quickly (R-type) to
$\approx 70$ pc within the first fraction of a Myr, stops for while
and then starts to expand again with the flow of the gas. Both the
speed and position compare well with \C2R{}. The initial speed in
\C2R{} has an apparent lag which is due to under-sampling in the front
positions \joki{at early times, as noted by \Ilb{}}. Other code
results which are better sampled in \Ilb{} show initial speeds that
are virtually identical to the \ramsesrt{} plot\joki{, especially
  those of the \rhod{} code}. The final front position in \ramsesrt{}
is slightly further out than that of \C2R{}, though very similar to at
least three of the codes in \Ilb{} (\flashhc{}, \licorice{} and
\rsph{}). It also appears that the \C2R{} front is starting to
accelerate slightly at the end, whereas the \ramsesrt{} front is about
to approach constant speed; \ramsesrt{} also agrees with most other
\Ilb{} codes on this point.

\Fig{Il6maps.fig} shows the overall structure of ionization and the
gas at $25$ Myr, here with a comparison between \ramsesrt{} (upper two
rows) and \TC2R{} (bottom row). \joki{The \CC2R{} version of this test
  is sensitive to so-called 'carbuncle' numberical instabilities (see
  Sec. 4.2 in \Ilb{}), so we compare here to the more stable and
  symmetric combination of \C2R{} coupled to the Eulerian TVD solver
  of \cite{{Trac:2004hp}} (used only in this test)}. In addition to
the default \ramsesrt{} run with on-the fly AMR refinement, we show
here in the middle row results from an identical \ramsesrt{} run with
the base resolution set to $128^3$ cells and AMR refinement turned
off. There are slight spherical asymmetries appearing in the top row
maps, in particular the $\xhii$, $T$ and Mach maps, and the middle row
maps are presented here to show that (the first) two of these are
purely artifacts of on-the-fly AMR refinement. The slightly square
shape of the inner region in the Mach map however does not seem to be
due to refinement and is likely rather a grid artifact which is
amplified by the radially decreasing density. It should also be noted
that the other plots produced for this test (I-front,
\Fig{Il6_Ifront.fig} and radial profiles, \Fig{Il6_Profs.fig}) are
absolutely identical regardless of whether on-the-fly refinement is
used or the full resolution applied everywhere, suggesting that AMR
refinement produces very robust results. \joki{The difference in
  runtime between the AMR and non-AMR runs is actually not much: the
  AMR run completes in about $2/3$ of the $\sim 640$ cpu hours taken
  for the non-AMR run. This lack of speedup is due to a combination of
  a large portion of the grid being refined ($\sim 60\%$ by volume
  when most), a shallow refinement hierarchy (one level of refinement)
  and overhead in refinement-related computations.}

As usual the I-front is considerably wider in \ramsesrt{} than in the
\C2R{} results, though we don't find the same discrepancy as in the
previous test between the photoheating intensity close to the source
(also, there is no such discrepancy here between \C2R{} and the other
codes in \Ilb{}). The two maps furthest to the right, of density and
Mach number, show the expanding shell of dense gas due to
photoheating. Here the shell appears considerably thinner in
\ramsesrt{} than in \TC2R{}, and indeed \TC2R{} appears to have
the thickest density shell of any of the codes in \Ilb{}
(\CC2R{} included, but here there are also severe
asymmetries). The \ramsesrt{} maps compare well with the \C2R{} ones,
and to most of the maps in \Ilb{}, and don't show any I-front
instabilities that seem to have a tendency to come up in this test
(and \Ilb{} do show that these are numerical and not physical
instabilities).

\Fig{Il6_Profs.fig} shows a comparison between \ramsesrt{} and \TC2R{}
for radially averaged profiles at $3$, $10$ and $25$ Myr of the
ionization state, pressure, temperature, density and Mach number. The
comparison is generally very good. The I-front (and corresponding
density shock) lag a little behind in \C2R{}, but it actually lags a
little behind all but one code in this test in \Ilb{}, and \ramsesrt{}
is spot-on compared with those others in every respect.

All in all, \ramsesrt{} thus performs well on this test, and
no problems appear that are worth mentioning.

\subsection{\Ilb{} test 7: \ \ Photo-evaporation of a dense clump}\label{iliev7.sec}
The setup of this test is identical to test 3 in \Ila{}, where UV
radiation is cast on a gas cloud, creating a shadow behind it and a
slowly-moving I-front inside it. Here however, since the hydrodynamics
are turned on, photo-heating causes the cloud to expand outwards and
simultaneously contract at the center. We recap the setup:

The box is $\Lbox=6.6$ kpc in width. A spherical cloud of gas with
radius $r_{\rm{cloud}}=0.8$ kpc is placed at
$(x_c,y_c,z_c)=(5,3.3,3.3)$ kpc from the box corner. The density and
temperature are $\nh^{\rm{out}}=2 \times 10^{-4}\, \cci$ and
$T^{\rm{out}}=8000$ K outside the cloud and $\nh^{\rm{cloud}}=200 \,
\nh^{\rm{out}}=4 \times 10^{-2} \, \cci$ and $T^{\rm{cloud}}=40$ K
inside it. From the $x=0$ boundary a constant ionizing flux of $F=10^6
\; \phflux$ is emitted towards the cloud. The simulation time is $50$
Myr, considerably longer than the $15$ Myr in the corresponding pure
RT test. The base resolution is $64^3$ cells, but on-the-fly
refinement on $\nh$, $\xhi$ and $\xhii$ gradients ensures the
prescribed effective resolution of $128^3$ cells at ionization and
shock fronts. In order to best capture the formation of a shadow
behind the cloud, we focus on a \ramsesrt{} run with the HLL solver,
but we also show some results with the usual GLF solver.

\begin{figure*}
  \centering
  \subfloat{\includegraphics[width=0.2\textwidth]
    {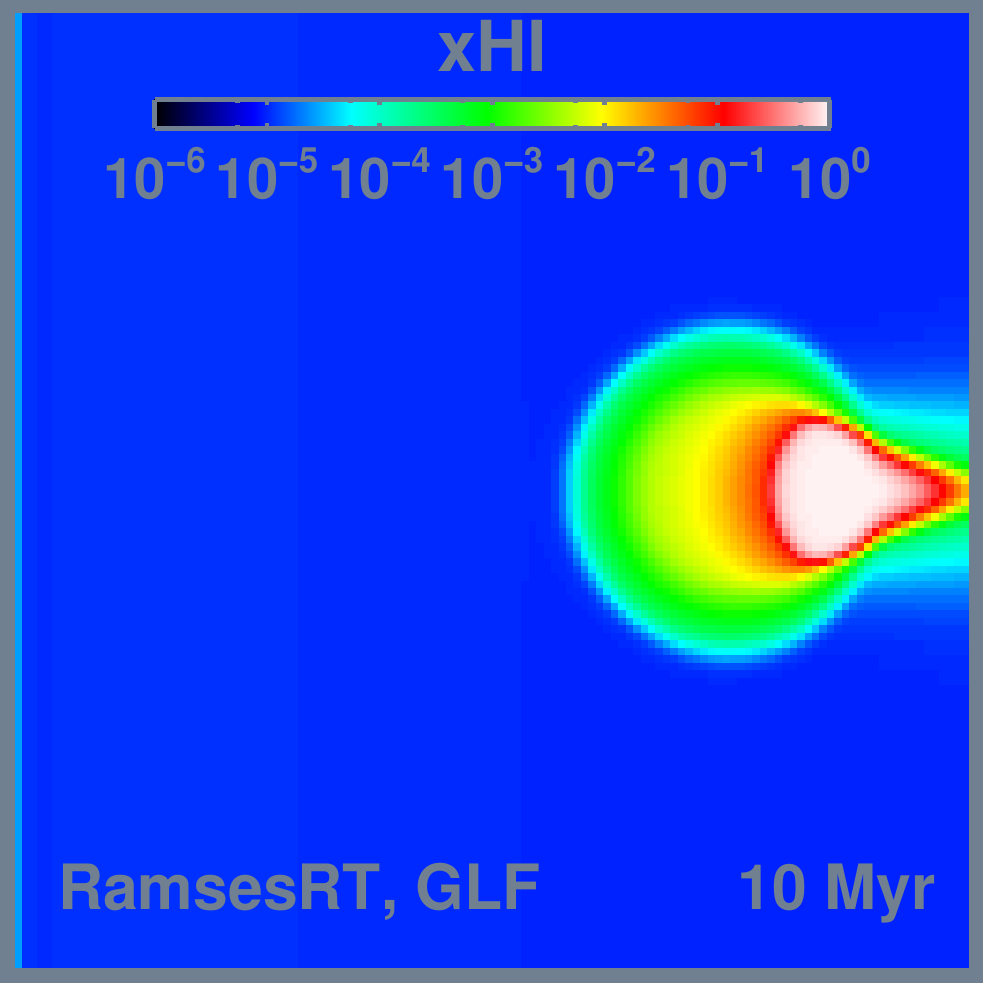}}\hspace{-1.2mm}
  \subfloat{\includegraphics[width=0.2\textwidth]
    {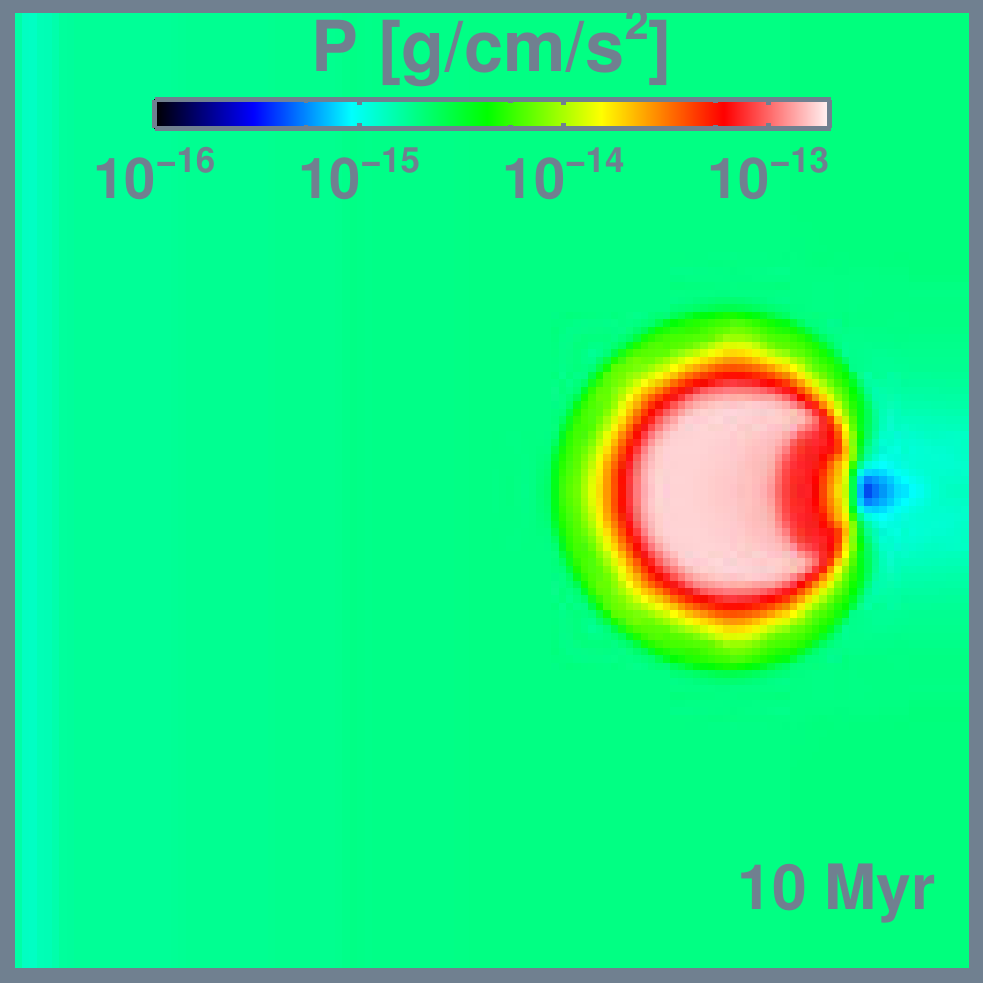}}\hspace{ -1.2mm}
  \subfloat{\includegraphics[width=0.2\textwidth]
    {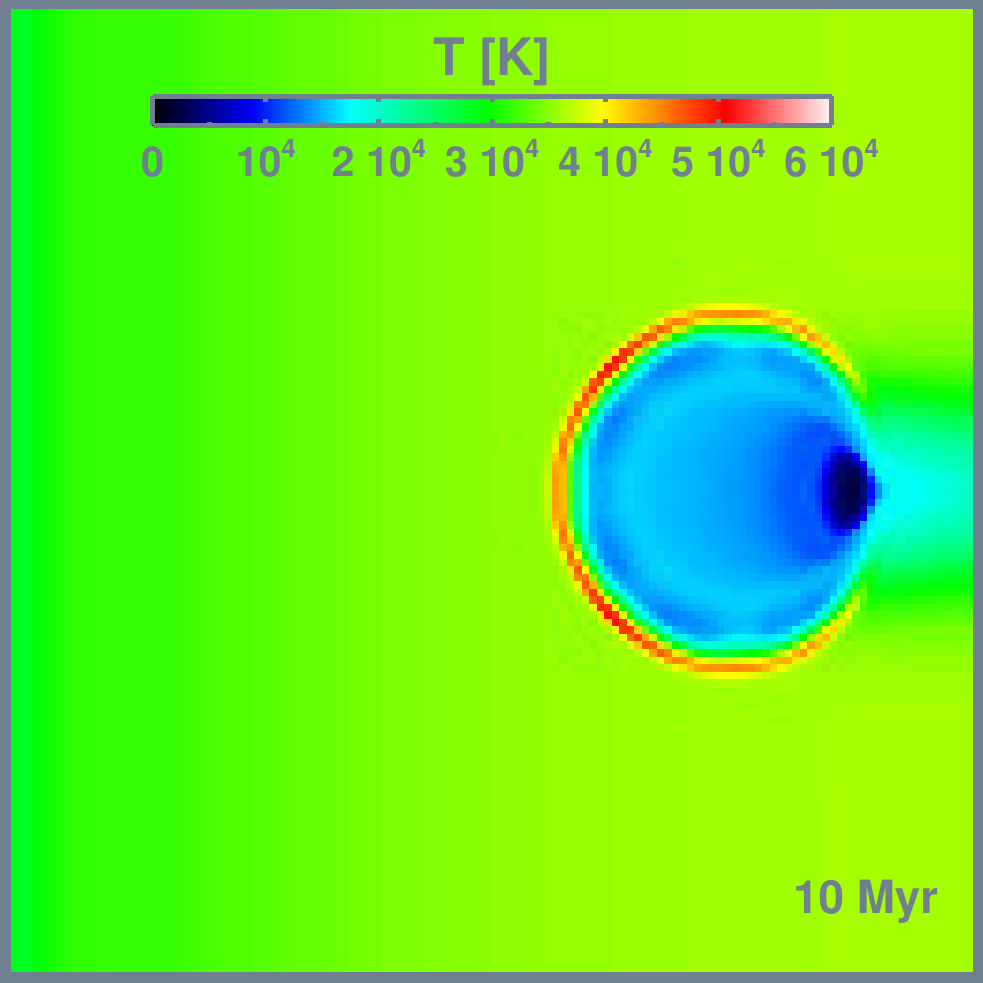}}\hspace{-1.2mm}
  \subfloat{\includegraphics[width=0.2\textwidth]
    {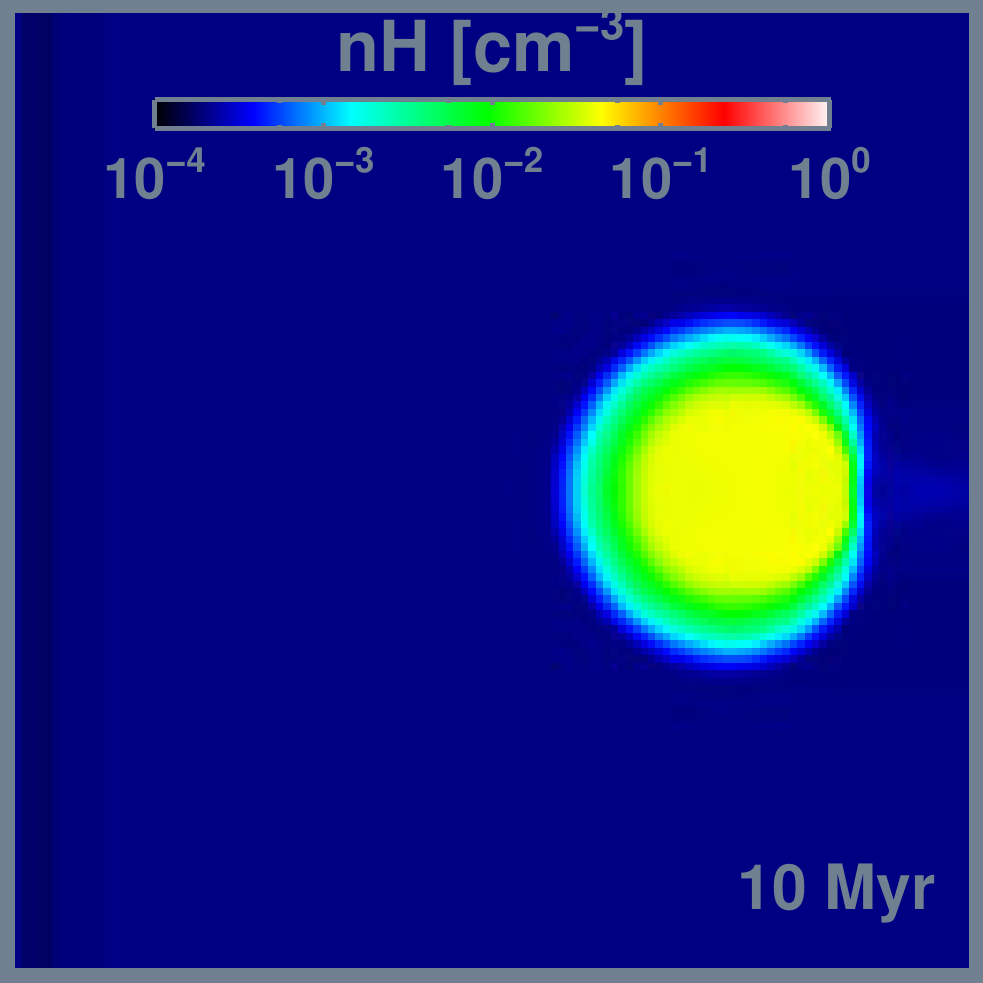}}\hspace{-1.2mm}
  \subfloat{\includegraphics[width=0.2\textwidth]
    {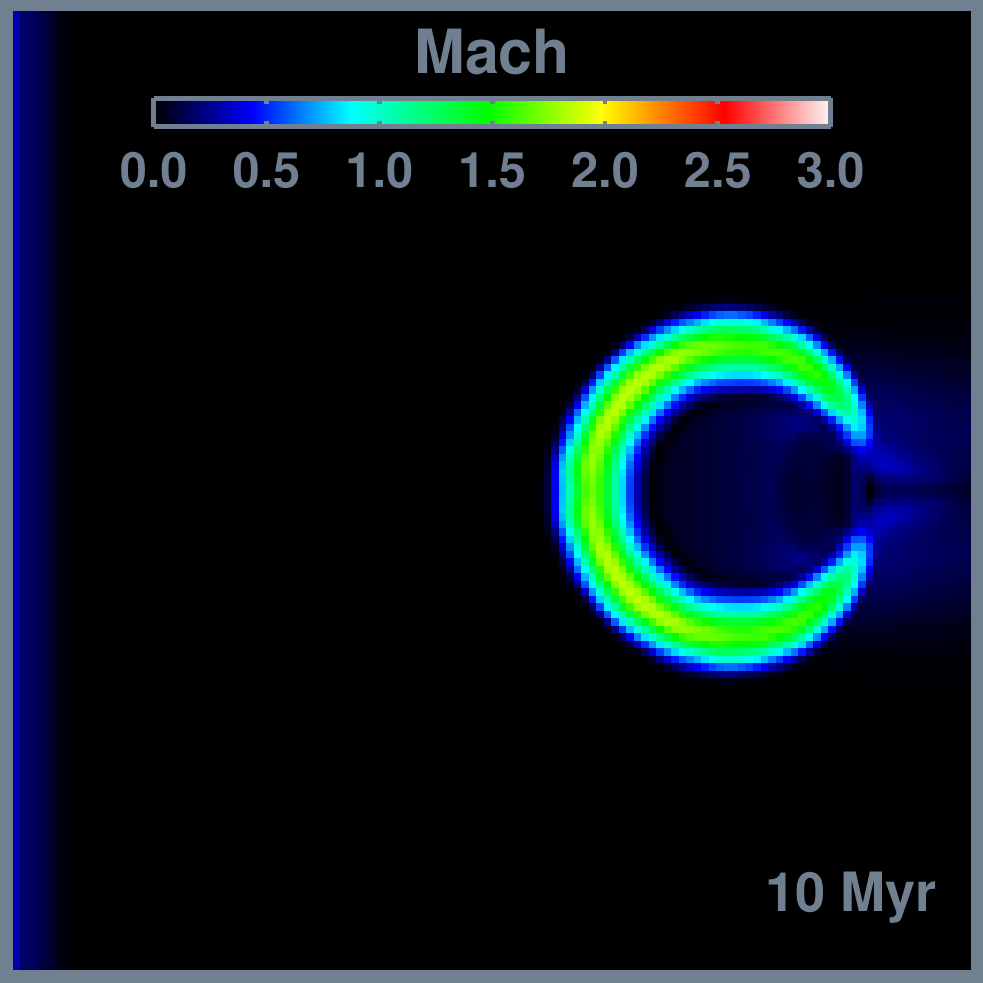}}\hspace{-1.2mm}
  \vspace{-4.mm}

  \subfloat{\includegraphics[width=0.2\textwidth]
    {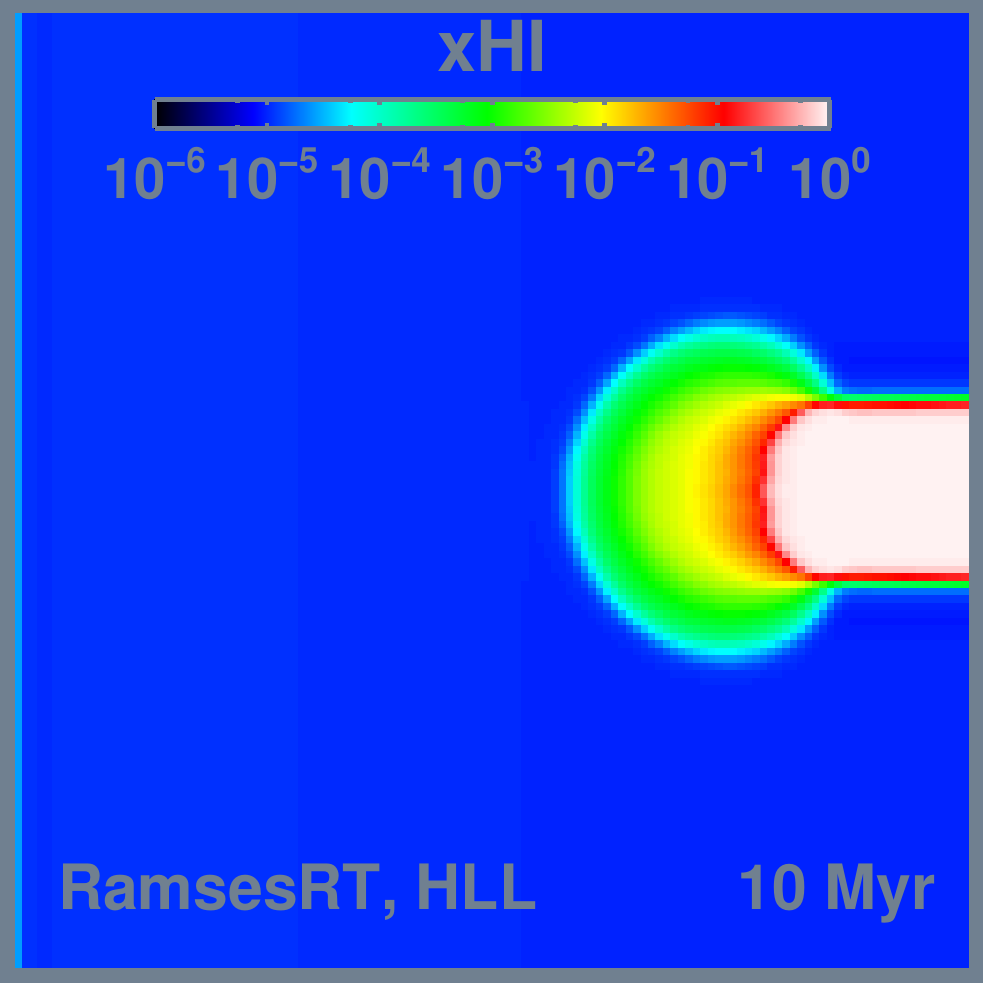}}\hspace{-1.2mm}
  \subfloat{\includegraphics[width=0.2\textwidth]
    {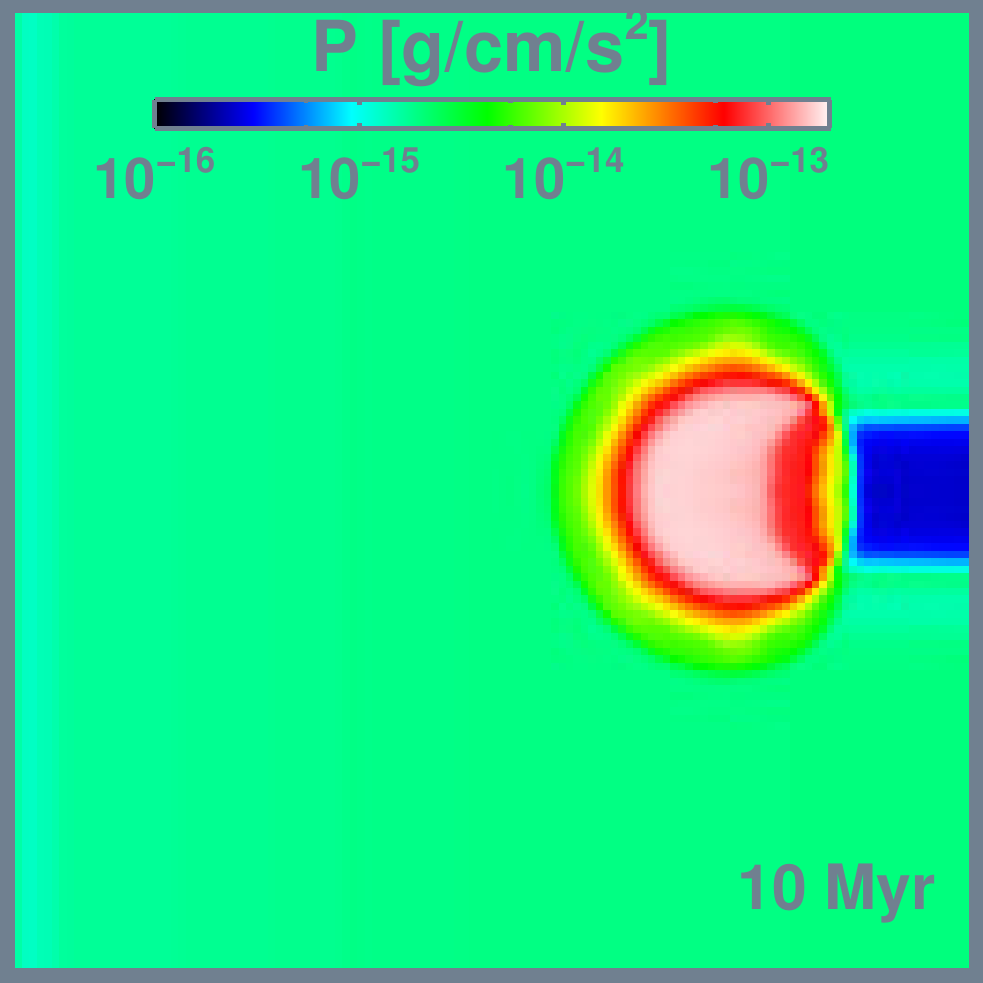}}\hspace{ -1.2mm}
  \subfloat{\includegraphics[width=0.2\textwidth]
    {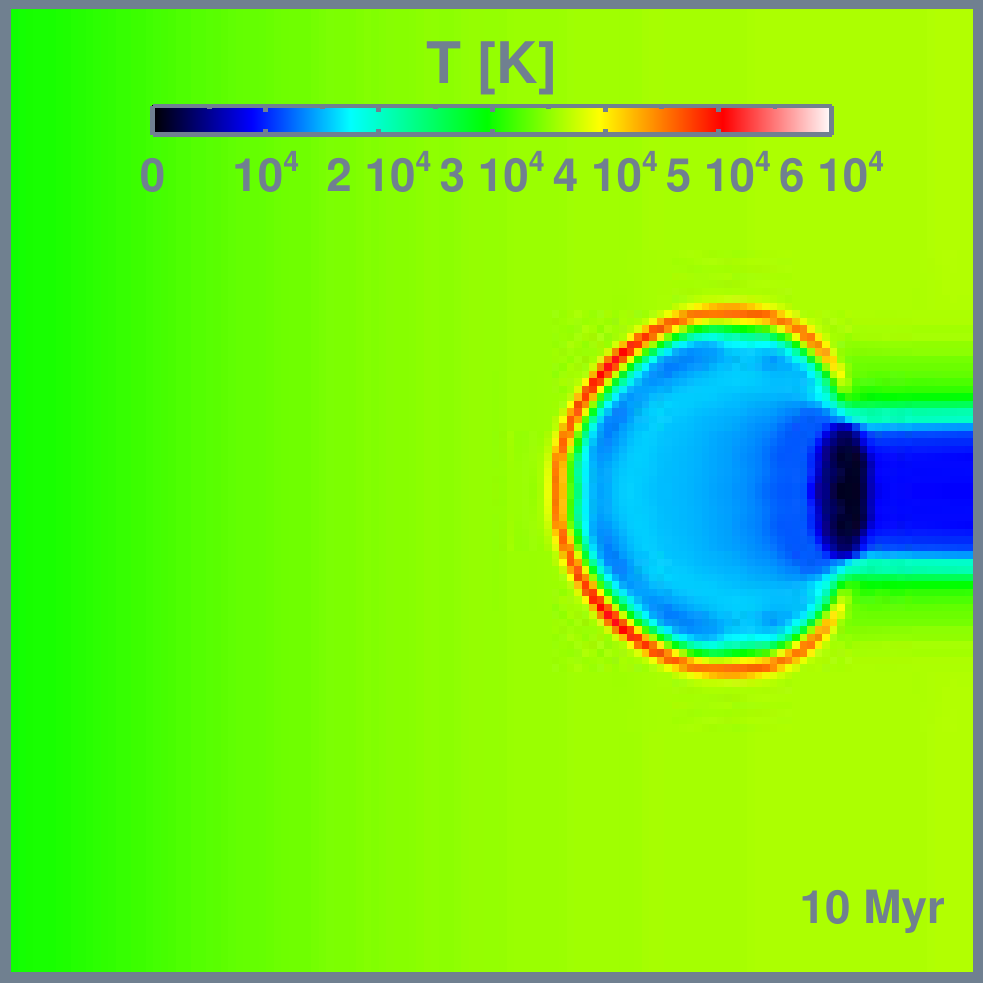}}\hspace{-1.2mm}
  \subfloat{\includegraphics[width=0.2\textwidth]
    {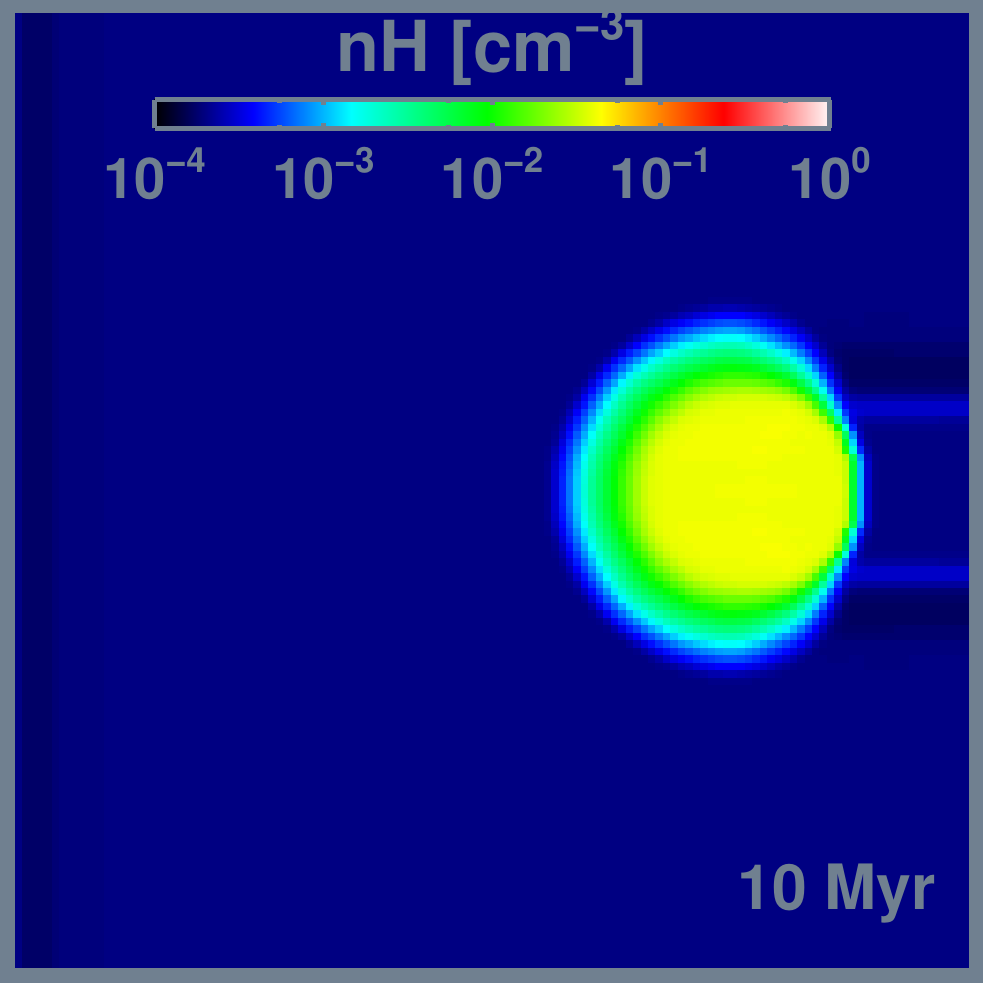}}\hspace{-1.2mm}
  \subfloat{\includegraphics[width=0.2\textwidth]
    {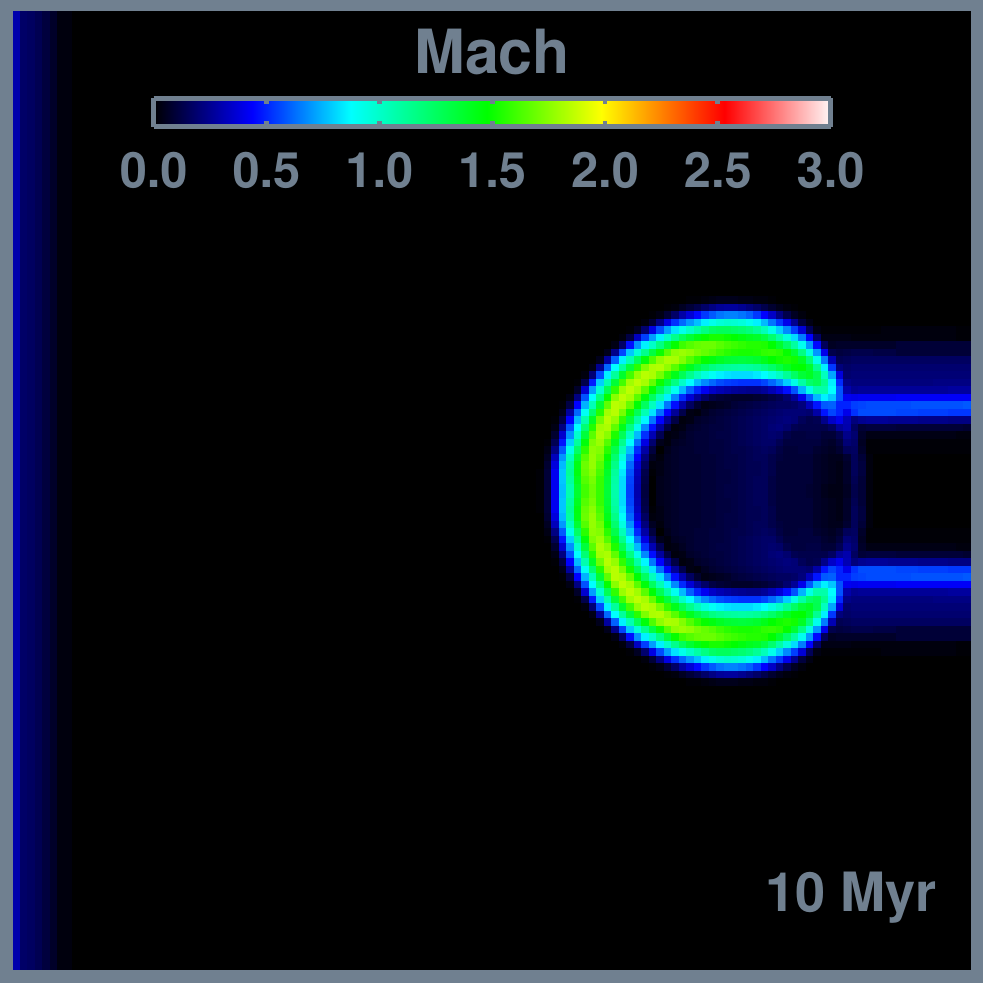}}\hspace{-1.2mm}
  \vspace{-4.mm}

  \subfloat{\includegraphics[width=0.2\textwidth]
    {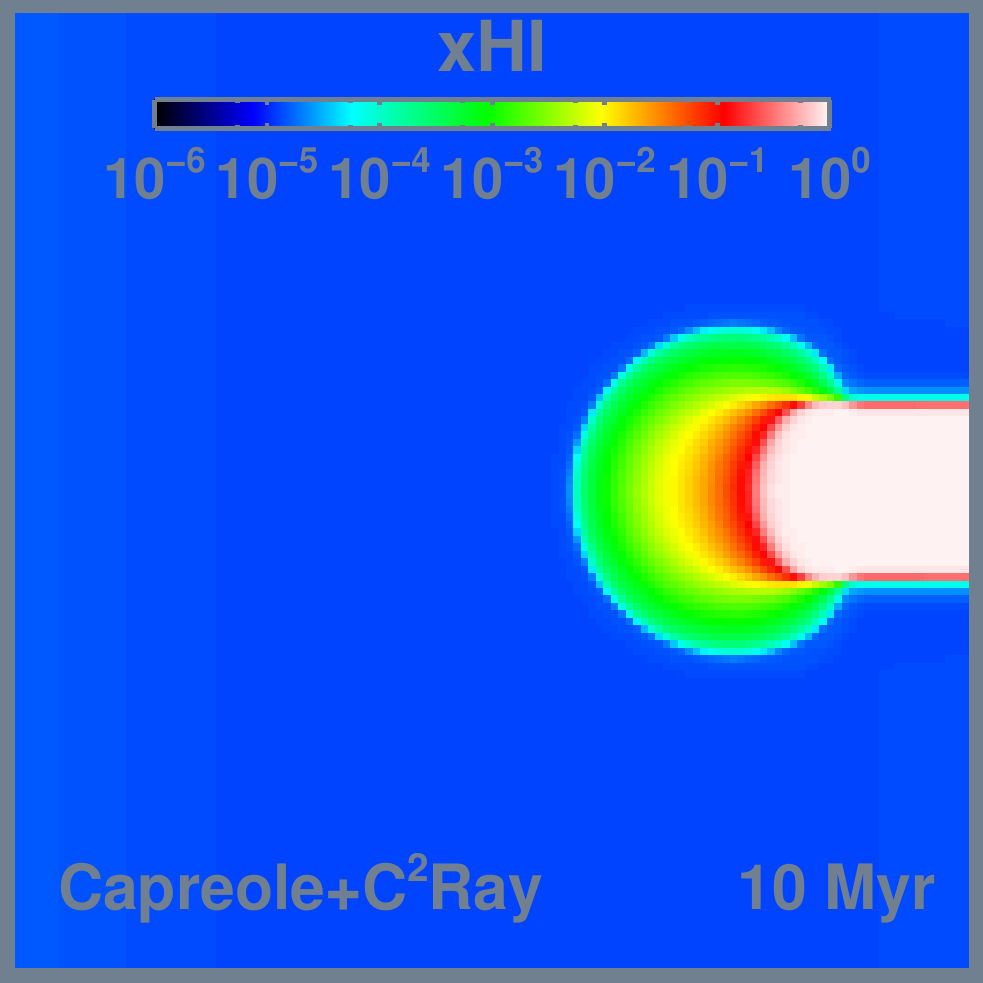}}\hspace{-1.2mm}
  \subfloat{\includegraphics[width=0.2\textwidth]
    {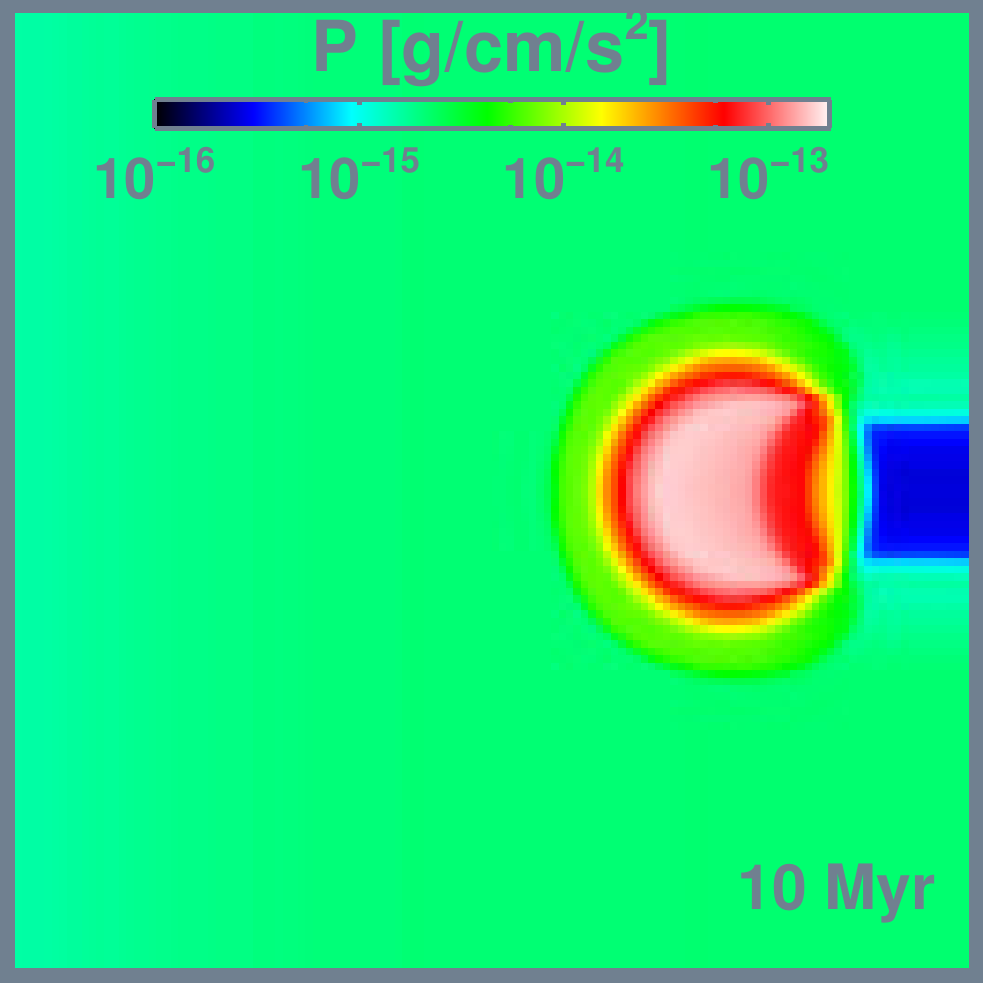}}\hspace{-1.2mm}
  \subfloat{\includegraphics[width=0.2\textwidth]
    {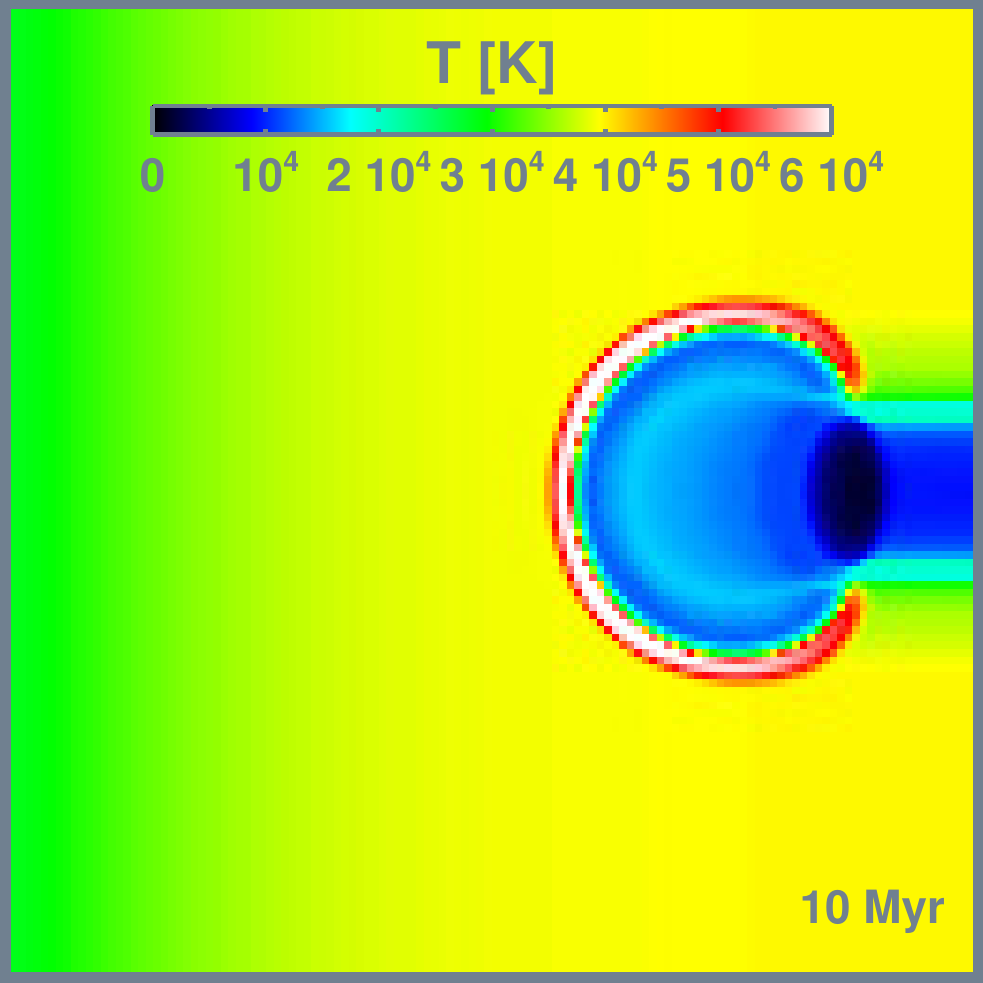}}\hspace{-1.2mm}
  \subfloat{\includegraphics[width=0.2\textwidth]
    {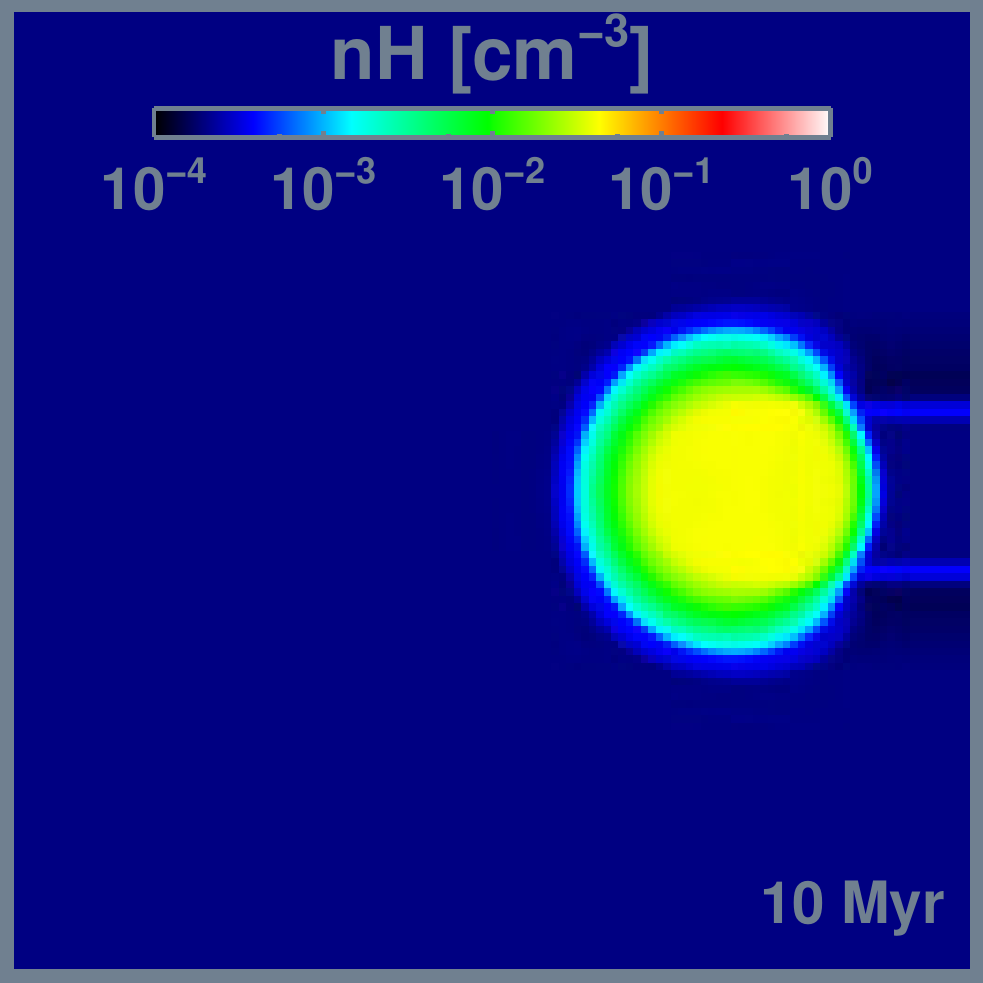}}\hspace{-1.2mm}
  \subfloat{\includegraphics[width=0.2\textwidth]
    {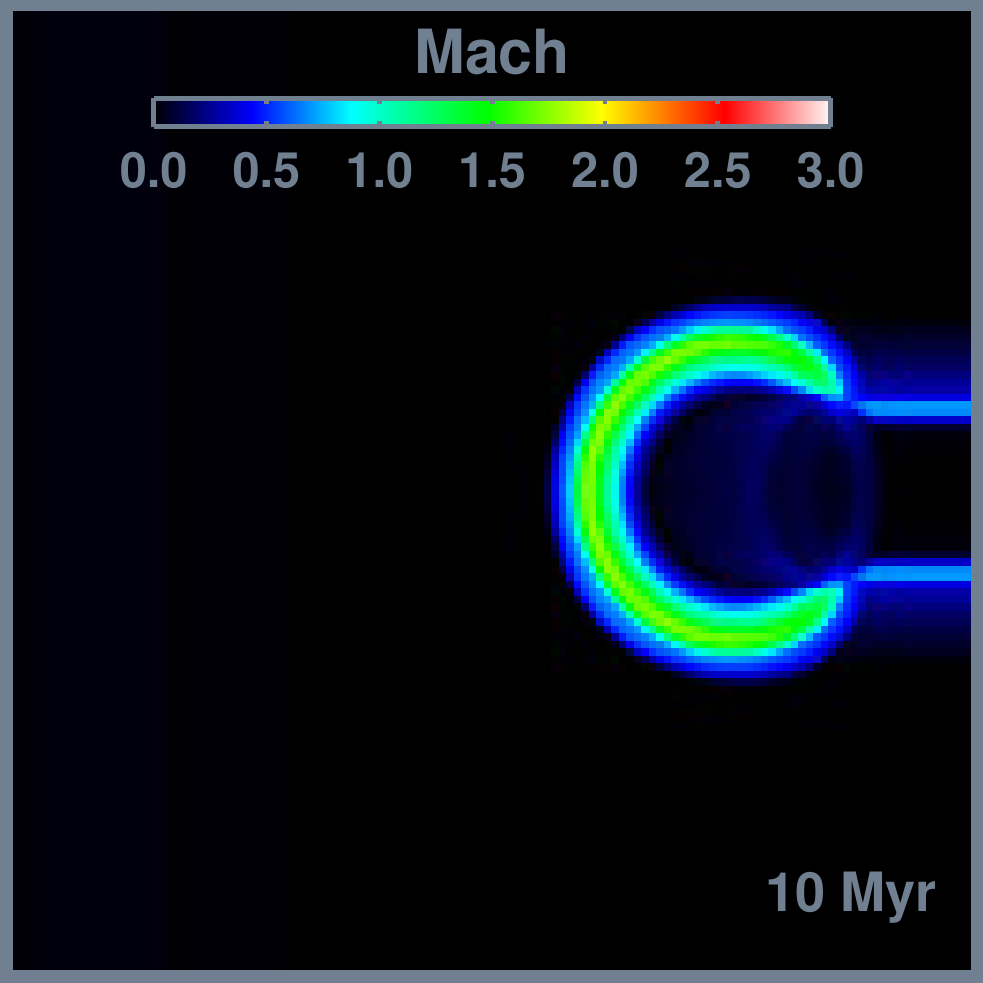}}\hspace{-1.2mm}
  \vspace{-2.5mm}

  \subfloat{\includegraphics[width=0.2\textwidth]
    {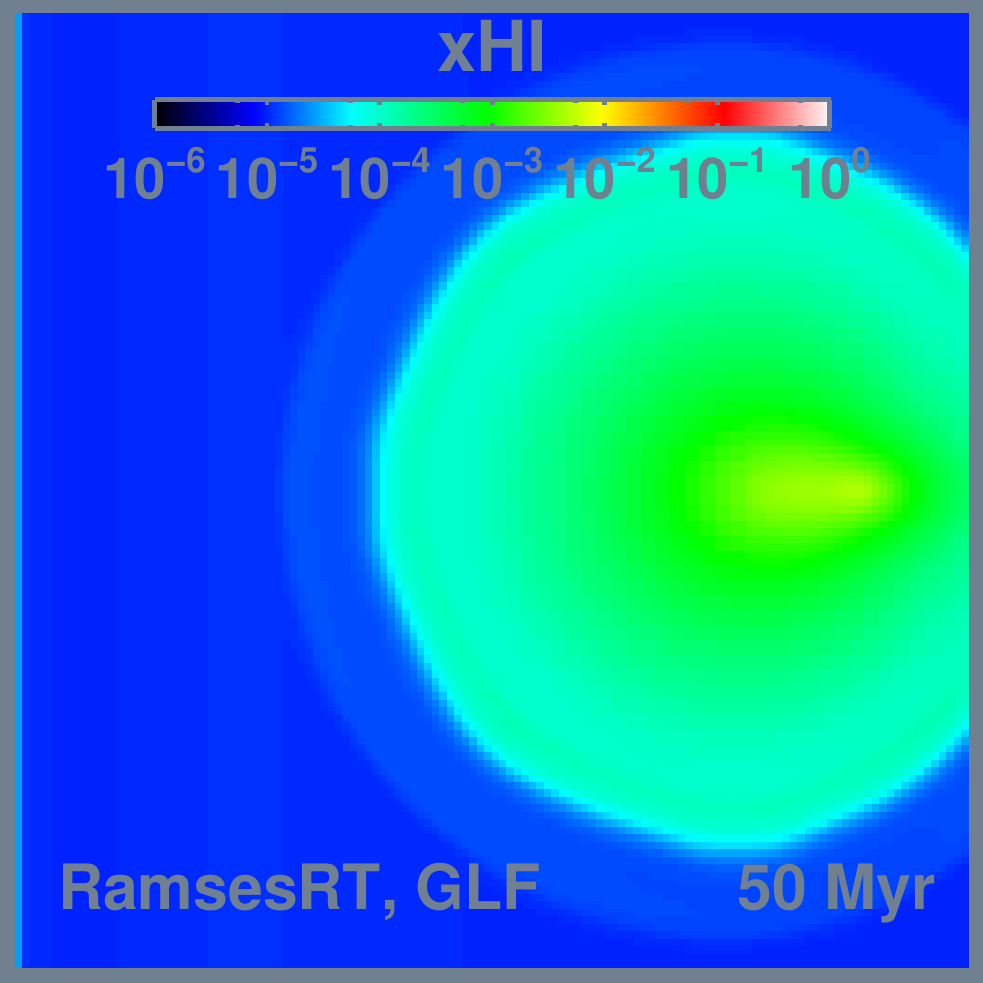}}\hspace{-1.2mm}
  \subfloat{\includegraphics[width=0.2\textwidth]
    {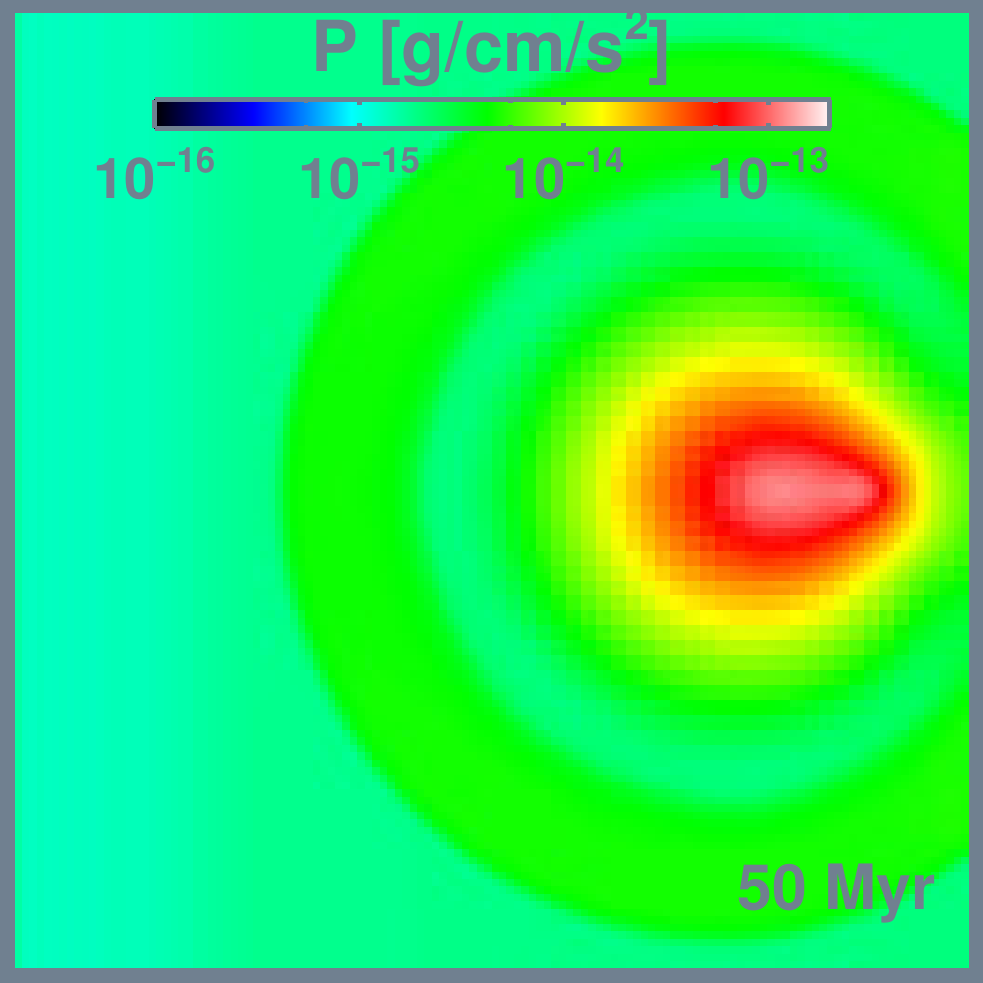}}\hspace{ -1.2mm}
  \subfloat{\includegraphics[width=0.2\textwidth]
    {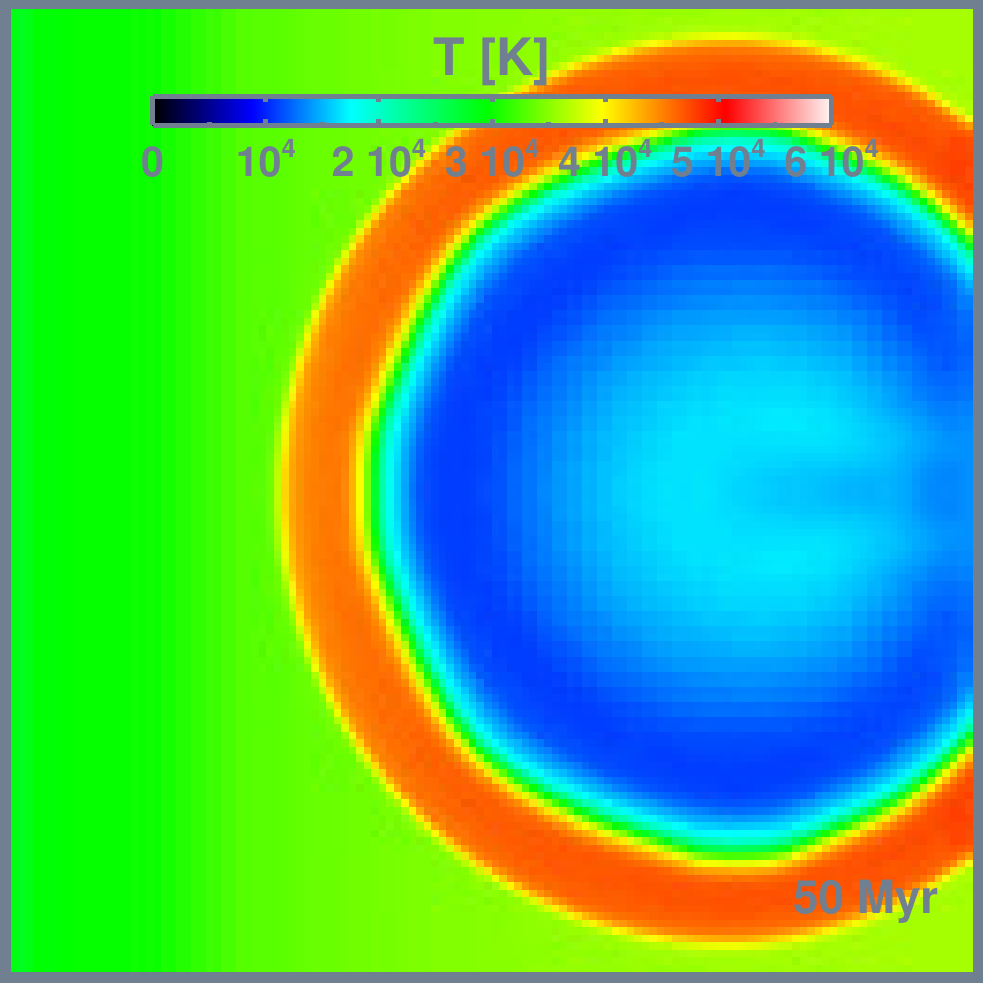}}\hspace{-1.2mm}
  \subfloat{\includegraphics[width=0.2\textwidth]
    {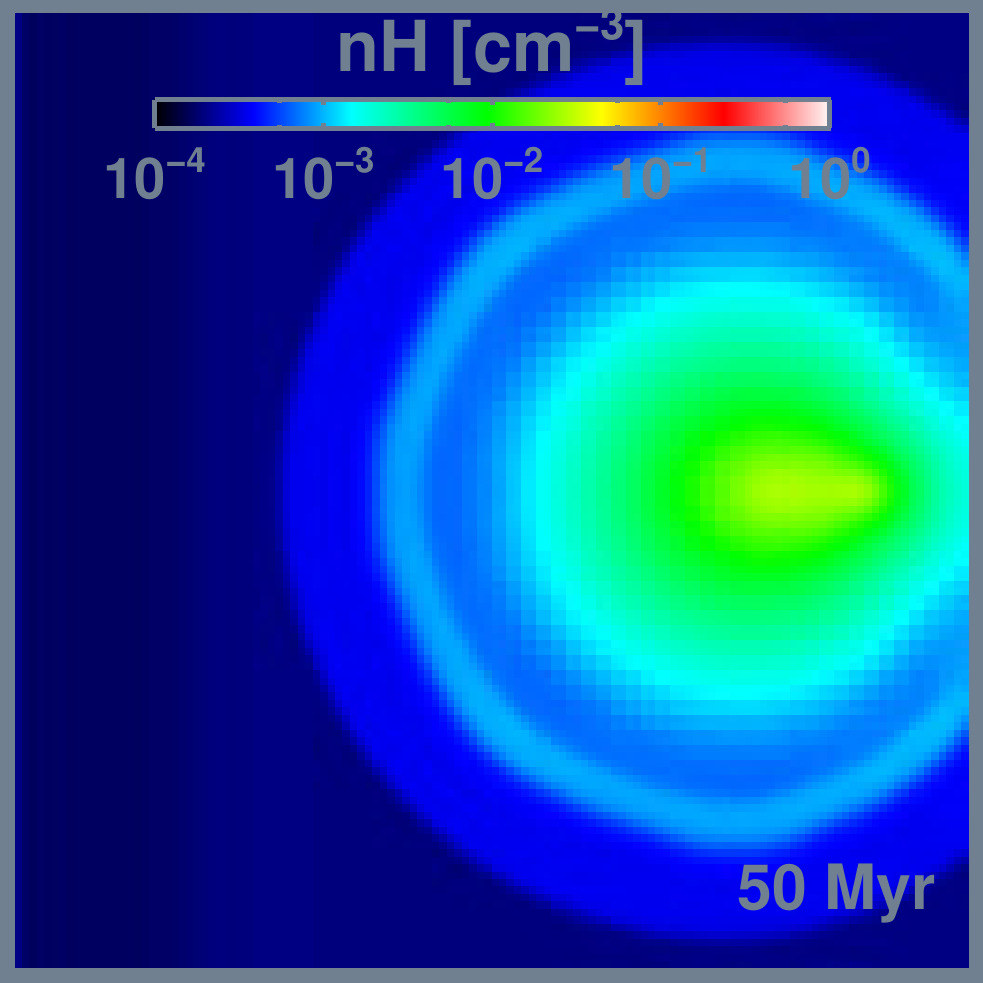}}\hspace{-1.2mm}
  \subfloat{\includegraphics[width=0.2\textwidth]
    {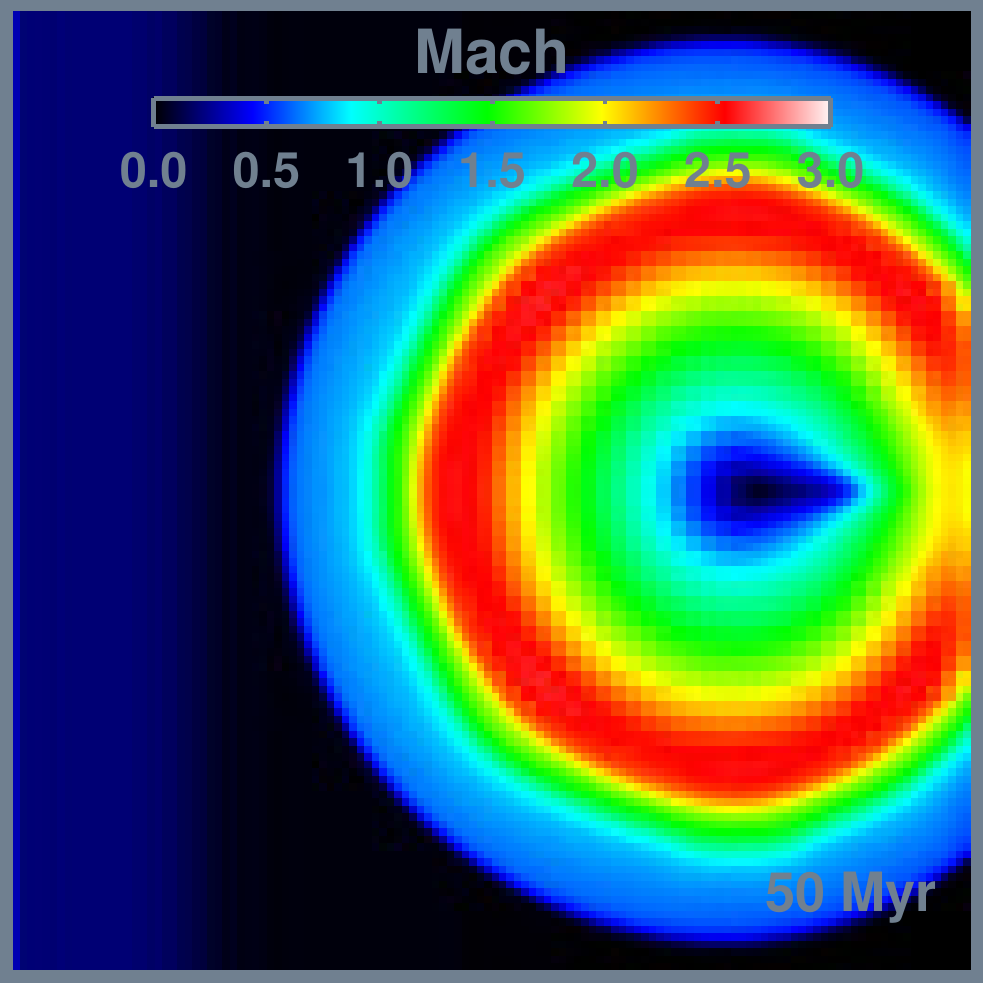}}\hspace{-1.2mm}
  \vspace{-4.mm}

  \subfloat{\includegraphics[width=0.2\textwidth]
    {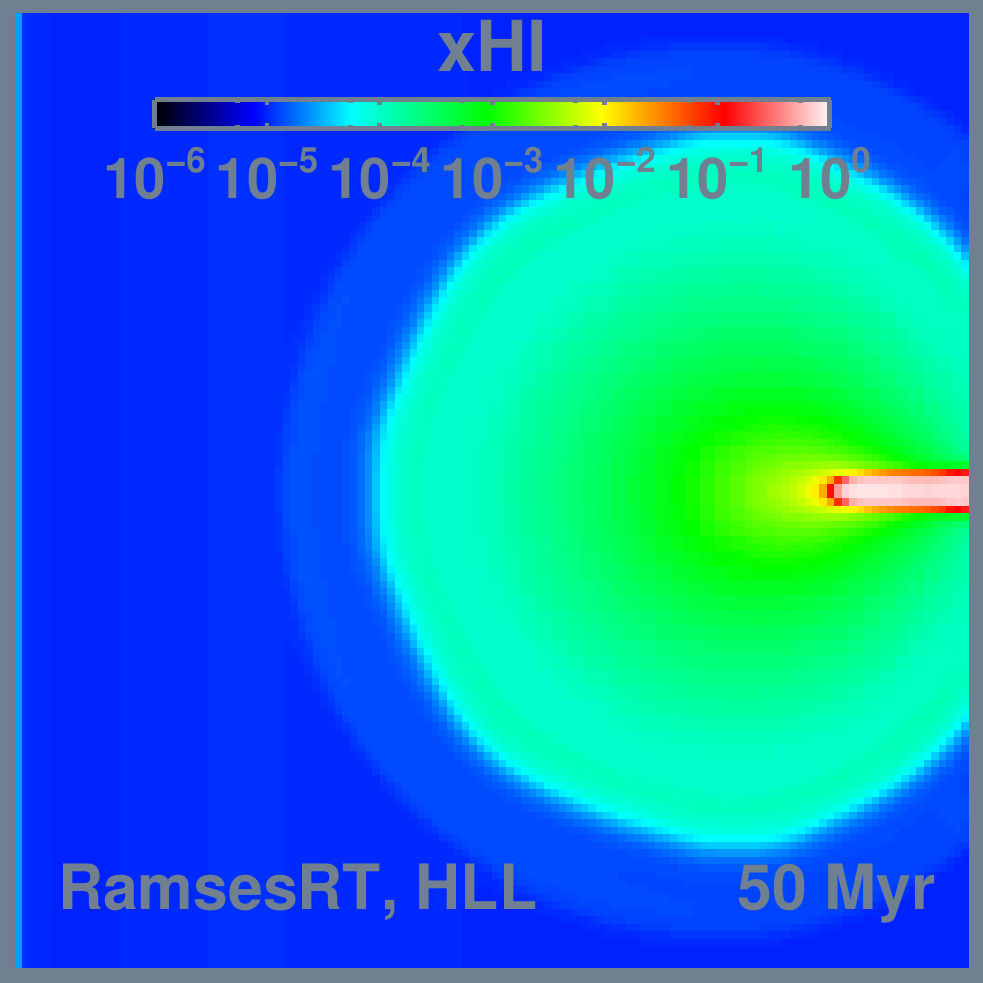}}\hspace{-1.2mm}
  \subfloat{\includegraphics[width=0.2\textwidth]
    {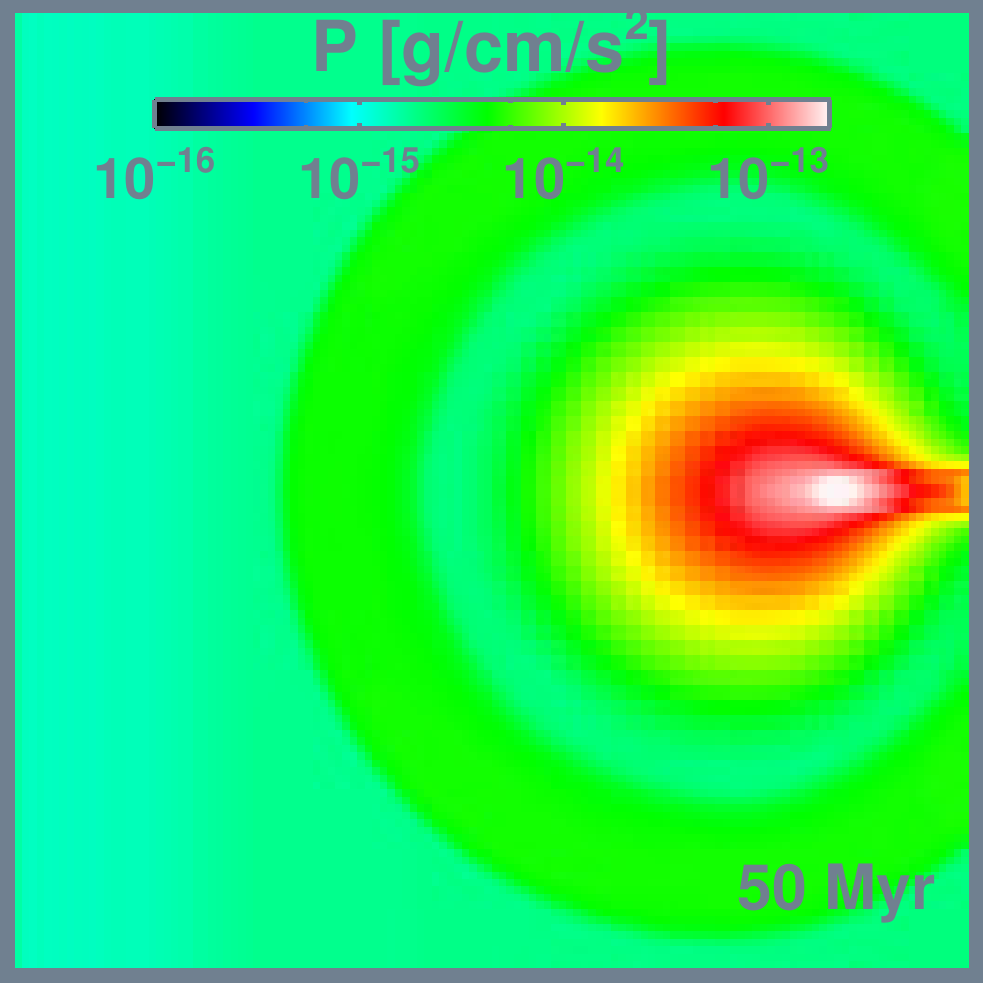}}\hspace{ -1.2mm}
  \subfloat{\includegraphics[width=0.2\textwidth]
    {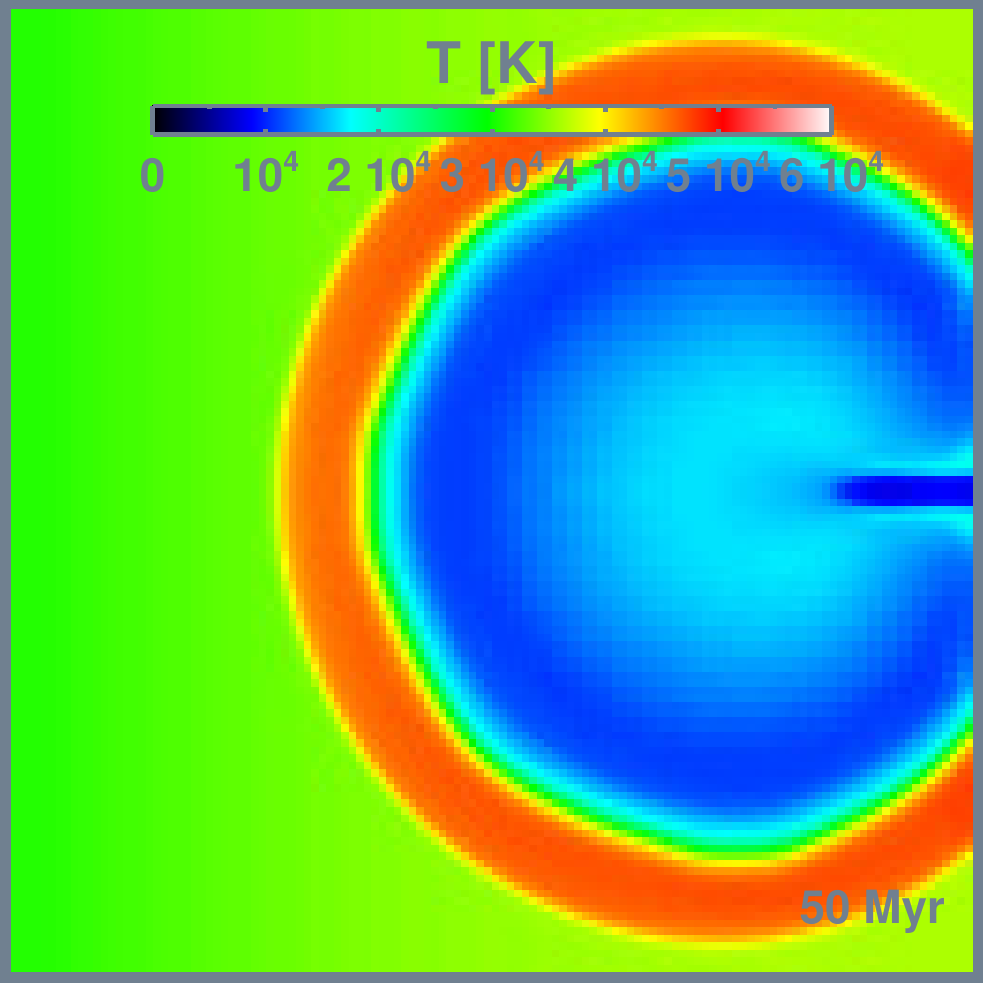}}\hspace{-1.2mm}
  \subfloat{\includegraphics[width=0.2\textwidth]
    {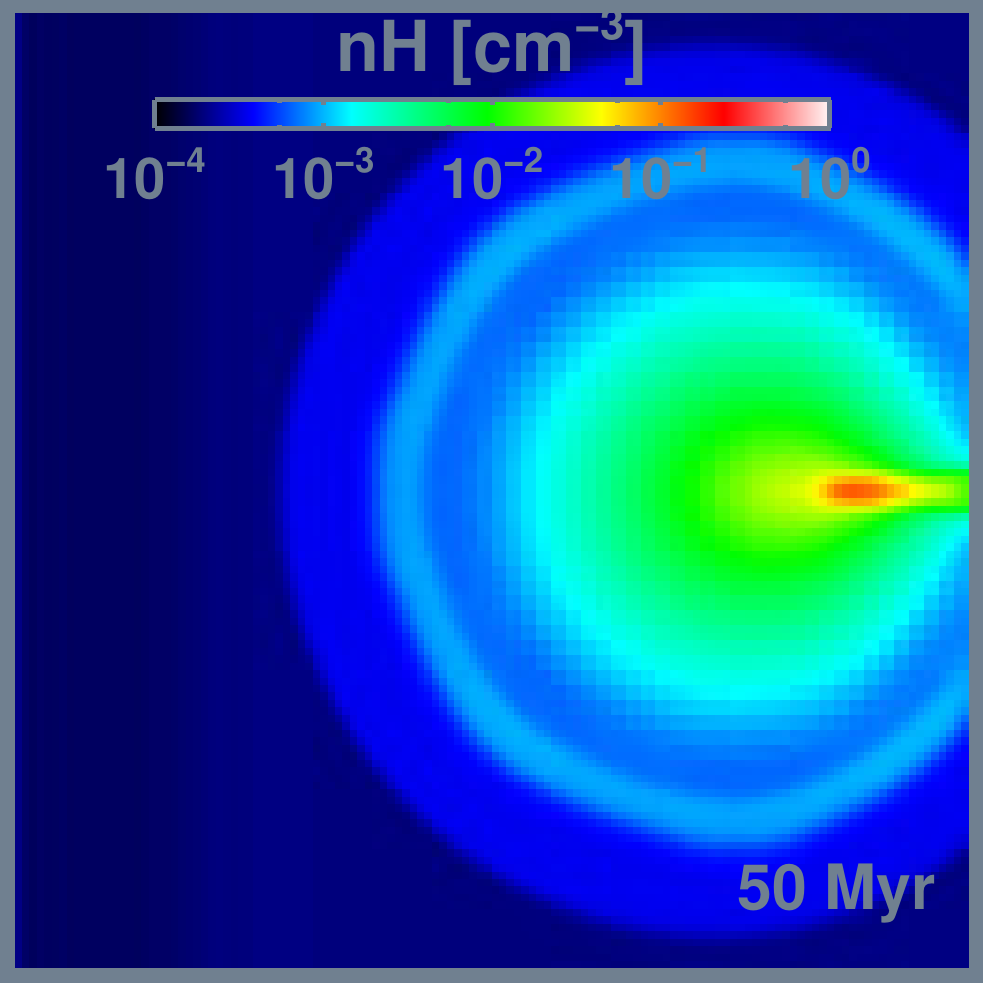}}\hspace{-1.2mm}
  \subfloat{\includegraphics[width=0.2\textwidth]
    {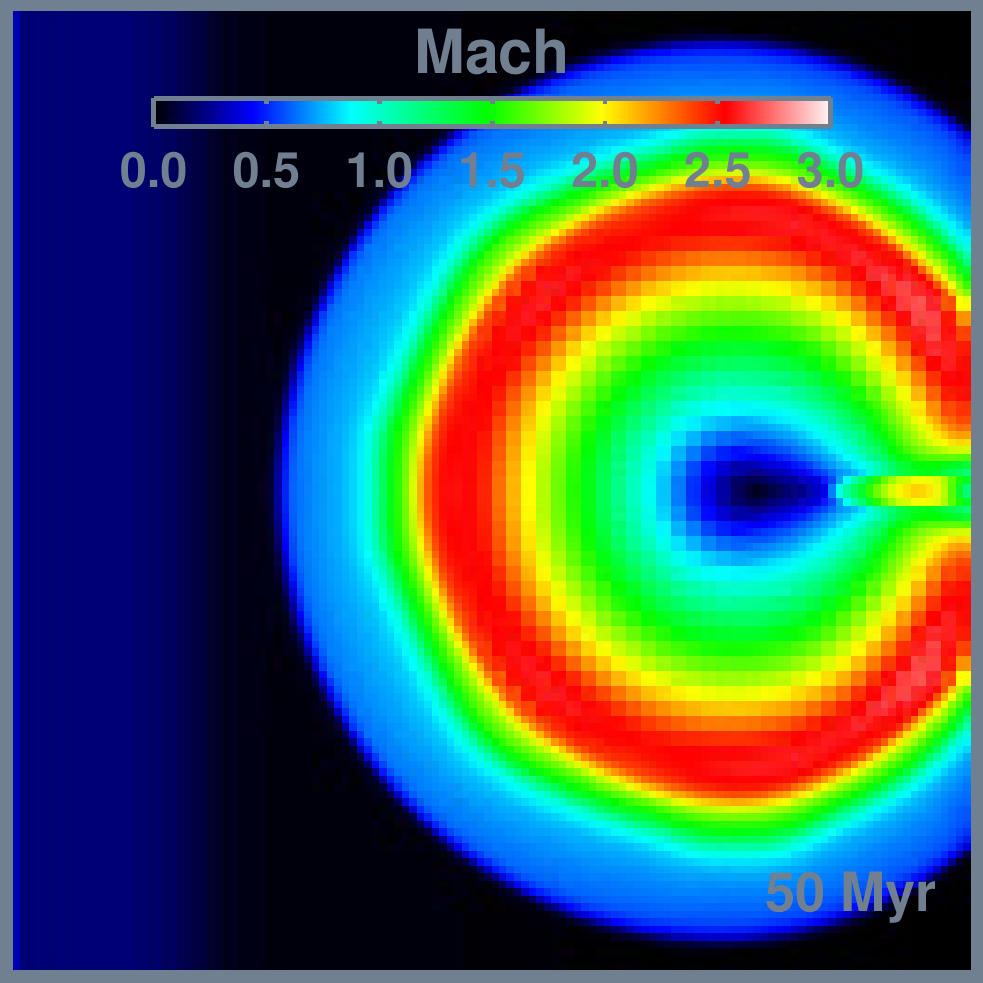}}\hspace{-1.2mm}
  \vspace{-4mm}

  \subfloat{\includegraphics[width=0.2\textwidth]
    {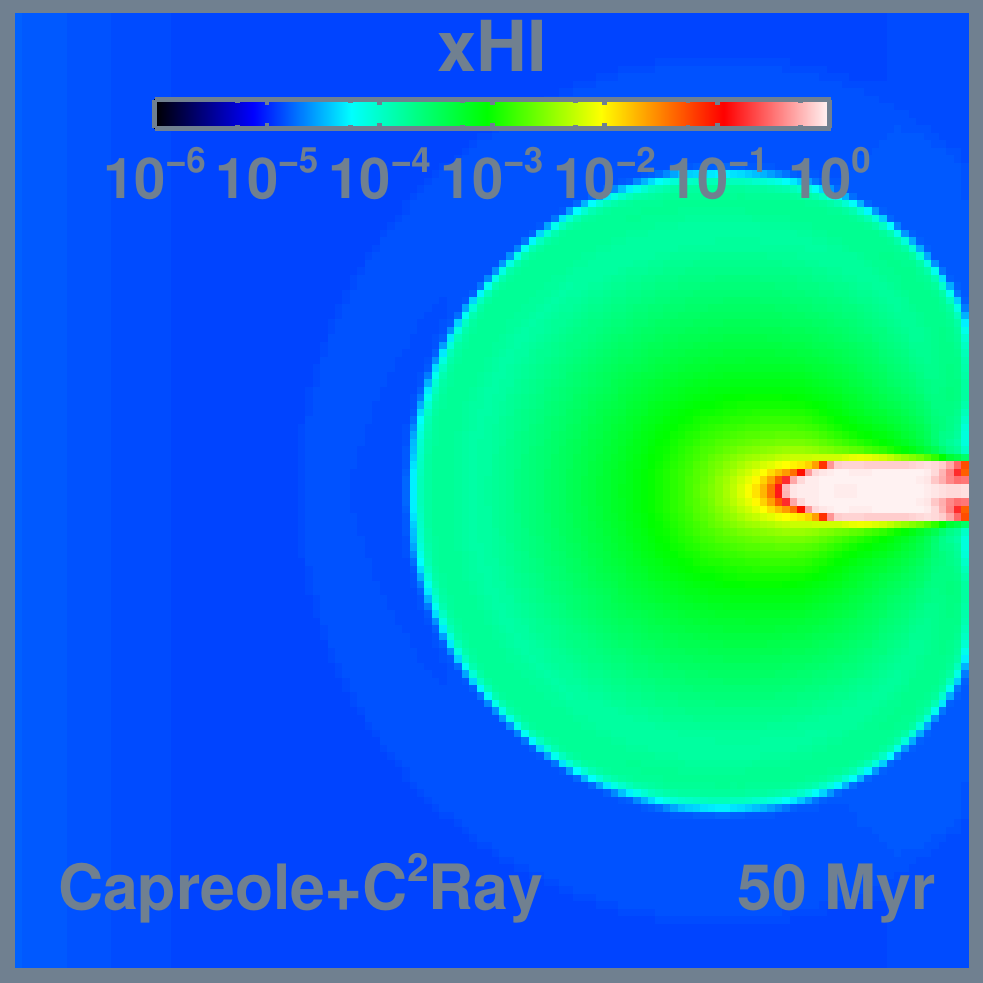}}\hspace{-1.2mm}
  \subfloat{\includegraphics[width=0.2\textwidth]
    {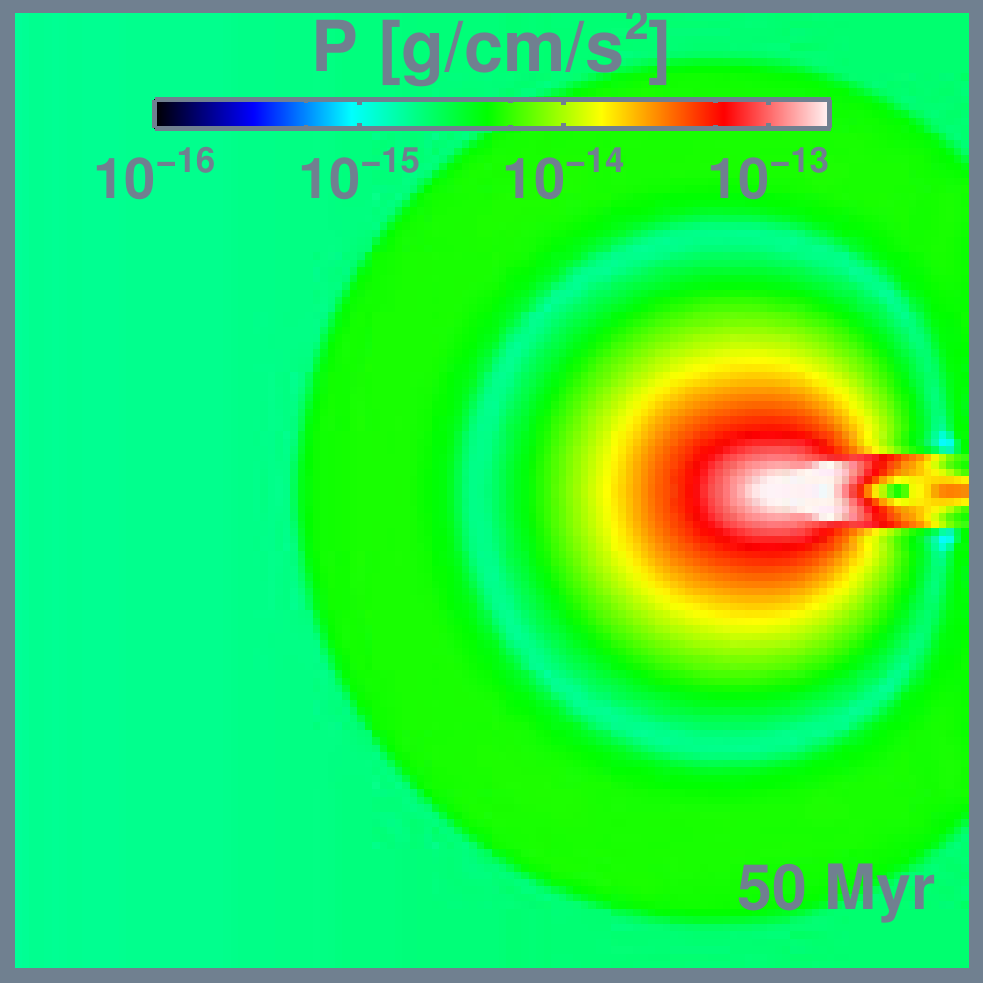}}\hspace{-1.2mm}
  \subfloat{\includegraphics[width=0.2\textwidth]
    {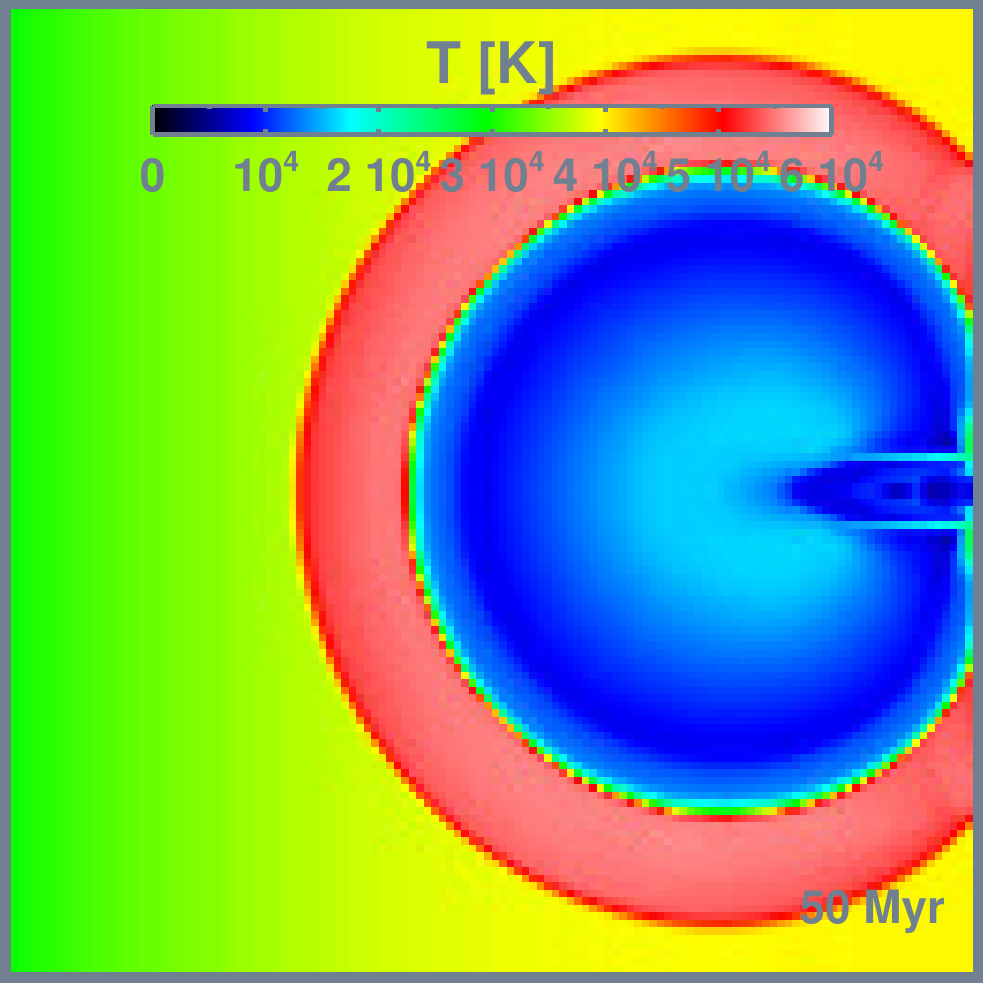}}\hspace{-1.2mm}
  \subfloat{\includegraphics[width=0.2\textwidth]
    {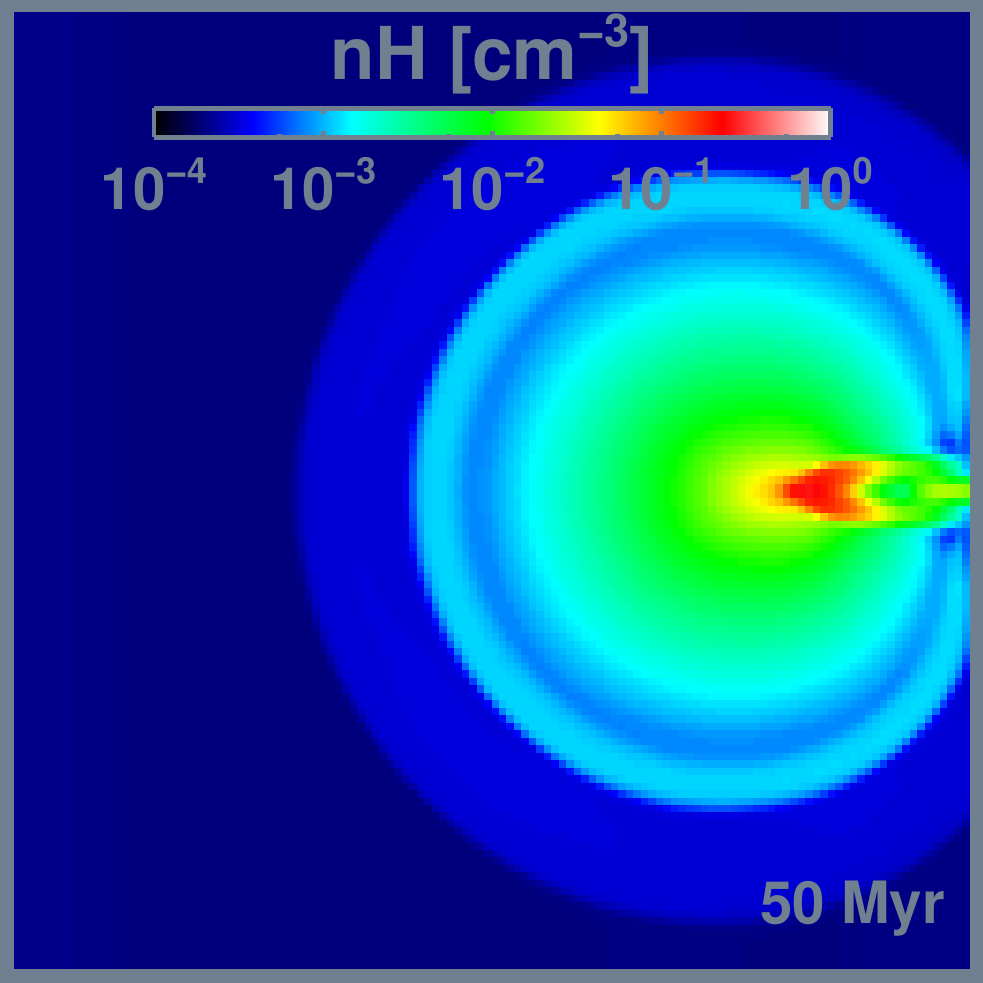}}\hspace{-1.2mm}
  \subfloat{\includegraphics[width=0.2\textwidth]
    {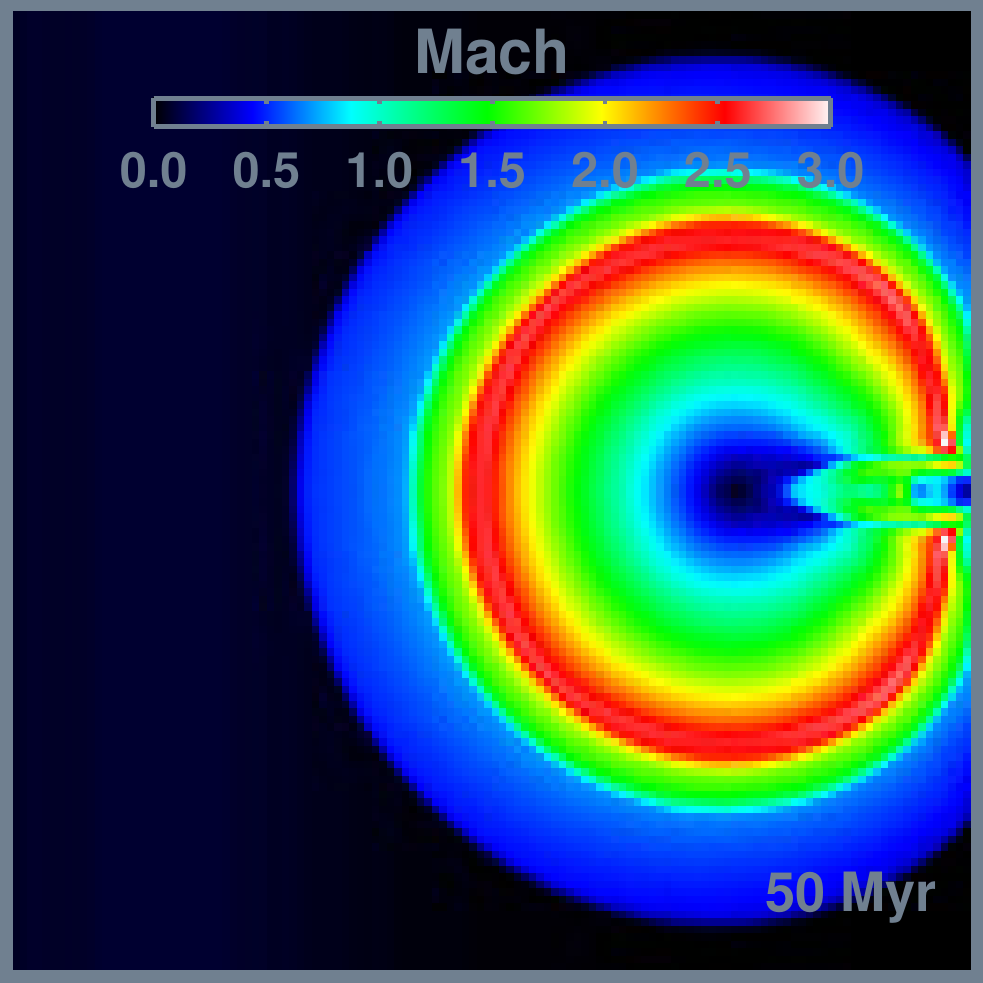}}\hspace{-1.2mm}
  \vspace{-2.5mm}
  \caption[\Ilb{} test 7 - maps]
  {\label{Il7maps.fig}\Ilb{} test 7. Maps showing slices at $z=0.5 \
    \Lbox$ of various quantities at 10 Myrs (top panel) and 50 Myrs
    (lower panel). In each panel, the top row shows the
    \ramsesrt{}+HLL results, the middle row shows \ramsesrt{}+GLF and
    the bottom row shows the \CC2R{} results.}
\end{figure*}

\begin{figure*}
  \centering
  \subfloat[]{\includegraphics[width=0.35\textwidth]
  {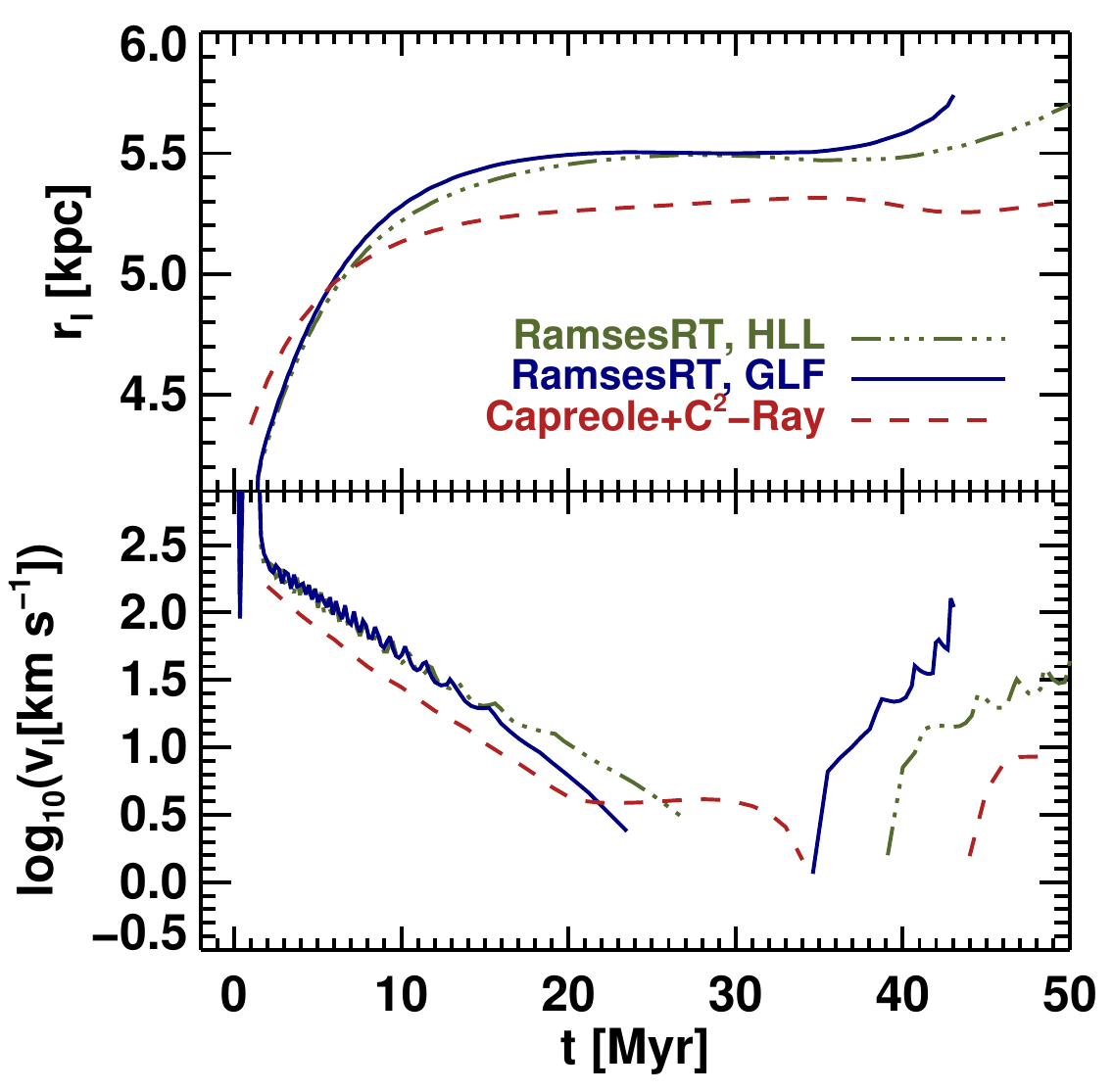}\label{Il7_Ifront.fig}}
  \subfloat[]{\includegraphics[width=0.35\textwidth]
  {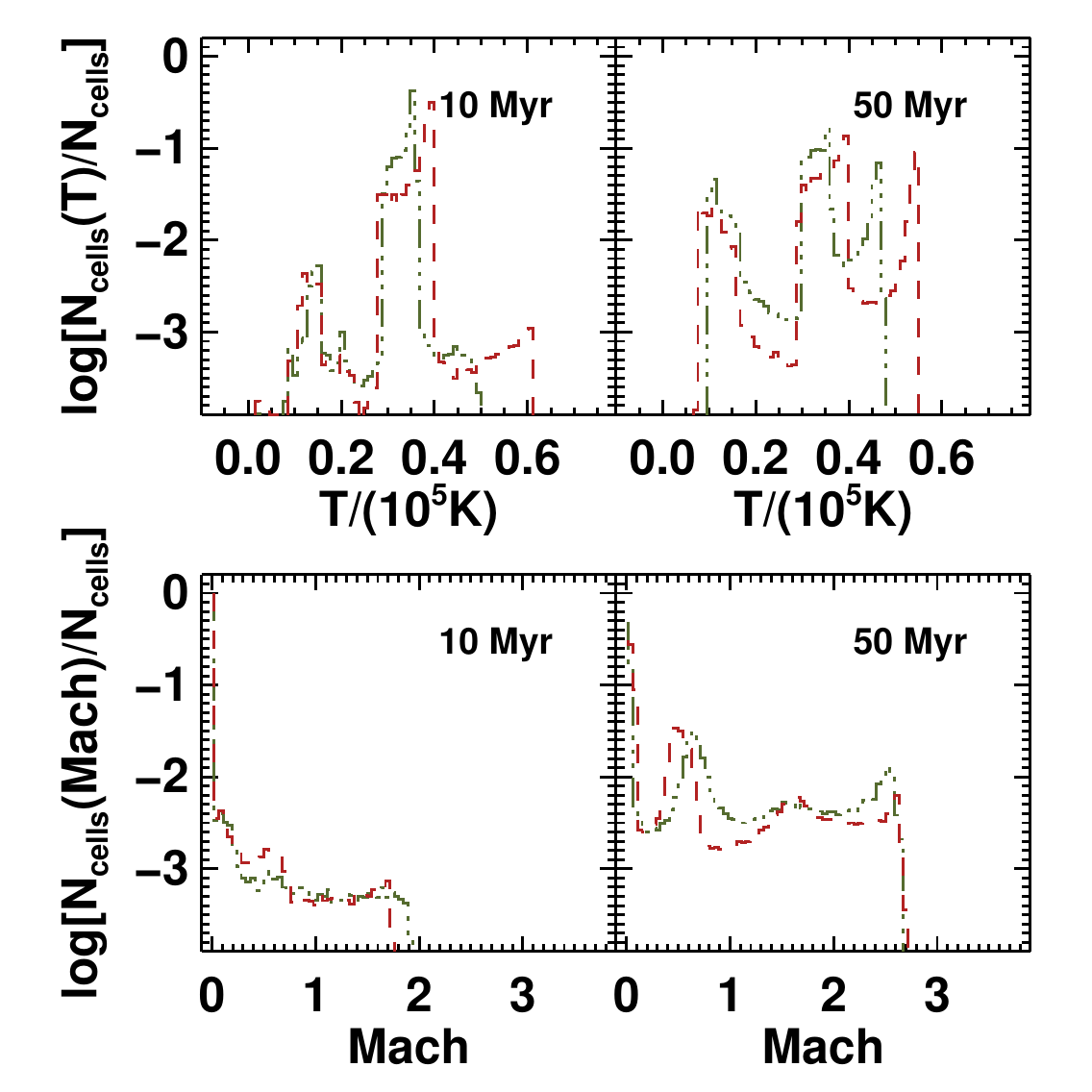}\label{Il7_Hists.fig}}
\caption[\Ilb{} test 7 - I-front and histograms]
{\Ilb{} test 7. \textbf{(a)} Time evolution of the position
  (top) and speed (bottom) of the ionization front along the x-axis of
  symmetry through the center of the box. \textbf{(b)} Histograms of
  the gas temperature (upper panel) and flow Mach number (lower panel)
  at 10 and 50 Myr for \ramsesrt{} and \CC2R{}.}
\end{figure*}

\Fig{Il7maps.fig} shows slices in the $xy$-plane through the middle of
the box of various quantities at $10$ and $50$ Myr, for the
\ramsesrt{} result and \C2R{} for comparison\joki{\footnote{\joki{In
      the official \C2R{} outputs from test 7 in \Ilb{}, the
      temperatures are too low and the densities too high by a factor
      1.3, which is a missing helium-based mean molecular weight
      (Garrelt Mellema, private communication). We have therefore
      adjusted the \C2R{} output temperatures and densities by this
      factor to retrieve their correct results. Making this change
      improves the agreement between temperature profiles from \C2R{}
      and other codes in Figures 40 and 43 in \Ilb{}, where \C2R{}
      otherwise stands out somewhat.}}}. As in the corresponding pure
RT test, it can be seen from the $\xhi$ maps that the shadow behind
the cloud is less conserved with \ramsesrt{} than with \C2R{}, though
the HLL solver does a much better job though than GLF. However, the
diffusion of photons doesn't have a large impact on the resulting
dynamics, or even the propagation of the I-front along the axis of
symmetry. The shadow becomes thinner towards the end of the run with
all codes in \Ilb{}, though it is thinner than most in
\ramsesrt{}+HLL, and it pretty much disappears in \ramsesrt{}+GLF. The
shadow thickness in \ramsesrt{}+HLL is still comparable at $50$ Myrs
to the results of \rsph{}, \zeusmp{} and \licorice{} in \Ilb{}. The
pressure maps of \ramsesrt{}+HLL, \C2R{} and other codes in \Ilb{} are
very similar both at $10$ and $50$ Myrs, though \C2R{}, and also to
some extent \flashhc{} and \licorice{} have a fork-like shape inside
what remains of the shadow at $50$ Myr. The other codes have the same
shape as \ramsesrt{}+HLL in this region. \joki{The temperature maps
  are similar as well, though the backwards-expanding cloud shell
  seems to be slightly less shock-heated in \ramsesrt{} than most
  other codes.} The shell expands in a very similar way for the two
codes, as can be seen in the density and Mach slices. The expansion
goes a bit further, though, in \ramsesrt{}. Also, the expanding cloud
seems to develop a slightly hexagonal shape in \ramsesrt{}, an effect
which is not apparent in any of the codes in this test in \Ilb{}
(though there is a hint of it in the \flashhc{} result). It can only
be speculated that this is a grid artifact. To be sure it doesn't have
to do with the on-the-fly refinement we ran an identical experiment
with a base resolution of $128^3$ cells and no refinement in
\ramsesrt{}+HLL. The \ramsesrt{}+HLL maps and plots presented here are
virtually identical to this non-refinement run, except of course for
graininess in the slice maps. None of these discussed effects
(hexagons and a slightly over-extended I-front compared to other
codes) are thus due to on-the-fly refinement.  \joki{As in the
  previous test, the speed-wise gain in using AMR is not a lot: the
  AMR run completes in about half of the $\sim 64$ cpu hours taken for
  the non-AMR run. Again the relatively modest speedup is due to a
  combination of a large portion of the grid being refined ($\sim
  30\%$ by volume when most), a shallow refinement hierarchy, and
  refinement-related overhead. With deeper refinement hierarchies in
  cosmological simulations, the speedup can be much greater, but a
  quantitative demonstration is beyond the scope of this paper.}

\begin{figure*}
  \centering
  \includegraphics[width=.85\textwidth]
  {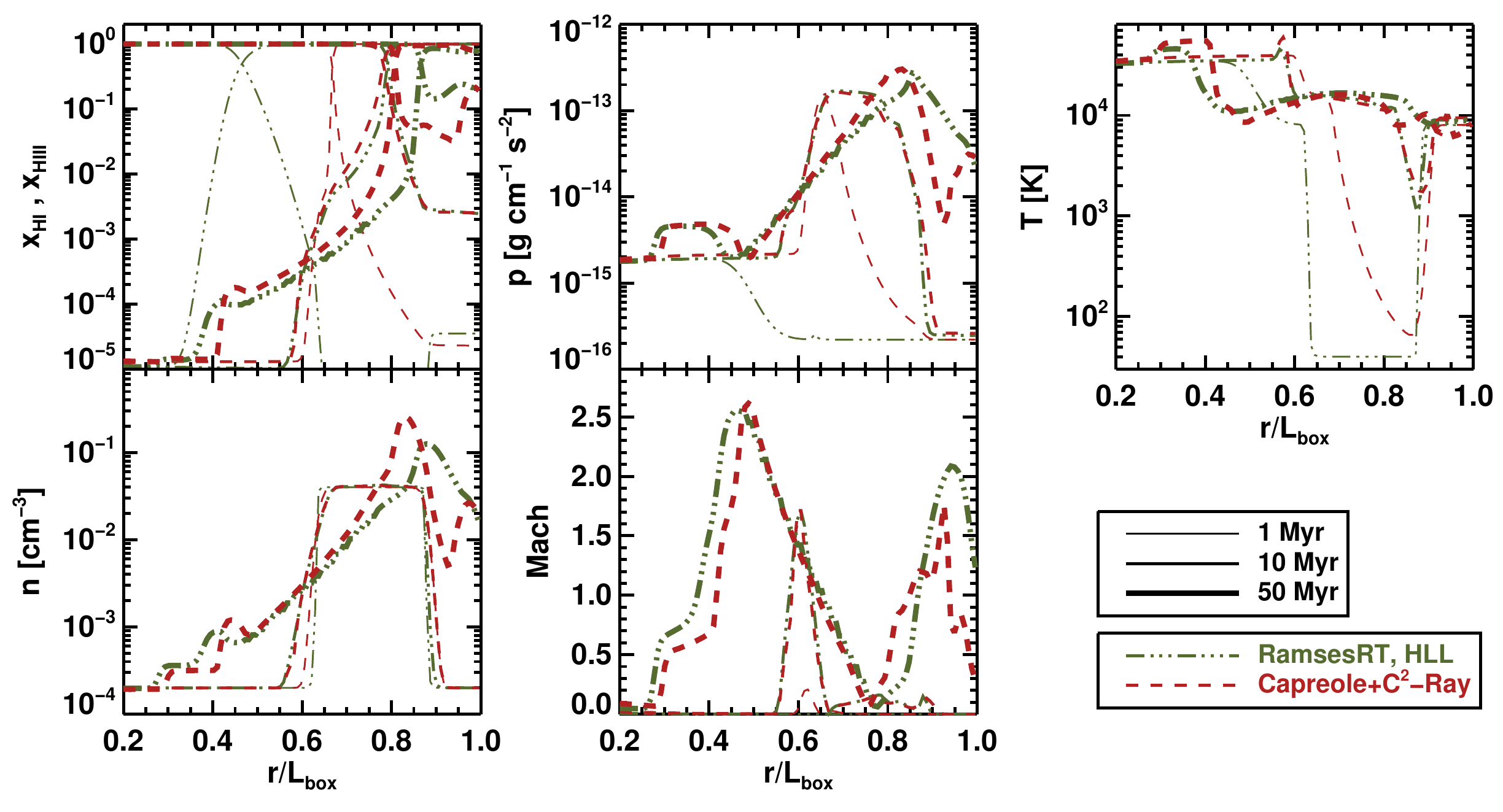}
  \caption[\Ilb{} test 7 - x-profiles]
  {\label{Il7_Profs.fig}\Ilb{} test 7. Profiles along the
    x-axis of symmetry through the center of the box, at 1, 10 and 50
    Myr for the \ramsesrt{} and \CC2R{} results. Clockwise from
    top left: ionization fractions, pressure, temperature, Mach
    number, atom number density.}
\end{figure*}

Next we turn our attention to the evolution of the position and speed
of the I-front along the x-axis of symmetry through the box. This is
presented for the \ramsesrt{} (HLL and GLF) and \C2R{} runs in
\Fig{Il7_Ifront.fig}. The I-front propagation is considerably
different between \ramsesrt{} and \C2R{}, but actually \C2R{}
considerably stands out here from other codes in \Ilb{}.  For the
first 7 Myrs or so, the \ramsesrt{} front lags behind that of \C2R{}
and in fact all the codes in \Ilb{}. This is due to the reduced speed
of light: before hitting the cloud, the photons have to travel from
the left edge of the box through a very diffuse medium -- so diffuse
that here the I-front speed apparently is approaching the speed of
light, or is at least considerably faster than the one-one-hundredth
of the light speed which is used in the \ramsesrt{} run. However, once
the I-front in the \ramsesrt{} run has caught up, the reduced light
speed should have a negligible effect on the results. After roughly 7
Myr, the \ramsesrt{} I-front overtakes \C2R{} front, and stays ahead
of it for the remainder of the run. This however is also the case for
most of the codes in \Ilb{}; their I-front is ahead of the \C2R{}
front, and four out of six codes end up with the I-front at $\sim 5.6$
kpc. The \ramsesrt{}+HLL front ends up at $\sim 5.7$ kpc, so slightly
ahead of what is typically found in \Ilb{}. Using the GLF solver
instead of HLL has the effect that the I-front disappears soon after
$40$ Myrs, which is due to diffusive photons eating into the shadow
from it's edges, but up to that point the I-front evolution is much
the same. \ramsesrt{} also reproduces the retreat of the I-front
between roughly 30 and 40 Myrs, which is seen in all runs in
\Ilb{}. This momentary negative speed is due to the expansion of the
cloud and the D-type movement of the I-front with the gas.

\Fig{Il7_Hists.fig} shows histograms of the gas temperature and Mach
number at $10$ and $50$ Myr in the \ramsesrt{}+HLL and \C2R{}
runs. The shapes of the histograms are very similar between the two
codes (and are also very similar to \ramsesrt{}+GLF, which is not
shown).

Finally, \Fig{Il7_Profs.fig} shows a comparison between \ramsesrt{}
and \C2R{} of profiles along the x-axis of symmetry of the various
quantities at $1$, $10$ and $50$ Myrs. The profiles compare badly at
$1$ Myr, but as already discussed this is simply due to the I-front
having not caught up at this early time when using the reduced speed
of light. At later times the profiles generally compare well, though
we see these effects which have already been discussed, of a further
expanding density-front out of the original cloud, and a further
progressed I-front. The \ramsesrt{} profile plots show a staircase
effect which is most obvious in the $50$ Myrs plot at the radial
interval $0.45\la r/L_{\rm{box}}\la 0.75$: this is simply due to the
grid being unrefined at this x-interval along the axis of symmetry,
i.e. at the effective base resolution of $64^3$ cells per box
width. The run with the full resolution and no AMR refinement shows no
staircases, but otherwise the results are identical to those shown
here.

We have made an alternative run with \ramsesrt{}+HLL with the
speed of light fraction set to $f_c=1/10$ rather than the default
$1/100$, and here the initial evolution of the I-front position and
radial profiles at $1$ Myr are almost identical to those of \C2R{}. At
later times the results are very much in line with those where
$f_c=1/100$, except the I-front position is slightly more advanced at
$50$ Myr, or at $5.78$ kpc rather than at $5.71$ kpc.

In summary, \ramsesrt{} performs well on this test with no apparent
problems. The reduced light speed ($f_c=1/100$) has very little effect
on the results and on-the-fly refinement gives results which are
identical to the fully refined simulation with a homogeneous $128^3$
cells grid. Even using the diffusive GLF solver retains much of the
results (I-front development, cloud expansion), except that the
I-front disappears a bit prematurely.

\subsection{RT test conclusions}
\ramsesrt{} performs very well on all the tests from \Ila{} and
\Ilb{}, with no discrepancies to speak of from expected results or
those from other codes. 

\joki{The most notable discrepancies clearly result from the reduced
  speed of light approximation, which leads to I-fronts that are
  initially too slow compared to full speed of light runs -- or
  \emph{infinite} speed, as is the case for many of the codes compared
  against from the RT comparison project. In test 4, the high-z
  cosmological field, we actually demostrated the reverse, where the
  codes we compared to had considerably \emph{premature} I-fronts as a
  consequence of their infinite light speed approximations.}  Our
shadows are considerably shorter lived with the GLF intercell flux
function than those of the other codes (most of which use ray-tracing
schemes). This can be fixed for problems involving shadows and
idealized geometries by using the HLL flux function instead, but as we
showed in \Sec{Transport.sec} the sacrifice is that isotropic sources
become anisotropic. Many codes in the RT comparison project show
various instabilities and asymmetries in ionization fronts; no such
features are manifested in the \ramsesrt{} results.


%% file: RamsesRT_chemtests.tex
To validate the non-equilibrium thermochemistry in \ramsesrt{} we ran
one-cell thermochemistry tests, that start at some initial state
(temperature, ionization state, photon flux) and evolve over roughly a
Hubble time.  We are interested here in verifying that our
implementation is correct and error free and also in comparing
equilibrium vs. non-equilibrium cooling -- e.g. \cite{{Cen:2006kh}}
report that the methods can produce significantly different
results. We compare against the equilibrium thermochemistry of
\ramses{} which has been modified to use the exact same heating,
cooling and interaction rates as \ramsesrt{}.

We test to see \textit{(i)} whether the thermochemistry of \ramsesrt{}
is stable, i.e. if the stiffness of the equations results in any
sudden divergence or `wiggles' in the evolution of the gas,
\textit{(ii)} whether \ramsesrt{} evolves the ionization fractions
towards the correct states predicted by the equilibrium solver of
\ramses{}, and \textit{(iii)} whether the \ramses{} and \ramsesrt{}
evolve the temperature towards the same final value.

There are four tests: first we disable cooling and evolve only the
ionization states of hydrogen and helium at different constant
temperatures in a zero UV radiation field, and see if we reach
equilibrium ionization states (predicted by \ramses{}). Then we turn
on a constant UV radiation field and again see if we reach equilibrium
states. Then we turn on cooling, and for two sets (zero, nonzero
radiation field) see if the temperature evolution is comparable to
\ramses{} equilibrium cooling from the same initial conditions.

\subsection{Ionization convergence at constant temperature and zero
  ionizing photon flux}
\begin{figure*}
  \centering
  \includegraphics[width=.9\textwidth]{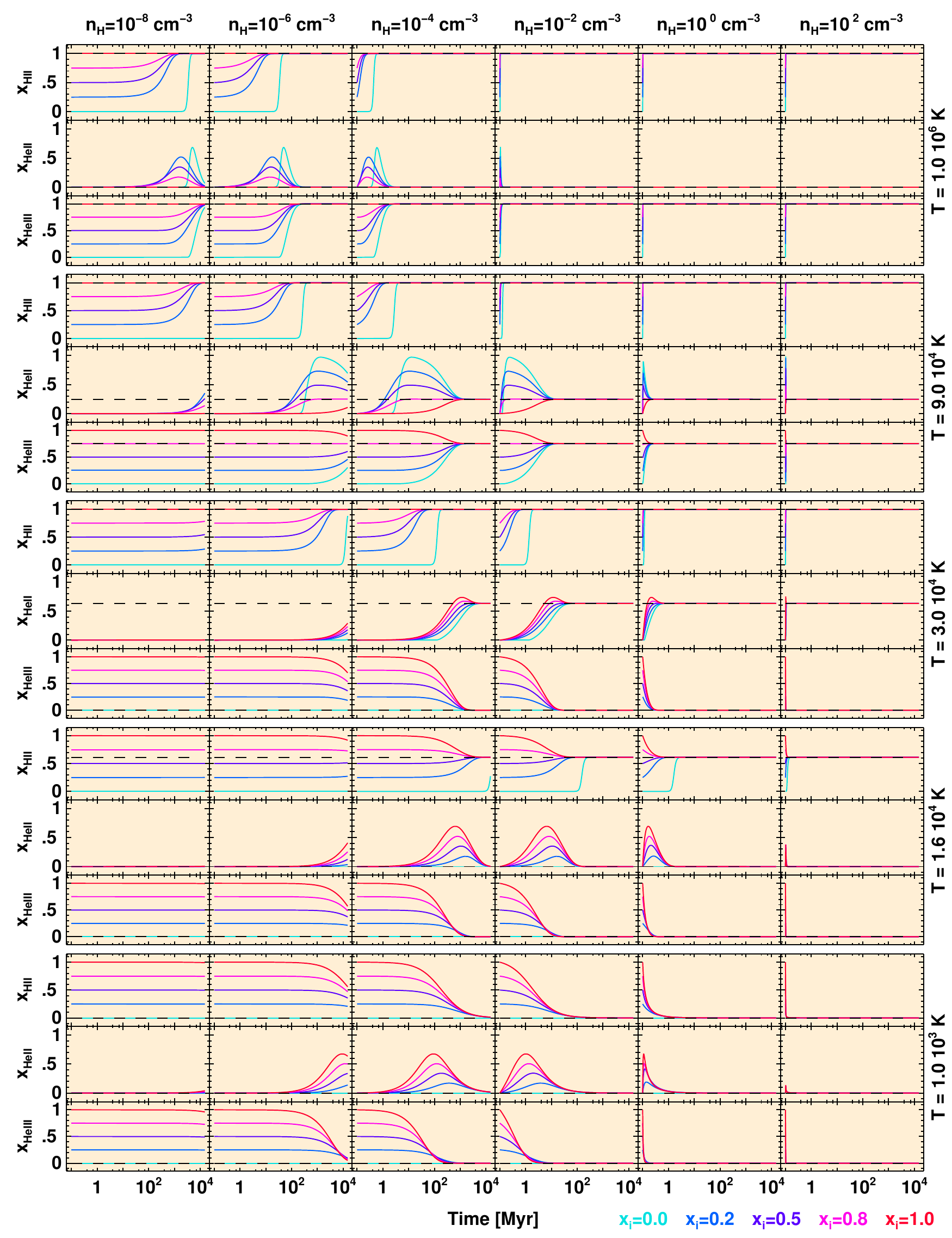}
  \caption[Ionization convergence test with zero ionizing flux]
  {\label{cTest1.fig}Ionization convergence test with
    constant $T$ and zero ionizing photon flux. Coloured lines show
    non-equilibrium evolution of the ionization fractions, given
    constant $T$ (right) and $\nh$ (top). Black dashed lines show the
    corresponding equilibrium ionization fractions as calculated in
    \ramses{}.}
\end{figure*}

In the first test, cooling is turned off and we check for a range of
densities, temperatures and initial ionization states whether we get a
convergence of the ionized fractions towards their equilibrium states,
as predicted by \ramses{}, assuming zero flux of ionizing photons.

Fig. \ref{cTest1.fig} shows the results. Each panel of $3\times6$
plots in the figure represents an evolution given the constant
temperature written to the right of the panel, and shows how the
ionized fractions, $\xhii$, $\xheii$ and $\xheiii$, evolve from
different (color-coded) starting states $x_i=\xhii=\xheiii$ (the HeII
fraction always starts at zero).  A black dashed line in each plot
shows the equilibrium ionization fraction for the given temperature
and species (which is gas density independent in the case of zero
ionizing flux). Each column of plots represents a (non-evolving)
hydrogen number density.

The non-equilibrium ionization fractions always evolve towards the
equilibrium ones, at a rate which depends on gas density, as
expected. It can even take longer than the age of the Universe to
reach equilibrium for the most diffuse gas ($\nh \la 10^{-6} \ \cci$),
which indeed is a significant difference from the equilibrium
assumption. If we zoom in around the equilibrium states we find a
difference between the calculated equilibrium state and the evolved
one which is typically around one in ten-thousand - this simply
corresponds to the allowed error in the iterative equilibrium
calculation, and can be decreased at will by reducing this error
margin.

\subsection{Ionization convergence at constant temperature and nonzero
  ionizing photon flux}
\begin{figure*}
  \centering
  \includegraphics[width=.9\textwidth]{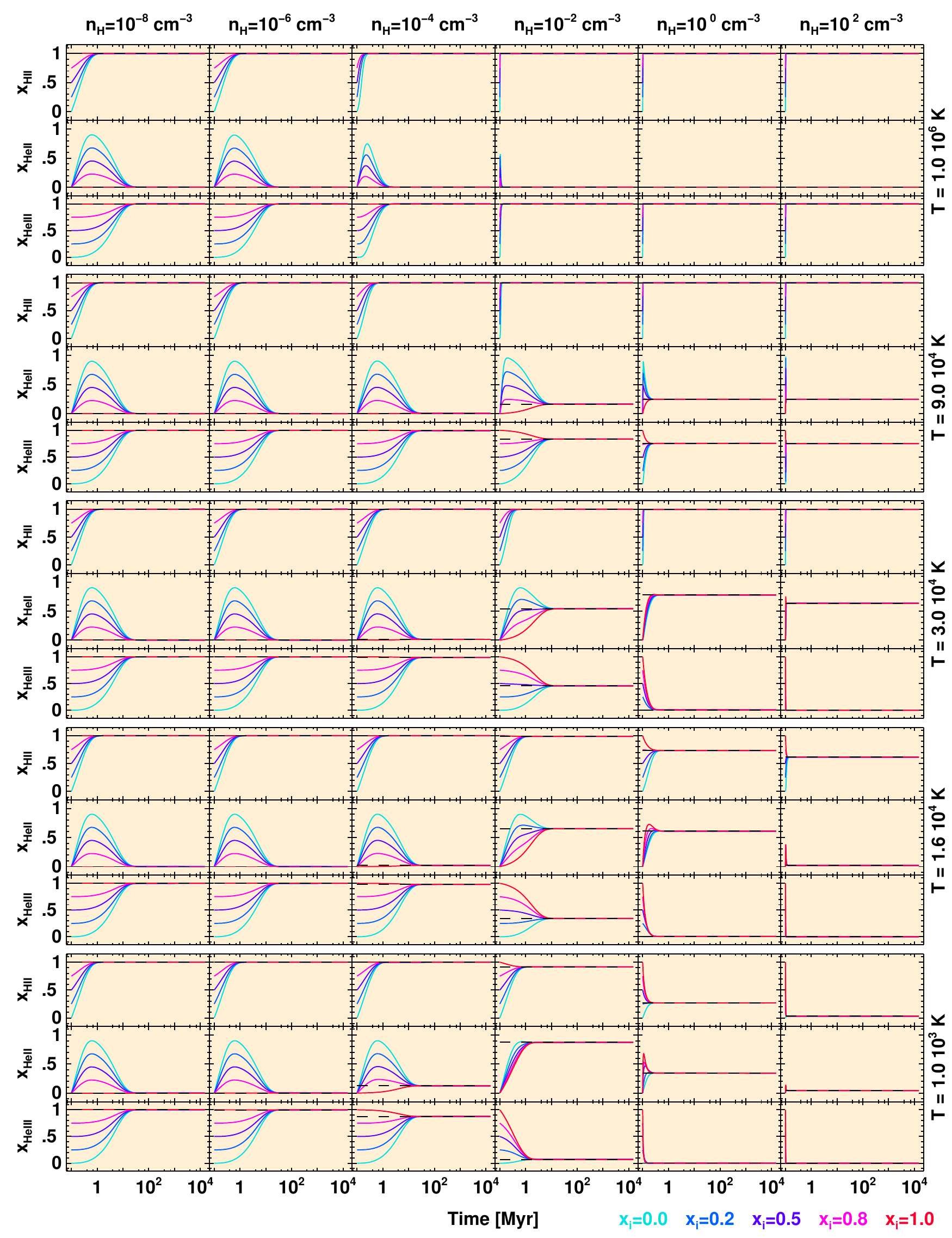}
  \caption[Ionization convergence test with nonzero ionizing flux]
  {\label{cTest2.fig}Ionization convergence test with constant
    temperature and an ionizing photon flux of $10^{5} \; \; \flux$.}
\end{figure*}
\begin{figure*}
  \centering
  \includegraphics[width=.86\textwidth]{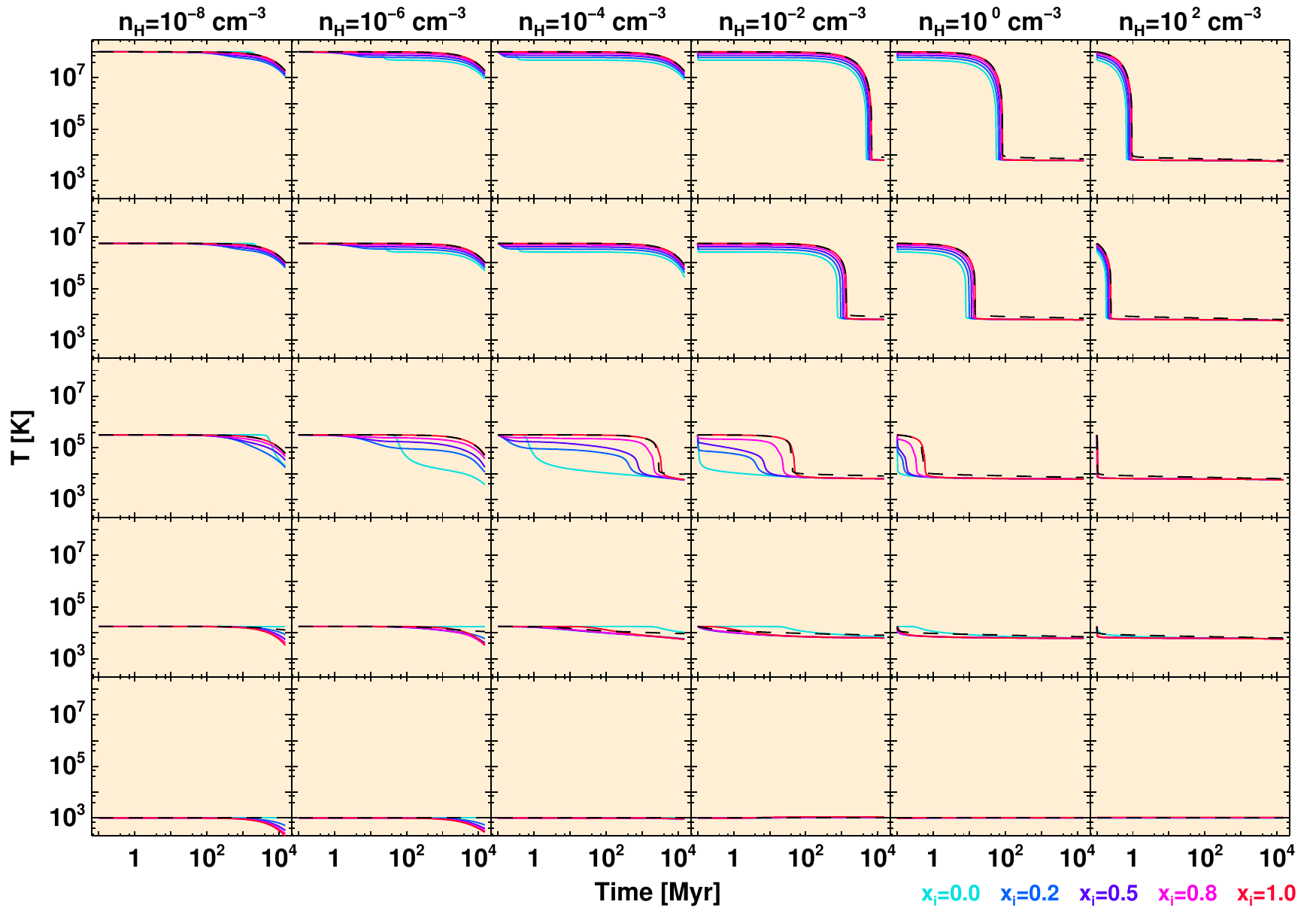}
  \caption[Temperature convergence with zero ionizing flux]
  {\label{cTest3.fig}Temperature convergence with zero ionizing
    flux. Color coded lines show different initial states of $\xhii$
    and $\xheiii$, as indicated by the color legend at bottom
    right. Black dashed curves show the equilibrium evolution from
    \ramses{}.}
\end{figure*}
\begin{figure*}
  \centering
  \includegraphics[width=.86\textwidth]{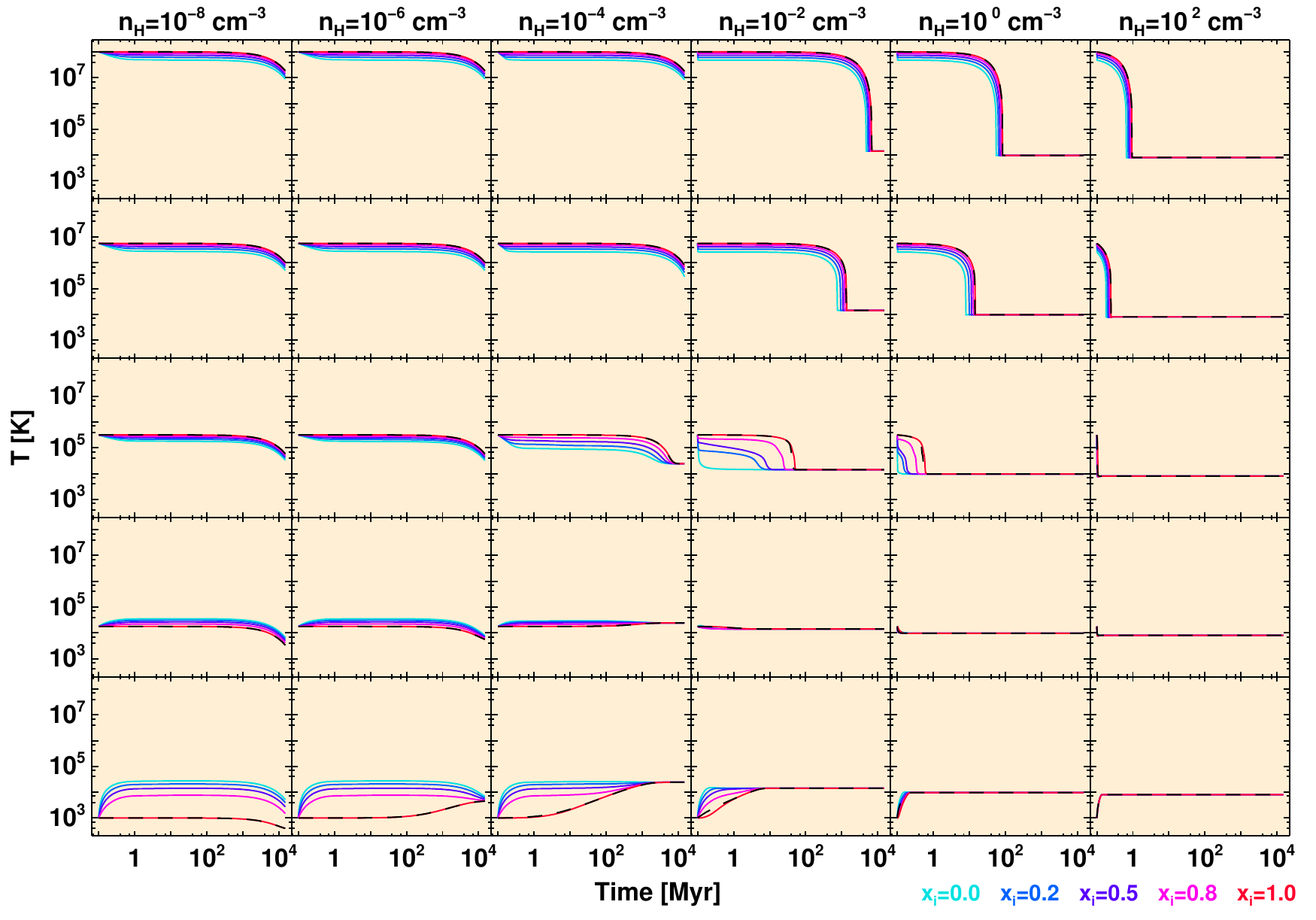}
  \caption[Temperature convergence with nonzero ionizing flux]
  {\label{cTest4.fig}Temperature convergence with an
    ionizing photon flux of $10^{5} \; \; \flux$.}
\end{figure*}

This is the same as the previous test, except now we apply a constant
flux of $10^5$ ionizing photons $\flux$ through the cell, assuming the
spectrum of a blackbody at $10^5$ K.

Fig. \ref{cTest2.fig} shows the results. The black dashed lines in
each plot show the equilibrium state which now is density dependent -
the denser the gas the harder it is for the radiation field to battle
against recombinations. Again the non-equilibrium ionized state always
evolves towards the equilbrium one at a gas density dependent rate,
though note that here it takes a maximum of $\sim 10$ Myr, which is
much shorter than it can take in the zero photon flux case.

\subsection{Temperature convergence with zero ionizing photon flux}
Now cooling is turned on, and we compare the \ramsesrt{}
non-equilibrium temperature evolution with that of equilibrium
\ramses{} (though keep in mind it has been adjusted to contain the
exact same cooling rates as \ramsesrt{}). Each of the 5 rows in
fig. \ref{cTest3.fig} shows cooling for a range of decreasing initial
temperatures, from top to bottom. The color-codings (initial
ionization states) and columns (hydrogen number densities) are the
same as before. The solid colored lines show non-equilibrium cooling
in \ramsesrt{} and the black dashed lines represent equilibrium
cooling in \ramses{} starting from the same temperature.

Clearly the temperature evolution is quite similar between
equilibrium/non-equilibrium cooling, especially if the initial
ionization fraction is 'correct', i.e. if it matches the equilibrium
one at the initial temperature.

The final temperature reached in the non-equilibrium case is usually a
bit lower than in the equilibrium case. This is independent of gas
density and initial temperature (as long as the initial temperature
allows for cooling to happen). The reason for this is that the
non-equilibrium ionization evolution lags behind the instantaneous
equilibrium one, so there is always a somewhat larger reservoir of
electrons in the non-equilibrium case. Electrons are the primary
cooling agents, and complete electron depletion completely stops
cooling, so it makes sense that if the electrons deplete more slowly,
cooling is more effective and can bring the gas to a lower final
temperature.

\subsection{Temperature convergence with nonzero ionizing photon flux}
This is the same as the previous test, except now we apply a constant
flux of $10^5$ ionizing photons $\flux$, assuming the spectrum of a
blackbody at $10^5$ K. The results are shown in
\Fig{cTest4.fig}. Things are much the same as before, except that the
non-equilibrium temperature seems to converge to a value which is much
closer to the the equilibrium one - because of the ionizing flux there
is always a reservoir of electrons both in the equilibrium and
non-equilibrium evolution, which makes for a much closer match in the
final temperature.

Although the final temperature reached is identical between the two
methods, the evolution towards that final temperature can be quite
different, depending on the initial ionization states.

\begin{figure}
  \centering
  \includegraphics[width=.37\textwidth]{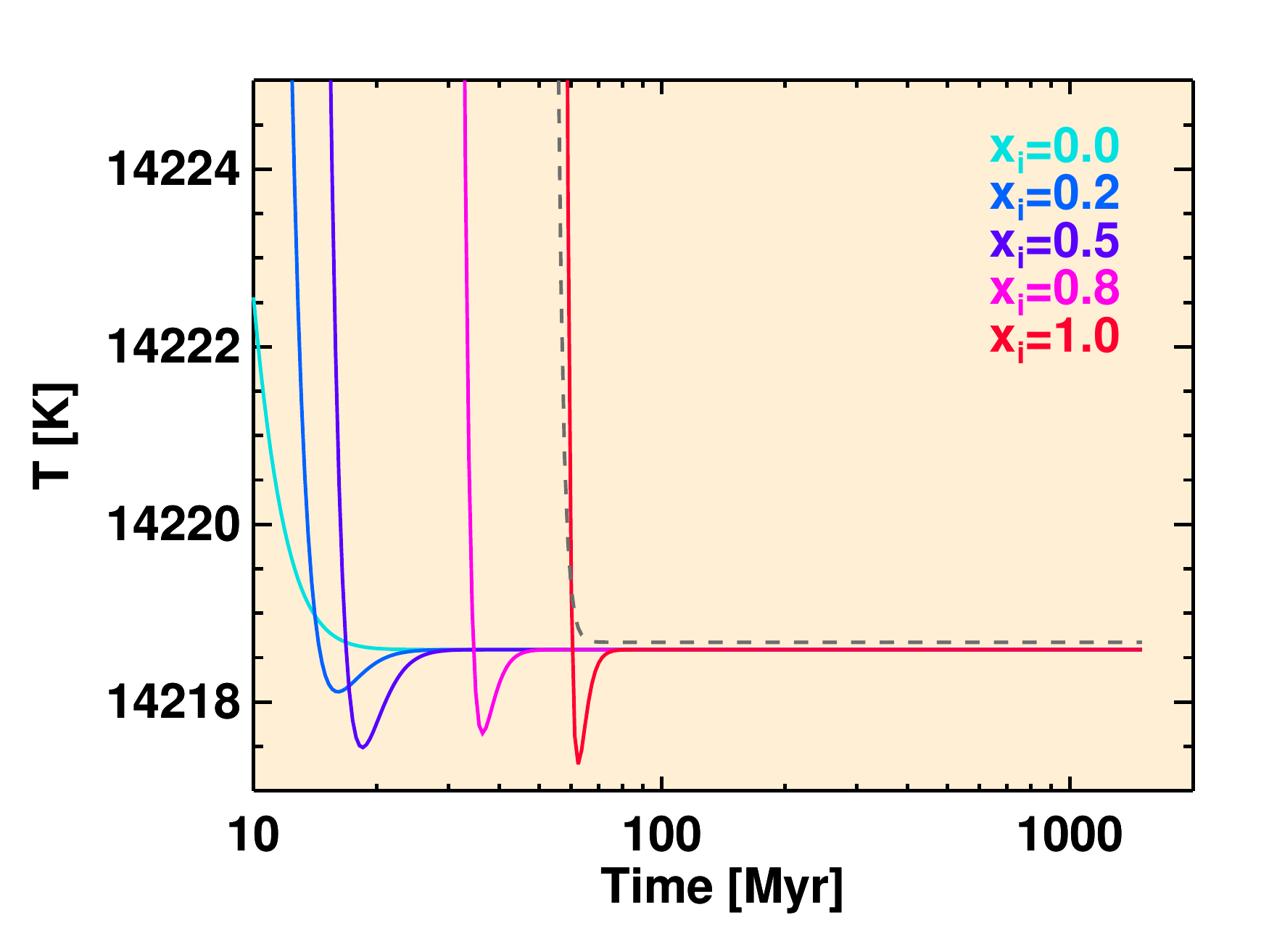}
  \caption[Temperature convergence in close-up]
  {\label{cTestClose.fig}Close-up of temperature convergence, for the
    UV inclusive test with initial temperature $T\approx10^5$ K and
    $\nh=10^{-2}$ $\cci$}
\end{figure}

A zoom-in on one of the plots is shown in \Fig{cTestClose.fig}, and
reveals that there is very little difference between the final
temperatures reached. The little difference there is results from
interpolation from cooling-rate tables in \ramses{} equilibrium
cooling and it can be decreased further by increasing the size of
these tables.

\subsection{Thermochemistry tests conclusions}

The main conclusions of the one-cell thermochemistry tests are:
\begin{itemize}
\item We always eventually reach the equilibrium ionization state with
  the non-equilibrium method...
\item ...but this can take a very long time to happen for diffuse gas,
  even more than a Hubble time.
\item Non-equilibrium temperature evolution of the gas is quite
  dependent on the initial ionization fraction of the gas at
  intermediate temperatures and low densities...
\item ...but in the end we reach the same or at least a very similar
  temperature as in the equilibrium case.
\item The convergence of the non-equilibrium solver towards the
  results of the equilibrium solver of \ramses{}, given the same
  cooling rate expressions, suggests that our thermochemistry solver
  is robust and correct.
\end{itemize}

